\documentclass[useAMS,usenatbib]{emulateapj}

\usepackage{pdflscape}
\usepackage{soul}
\usepackage{times}
\usepackage{url}
\usepackage{graphicx}
\usepackage[usenames]{color}
\DeclareGraphicsExtensions{.pdf,.png,.jpg,.mps,.eps,.ps}
\usepackage{amsmath}
\usepackage{natbib}
\usepackage{tablefootnote}
\usepackage{enumitem}
\usepackage{soul}

\def\unit #1{\,{\rm #1}}

\newcommand\cmsqi{\rm \,\unit{cm^{-2}}}
\newcommand\cm{\rm \,\unit{cm}}

\newcommand\kev{\rm \,\unit{keV}}

\newcommand\funit{\rm \,erg\,cm^{-2}\,s^{-1}}

\newcommand\lunit{\rm \,erg \,s^{-1}}

\newcommand\ledd{L_{\rm \, Edd}}

\newcommand\lbol{L_{\rm \, bol}}

\newcommand\msol{M_{\odot}}

\newcommand\mbh{M_{\rm BH}}

\newcommand\nh{ N_{\rm H}}

\newcommand\ks{\, \rm ks}

\newcommand\dc{\, \Delta\chi^2}

\newcommand\sax{{\it BeppoSAX}}
\newcommand\chandra{{\it Chandra}}
\newcommand\suzaku{{\it Suzaku}}

\newcommand\rxte{{\it RXTE}}

\newcommand\nustar{{\it NuSTAR}}
\newcommand\xmm{{\it XMM-Newton}}

\def\aap{A\&A} \def\apjl{ApJ}  
 \def\apjs{ApJS}

\usepackage{graphicx}

\begin{document}

\title[The variable and non-variable X-ray absorbers in Compton-thin type-II AGN]{{ The variable and non-variable X-ray absorbers in Compton-thin type-II Active Galactic Nuclei}}

\author{Sibasish Laha\altaffilmark{1,2,3}, Alex G. Markowitz\altaffilmark{4,1}, Mirko Krumpe\altaffilmark{5}, Robert Nikutta\altaffilmark{6}, Richard Rothschild\altaffilmark{1}, \\ \& Tathagata Saha \altaffilmark{4}}

\altaffiltext{1}{{University of California, San Diego, Center for Astrophysics and Space Sciences, 9500 Gilman Dr, La Jolla, CA 92093-0424, USA.} ; {\tt email:sib.laha@gmail.com, sibasish.laha@nasa.gov}}
\altaffiltext{2}{ Astroparticle physics laboratory, NASA Goddard Space Flight Center, Greenbelt, MD 20771, USA}
\altaffiltext{3}{Center for Research and Exploration in Space Science and Technology (CRESST) and Department of Physics, University of Maryland, Baltimore County, 1000 Hilltop Circle, Baltimore, MD 21250, USA}
\altaffiltext{4}{Nicolaus Copernicus Astronomical Center, Polish Academy of Sciences, Bartycka 18, PL-00-716 Warszawa, Poland.}
\altaffiltext{5}{Leibniz-Institut f\"ur Astrophysik Potsdam (AIP), An der Sternwarte 16, D-14482 Potsdam, Germany.}
\altaffiltext{6}{National Optical Astronomy Observatory, 950 N Cherry Ave, Tucson, AZ 85719, USA.}

\begin{abstract}
	{ We have conducted an extensive X-ray spectral variability study of
  a sample of 20 Compton-thin type II galaxies using broad band
	spectra from \xmm{}, \chandra{}, and \suzaku{}. The aim is to study
  the variability of the neutral intrinsic X-ray obscuration along the
  line of sight and investigate the properties and location of the
	dominant component of the X-ray-obscuring gas. The observations are
  sensitive to absorption columns of $\nh \sim 10^{20.5-24} \cmsqi$ of
  fully- and partially-covering neutral and/or lowly-ionized gas on
  timescales spanning days to well over a decade.  We detected variability in the column density of the full-covering
absorber in 7/20 sources, on timescales of months-years, indicating a
component of compact-scale X-ray-obscuring gas lying along the line of sight of each of these objects.
Our results imply that torus models incorporating clouds or overdense
regions should account for line of sight column densities as low as
$\sim$ a few $\times 10^{21}$ cm$^{-2}$. However, 13/20 sources yielded no detection of significant variability
in the full-covering obscurer, with upper limits to ${\Delta}N_{\rm H}$
spanning $10^{21-23}$ cm$^{-2}$.
The dominant absorbing media in these systems could be distant, such
as kpc-scale dusty structures associated with the host galaxy, or
a homogeneous medium along the line of sight.
Thus, we find that overall, strong variability in full-covering
obscurers is not highly prevalent in Compton-thin type IIs, at least
for our sample, in contrast to previous results in the literature. Finally, 11/20
sources required a partial-covering, obscuring component in all or
some of their observations, consistent with clumpy near-Compton-thick
	compact-scale gas.

	}



 \end{abstract}



\section{INTRODUCTION}

{ It is now generally agreed that the main source of
  energy of an active galactic nucleus (AGN) is the accretion of
  matter onto a supermassive black hole (SMBH). However, it is still
  unknown how gas located at kpc scales in the host galaxy loses
  its angular momentum and falls into the gravitational potential
  well of the SMBH at sub-pc scales and thereby powers the
  central engine.  Galactic-scale bars, circumnuclear disks at scales
  of a few hundred parsecs, and circumnuclear gas structures at scales
  of parsecs, in the near vicinity of the SMBH, are each believed to
  play roles in transferring matter ultimately from large distances
  into the SMBH accretion disk.}

{ The observed type 1/2 Seyfert dichotomy in the optical band led to
orientation-dependent unification schemes:
all AGN function similarly, and the different spectral classifications
of AGN arise only due to the different lines of sight toward the central
engine \citep{1985ApJ...297..621A}.
When we have a direct unobscured view of the central engine, then the optical-UV spectra exhibit broad as well as narrow
emission lines and the source is classified as a type 1--1.8 (collectively
hereafter referred to as type I). On the other
hand, if our line of sight to the central engine cuts across a dusty
structure popularly known as a ``torus", the central engine is no longer
visible directly and the optical-UV spectra we observe are
characterized only by narrow emission lines. In such a case, the source
is regarded as a type 1.9--2 (hereafter type II) AGN.}
{ Classically, the dusty torus was expected to extend to $\sim$pc scales ---
larger than the BLR but smaller than the NLR \citep[e.g., ][]{1988ApJ...329..702K}.  The simplest configuration is an axisymmetric donut-shaped torus, but this notion was effectively a starting point for more
complex models, and in recent decades the community has been probing
the morphology, content, and radial extent of the circumnuclear gas
\citep[see e.g., the reviews by ][]{2012AdAst2012E..17B,2017NatAs...1..679R}}

{ Firstly, the community has been accumulating evidence for optical-reddening
dust and X-ray-obscuring gas (which can potentially be dusty or
non-dusty) to exist across multiple distance scales from the SMBH.
Inside the dust sublimation radius, and commensurate with the BLR,
temporary X-ray obscuration can occur due to individual clouds
(possibly BLR clouds themselves) transiting the line of sight \citep[e.g., ][]{2009ApJ...696..160R,2011MNRAS.410.1027R}. In addition, observations of ratios of
$N_{\rm H}$ (as probed by X-rays) to V-band extinction $A_{\rm V}$ are found to
be much greater --- sometimes a couple orders of magnitude --- than
the Galactic ratio \citep{2001A&A...365...37M}.
Farther out, optical/IR reverberation monitoring indicates
thermally-emitting dust on scales of light-weeks to light-months \citep[see e.g., ][]{2006ApJ...639...46S}.
In addition, dusty gas on scales of parsecs to tens--hundreds of parsecs
is revealed by IR interferometry \citep[e.g., ][]{2009A&A...507L..57K,2009A&A...502...67T};
sub-mm observations also indicate dense
molecular gas at these distance scales \citep[e.g., ][]{2000ApJ...533..850S,2011A&A...525A..18B,2016ApJ...829L...7G,2016ApJ...822L..10I,2018ApJ...853L..25I,2016ApJ...823L..12G,2019A&A...623A..79C}.
These radial structures may potentially be connected: for example \citep{1993ApJ...404L..51N} and \citep{2007ASPC..373..415E} posit that material spanning
both the dusty torus and (non-dusty) BLR forms a radially-continuous
component (Toroidal Obscuring Region/BLR-Obscuring Region, or
"TOR/BOR").
{Henceforth, in this paper, for simplicity, we refer to the ``torus''
as a synonym for ``compact scale (less than 10 pc) X-ray-obscuring
gas,'' with the exact morphology and extent to be determined.
Specifically, we focus on all X-ray-obscuring gas along the line of sight
both inside and outside the dust sublimation radius, regardless of morphology.}
}


{ Secondly, some components of circumnuclear gas may contain discrete
clumps or filaments, and/or overdensities embedded in a continuous, lower-density
medium, as opposed to having a one-component continuous, homogeneous structure;
clumpy-torus models positing extended distributions of clouds \citep[e.g., ][]{2006ApJ...648L.101E,2007ApJ...659L.111R,2008ApJ...685..160N,2013ApJ...771...87H}
 are consequently finding observational support, particularly from X-ray
spectral studies.
For example, \citet{2002ApJ...571..234R} studied variability
of line-of-sight, neutral, X-ray-obscuring column density $N_{\rm H}$
in a sample of Compton-thin and moderately Compton-thick type IIs.
They detected almost ubiquitous (22/25 objects) variability in $N_{\rm
H}$ on timescales of months to several years, with typical variations
up to factors of $\sim1.5-3$.  Their analysis combined multiple
single-epoch observations across a range of different X-ray missions.
For a subsample of 11 sources the authors could detect relatively
rapid variations ($\leq1$ year), with obscuring columns typically
varying by $10^{22}-10^{23}$ cm$^{-2}$.
More recently, the community has used more continuous X-ray monitoring data
(e.g., from \textit{Rossi X-ray Timing Explorer}; \textit{RXTE})
or single-epoch X-ray long-looks (with
e.g., \xmm{} or \suzaku{}) to track ingress/egress
of individual clouds, finding support for clouds existing at radii
spanning both inside and outside the dust sublimation radius \citep[][]{2003MNRAS.342L..41L,2007MNRAS.377..607P,2009ApJ...696..160R,2011MNRAS.410.1027R,2010A&A...517A..47M,2013MNRAS.436.1588S}.
\citet{2014MNRAS.439.1403M} (2014; hereafter MKN14)
provided the first X-ray-based
statistical support for the clumpy-torus model of \citet{2008ApJ...685..160N} by studying the obscuration variability of a sample of 55 type Is and Compton-thin type IIs using long-term \textit{RXTE} monitoring. This
variability database yielded a total of 12 full-covering eclipse
events across eight objects. The event durations spanned hours to $>$
a year, with clouds' column densities typically ($4-26$) $\times
10^{22}$~cm$^{-2}$, i.e., no full-covering Compton-thick eclipse
events were observed.  In seven objects, the clouds were inferred to
be located at radial distances commensurate with the outer BLR or the
inner dusty torus. MKN14 also provided the first X-ray-based
probability estimates for witnessing eclipses in type I/II objects.
Finally, infrared studies probing the dusty part of the obscurer also support
clumpy-torus models, via spectral energy distribution modeling \citep[][]{2011ApJ...731...92R,2014MNRAS.439.3847R}, the co-existence of relatively hotter and
cooler dust components in nearby AGN \citep{2004Natur.429...47J,2009MNRAS.394.1325R}, and the range of 9.7~$\mu$m Si emission/absorption features
spanned by type I and II Seyferts \citep{2009ApJ...707.1550N}.}


{ Both the dusty and non-dusty components of the torus are
believed to play an active role in SMBH accretion, and
hence, understanding the structure of the torus is essential for
understanding both disk/SMBH fueling and orientation-dependent
unification schemes.  However, there are additional complications that
simple orientation-dependent unification cannot easily explain.  There
is likely a dependence of the torus covering factor on luminosity or
$\lambda_{\rm Edd} \equiv L_{\rm Bol}/L_{\rm Edd}$; relatively
stronger radiation fields from the nucleus can clear out more
obscuring material \citep{2017Natur.549..488R}.
In addition, there is support for the BLR to disappear towards low
values of AGN bolometric luminosity, forming the "true type 2" objects \citep{2009ApJ...701L..91E,2012MNRAS.426.3225B}.
One might therefore refer to \textit{the} "torus" or "TOR/BOR"
component, but it is likely the case that its morphology and/or spatial
extent do not remain the same from one object to the next.
We reiterate that in this paper, we refer to the "torus" just to
indicate compact-scale circumnuclear gas, with the precise morphology
and spatial extent still to be determined by the community (e.g., a TOR/BOR is just one
possibility), the content (smooth, clumpy, or mixed) also to be
determined, and with the assumption that even if it is present in all
AGN, its morphology and extent are not guaranteed to be the same
universally.}


{ Yet another major complication for unification schemes is potential
optical extinction and X-ray obscuration originating at length scales
much greater than the compact torus, at 100s of pc to kpc, and due
to dusty structures or lanes associated with the host galaxy.
Optically-selected samples of Seyferts tend to
yield a systematic dearth of type Is in relatively more edge-on systems,
\citep{1995ApJ...454...95M,2011MNRAS.414.2148L}. Edge-on systems also tend to exhibit relatively stronger optical extinction \citep{2007MNRAS.379.1022D,2007ApJ...659.1159S}. 
The expansive \textit{Hubble Space Telescope} (\textit{HST})
snapshot survey of over 250 nearby Seyfert
and starburst galaxies performed by \citet{1998ApJS..117...25M} (1998;
hereafter MGT98) revealed an array of fine-scale dusty structures in
galaxies' centers. They \citep[and others such as ][]{2014MNRAS.442.2145P} concluded that type II Seyferts are intrinsically more likely to
be hosted in galaxies with nuclear dust structures crossing the line
of sight, potentially alleviating the requirement for a compact
torus to explain extinction of BLR lines (a fundamental component of
orientation-dependent unification schemes).}

{ X-ray studies yield a similar picture:
in some high-spatial resolution X-ray images of nearby AGN,
we can resolve where dust lanes directly obscure soft X-ray diffuse
emission \citep[e.g., NGC~7582 and Cen~A: ][]{2007MNRAS.374..697B,2008ApJ...677L..97K}.
Moreover, \citet{2001MNRAS.327..323G} and \citet{2005A&A...444..119G} have
compared X-ray obscuring columns with Balmer decrements or
nuclear dust morphology in samples of Compton-thin and Compton-thick
Seyferts. Their results support the notion \citep[put forth by e.g.,][]{2000A&A...355L..31M} that Compton-thin type IIs tend to reside preferentially in galaxies with dusty nuclear environments on scales of $>$$\sim$ 0.1 kpc.
However, Compton-thick obscuration does not seem highly affected
by nuclear dust content and is likely due to a compact torus
instead.}\footnote{On the other hand, \citet{2012ApJ...755....5G}
find evidence from mid-IR
spectroscopy of nearby Compton-thick AGN that in at least some of these
sources, the dominant dust extinction is associated with the host galaxy
instead.}
{ Host galaxy characteristics --- namely the chance of having or not
having a dusty filament along the line of sight to the nucleus ---
can therefore potentially impact both optical spectral type and
whether or not a source is perpetually Compton-thin obscured. Searching for time variability in X-ray obscuration
can potentially provide clues to distinguish between
obscuration due to a compact torus versus that from host galaxy
structures. Therefore, the main goal of the present paper
is to test this simplified model with X-ray monitoring data,
wherein the location of the X-ray-obscuring gas can be discerned by
the extent of the variability in the X-ray obscurer.  A detection of
variability on timescales of years or shorter would point to a clumpy
structure very likely associated with the torus. A lack of
variability on timescales of years and longer in a given object
supports the notion that the dominant obscuring gas is more likely
associated with host galaxy dusty structures.}

{ The present paper is motivated in part by MKN14's
  results on a subsample of eight type IIs monitored with
  \textit{RXTE}: these objects' X-ray column densities remained
  constant over timescales from 0.6 to 8.4 years.  However,
  \textit{RXTE}'s limited bandpass (no coverage $<2$~keV) meant that
  sensitivity in ${\Delta}N_{\rm H}$ in these objects was limited,
  with limits on variability spanning $0.6 - 9 \times10^{22}$
  cm$^{-2}$. In contrast, \textit{XMM-Newton}, \textit{Suzaku}, and
  \textit{Chandra} observations of type II Seyferts can provide
  comparably stronger sensitivity in ${\Delta}N_{\rm H}$, courtesy of
  their soft X-ray coverage.}


{ In this paper we investigate the variability of the X-ray
obscuration column density of a sample of perpetually X-ray-obscured
type II AGN in the local Universe
to address the question of "What is the origin of perpetual Compton-thin
X-ray obscuration in optical type IIs?" 
The rest of this paper is as follows: We present the sample
and data reduction in Section 2, the
spectral analysis in Section 3, 
the results for X-ray obscuration and its variability in Section 4, and
we discuss physical interpretations in Section 5.
Section 6 contains our main conclusions.}


\section{Sample selection, Observations and data reduction}\label{Sec:obs}

\subsection{The Sample selection and properties}\label{subsec:sample}

\subsubsection{Sample selection}
{ The sample of Compton-thin type II (X-ray classification) sources was selected from the existing literature, with the constraint that each source must have at least two observations for a given instrument with a minimum time separation of two days (observations executed within two days of each other are almost always part of the same proposal/long-look). The X-ray spectra for the sources in the sample are obtained from \xmm{}, \chandra{}, and \suzaku{} observatories that are in the HEASARC public archives as on 1 July 2017. Only the  Seyfert sub-types 1.9-2 (as listed in NED), referred to as Compton-thin type II, are considered in this work. We obtained a final list of 20 sources (See Table \ref{Table:sources} for details). We focus on optical type II Seyferts because they are more likely to
be perpetually obscured in X-rays compared to type Is, and we reviewed
the literature to ensure that each source in our sample is indeed
perpetually X-ray-obscured. 
Our sample is not intended to be a complete sample \citep[for instance, very
roughly $10\%$ of type IIs have values of $N_{\rm H}$ below $10^{21}$
cm$^{-2}$, e.g., ][]{1999ApJS..121..473B}, but it is an exploratory 
sample for expanding our knowledge on variability or lack thereof
in the X-ray obscurers of type IIs.
We focus on relatively X-ray bright objects (average observed 2--10 keV
fluxes brighter than typically a few $\times$$10^{-12}$ erg cm$^{-2}$ s$^{-1}$
to ensure adequate signal-to-noise within each observation).
We exclude Compton-thick-obscured AGN, as their X-ray
spectra are best studied with bandpasses extending above 10 keV, such
as \textit{NuSTAR}.  However, we are clearly sensitive to potentially
detecting any Compton-thin to -thick transitions (or vice versa),
though as we note below, none were observed. }


\subsubsection{Activity properties}
{ Optical spectral classifications are listed in Table~1.  Some of the
sources are not Seyfert galaxies, such as Cen A or Cyg A, which host
radio jets, or NGC 1052, which is an X-ray-obscured LINER exhibiting
broad polarized lines.
We note as a caveat that even if the torus exists in all objects, its
morphology and spatial extent (scaled relative to $M_{\rm BH}$) may
very well likely differ between different activity classes, e.g.,
radio-loud versus radio-quiet objects; a detailed discussion of the
impact of the presence/lack of a jet on torus morphology is beyond the
scope of the current paper.
Ten sources have been confirmed to harbor hidden BLRs, five using
scattered polarized emission (denoted by ``$1H$'' in Table~1) in which
case a compact torus is likely to exist along the line of sight.
The other five (denoted by ``$1I$'') have detections of broad
recombination lines in the IR \citep[e.g., ][]{2002A&A...392...53N}; on that
basis alone, it is not clear where the absorbing gas lies or how much of the total observed column
is due to the host galaxy versus any putative torus; a potential observation of
variability in $N_{\rm H}$ could confirm the existence of the compact
torus along the line of sight in such cases.}

\subsubsection{Host galaxy properties of the sample}

The objects in our sample are known to span a range of host galaxy properties. A majority of our sample have been studied in the snapshot
    survey by \citet{1998ApJS..117...25M}: eight have dust lanes crossing the
    line of sight to the nucleus or just offset from it; four have
    filamentary/wispy or irregular dusty structures. Sixteen of
    our objects are hosted in spirals, with semi-major/minor axis
    ratios (as listed on NED) indicating disk inclinations spanning
    roughly 30--75 degrees from the plane of the sky.  The other four are hosted
    in ellipticals (Cen~A, Cyg~A, NGC~1052, and NGC~6251).



\subsection{Observations and data reduction}

To effectively detect any Compton-thin variable obscuration in the
X-rays, the best instruments to use are the EPIC cameras aboard \xmm{}, the
XIS detectors aboard \suzaku{} and the ACIS detectors aboard
\chandra{}, as they each provide a broadband spectral view in the energy
range $\sim 0.5 - 10\kev$, crucial for tracking the neutral
absorption roll-over. We describe below the methods employed to
reprocess and clean the X-ray spectral data obtained from these
telescopes.


\subsubsection{XMM-Newton}

The EPIC-pn data from \xmm{} were reduced using the scientific
analysis system (SAS) software (version 15) with the task {\it
  epchain} and using the latest calibration database available at the
time we carried out the data reduction. We used EPIC-pn data because
of its higher signal to noise ratio as compared to MOS. We filtered
the EPIC-pn data for particle background counts using a rate cutoff of
$< 1$ ct s$^{-1}$ for photons $>10\kev$, and created time-averaged
source and background spectra, as well as the response
matrix function (RMF) and auxiliary response function (ARF) for each
observation using the {\it xmmselect} command in SAS. The source
regions were selected with a circle radius of $40"$ centred on the
centroid of the source. The background regions were selected with a
circle of $40"$ located on the same CCD, but located a few arc-minutes away from
the source and avoiding X-ray-emitting point sources.
Spectra were accumulated using pattern $0-4$.  We found
that the sources NGC~4258 and Cyg~A are extended in the EPIC-pn CCD
image, possibly due to the resolved, diffuse stellar emission in the
former, and due to diffuse X-ray emission from intercluster gas in the
latter.

We checked for possible pile up in the sources using the command {\it
  epatplot} in SAS, and found that the spectra of the source Cen~A are
piled up. For Cen~A we thus used an annular extraction region for the
source, with an inner radius of $20"$ and outer radius $40"$ to
minimize pile-up. We did not detect significant pile up in the EPIC-pn
spectra of the other sources.


\subsubsection{Chandra}

We considered Advanced CCD Imaging Spectrometer (ACIS-I and ACIS-S) data
as well as 0th-order High-Energy Transmission Grating Spectrometer (HETGS) data.
All \chandra{} data were reprocessed using the command {\it
chandra\_repro} in the {\it CIAO} software (version 4.7.1) and using
the latest calibration database. Source regions were selected using a
circle of radius 4.0". The background regions were selected using a
circle of radius of 4.0" on the same CCD as the source, but away from
the source. We detected pileup (ranging from severe to mild) in the
ACIS CCD spectra for the sources IRAS~F05189$-$2524, NGC~1052,
NGC~5252, NGC~5506, NGC~6251, NGC~6300, NGC~7172,
NGC~7582, Mkn~348, NGC~4507 and MCG$-$5-23-16.
Excluding a central circular region from the source image, as is typically done to exclude
piled up data in \xmm{} EPIC spectra,
may lead to issues with the ARF in the \chandra{} spectra, and so that method was avoided.
We instead use the
pile-up kernel in the spectral fitting codes to model the pile
up. In those cases where the pile-up is too
  severe to be modeled by such a kernel (such as in Cen~A), we excluded those 
  observations from our study.


\subsubsection{Suzaku}

The \suzaku{} observations were performed using the X-ray Imaging
Spectrometer (XIS) \citep{2007PASJ...59S..23K} and Hard X-ray Detector
(HXD) \citep{2007PASJ...59S..35T}. The XIS observations were obtained
in both the $3 \times 3$ and $5 \times 5$ data modes. The {\sc
  aepipeline} tool was used to reprocess and clean the unfiltered
event files and to create the cleaned event files. In all
observations, for both the XIS0 and XIS3 (front-illuminated CCD) and
for XIS1 (back-illuminated CCD), we extracted the source spectra for
each observation from the filtered event lists using a $120$" circular
region centered at the source position. We also extracted the
corresponding background spectral data using four circular regions of
$120{\rm~arcsec}$ radii, a few arcminutes away from the source region
and avoiding X-ray-emitting point sources. There are only a few cases
of pile-up in \suzaku{} observations; we excluded those
centrally-located pixels for which pile-up exceeded a threshold of
$4\%$.  During spectral fits, we did not co-add the front-illuminated
XIS spectra, instead fitting them separately.


\section{Spectral analysis}\label{Sec:spec}

We used ISIS (Interactive Spectral Interpretation System) software
\citep{2000ASPC..216..591H} for spectral fitting carried out in this
work. The \xmm{} spectra were grouped by a minimum of 20 counts per
channel and a maximum of five resolution elements using the command
{\it Specgroup} in SAS. The \chandra{} and \suzaku{} spectra were
grouped by a minimum of 20 counts per channel in ISIS. As described below, we have carried
out iterative steps to systematically fit all the X-ray spectra and account for
both soft- and hard-band components while obtaining 
precise estimates of the intrinsic neutral absorption column densities. 
In this section we first describe the models we used to fit the
spectra and then elucidate stepwise the fitting procedure that we
employed.

We started from a ``baseline'' model that follows (using ISIS notation)\\

\noindent {\it TBabs(1) $\times$ (apec(1)+ apec(2) + powerlaw(1) + zTBabs(1) $\times$ (powerlaw(2) + pexmon(1)+ zgauss))}. \\

If an additional partial-covering absorption component is required by the data, then our model became \\

\noindent {\it TBabs(1) $\times$ (apec(1) + apec(2)+ powerlaw(1) + zTBabs(1) $\times$ zpcfabs(1) $\times$ (powerlaw(2) + pexmon(1) + zgauss ))}. \\

\noindent The {\it TBabs} and {\it zTBabs} components model the
Galactic and intrinsic fully-covering neutral absorption column,
respectively. {\it zpcfabs} models the partial covering absorption
component, if significantly detected.  The primary, hard power law
({\it powerlaw(2)}) models the Compton-upscattered emission from a hot
optically-thin corona in the central AGN. In addition we have tested for the possible presence of warm ionized absorbers \citep{2005A&A...431..111B,2014MNRAS.441.2613L} using {\it warm-abs} model \citep{2001ApJS..133..221K} in ISIS, but did not detect any statistically significant warm absorption in any of the sources in the sample. The best fit models and details of the analysis for every source have been reported in Appendix A.

The soft band may contain emission from thermal plasma, which could be
due to star formation \citep[e.g.,][]{1997ApJS..113...23T}. 
Continuum emission due to scattering of the primary X-ray emission in
Compton-thin circumnuclear gas out to $\sim$ a kpc \citep[e.g.,][]{2006A&A...446..459C,2007ApJ...664L..79U,2008PASJ...60S.293A,2017ApJS..233...17R} is also expected. There also likely exist signatures of gas being photo-ionized and
photo-excited, namely soft emission lines and radiative recombination
continuum (RRC) features, likely originating in the AGN-illuminated
regions of the Narrow Line Region \citep[e.g.,][]{2006A&A...448..499B,2007MNRAS.374.1290G}.  Indeed, 12 of our sources are contained in the CIELO-AGN
sample of \citet{2007MNRAS.374.1290G}; however, such features are
typically identified by gratings observations and will be blurred at
CCD resolution, so we do not explicitly model them here.

We use one (or two, if necessary) {\it apec} component(s) to model any thermal
emission.  {\it powerlaw(1)} denotes the secondary (soft) power law to model the scattered
emission, with the expectation that the normalization of the soft
power law will be of the order $\lesssim1\%$ that of the hard power
law, approximately. We first attempt to fit with the value of soft photon index
$\Gamma_{\rm SX}$ tied to that of the primary (hard) X-ray power law
but only if a significant improvement in fit results from thawing
$\Gamma_{\rm SX}$ then we do so. As mentioned earlier, the soft-band power-law is expected to
model scattered nuclear emission and in the ideal case, the photon
indices of the soft and hard power laws should match.  However, there can be
numerous potential reasons for a mismatch, including that the current
value of the hard X-ray power-law photon index may be different from
the long-term averaged photon index scattered off extended diffuse
gas, or there may be blending with emission from other components,
such as unresolved point sources (ULXs).

Although the baseline model gives a reasonable fit in most cases, it is definitely not the case that a common baseline
model can be applied equally to all objects/observations. For a given instrument
(e.g., {\it XMM-Newton}), some objects' spectra require only one {\it
  apec} component; others require two.  Relativistically broad Fe~K
emission lines were detected only in two sources, MCG--5-23-16 and
Fairall~49, and for simplicity we have used {\it diskline} to model
them (See Appendix A for details of the fits). 

In addition, for a given object, different instruments have different
apertures, different responses/effective areas, etc., so some
components ({\it apec}, narrow Fe K line) detected in one instrument
for a given object are not detected in other instruments. As one
example, the {\it Chandra} observations for the sources NGC~2992 and
NGC~7314 did not require any {\it apec} components, but the \xmm{} and
\suzaku{} observations for the same sources required them. The goodness of fit upon adding a new model component has been 
tested both using $\dc$ and Ftest, requiring a $>5\sigma$ improvement in statistics to consider the new model component as required in the fit.



Ideally, we would have liked to perform, for each object/instrument
combination, joint fitting in which we can have certain parameters
freed but tied across all spectral fits (power-law photon indices,
{\sc apec} temperatures, etc.), but for practical reasons that could not
be done due to the huge computational power required (especially for
those objects with multiple {\it Suzaku} datasets). We also note that not all objects in the sample adhere
to a common ``baseline" model, and not all objects follow the same spectral
variability behavior in soft and hard x-ray bands. For example, 
given that soft-band emission likely originates in diffuse gas, we do not expect it to exhibit
variability on timescales of $\sim$ years and shorter.  However in
several object/instrument cases we did find strong evidence for
soft-band variability; keeping soft-band parameters frozen resulted in
poor fits in these cases: the {\it XMM-Newton} spectra of
MCG--5-23-16, NGC~526a, NGC~2992, and NGC~7314, and the {\it Chandra}
spectra of NGC~526a 
(See the spectral overplots in Appendix Fig \ref{Fig:CenA}-\ref{Fig:NGC7582}).  For example,
in the first {\it XMM-Newton} observation of NGC 2992 (denoted as
X-1), the entire continuum (except for the narrow Fe line flux, which
is $\sim$ constant) is higher than for all the subsequent 
{\it XMM-Newton} observations.  There could be several possible reasons:
a leaky, patchy absorber which obscures the AGN and which has changed
its covering fraction, a sudden spurt in stellar emission, or flaring
emission from a transient point source such as a ULX. A detailed study
of the causes behind each of these soft-band spectral variations would
require high spatial resolution to separate out the AGN, stellar
emission, other point sources, etc., and is therefore beyond the scope
of this paper.
%
At any rate, consequently, we adopt a two-step system: We first fit each
observation separately, then for each object/instrument combination,
we adopt the average values of all soft-band component parameters and
freeze them during a second round of fits. While adopting this process we note as caveats that 1. different spectra can have different statistical weights
and 2. freezing some parameters may shrink some error bars on fitted $\nh$ values.


There is also the issue of the Compton reflection hump (hereafter
CRH). The CRH is detected and mostly well constrained in 
individual \suzaku{} observations due to its broadband 
$0.3-40\kev$ coverage, but remains unconstrained for the {\it XMM-Newton} or {\it
  Chandra} observations --- meaning
that CRH reflection strength and hard X-ray power-law parameters
(which can in turn impact modeling of $N_{\rm H}$) cannot be
unambiguously and independently constrained in $80\%$ of the
observations involved in this work.
We thus started our analysis for each object with \suzaku{}
spectra and then applied that model to \xmm{} and \chandra{}. We note here that for all the 20 sources in our sample we have at least one \suzaku{} observation, and hence we could use this approach for all of the sources. 

We use {\it pexmon} to model the CRH and narrow Fe K emission line
simultaneously.  However, we need to understand how the CRH has varied
with time for each object, in order to know which parameter values of
the {\it pexmon} component to use for cases with multiple \suzaku{}
observations. For simplicity, we consider two scenarios: 1. The CRH
remains constant in absolute normalization with time, irrespective of
the hard X-ray power-law $\Gamma$ and flux, and 2. The CRH responds
instantaneously to the hard X-ray powerlaw variations (relative
normalization constant). There are 8 sources for which there are
multiple \suzaku{} observations and 12 sources with only one.
For the 12 sources with only one \suzaku{} observation we have used
the best fit {\it pexmon} values from that observation and assumed it
to remain constant in absolute normalization in the \xmm{} and
\chandra{} observations, as there is no way to rule out or vindicate any of the above 
scenarios with the {\it XMM-Newton} or {\it Chandra} data.
For the seven sources with multiple \suzaku{} observations (excluding
Cen~A, which lacks any CRH detected to date), we investigated
potential CRH variability.  After obtaining a broad-band best fit to
each \suzaku{} observation, we calculated the absolute normalization
of the {\it pexmon} component (simply the product of the model
normalization and the reflection fraction $R$).  For the 
sources MCG--5-23-16, NGC~4258, NGC~7314, and NGC~7582  
 the absolute normalization is consistent with being constant in time. This is
consistent with the notion that at least in these objects, the CRH
arises from a distant medium and does not vary over the timescales of
our observations. For NGC~2110 the absolute normalization 
tracks the hard X-ray power-law normalization, suggesting that CRH flux tracks 
that of the coronal power law closely. For the sources NGC~2992 and NGC~5506, 
insufficient SNR or lack of significant variability in the hard X-ray powerlaw
did not allow us to conclude the nature of CRH variability. Given the fact that 
a majority of these sources with multiple \suzaku{} observations are consistent with
a constant-absolute normalization CRH, and since we lack information on the rest 
of the 12 objects, for simplicity and uniformity we assumed a constant-absolute normalization CRH 
for all the sources in our sample. In other words, we used the {\it pexmon} parameter values 
from the best-fit to S-1 (for a given object) and held those frozen when fitting the \xmm{} and
\chandra{} datasets for each of these objects.

In addition, we note that we did not detect any significant variability in the narrow
Fe~K$\alpha$ emission line (at $\sim 6.4\kev$) flux in any of the objects.
If we assume that the line arises from the same reprocessing medium as
that responsible for the CRH, as is implicitly assumed when using 
{\it pexmon}, then this further supports the notion that reflected emission (CRH + Fe K line) is constant with time.
We note however, that for the high SNR {\it XMM-Newton} observations of
MCG--5-23-16, we found that the narrow Fe K emission line and the CRH
could not be simultaneously modeled by {\it pexmon}, implying e.g.,
that they arise from different reprocessing media, or there is a
non-solar Fe abundance, In fact, we had to thaw the Fe abundance relative to solar, $Z_{\rm Fe}$,
in {\it pexmon} to sub-solar values to obtain good fits: 
$Z_{\rm Fe}$ typically falls to $\sim 0.19$ and $\chi^2$ drops by at least $200$ in the X-2, X-3 and X-4 spectra.

The ``second round'' of fits are our best fits, listed in Tables
\ref{Table:Xray} and \ref{Table:Xray2}.  The error quoted on each
parameter is the $90\%$ confidence interval for one free parameter. Note that we have only reported the errors for the soft X-ray parameters in Table \ref{Table:Xray2}
when they are kept free in the second round of fits, that is, when the fit requires a different value of these parameters than those of the average values.


\section{Results}\label{Sec:results}


  %
  Table 3 lists the best fit parameters for the full and partial-covering line of sight
  absorbers, along with the $90\%$ confidence uncertainties.
  %
  %
  We first discuss the characteristics of the full-covering absorbers.
  The column densities of the full covering absorber in our sources
  have values spanning three orders of magnitude ($N_{\rm H} \sim
  10^{20.5 - 23.5}$ cm$^{-2}$).  We note that the distribution of mean
  values of $N_{\rm H}$ is roughly uniform, and does not show any
  clustering towards low or high values.  We present light curves of
  $N_{\rm H}$ for all sources, shown in Fig. \ref{Fig:lightcurve}.

%
%
For a given source, we can search for variability in $N_{\rm H}$ by
examining a single instrument only (to eliminate cross-instrument
systematic effects), and in parallel, across different missions. The
latter, however, is subject to cross-instrument calibration issues
which are not straightforward to quantify and may depend on e.g.,
intrinsic spectral shape, the effect of differing apertures, etc., so
cross-instrument comparisons of a given parameter must be taken with a
grain of salt, and is discussed further in Section \ref{subsec:crossinstrument}.  
Nonetheless, we note firstly that none of the sources exhibits any
Compton-thin to -thick (or vice versa) transitions, considering both
single instruments and across missions.  Furthermore, variations in
best-fit values of $N_{\rm H}$ are usually modest even over timescales of years: for any given
object except Fairall 49, the maximum/minimum best-fit values of
$N_{\rm H}$ typically never vary in ratio by more than $\sim$ 1.5--1.8
(within one instrument) or more than $\sim$ 2--5 (across all
instruments for a given source).
%
%
Fairall 49 is the standout exception, displaying an order of magnitude increase in
$N_{\rm H}$ (this source is discussed further below).
%
%
 %

   As a caveat we remind the reader that in this work we are
     limited by the relatively sparse time sampling of the data, and we are not
exploring variability on timescales less than $\sim$1 day in this work.
     This lack of sustained sampling means we are not as sensitive
     compared to \rxte{} in detecting complete (ingress-to-egress) eclipse events
as detected by e.g., MKN14, \citet{2011MNRAS.410.1027R}.
Nonetheless we can probe up to timescales of nearly two decades, so we are probing a spatial extent
similar to that of MKN14, although here we have greater sensitivity to smaller variations in $N_{\rm H}$,
covering the range $\log(\nh/\cmsqi)$ $\sim 20.5-23.5$.

\subsection{Candidates for variability in $N_{\rm H}$ using single instruments}\label{subsec:variablesources}

 We select candidates for sources exhibiting variability in
  $\nh$ (henceforth ``variable-$\nh$ sources"), but we first
  concentrate only on using single instruments for a given object.  
To be classified in this category, a given object/instrument combination must exhibit
variability as follows:

\begin{itemize}

\item As a first cut, values of $N_{\rm H}$ between any two
  observations must differ by at least 3 times the 90\% error in one
  parameter obtained from ISIS spectral fit (adopting a conservative
  criterion).

	\item A simple $\chi^2$ fit of the $N_{\rm H}$($t$) light curve against a constant
must satisfy $\chi^2$/$dof$ $> 5$.

	\item The X-ray spectra must be checked for possible model degeneracies
that could influence $N_{\rm H}$ values; as described below, we perform Bayesian analysis with MultiNest
to vet candidate-variable objects.

\end{itemize}

The first criterion lead to eight object/instrument combinations as candidates
(here, X, S, and CA denote {\it XMM-Newton}, {\it Suzaku}, and {\it Chandra}-ACIS, respectively):
Fairall~49/X, Cen~A/S, MCG--5-23-16/X,  MCG--5-23-16/S, NGC~2992/X, NGC~5252/CA, NGC~5506/X, and NGC~7582/X. We note that the \xmm{} and \suzaku{} events of MCG--523-16 are different and not overlapping in time. All eight of the above events pass the second criterion, as testing against a constant
yielded $\chi^2$/$dof$ $> 5$.
Two additional objects (NGC~4258/S and NGC~7582/X) pass this criterion
but fail the first criterion (and partial coverers and/or low
signal/noise may be at play), and hence we do not consider them further. Choosing a much less strict threshold for the first criterion, say 2 times the
90$\%$ errors, would have allowed only three more object/instrument
combinations to pass this criterion. Similarly, choosing a lower threshold for $\chi^2/{dof}$ would not have
significantly increased the number of objects passing the second
criterion; lowering the threshold to 2.5, for example, would have
allowed only two more object/instrument combinations to pass this criterion. We are thus confident that these two criteria are each reasonable in
terms of separating outlying variability from the bulk of the
distribution in which variability is not detected.

We then conducted Bayesian analysis to vet
these candidates and verify that modest variations in $N_{\rm H}$ are not
the result of degeneracies with other spectral component parameters.
Specifically, we use the MultiNest nested-sampling algorithm \citep{2004AIPC..735..395S,2009MNRAS.398.1601F} via the Bayesian X-ray Analysis (BXA) 
and PyMultiNest packages 
\citep{2014A&A...564A.125B}\footnote{\url{https://github.com/JohannesBuchner/BXA}} 
for XSPEC version 12.10.1f.
Standard Markov Chain Monte Carlo (MCMC) algorithms form ``chains" by
comparing the likelihood of a test point against that of a new point
randomly chosen from the prior distribution, and moves to the new
point with a probability determined by the likelihoods.  However,
there may be convergence issues, in that parameter sub-spaces with
non-negligible probabilities can potentially be under-explored by such
chains. Nested sampling algorithms, including MultiNest, attempt to map out
all of the most probable regions of parameter sub-space: it maintains
a set of parameter vectors of fixed length, and removes the
least-likely point, replacing it with a point with a higher
likelihood, and thus shrinking the volume of parameter space.
We use MultiNest version 3.10 with default arguments (400 live points, sampling efficiency of 0.8)
set in BXA version 3.31.
We paid particular attention to potential degeneracies between 
$N_{\rm H}$ and each of partial-covering parameters, photon indices of the
power laws, and \textsc{apec} component normalizations.
For all candidates, \textsc{pexmon} and emission line parameters were all kept frozen at best-fit values;
additional details for individual objects' MultiNest runs are listed in Appendix \ref{Sec:BXA}.


Given the $90\%$ distribution of the posterior distributions on $N_{\rm H}$, we conclude that
model degneracies do not significantly impact 
and that the observed variations in $N_{\rm H}$ are intrinsic to the objects. For brevity, we defer presentation of the confidence contours obtained
from the MultiNest runs to  Appendix \ref{Sec:BXA}.




From Table 3, and taking into account the model degeneracies, we conclude that 
variations in $N_{\rm H}$ are robust for the following objects; here, $N_{\rm H22}$ denotes 
$N_{\rm H}$ / ($10^{22}$ cm$^{-2}$):

$\bullet$ Cen~A/Suzaku, S-3 to S-5: $N_{\rm H22}$ dropped from $11.03
\pm 0.08$ (S-3) to $9.98 \pm 0.14$ (S-5); the 90$\%$ confidence
intervals from the posterior distribution in MultiNest was $0.29$.

Considering the times and 
column densities of the other Suzaku observations as well, we infer
that $N_{\rm H22}$ dropped from $\sim$10.9 in July--August 2009
(S-2--4) to 9.98 by August 2013 (S-5; $\Delta$$N_{\rm H22} = -1.05 \pm 0.29$.)
(Unfortunately, there were no \textit{XMM-Newton} or \textit{Suzaku}
observations during the two spikes in $N_{\rm H}$ obtained from
\textit{RXTE} monitoring, in 2003--4 and 2010--1; Rothschild et
al.\ 2011; Rivers et al.\ 2011; MKN14.)

$\bullet$ Fairall~49, X-1 to X-2: 
$N_{\rm H22}$ increased from $0.067 \pm 0.017$ in 2001
to $0.75 \pm 0.05$ in 2013. 
We ran MultiNest for both X-1 and X-2 separately, given the large
difference in measured columns, the 90$\%$ confidence intervals from
the posterior distribution spanned 0.03 and 0.05, respectively
(we adopt $\Delta$$N_{\rm H22} = +0.68 \pm 0.04$).

$\bullet$ MCG--5-23-16/{\it XMM-Newton}, X-1 to X-2:
$N_{\rm H22}$ decreased from $1.020^{+0.019}_{-0.011}$ in
Dec.\ 2001 to $1.366\pm0.012$ in Dec.\ 2005. 
The MultiNest 90$\%$ confidence interval on
X-1's $N_{\rm H22}$ was 0.01;
we adopt ${\Delta}N_{\rm H22} = +0.35 \pm 0.01$.

$\bullet$ MCG--5-23-16/Suzaku, S-1 to S-2: $N_{\rm H22}$ dropped from
$1.44 \pm 0.01$ in Dec.\ 2005 (S-1) to $1.34 \pm0.02$
in June 2013 (S-2); the MultiNest 90$\%$ confidence interval was 0.02;
we thus adopt $\Delta$$N_{\rm H22} = -0.10\pm0.02$. \\

$\bullet$  NGC~2992, X-1 to X-2: $N_{\rm H22}$ increased from $0.60\pm0.01$ to 
$0.82\pm 0.03$ from 2003 to 2010. The MultiNest 90$\%$ uncertainty
for X-1 was 0.01; $\Gamma_{\rm soft}$ and $kT_1$ are left free during the MultiNest runs
(we adopt $\Delta N_{\rm H22} = +0.22 \pm 0.01$).
Curiously, X-1 corresponds to the highest flux state, both in the hard and soft X-ray bands.
That is, the soft-band emission seems to track the decrease in hard power-law flux from
2003 to 2010.

$\bullet$ NGC~5252/{\it Chandra}-ACIS, CA-1 to CA-2, and CA-3 to CA-4:
$N_{\rm H22}$ increased from $2.84\pm0.07$ in Aug.\ 2003 (CA-1) to
$4.51\pm0.11$ in Mar.\ 2013 (CA-2).
Values for CA-2 and CA-3 are consistent with each other; this is not
surprising since the observations occurred only a few days apart.
However, $N_{\rm H22}$ had dropped to $3.51\pm0.10$ by May 2013.
Given the MultiNest $90\%$ confidence intervals, we adopt
$\Delta N_{\rm H22} = +1.67\pm0.07$ from 2003--2013 and
$ \Delta N_{\rm H22} = -1.07\pm0.10$ from March to May 2013.

$\bullet$  NGC~5506, X-1 \& X-2 to X-3: $N_{\rm H22}$ increased from $2.77
\pm 0.05$ and $2.80 \pm 0.05$ in 2001 and 2002, respectively, to $3.02
\pm 0.05$ in 2014; given the MultiNest uncertainty on X-1,
we adopt $\Delta$$N_{\rm H22} = +0.25\pm0.07$ over a period of
3.5 years. In addition, as can be seen in Fig \ref{Fig:overplot} the column densities increase 
from 2001 to 2004 but then remain consistent with being constant from 2004-2015

$\bullet$ NGC~7582, X-1 to X-4: $N_{\rm H22}$ increased from $15.5 \pm
1.7$ to $29.8 \pm 0.5$ over $\Delta t = 14.2$ years; $\Delta$$N_{\rm H22} = +14.3$, with a
combined error from MultiNest runs of 2.7.

 All of the above eight cases of $\nh$ variability, and the corresponding  event durations are listed in Table \ref{Table:NHsumm}. We present spectral overplots for these sources in Fig. \ref{Fig:overplot}.
For most of them, the variations in $N_{\rm H}$ are modest enough that
the change in spectral curvature is not always visually obvious, although the
variation in Fairall~49's absorber is quite apparent.

For the other objects in the sample, where there exist
multiple observations per telescope, we can rule out variations in
$N_{\rm H22}$ down to approximately
\begin{itemize}
	\item $0.1$ (NGC~7314/X), $0.2$ (NGC~2992/X, excluding X1; NGC~7314/CA),
	\item $0.25-0.40$ (NGC~526A/CA; NGC~2110/S; NGC~2992/S; NGC~5506/S; NGC~7314/CH),
	\item $0.5-0.8$ (Fairall~49/CA; NGC~526A/CH; NGC~6251/CA),
	\item $1.0-1.4$ (MCG--5-23-16/CA; NGC~526A/X; NGC~2110/CH; NGC~7172/X),
	\item $1.9-2.3$ (IRAS~05189/X; NGC~4258/CH),
	\item $2.7-4.2$ (Cen~A/X; IRAS~00521/X; IRAS~05189/CA; NGC~1052/X; NGC~4258/CA; NGC~4258/S; NGC~6300/CA),
	\item $ 7$ (Mkn~348/X; NGC~4258/X), 10 (Cyg~A/CA),
	\item $17$ (NGC~4507/X), and
	\item $20$ (NGC7582/S).

\end{itemize}
 However, the reader is reminded that these limits are based on the
statistical error on $N_{\rm H}$ only and do not take into account
potential model degeneracies with other parameters such as
partial-covering parameters.
In addition, the various object/instrument combinations do not have
equal numbers of points nor cover the same durations, so these limits
cannot be considered to be uniformly derived in those senses. We present the overplots of the spectra of these sources ($\nh$ not varied) in Appendix Fig. \ref{Fig:CenA}-\ref{Fig:NGC7582}.

\subsection{Variability in $N_{\rm H}$ across multiple instruments}\label{subsec:crossinstrument}

Across the full sample, we would like to be able to, ideally,
cross-calibrate values of full-covering column density 
between different instruments from different
missions, and thus derive systematic differences in $\nh$, which can
enable us to not only create one combined $N_{\rm H}$ lightcurve for each
object, but to interpret it as well.
However, doing so is particularly difficult for this sample of
absorbed type IIs, for multiple reasons. Any offset value in $\nh$ we
try to compute (e.g., $\nh$(ACIS) -- $\nh$(XMM)) would likely
have strong object-to-object and/or telescope-to-telescope variance
due to:
(1) differing soft band spectra --- even for the same object --- as
different extraction regions and effective areas/responses can lead to
differing modeled contributions from extended thermal emission; (2) intrinsically
variable hard X-ray power-law slope values from non-simultaneous
observations of the same object; and (3) in a few objects,
partial-covering components are detected only in a fraction of the observations.
Finally, the location of the continuum rollover due to absorption will be quite
different from one object to the next, given the wide range of column values
and given how differences in response
and effective area between any two telescopes evolve with energy; 
comparing systematic offsets between missions for objects with
$N_{\rm H}$ $\sim 10^{21}$~cm$^{-2}$ to those obtained for $\sim
10^{23}$~cm$^{-2}$ thus may not be highly fruitful.  Consequently, a
detailed analysis of the full range of potential systematic differences in
$N_{\rm H}$ (or other parameters) is beyond the scope of the current paper.


 Nonetheless, we can still consider simultaneous observations of the
same object as an initial exploration of such systematic differences,
and derive approximate thresholds for detecting gross changes in
full-covering $N_{\rm H}$.  The only quasi-simultaneous observation of
a source with all three missions is that of MCG--5-23-16 (observation
IDs: CH2, CH3, X2, S1), which occurred on 7--10 December 2005, and
analyzed by \citet{2007PASJ...59S.301R}; S1, X2, and CH2 were in fact
directly overlapping from 8 December 2005 $\sim$ 21 UTC until 9
December 2005 $\sim$ 2 UTC; S1, X2, and CH3 were directly overlapping
from 9 December 2005 $\sim$ 21 UTC until 10 December 2005 $\sim$ 3
UTC.  CH2 and CH3 did not yield any significant spectral variability,
so we henceforth average the best-fit model parameters.
We find 
$N_{\rm H22}$ (CH) -- $N_{\rm H22}$ (X) = $0.00 \pm 0.25$,
$N_{\rm H22}$ (S) -- $N_{\rm H22}$ (CH) = $0.07 \pm 0.25$, and
$N_{\rm H22}$ (S) -- $N_{\rm H22}$ (X) = $0.07 \pm 0.02$
(CH, X, and S denote HETG, \textit{Suzaku}, and \textit{XMM-Newton}).
That is, one can conclude that
values of $N_{\rm H22}$ measured from \textit{Suzaku} are will be $0.07 \pm 0.02$ higher
than those for \xmm{} in the absence of intrinsic variability in column density;
However, such a conclusion would only be reasonably applicable to
those sources with a spectral shape very similar to that of
MCG--5-23-16: full-covering $N_{\rm H} \sim 1.0-1.8 \times 10^{22}$ cm$^{-2}$,
no partial-covering component, and extremely low amounts of soft
thermal emission and scattered power-law emission below 1 keV
\citep[see e.g, Fig.~4 of ][]{2007PASJ...59S.301R}.
In the \textit{Suzaku} XIS spectrum for instance, the 
value of spectral counts in counts s$^{-1}$ keV$^{-1}$ drops by
well over an order of magnitude from $\sim2$ keV to $\sim0.7$ keV).
Across our sample, only NGC~526A has a similar spectral shape.
Considering observations taken two years apart, values of best-fit
full-covering $N_{\rm H}$ for NGC~526A's S1 observation (in 2011) and
X3 (in 2013) yield $N_{\rm H22}$ (S) -- $N_{\rm H22}$(X) = $0.19 \pm
0.09$, a bit higher than for the (simultaneous) observations of MCG--5-23-16.  Similarly, offsets to $N_{\rm H22}$ (CH) are
consistent with the upper limits derived for MCG--5-23-16.
We conclude that the measured differences in $N_{\rm H}$ between
various missions for NGC~526A are consistent with inter-mission
systematic offsets, and there is no evidence for variability in $\nh$ here.

Cen~A also has a pair of simultaneous observations (X-5 and S-6)
and a pair separated by eight days (X-4 and S-5).
Assuming that the column does not vary on timescales less than eight days,
these pairs of observations would imply that $N_{\rm H22}$ (X) is
roughly 1.1--1.4 higher than $N_{\rm H22}$ (S) for objects with a
spectral shape similar to that of Cen~A.
However, most of the other objects with columns similar to that of
Cen~A have very strong soft-band emission (NGC~4258) and/or partial
coverers (e.g., Mkn~348, NGC~4507), so a straightforward application
is not possible.

Across the sample, excluding those sources where we have claimed
variability in full-covering $N_{\rm H}$, we find 
that ratios of $\nh$ for the
following instrument pairs typically span:
$N_{\rm H}$(CA)/$N_{\rm H}$(X) $\sim 0.3-1.2$,
$N_{\rm H}$(CH)/$N_{\rm H}$(X) $\sim 1.0-1.7$,
$N_{\rm H}$(CH)/$N_{\rm H}$(CA) $\sim 0.8-2.3$,
$N_{\rm H}$(S)/$N_{\rm H}$(X) $\sim 0.7-1.0$.
These ratios show that there is no general trend of any instrument
consistently detecting higher/lower values of $\nh$ for the same
source compared to other instruments.
In addition, under the assumption that $N_{\rm H}$ is intrinsically
non-varying in these sources, these ratios demonstrate the approximate
level of sensitivity required to claim variability in $\nh$ between
different telescopes.
While comparing the values of $\nh$ for a given object from different
instruments, we conservatively consider differences in $\nh$ to be
significant only if their ratio is greater than $\sim$ 2; to that
effect, we do not find any object to display significant variability
in full-covering $\nh$ up to this factor between different
instruments.



At this point it is worth noting a few important differences
between the results found by \citet{2002ApJ...571..234R} (hereafter REN02) and our
work. 15 out of 20 sources in the sample by
REN02 overlap with our sample. However, we do
not detect $\nh$ variability with the same frequency as detected by
REN02. Possible causes include: 1)
REN02 considered variability in $\nh$ as
obtained from different missions to be bona fide. However, relative
flux and energy cross-instrument calibration issues likely play a
role, different instruments had different apertures and/or energy bands, and in addition,
model degeneracies play an important role in estimating the errors on
the measured $\nh$, which the authors have not considered. 2) The
various sources were analysed by different authors using different
techniques and models, thus introducing an unknown amount of scatter
in the errors derived on the measured parameters. 3) The data were
obtained using missions which sometimes had poorer energy resolution
and/or narrower bandpass compared to our work.  4) The data quality
did not allow REN02 to detect and constrain any
partial covering absorption as we could do in our work.


\subsection {Note on partial covering absorbers}

  For four sources in our sample (Fairall~49, IRAS~F00521--7054, NGC~1052, and NGC~4507), we
  consistently detected partial-covering absorption components in all
  observations; in addition in Mkn~348 
  we detected partial-covering components in all but one observation. We have seven sources in which we 
  detected partial-covering absorption in some of their observations (Cen~A, IRAS~F05189--2524, Mkn~348, NGC~2110, NGC~5252, NGC~7172, NGC~7582).
  Among these 11 sources (in total), partial-covering column densities are
  typically $N_{\rm H}^{\rm pc} \sim 10^{23-24}$ cm$^{-2}$ and with
  covering fractions $f_{\rm pc}$ typically spanning $\sim
  0.3-0.9$. The detection of a partial coverer is independent of the
  value of the column density of the full coverer or the spectral
  index $\Gamma$, implying that the detection of the partial coverer
  is bona fide in these cases.  In virtually all cases, the errors on
  both $N_{\rm H}^{\rm pc}$ and/or $f_{\rm pc}$ are large and impacted
  by some degree of model degeneracy, preventing us from making any
  statement about variability or constancy in these partial-covering model parameters
  as a function of time, though the partial-covering model component is statistically required in the fits in these cases.
 Future broad-band ($0.3-50\kev$) high SNR observations can distinguish between the following scenarios 1) if the
partial-covering components are intrinsically variable in terms of
crossing the line of sight 2) or if they are not
detected due to the complexity of the strongly absorbed spectrum
 and/or the lack of spectral coverage above 10 keV for
\xmm{} and \chandra{}, 3) Or simply due to a lack of SNR.


\section{Discussion}

In this work, we have conducted a systematic study of variations in
line of sight absorption column density $N_{\rm H}$ across a sample of
perpetually-absorbed Compton-thin type II AGN.  We have improved upon
the \textit{RXTE}-based study of MKN14 by using \textit{XMM-Newton},
\textit{Chandra}, and \textit{Suzaku}, which yield comparatively
greater sensitivity to smaller variations in ${\Delta}N_{\rm H}$ (by
roughly an order of magnitude) as well as greater sensitivity to
partial-covering absorbers.
We have classified the full-covering absorbers in each source into
variable or non-variable (down to sensitivity levels of roughly
$3-10\%$ when considering a single telescope, or factors of
very roughly 2 when comparing inter-telescope data). We find evidence for variability in the full-covering obscuration components in seven sources (Cen~A, Fairall~49,
MCG--5-23-16, NGC~2992, NGC~5506, NGC~5252, and NGC~7582) to vary on timescales of 2 months to 14.5 years,
with values of $\vert{\Delta}N_{\rm H}\vert$ spanning $\sim$ 0.1 to 1.9 $\times
10^{22}$ cm$^{-2}$; in all cases, the variability is at the $\geq3 \sigma$ level.
We also find that almost half the sources in our
sample (9/20) require a partial-covering absorber in all or almost all
of their observations.

Below, we discuss the nature and location of the various absorbing
components: In short, variable full-covering X-ray-obscuration
components likely delineate compact-scale gas (less
than $\sim 1-10$ pc) which could be associated with the dusty or
non-dusty components of the ``torus.''
Meanwhile, non-variable columns could potentially indicate either 
distant material residing at scales of 0.1 kpc to kpcs, such as dust lanes,
though smooth (non-clumpy) homogeneous compact-scale gas is also a possible explanation.


\subsection{The full-covering X-ray-obscuring gas}

\subsubsection{Variable full-covering $\nh$, and implications for compact-scale gas}

{
 As stated above, we detected eight  occurrences of variable full-covering $N_{\rm H}$
 across seven sources. 
 We discuss three physical models below, although the sparse sampling
 makes it impossible to fully distinguish between these models and thus
 discern the true nature of the variable-$N_{\rm H}$ gas in each of the
 seven objects where variability in $N_{\rm H}$ was detected.
We did not,
 for example, detect any new complete eclipse events, with egress and
 ingress, for which sustained monitoring is usually necessary, e.g.,
 as \textit{RXTE} provided for variability on timescales of
 days--years, or as long-looks from \textit{XMM-Newton} provided for
 timescales $\lesssim$ a day.
Nonetheless, even establishing variability in full-covering $N_{\rm
  H}$ is a rudimentary first step because it establishes the presence
of relatively compact-scale gas contributing to the total observed
value of $N_{\rm H}$.

 Model A: {\it All} full-covering obscuration is due to discrete
 clumps only, e.g., in the torus, following \citet{2008ApJ...685..160N},
 and an observed increase (decrease) in $N_{\rm H}$ indicates the
 number of clouds along the line of sight increasing from $N$ to $N+1$
 (decreasing from $N+1$ to $N$), where $N$ cannot be zero for our
 perpetually-absorbed sample. 
Ingress/egress of individual clouds should cause a step-like behavior
in $N_{\rm H}$ if the whole cloud enters the line of sight faster than
the observation sampling.  However, if a cloud ingress occurs very
slowly relative to the observation sampling, then a slow
increase/decrease in $N_{\rm H}$ could be observed, depending on
the cloud's transverse density profiles.

 We do not have sufficient data to determine what the ``average''
 value of $N_{\rm H}$ corresponding to $N$ clumps is for any source.
 We thus assume for simplicity that the lowest measured values of $N_{\rm H}$
 correspond to $N$ clouds. We also use the  
 simplifying assumption that all individual clouds have identical column densities.
 If the observed values of ${\Delta}N_{\rm H}$ correspond to ingress
 (egress) of one cloud into (out of) the line of sight, then, for
 example, the observed increases in $N_{\rm H}$ in both NGC~2992 and
 NGC~5252 could each correspond to an increase from $\sim$ 3 to 4
 clouds.

 Model B: Full-covering obscuration is explained by the sum of a time-constant component
 (e.g., kpc-scale dust structures, as discussed below) 
plus some number of compact-scale discrete clumps.
 Here, the extremely sparse sampling of our data precludes us from being able to
 cleanly separate the light curve into ``eclipsed'' versus ``non-eclipsed'' periods
 (in contrast to the sustained monitoring provided by \textit{RXTE}).

We do not observe both ingress and egress for any variability event,
so we would not be able to estimate radial distance to the occulting
structure \citep[e.g., following][eqn.\ 3, which assumes Keplerian motion]{2002ApJ...571..234R} invoking
assumptions on full eclipse duration and in particular cloud density.

Such constraints are necessary to obtain accurate distances and thus
meaningful insights into the physical processes that create and sculpt
clouds. For example, it would help to know if clouds are inside or
outside the dust sublimation region, since the presence of dust can
play a crucial role in the physical processes that form, shape, and
drive compact structures e.g., via radiation pressure on dust to drive
winds \citep{2011A&A...525L...8C,2012ApJ...761...70D,2018MNRAS.474.1970B}.

 
In a third model, Model C, the variable component of the X-ray
obscuration is due to a non-clumpy, volume-filling (contiguous),
compact-scale medium, which contains inhomogeneities that transit the
line of sight.
MKN14 discuss a similar interpretation of the
observed $\nh$($t$) light curve derived from \textit{RXTE}
monitoring of Cen~A during 2010--2011. In addition to sharp increases
in column density, interpreted as transits by discrete clumps, they
detected a smooth decrease then increase by $\gtrsim10\%$ over an 
80-day span\footnote{We should note that since Cen~A is a radio galaxy and no BLR
has been confirmed yet, it may not represent a standard Seyfert
galaxy; nonetheless, searches for such non-clumpy, contiguous
components of the torus are important for testing the applicability of
clumpy torus models across AGN.}.
In the current study, we observe an increase in $\nh$ in the \xmm{} observations
of NGC~5506 over a period of 3.5 years, followed by $\nh$ remaining constant for an additional
$11$ years. While we cannot completely rule out that this trend is due to ingress by a single cloud, it is unlikely 
unless the cloud has a rather contrived transverse density profile. Such smooth trends argue against the clumpy-torus model being
able to explain all of the observed absorption in these two objects:
ingress/egress of individual clouds would produce sharp step functions
in the $\nh$ light curve, but the observed smooth trends 
(particularly in the \textit{RXTE} data for Cen~A) argue against such
an interpretation.  One possibility is that these variations are due
to the line of sight's passing through a contiguous component of the
torus \citep[i.e., possibly an intercloud medium;
][]{2012MNRAS.420.2756S}, and relative over- or under-dense regions
transit the line of sight.  That is, during 2001-2004, the line of sight in NGC~5506 was transited by a relatively underdense region (by$\ge 8\%$ relative to the long-term average), thus causing the observed ``dip" in $\nh$. Our observations thus provide constraints for 
column density ratios in such media for these cases.

\subsubsection{Constant-$\nh$ sources: Origin?}\label{subsubsection:constantnh}  

For 13 objects in our sample, the full-covering obscurer's column
density is consistent with being constant in time, down to sensitivity levels of $\Delta$$N_{\rm H}$
ranging from $\sim$ 0.1 to 17 $\times10^{22}$ cm$^{-2}$. Could such obscuration be due to a single discrete cloud?
At a distance of a few pc and more, typical velocities are of order hundreds of km s$^{-1}$.
To obscure for $\sim$ a decade, its transverse diameter must be at least 
of order light-days. This is a very unrestrictive limit, not much larger
than the inferred sizes of X-ray clumps so far. Furthermore,
some models posit large-scale structures at tens of parsecs
comprised of filaments of order a parsec thick \citep[e.g.,][]{2012ApJ...758...66W}.
However, a decade-long eclipse by a single cloud would require a near-uniform
cloud density in the transverse direction, which is a somewhat contrived scenario.
It is also highly unlikely that the bulk of the objects in our sample {\it each} have such a 
cloud along their lines of sight. For these objects, and/or to explain any potential non-variable
component in the variable-$N_{\rm H}$ objects, we therefore consider the
following three interpretations:

(a) In the context of the clumpy-torus model of
\citet{2008ApJ...685..160N}, there could potentially exist a large
number of clouds $N$ along the line of sight, each with a very low
value of $N_{\rm H}$, such that ingress/egress of individual clouds
does not change $N$ or the observed value of $N_{\rm H}$ by
perceptible amounts, giving us an impression of a non-varying column.
For a fiducial total column of, say, $10^{22}$ cm$^{-2}$, and limits on
sensitivity of ${\Delta}N_{\rm H} \sim 10^{21}$ cm$^{-2}$, there would
typically have to be at least $\sim10$ clouds with columns $\leq$ this limit
in order to give us the impression of a non-varying $\nh$.
However, from theoretical considerations, \citet{2008ApJ...685..160N} posit that individual clouds
each typically have visual optical depths of $\sim30-100$, corresponding to $N_{\rm H} \sim 8 - 20 \times 10^{22}$ cm$^{-2}$  
for typical Galactic dust/gas ratios \citep[e.g., ][]{2012ApJ...759...95N}, so a
large number of clouds each with column $\sim 10^{21}$ cm$^{-2}$ is unlikely.

(b) {\it A smooth, contiguous, compact torus or inter-cloud medium:}
To model the IR emission of dusty tori, \citet{2012MNRAS.420.2756S} and \citet{2015A&A...583A.120S} 
assumed the torus to exist in a two-phase medium, with high-density
clouds and low-density gas filling the space between the cloud.
%
%
In our study, the constant level of $\nh$ observed in X-rays may
denote the intercloud medium, while the higher column density partial
coverers and/or variable absorbers detected in X-rays denote the high
density clumps.
In this case, the limits on column density between relative over- or
under-dense regions must be $\lesssim 10^{21}$ cm$^{-2}$ for the relatively less-absorbed sources in our sample.

}


{(c) {\it Host galaxy dusty structures, e.g., lanes or filaments:} A constant level of full-covering
  X-ray obscuration  could also be
  attributed to dusty gas residing along the line of sight at scales
  $\gtrsim 0.1$ kpc to several kpcs.  As noted in the Introduction, there
  are multiple indications that the host galaxies of optical type II
  sources themselves may play a role in the observed X-ray obscuration
  and optical extinction.

Ideally, we would like to go through each source on a case-by-case
basis, and compare the observed value of $N_{\rm H}$ to values of $N_{\rm H}$ estimated from both
1) $A_{\rm V}$ from known sources of dust residing at kpc scales
and 2) $A_{\rm V}$ from the dust residing in the pc-scale torus at radial distances outside $R_{\rm dust}$.
If component 1) alone can fully account for $N_{\rm H}$, it would
minimize the need to invoke a torus intersecting the line of sight (at
least in that given object).
If components 1) and 2) both exist and cannot account for $\nh$, it
would indicate a significant amount of \textit{non}-dusty gas in a
given object, likely residing inside $R_{\rm dust}$.
There are various known sources of dust extinction for many of our sources, as measured by
Balmer decrements to narrow lines  (e.g., several of our sources are contained in
the samples of \citet{2001A&A...365...37M} and \citet{2017ApJ...846..102M}),
high-spatial resolution color-color maps \citet[e.g., with \textit{HST}, ][]{1994ApJ...436..586M,1996ApJ...459..535S,2014MNRAS.442.2145P}
and NIR-MIR spectral fits \citep[see e.g., ][]{2016A&A...586A..28B}.
We could also consider 9.7~$\mu$m absorption as studied by \citet{2010ApJS..187..172G} using \textit{Spitzer}: the Si-containing gas
absorbs 9.7~$\mu$m continuum from warm dust, and must be due to gas more extended than that warm dust.

However, there are multiple obstacles to this goal:

1) The above methods to determine $A_{\rm V}$ cannot cleanly separate
dust extinction along the total line of sight due to kpc-scale dust
lanes versus that due to a compact torus: one simply gets the total
extinction along the line of sight.

2) Certain methods (color color maps, spectral fits) may lack the
spatial resolution to guarantee that all optical extinction along the
line of sight to the AGN is indeed accounted for; there might,
potentially, be some compact giant molecular cloud lying along the line of sight that would be
missed by the above methods, but would contribute to $\nh$. It is even possible that $N_{\rm H}$ along the line of sight could be overestimated if there exists a “hole” not picked up by the above methods.

3) In those cases where individual kpc-scale dust structures are
resolved and noted to cross the line of sight to the nucleus (``DC'' in
\citet{1998ApJS..117...25M}), but where dust extinction maps (from
color-color maps) have not yet been made, we could \textit{attempt} to
assign a ``canonical'' or ``generic'' value of $A_{\rm V}$ to all dust lanes.  For
example, based on color-color maps made with \textit{HST} for nearby Seyferts,
$A_{\rm V}$ is typically $\sim 0.5-2$ magnitudes \citep[e.g., ][]{1994ApJ...436..586M}, or $A_{\rm V}
\sim 3-6$ in the case of Cen~A's famous dust lane \citep{1996ApJ...459..535S}.  Applying this to all galaxies, however, is dangerous: there is very strong dispersion
from one dust lane to the next and from one line of sight to the next.

We found 45 total estimates of either V-band exinction or 9.7~$\mu$m Si
line optical depth for our 20 sources in the literature, from the
aforementioned references; see Fig.~\ref{Fig:NHcompare}. In estimating the corresponding values of
$N_{\rm H}$, we assume the Galactic dust/gas conversion of Nowak et
al.\ (2012): $N_{\rm H} = A_{\rm V} \times 2.7 \times10^{21}$
cm$^{-2}$ mag$^{-1}$.  For the Si line optical depths in \cite{2010ApJS..187..172G}, we multiply by 10 to obtain estimates of $A_{\rm V}$.

The median value (in linear space) of all these
estimates is $N_{\rm H} = 0.84\times 10^{22}\cmsqi$, and the 16th/84th percentiles are
0.40 and 2.1 $\times 10^{22}\cmsqi$ respectively. There are some individual cases for which various measurements of
$A_{\rm V}$ in the literature imply values of $N_{\rm H}$ that are
roughly equal to or greater than our measured values, raising the possibility that
all dusty gas (kpc + pc scale, in total) can indeed account for with
all X-ray obscuration, and that there is no need to invoke non-dusty gas
inside $R_{\rm dust}$.  However, other measurements (sometimes for the
same object) yield estimates of $N_{\rm H}$ that fall short.

We can only make the very general conclusion that when $N_{\rm H}$ is of order of magnitude
$10^{22}$ cm$^{-2}$ or higher,
there is a relatively increased likelihood that a component of {\it non}-dusty gas
(likely inside $R_{\rm dust}$ and thus part of the innermost compact torus) exists.
For smaller columns, there is a relatively increased likelihood
that dust-containing structures intersecting the line of sight (sum of kpc-scale
and dusty pc-scale structures) can explain $N_{\rm H}$. 
Our conclusions are generally consistent with those of 
several early and recent studies
aiming to separate
the contributions of galaxy-scale dust lanes and nuclear obscuration 
such as
\citet{2000A&A...355L..31M, 2001MNRAS.327..323G,2005A&A...444..119G}
and \citet{2017MNRAS.465.4348B}.

Although subject to very low number statistics, a Kolmogorv-Smirnov
(KS) test indicates that the distributions of values of $N_{\rm H}$ in
the $\nh$-variable and the $N_{\rm H}$-non-variable subsamples
are consistent with arising from the same parent population (the null
hypothesis in the KS test cannot be ruled out at a confidence of even
merely $50\%$). See Fig. \ref{Fig:hist} right panel for the two distributions. This finding would suggest that in Compton-thin obscured type IIs, neither
the structures that comprise non-homogeneous tori (and thus $\nh$-variable)
nor the structures comprising constant-$N_{\rm H}$ media
(be they due to host-galaxy structures or a homogeneous compact torus)
have a preference for relatively high or low columns.


\subsection{Sources with partial covering absorption}     

 As mentioned earlier, previous sample studies on X-ray absorption in
 Seyfert-2 galaxies, such as in \citet{2014MNRAS.439.1403M}, used
 \rxte{}, which was not highly sensitive to partial covering and lower column density ($\nh\le 10^{21}\cmsqi$) absorbers. However, \xmm{}, \chandra{} and \suzaku{} are, and
 hence we can additionally constrain partial coverers apart from full
 coverers. 11 out of 20 sources in the sample show signatures of
 partial-covering (hereafter PC) absorption. Many similar PC absorption features have been identified
 in other observations of Seyfert galaxies such as NGC~1365
 \citep{2009ApJ...696..160R}, Mkn~766 \citep{2011MNRAS.410.1027R},
 NGC~3227 \citep{2018MNRAS.481.2470T}, including previous observations
 of objects in our sample, e.g., NGC~7582
 \citep{2009ApJ...695..781B}. However, our small sample spans a
 relatively small range in system parameters such as
 $L_{2-10\kev}$ and $\lbol/\ledd$, and thus extrapolation to
 determining the fraction of sources hosting sustained PC components
 across all Compton-thin and/or optically-identified type IIs in the
 local Universe is not straightforward. In our sample we find the
 best fit covering fractions spanning typically $30-90\%$ and column
 densities spanning $1-80\times 10^{22}\cmsqi$ (assuming a neutral
 absorber in our model). We must note as a caveat that we do not have
 strong data constraints on PC model parameter values, given the CCD
 energy resolution and model complexity. We thus caution the reader not to
 interpret measured changes in partial covering $\nh$ and/or covering fraction
 too literally. There may exist multiple clouds residing and partially
 covering the line of sight, but we cannot discern ingress/egress of
 individual clouds; current data thus prevent us from confirming or rejecting this notion.  Constraints
 on the sizes and the location of PC clouds from our data alone are
 not strong. If the clouds partially cover the corona, then the clouds
 must be smaller, so a corona size of say, $10-30 \, R_{\rm g}$
 provides an upper limit on the size of the cloud. For example, for a
 $10^8\msol$ black hole, $30\, R_{\rm g}=4.5\times 10^{14}\cm$. Such
 sizes are consistent with estimates using occultations by individual
 clouds \citep[e.g., NGC~1365, ][]{2009ApJ...696..160R}.
 Since we detected only neutral absorbers in our fits (and no ionized absorbers), we do not have a good
 handle on the ionization parameter of these clouds. For that matter,
 any value of the ionization parameter $\xi$ that yields strong
 continuum curvature at $\sim 6\kev$ or below is plausible.
 Constraints based on ionization parameter generally thus only provide a
 rough lower limit to the radial distance of (order of magnitude) a light day in most cases.

 The consistency of the PC components across over a decade could indicate that there exists a population of clumps
 that are long-lived and orbiting mostly in Keplerian motion, with
 clouds either too dense to be tidally sheared by the SMBH, or else
 confined eternally by the ambient gas and pressure or a magnetic
 field \citep{1987MNRAS.228P..47R,1988ApJ...329..702K}.
 Another possibility a mechanism which continuously
 produces clumps and deposits them along the line of sight, and which is
 both active and stable over timescales of at least $1-2$
 decades. Potential mechanisms include magnetohydrodynamic-driven
 winds
 \citep{1982MNRAS.199..883B,1994ApJ...429..139C,1994ApJ...434..446K,2010ApJ...715..636F},
 or a turbulent dusty disk wind as proposed by
 \citet{2011A&A...525L...8C}. The PC column density from our sample
 are mostly consistent with those derived by
 \citet{2010ApJ...715..636F}. If the physical conditions in the disk
 remain stable over timescale of years, then it's not hard to envision
 a persistent wind process.
  
 Using a Kolmogorv-Smirnov (KS) test, we find that the distributions of
 the values of full-covering $\nh$ of sources with and without partial
 coverers are consistent, (i.e., the null hypothesis in the KS test
 cannot be ruled out at a confidence of even merely $\sim 60\%$ implying that these
 samples have been likely derived from the same parent sample), and
 suggesting that the full and partial coverers are two independent
 components. See Fig. \ref{Fig:hist} left panel for the two distributions.


\section{Conclusions}

We carried out an extensive X-ray spectral variability study of a
sample of 20 Compton-thin Seyfert-2 galaxies to investigate the nature
of the variability of the neutral intrinsic absorption in X-rays along
the line of sight, and derive constraints on the location and properties
of the X-ray obscurer. We are sensitive to absorption column density
of $\nh\sim 10^{20.5-23.5}\cmsqi$ of fully- and partially-covering,
neutral and/or lowly-ionized clouds transiting along the line of sight on
timescales of days to decades. We list below the main conclusions
from our study:

\begin{itemize}

\item{We detected variability in full-covering absorption column $\nh$ in
	X-ray spectra of seven out of 20 objects at $\geq3\times$ the $90\%$ confidence level (obtained from spectral fits),
  implying compact-scale, non-homogeneous gas
		along our line of sight in those objects. We detected variations as small as $\sim1-2\times10^{21}\cmsqi$ in some objects (See Table \ref{Table:NHsumm}). Models that explain torus geometry by invoking discrete clouds or other compact structures thus must include the possibility of structures with values of column density as small as these. }

\item{For most of these seven objects, due to their 
  sparse sampling, we cannot distinguish
  between variability due to discrete clouds transiting the line of
  sight or a contiguous (volume-filling) inhomogeneous medium. 
		 An exception, though, is NGC~5506, in which we observe an increase in $\nh$ over 3.5 years, followed by $\nh$ remaining constant  for an additional 11 years. Such a trend is qualitatively similar to the ``dip" in $\nh$ in Cen~A noted by MKN14. These trends are difficult to explain in the context of clumpy-torus models; one possible explanation is 
		that the variable component of its column density originates in a non-homogeneous contiguous medium. That is, we observed a relatively underdense region (by $\ge 8\%$ relative to the long-term average) transit the line of sight in NGC~5506 before 2004.}

\item{We do not detect any significant $\nh$ variability for 13/20
  sources. Nuclear $\nh$ variability
  of Compton-thin type IIs is thus far less prevalent than previously
  reported in the literature. The X-ray obscurers in these sources may be associated with
  a contiguous, highly homogeneous (column density variations
  typically $< \sim 10^{21}$ cm$^{21}$) compact scale medium. They
  could instead be associated with large-scale dusty structures or filaments
  intersecting the line of sight at distances of $\gtrsim$ 0.1 kpc to
  kpcs, consistent with previous studies.}

\item{We detected partial covering absorption in 11/20 sources over 1--2 decades, suggesting a
  long-lived population of clumpy clouds or a long-lived mechanism for
		producing such clouds.  The distributions of the values of
  full-covering $\nh$ of sources with and without partial coverers are
  consistent, suggesting that the full and partial coverers are two
  independent components. There are six sources for which we detected partial-covering absorption in some of their observations, but we refrain from commenting on the variability
		and/or the properties of the partial coverers due toi lack of signal-to-noise and lack of broad band pass ($0.3-50\kev$) in $80\%$ of our observations (\xmm{} and \chandra{}). Future broad-band ($0.3-50\kev$) high SNR observations can distinguish between the scenarios 1. if the
partial-covering components are intrinsically variable in terms of
crossing the line of sight 2. or if they are not
detected due to the complexity of the strongly absorbed spectrum
 and/or the lack of spectral coverage above 10 keV for
\xmm{} and \chandra{}, 3. Or not detected simply due to lack of SNR.}

\item{We do not observe any Compton-thin to -thick transitions, or vice versa, in our sample.}

\item  {The distributions of average values of $\nh$ in
the $\nh$-variable and the $\nh$-non-variable subsamples
are consistent with arising from the same parent population suggesting that in Compton-thin obscured type IIs, neither
the structures that comprise non-homogeneous tori (and thus $\nh$-variable)
		nor the structures comprising constant-$\nh$ media (be they due to host-galaxy structures or a homogeneous compact torus) have a preference for relatively high or low columns. We are however limited to small number statistics (See Fig. \ref{Fig:hist}).}

\end{itemize}

    Future X-ray observations of larger samples of Compton-thin-obscured
Seyferts can yield additional insight into compact-scale X-ray
obscurers the applicability of clumpy-torus models, and the potential
presence of compact-scale non-clumpy gas such as an intercloud medium
by further quantifying the fractions of sources with variable
full-covering $N_{\rm H}$.  Specifically, the community needs the
combination of sustained multi-timescale monitoring (to probe spectral
variability on timescales from days to years), as \textit{RXTE}
provided, plus soft X-ray coverage with at least CCD-quality
resolution, as provided by \xmm{}, \chandra{}, and \suzaku{}, to build
a new database of $N_{\rm H}$ variations, and distinguish among the
various physical explanations for variations in $\nh$.


{\bf ACKNOWLEDGEMENTS:} {S.L.\ and A.G.M.\ acknowledge financial support
  from NASA via NASA-ADAP Award NNX15AE64G. S.L.\ and A.G.M.\ thank Matt
  Malkan for insightful discussions. A.G.M.\ and T.S. both acknowledge partial funding
  from Narodowy Centrum Nauki (NCN) grant
  2016/23/B/ST9/03123. M.K.\ acknowledges support from DLR grants
  50OR1802 and 50OR1904. The authors thank Johannes Buchner for assistance in setting up and running
BXA and MultiNest. This research has made use of data obtained from the
  \chandra{}, \xmm{}, and \suzaku{} missions by NASA, ESA and
  JAXA. This work has made use of HEASARC online services, supported
  by NASA/GSFC, and the NASA/IPAC Extragalactic Database, operated by
  JPL/California Institute of Technology under contract with NASA.}


\begin{table*}

\caption{Source properties.} \label{Table:sources}
  \begin{tabular}{lccccccccccccccc} \hline\hline 
Source		        &R.A.		&Dec.	        &Redshift  &$\mbh$	        &Ref$^a$ &Method$^b$	&$\nh^{\rm Gal}$		& Optical\\		 
		        &(J2000)	&(J2000)        &	   &$\log(\mbh/\msol)$	&        &		&$\times 10^{20}\cmsqi$	& Classification$^c$\\
(1)		        &(2)	        &(3)	        &(4)	   &(5)			&(6)     &(7)		&(8)			&(9)	\\ \hline
1. Cen~A		&13h25m27.6s	&--43d01m09s 	&0.0018	   &$7.7_{-0.3}^{+0.2}$	& C09	 & stellar	&$8.09$			&RG\\
2. Cyg~A		&19h59m28.3s	&+40d44m02s	&0.0561	   &$9.40_{-0.14}^{+0.11}$	& T03	 & gas		&$2.72$			&RG	\\
3. Fairall~49		&18h36m58.3s	&--59d24m09s	&0.0200	   &$6.3$		& I04	 & X var	&$6.47$			&Sy$2\longrightarrow 1H$ \\
4. IRAS~F00521--7054	&00h53m56.1s	&--70d38m04s	&0.0689	   &--			&	 &		&$5.26$			&Sy$2$	\\
5. IRAS~F05189$-$2524	&05h21m45s	&--25d21m45s	&0.0426	   &$8.6$		& X17	 & stellar	&$1.66$			&Sy$2\longrightarrow 1H$\\
6. MCG--5-23-16	        &09h46m48.4s	&--33d36m13s	&0.0081	   &$7.31\pm1.00$	& P12	 & X var	&$8.70$			&Sy$2\longrightarrow 1I$\\
7. Mkn~348		&00h48m47.1s	&+31d57m25s	&0.0150	   &$7.21$		& WU02	 & stellar	&$5.79$			&Sy$2\longrightarrow 1H$\\
8. NGC~526A		&01h23m54.4s	&--35d03m56s	&0.0199	   &$8.02$		& W09	 & K lum.	&$2.31$			&$1.9$	\\
9. NGC~1052		&02h41m04.8s	&--08d15m21s	&0.0050	   &$8.19$		& WU02	 & stellar	&$2.83$			&LINER	\\
10. NGC~2110		&05h52m11s	&--07d27m22s	&0.0077	   &$8.3\pm0.2$		& M07	 & stellar	&$1.66$			&Sy$2\longrightarrow 1I$\\	  
11. NGC~2992		&09h45m42.0s	&--14d19m35s	&0.0077	   &$7.72$		& WU02	 & stellar	&$4.87$			&Sy$2\longrightarrow 1I$\\
12. NGC~4258		&12h18m57.5s	&+47d18m14s	&0.0015	   &$7.59\pm0.01$	& H99    & maser	&$1.60$			&Sy$2$	\\
13. NGC~4507		&12h35m36.6s	&--39d54m33s	&0.0118	   &$8.39$		& W09	 & K lum.	&$7.04$			&Sy$2\longrightarrow 1H$\\
14. NGC~5252		&13h38m15.9s	&+04d32m33s	&0.0229	   &$8.04$		& WU02	 & stellar	&$2.14$			&Sy$2$	\\
15. NGC~5506		&14h13m14.9s	&--03d12m27s	&0.0061	   &$8\pm1$		& O99	 & stellar	&$4.08$			&Sy$1.9\longrightarrow 1I$\\
16. NGC~6251		&16h32m32s	&+82d32m16s	&0.0247	   &$8.8^{+0.2}_{-0.1}$	& FF99	 & gas	        &$5.57$			&Sy2	\\
17. NGC~6300		&17h16m59.5s	&--62d49m14s	&0.0037	   &$6.7$		& V10	 & K lum.	&$7.79$			&Sy$2$\\	  
18. NGC~7172		&22h02m01.9s	&--31d52m11s	&0.0087	   &$8.31$		& W09	 & K lum.	&$1.95$			&Sy$2$	\\
19. NGC~7314		&22h35m46.2s	&--26d03m02s	&0.0048	   &$7.84$		& W09	 & K lum.	&$1.50$			&Sy$1.9\longrightarrow 1H$\\  
20. NGC~7582		&23h18m23.5s	&--42d22m14s	&0.0053	   &$8.31$		& W09	 & K lum.	&$1.33$			&Sy$2\longrightarrow 1I$\\
\hline \hline
\end{tabular}
     
	{$^a$References for $\mbh$:
     C09=\citet{2009MNRAS.394..660C},
     FF99 =\citet{1999ApJ...515..583F},
     H99 =\citet{1999Natur.400..539H},
     I04=\citet{2004MNRAS.347..411I},
     M07=\citet{2007ApJ...668L..31M},
     O99=\citet{1999A&A...350....9O},
     P12=\citet{2012A&A...542A..83P},
     T03=\citet{2003MNRAS.342..861T},
	V10 = \citet{2010MNRAS.402.1081V},
     W09=\citet{2009ApJ...690.1322W},
     WU02 = \citet{2002ApJ...579..530W},
     X17=\citet{2017ApJ...837...21X}          }\\
  {$^b$ Methods for black hole mass estimate:
    gas = gas dynamics;
    K lum.\ = estimated from K-band bulge stellar luminosity;
    maser = water masers; stellar = stellar velocity dispersion;
    X~var.\ = from short-term X-ray variability amplitude.}\\
	{$^c$ Optical classification: To the left of the arrow is the optical classification from NED,
          while to the right are either: $1H$, denoting that the source contains a Type-1 hidden BLR
          observed in polarized optical emission, or $1I$, denoting a Type-1 hidden BLR identified via IR emission lines.}\\
	{The Galactic column densities (column 8) are obtained from the LAB survey of \citet{2005A&A...440..775K}.}


\end{table*}


\begin{table*}

\centering
\caption{List of X-ray observations of the sources in the sample.} \label{Table:obs}
  \begin{tabular}{lccccccccccccccc} \hline\hline 
	
	  Number&Source		& Telescope	&Observation	&Observation 	&Exposure	&Short obs-id 	\\ 
		&		&		&ID		&Date		&		&		\\	\hline\\

	  1.	& CenA	&\xmm{}		&0093650201	&2001-02-02		&24	&X-1		&		\\
	  &		&\xmm{}		&0093650301	&2002-02-06		&15	&X-2		&		\\
	  &		&\xmm{}		&0724060501	&2013-07-12		&12	&X-3		&		\\  
	  &		&\xmm{}		&0724060601	&2013-08-07		&12	&X-4		&		\\
	  &		&\xmm{}		&0724060701	&2014-01-06		&27	&X-5		&		\\
	  &		&\xmm{}		&0724060801	&2014-02-09		&23	&X-6		&		\\

	  &		&\suzaku{}	&100005010	&2005-08-19		&65	&S-1		&		\\
	  &		&\suzaku{}	&704018010	&2009-07-20		&62	&S-2		&		\\
	 &		&\suzaku{}	&704018020	&2009-08-05		&51	&S-3		&		\\
	   &		&\suzaku{}	&704018030	&2009-08-14		&56	&S-4		&		\\

	  &		&\suzaku{}	&708036010	&2013-08-15		&11	&S-5		&		\\
	  &		&\suzaku{}	&708036020	&2014-01-06		&7	&S-6		&		\\  \hline

2.	&CygA$^*$	&\xmm{}					&0302800101	&2005-10-14	&23	&X-1	\\

	  &		&\suzaku{}				&803050010	&2008-11-15	&45	&S-1	\\\hline

3.	&Fairall 49	&\chandra{} HETG	&3148		&2002-03-20	&57	&CH-1		\\
	  &		&\chandra{} HETG	&3452		&2002-03-23	&51	&CH-2		\\
	
	  &		&\xmm{}			&0022940101	&2001-03-05	&75	&X-1		\\
	  &		&\xmm{}			&0724820101	&2013-09-04	&110	&X-2	\\
	  &		&\xmm{}			&0724820201	&2013-10-15	&107	&X-3	\\

	  &		&\suzaku{}		&702118010	&2007-10-26	&78	&S-1	\\ \hline

4.	&IRAS F00521--7054 	&\xmm{}		&0301150101	&2006-03-22	&17	&X-1		\\
	  &			&\xmm{}		&0301151601	&2006-04-22	&14	&X-2	\\
	  &			&\suzaku{}	&708005010	&2013-05-19	&103	&S-1	\\ \hline

5. 	&IRAS F05189--2524	&\chandra{} ACIS-S	&2034		&2001-10-30	&20	&CA-1		\\
	  &			&\chandra{} ACIS-S 	&3432		&2002-01-30	&15	&CA-2		\\

	  &			&\xmm{}			&0085640101	&2001-03-17	&12	&X-1	\\
	&			&\xmm{}			&0722610101	&2013-10-02	&38	&X-2		\\

	  &			&\suzaku{}		&701097010	&2006-04-10	&78	&S-1	\\ \hline

6.	&MCG--5--23--16	&\chandra{} HETG	&2121		&2000-11-14	&76	&CH-1		\\
	  &		&\chandra{} HETG	&6187		&2005-12-08	&30	&CH-2		\\
	  &		&\chandra{} HETG	&7240		&2005-12-09	&20	&CH-3		\\
	
	  &		&\xmm{}			&0112830401	&2001-12-01	&25	&X-1		\\
	&		&\xmm{}			&0302850201	&2005-12-08	&132	&X-2		\\
	&		&\xmm{}			&0727960101	&2013-06-24	&138	&X-3		\\
	&		&\xmm{}			&0727960201	&2013-06-26	&139	&X-4		\\

	&		&\suzaku{}		&700002010	&2005-12-07	&96	&S-1		\\ 
	&		&\suzaku{}		&708021010	&2013-06-01	&160	&S-2		\\
	&		&\suzaku{}		&708021020	&2013-06-05	&139	&S-3		\\ \hline

7. 	&Mkn~348	&\chandra{} ACIS-S	&12809		&2010-10-13	&95	&CA-1			\\
	  &		&\xmm{}			&0067540201	&2002-07-18	&49	&X-1			\\
	  &		&\xmm{}			&0701180101	&2013-01-04	&13	&X-2			\\
	  &		&\suzaku{}		&703029010	&2008-06-28	&87	&S-1			\\ \hline

8.	&NGC~526A	&\chandra{} ACIS-S	&342		&2000-02-07	&9	&CA-1		\\
	&		&\chandra{} ACIS-S	&442		&2000-04-23	&5	&CA-2		\\
	&		&\chandra{} HETG	&4437		&2003-06-21	&29	&CH-1		\\
	&		&\chandra{} HETG	&4376		&2003-06-21	&29	&CH-2		\\

	&		&\xmm{}			&0109130201	&2002-06-30	&12	&X-1		\\
	&		&\xmm{}			&0150940101	&2003-06-21	&48	&X-2		\\
	&		&\xmm{}			&0721730301	&2013-12-21	&56	&X-3		\\
	&		&\xmm{}			&0721730401	&2013-12-22	&46	&X-4		\\

	&		&\suzaku{}		&705044010	&2011-01-17	&73	&S-1		\\ \hline

\end{tabular}

{$^*$ Cyg A Chandra observations and the corresponding best fit parameters are listed in Table \ref{Table:cygA}.}


\end{table*}


\begin{table*}

\centering
\setcounter{table}{1}
\caption{List of X-ray observations of the sources in the sample.}
  \begin{tabular}{lccccccccccccccc} \hline\hline 
	
 Number&Source		& Telescope	&Observation	&Observation 	&Exposure	&Short obs-id 	\\ 
	&		&		&ID		&Date		&($\ks$)	&		\\	\hline\\

9.	&NGC~1052&\chandra{} ACIS-S	&5910		&2005-09-18	&60	&CA-1		\\
	  		
	  &		&\xmm{}			&0093630101	&2001-08-15	&16	&X-1		\\
	  &		&\xmm{}			&0306230101	&2006-01-12	&55	&X-2		\\
	  &		&\xmm{}			&0553300301	&2009-01-14	&52	&X-3		\\
	  &		&\xmm{}			&0553300401	&2009-08-12	&59	&X-4		\\
	  &		&\suzaku{}		&702058010	&2007-07-16	&101	&S-1		\\ \hline

10.	&NGC~2110 			&\chandra{} ACIS-S 	&883		&2000-04-22	&50	&CA-1		\\
			&		&\chandra{} HETG	&3143		&2001-12-19	&34	&CH-1	\\
			&		&\chandra{} HETG	&3418		&2001-12-20	&76	&CH-2	\\
			&		&\chandra{} HETG	&3417		&2001-12-22	&33	&CH-3	\\
			&		&\chandra{} HETG	&4377		&2003-03-05	&96	&CH-4	\\

			&		&\xmm{}			&0145670101	&2003-03-05	&60	&X-1		\\

			&		&\suzaku{}		&100024010	&2005-09-16	&102	&S-1		\\
			&		&\suzaku{}		&707034010	&2012-08-31	&103	&S-2		\\
			&		&\suzaku{}		&709011010	&2015-03-20	&46	&S-3		\\ \hline

	  11. 	&NGC~2992&\chandra{} HETG		&11858		&2010-02-09	&96	&CH-1		\\

	  &		&\xmm{}				&0147920301	&2003-05-19	&29	&X-1		\\
	  &		&\xmm{}				&0654910301	&2010-05-06	&59	&X-2		\\
	  &		&\xmm{}				&0654910401	&2010-05-16	&61	&X-3		\\
	  &		&\xmm{}				&0654910501	&2010-05-26	&56	&X-4		\\
	  &		&\xmm{}				&0654910601	&2010-06-05	&56	&X-5		\\
	  &		&\xmm{}				&0654910701	&2010-11-08	&56	&X-6		\\
	  &		&\xmm{}				&0654910801	&2010-11-18	&56	&X-7		\\
	  &		&\xmm{}				&0654910901	&2010-11-28	&56	&X-8		\\
	  &		&\xmm{}				&0654911001	&2010-12-08	&61	&X-9		\\
	&		&\xmm{}				&0701780101	&2013-05-11	&13	&X-10		\\

	&		&\suzaku{}			&700005030	&2005-12-13	&47	&S-1		\\
	&		&\suzaku{}			&700005010	&2005-11-06	&38	&S-2		\\
	&		&\suzaku{}			&700005020	&2005-11-19	&37	&S-3	\\  \hline

12.	&NGC~4258	&\chandra{} ACIS-S		&350		&2000-04-17	&14		&CA-1	\\
	  &		&\chandra{} ACIS-S		&1618		&2001-05-28	&21		&CA-2	\\
	  &		&\chandra{} ACIS-S		&2340		&2001-05-29	&8		&CA-3	\\
	&		&\chandra{} HETG		&7879		&2007-10-08	&153		&CH-1		\\
	  &		&\chandra{} HETG		&7880		&2007-10-12	&60		&CH-2	\\
	&		&\chandra{} HETG		&9750		&2007-10-14	&107		&CH-3			\\

	  &		&\xmm{}				&0110920101	&2000-12-08	&23		&X-1		\\
	  &		&\xmm{}				&0059140101	&2001-05-06	&13		&X-2	\\
	  &		&\xmm{}				&0059140201	&2001-06-17	&13		&X-3	\\
	  &		&\xmm{}				&0059140401	&2001-12-17	&15		&X-4	\\
	  &		&\xmm{}				&0059140901	&2002-05-22	&17		&X-5	\\
	  &		&\xmm{}				&0203270201	&2004-06-01	&49		&X-6	\\
	  &		&\xmm{}				&0400560301	&2006-11-17	&65		&X-7	\\
	  
	  &		&\suzaku{}			&701095010	&2006-06-10	&100		&S-1	\\  
	  &		&\suzaku{}			&705051010	&2010-11-11	&104		&S-2	\\\hline

13.	&NGC~4507	&\chandra{} HETG	&2150		&2001-03-15	&140	&CH-1	\\
	  &		&\chandra{} ACIS-S	&12292		&2010-12-02	&44	&CA-1		\\
	
	  &		&\xmm{}			&0006220201	&2001-01-04	&46	&X-1	\\
	  &		&\xmm{}			&0653870201	&2010-06-24	&20	&X-2	\\
	  &		&\xmm{}			&0653870301	&2010-07-03	&17	&X-3	\\
	  &		&\xmm{}			&0653870401	&2010-07-13	&17	&X-4	\\
	  &		&\xmm{}			&0653870501	&2010-07-23	&17	&X-5	\\
	  &		&\xmm{}			&0653870601	&2010-08-03	&22	&X-6	\\

	  &		&\suzaku{}		&702048010	&2007-12-20	&104	&S-1\\\hline

\end{tabular} 


\end{table*}


\begin{table*}

\centering
\setcounter{table}{1}
\caption{List of X-ray observations of the sources in the sample.}
	\begin{tabular}{lccccccccccccccc} \hline\hline 
	
 Number&Source		& Telescope	&Observation	&Observation 	&Exposure	&Short obs-id	& 	\\ 
	&		&		&ID		&Date		&($\ks$)	&		&		\\	\hline\\

14.	&NGC~5252	&\chandra{} ACIS-S	&4054		&2003-08-11	&63	&CA-1		\\
	  &		&\chandra{} ACIS-S	&15618		&2013-03-04	&42	&CA-2	\\
	  &		&\chandra{} ACIS-S	&15022		&2013-03-07	&71	&CA-3	\\
	  &		&\chandra{} ACIS-S	&15621		&2013-05-09	&65	&CA-4	\\
		
	  &		&\xmm{}			&0152940101	&2003-07-18	&67	&X-1	\\
	
	  &		&\suzaku{}		&707028010	&2012-12-26	&50	&S-1	\\ \hline

15.	&NGC~5506	&\chandra{} HETG		&1598		&2000-12-31	&90		&CH-1		\\

	  &		&\xmm{}				&0013140101	&2001-02-02	&20		&X-1		\\
	  &		&\xmm{}				&0013140201	&2002-01-09	&14		&X-2		\\
	  &		&\xmm{}				&0201830201	&2004-07-11	&22		&X-3		\\
	  &		&\xmm{}				&0201830301	&2004-07-14	&20		&X-4		\\
	  &		&\xmm{}				&0201830401	&2004-07-22	&22		&X-5		\\
	  &		&\xmm{}				&0201830501	&2004-08-07	&20		&X-6		\\
	  &		&\xmm{}				&0554170201	&2008-07-27	&91		&X-7		\\
	  &		&\xmm{}				&0554170101	&2009-01-02	&89		&X-8		\\
	  &		&\xmm{}				&0761220101	&2015-07-07	&132		&X-9		\\

	  &		&\suzaku{}			&701030020	&2016-05-27	&53		&S-1	\\
	  &		&\suzaku{}			&701030010	&2016-05-27	&48		&S-2	\\ 
	  &		&\suzaku{}			&701030030	&2016-05-28	&57		&S-3		\\  \hline

16.	&NGC~6251	&\chandra{} ACIS-I			&847		&2000-09-11	&37	&CA-1		\\ 
	  &		&\chandra{} ACIS-S			&4130		&2003-11-11	&49	&CA-2	\\

	  &		&\xmm{}					&0056340201	&2002-03-26	&50	&X-1		\\
	
	&		&\suzaku{}				&705039010	&2010-12-02	&87	&S-1		\\   
	 &		&\suzaku{}				&806015010	&2011-11-20	&100	&S-2	\\\hline

17.	&NGC~6300	&\chandra{} ACIS-S	&10289		&2009-06-03	&10	&CA-1		\\
	  &		&\chandra{} ACIS-S	&10290		&2009-06-07	&10	&CA-2	\\
	  &		&\chandra{} ACIS-S 	&10291		&2009-06-09	&10	&CA-3	\\
	  &		&\chandra{} ACIS-S	&10292		&2009-06-10	&10	&CA-4		\\
	  &		&\chandra{} ACIS-S	&10293		&2009-06-14	&10	&CA-5	\\

	  &		&\xmm{}			&0059770101	&2001-03-02	&47	&X-1	\\

	  &		&\suzaku{}		&702049010	&2007-10-17	&83	&S-1	\\ \hline

18.	&NGC~7172	&\chandra{} ACIS-I	&905		&2000-07-02	&50	&CA-1	\\
		
	&		&\xmm{}			&0147920601	&2002-11-18	&17	&X-1\\
	&		&\xmm{}			&0202860101	&2004-11-11	&59	&X-2\\
	&		&\xmm{}			&0414580101	&2007-04-24	&58	&X-3	\\
	&		&\suzaku{}		&703030010	&2008-05-25	&82	&S-1	\\ \hline

19.	&NGC~7314	&\chandra{} HETG			&3016		&2002-07-19	&29	&CH-1	\\
	&		&\chandra{} HETG			&3719		&2002-07-20	&68	&CH-2\\
	&		&\chandra{} ACIS-S			&6976		&2006-09-10	&25	&CA-1\\
	&		&\chandra{} ACIS-S			&7404		&2006-09-15	&15	&CA-2\\

	&		&\xmm{}					&0111790101	&2001-05-02	&45	&X-1\\
	&		&\xmm{}					&0311190101	&2006-05-03	&84	&X-2\\
	&		&\xmm{}					&0725200101	&2013-05-17	&140	&X-3\\
	&		&\xmm{}					&0725200301	&2013-11-28	&132	&X-4\\
	&		&\xmm{}					&0790650101	&2016-05-14	&65	&X-5\\
	
	&		&\suzaku{} 				&702015010	&2007-04-25 	&109	&S-1\\
	&		&\suzaku{}				&806013010	&2011-11-13	&101	&S-2\\ \hline

20.	&NGC~7582	&\chandra{} ACIS-S	&436		&2000-10-14	&14	&CA-1	\\
	&		&\chandra{} ACIS-S	&2319		&2000-10-15	&6	&CA-2	\\

	&		&\xmm{}			&0112310201	&2001-05-25	&23	&X-1	\\
	&		&\xmm{}			&0204610101	&2005-04-29	&102	&X-2	\\
	&		&\xmm{}			&0405380701	&2007-04-30	&45	&X-3	\\
	&		&\xmm{}			&0782720301	&2016-04-28	&101	&X-4		\\

	&		&\suzaku{}		&702052010	&2007-05-01	&24	&S-1		\\
	&		&\suzaku{}		&702052020	&2007-05-28	&29	&S-2		\\
	&		&\suzaku{}		&702052040	&2007-11-16	&32	&S-3		\\
	&		&\suzaku{}		&702052030	&2007-11-09	&29	&S-4		\\\hline

\end{tabular} 


\end{table*}


\clearpage
\begin{table}
{\scriptsize
\centering
	\caption{The best fit parameters obtained from X-ray spectral fits. \label{Table:Xray}}
  \begin{tabular}{lllllllllllllll} \hline\hline 

Source	&obsid(year)$^{\rm A}$	&$\nh$			&$\nh^{\rm pc}$			&$f_{\rm pc}$	&$\Gamma_{\rm Hard}$	&PL-norm$^{\rm B}$	&$\chi^2/{\chi^2_{\nu}}$	&$2-10\kev$ flux$^{\rm C}$	 \\
	  &			&($10^{22}\cmsqi$)	&($10^{22}\cmsqi$)		&		&		&	&			& $10^{-11}\funit$	\\ \hline \\

1.Cen~A	&X-1(01)		&$11.9\pm0.4$		&$--$				&$--$		&$2.17\pm0.07$		&$0.167\pm0.021$	&$1612/1.14$			&$33.10$	\\	
	&X-2(02)		&$12.9\pm0.6$		&$--$				&$--$		&$2.18\pm0.09$		&$0.177\pm0.032$&$865/0.95$			&$32.31$	\\	
	  &X-3(13)		&$10.9\pm0.3$		&$--$				&$--$		&$1.96\pm0.05$		&$0.292\pm0.031$&$1402/1.05$			&$83.12$	\\  	
	  &X-4(13)		&$11.1\pm0.3$		&$--$				&$--$		&$2.04\pm0.05$		&$0.309\pm0.033$&$1319/0.998$			&$70.71$	\\	
        &X-5(14)		&$12.0\pm0.3$		&$--$				&$--$		&$1.98\pm0.05$		&$0.125\pm0.330$&$1418/1.003$			&$55.31$	\\	
	 &X-6(14)		&$10.8\pm0.8$		&$19_{-4}^{+7}$			&$0.51\pm0.08$	&$2.44\pm0.09$		&$0.424\pm0.156$&$1514/1.023$			&$37.11$	\\	
        &S-1(05) 		&$10.76\pm0.15$		&$26\pm 8$			&$0.157\pm0.022$&$1.83\pm0.02$		&$0.132\pm0.007$&$10388/1.05$			&$63.11$	\\  	
      	&S-2(09) 		&$10.86\pm0.07$		&$--$				&$--$		&$1.82\pm0.01$		&$0.169\pm0.003$&$8226/1.09$			&$58.81$	\\

      	&S-3(09) 		&$10.93\pm0.07$		&$--$				&$--$		&$1.80\pm0.01$		&$0.175\pm0.004$&$7851/1.04$			&$70.77$	\\
	&S-4(09)		&$11.03\pm0.08$		&$--$				&$--$		&$1.79\pm0.01$		&$0.149\pm0.004$&$7775/1.06$			&$63.12$	\\ 
      	&S-5(13) 		&$9.98\pm0.14$		&$--$				&$--$		&$1.79\pm0.02$		&$0.143\pm0.006$&$5305/0.99$			&$51.22$	\\  	
      	&S-6(14)		&$10.59\pm0.27$		&$--$				&$--$		&$1.79\pm0.04$		&$0.095\pm0.008$&$3256/0.98$			&$72.42$	\\ \hline\\


	  2.{Cyg~A}$^{\rm D}$&X-1(05) 	&$21.35\pm1.12$		&$--$				&$--$		&$1.7^*$	&$0.0069\pm0.0003$	&$297/1.18$			&$3.01$	\\
      			     &S-1(08) 	&$22.52\pm2.75$		&$--$				&$--$		&$1.92\pm0.02$	&$0.0106\pm 0.0001$	&$5936/1.05$			&$5.91$	\\\hline\\


3.{Fairall 49}	&CH-1(02)	&$0.81\pm0.13$		&$2.59\pm0.68$			&$0.68\pm 0.08$		&$2.46\pm0.02$	&$0.016\pm0.002$		&$388/0.97$			&$2.23$		\\	
		&CH-2(02)	&$0.55\pm0.40$		&$<50$				&$<0.95$		&$1.96\pm 0.12$	&$0.012$		&$417/0.98$			&$2.45$		\\	
		&X-1(01)	&$0.067\pm0.017$	&$3.98\pm0.52$			&$0.92\pm0.01$		&$2.60\pm 0.16$	&$0.012\pm0.004$	&$226/1.45$			&$1.07$		\\	
		&X-2(13)	&$0.75\pm0.05$		&$0.87\pm 0.08$			&$0.62\pm 0.07$		&$2.14\pm0.02$	&$0.0159\pm0.0005$	&$381/1.57$			&$3.46$		\\	
		&X-3(13)	&$0.50\pm0.06$		&$0.71\pm0.07$			&$0.77\pm 0.05$		&$2.07\pm 0.01$	&$0.0099\pm0.0003$	&$462/1.90$			&$2.45$		\\	
		&S-1(07)	&$1.09\pm0.04$		&$4.04\pm 0.89$			&$0.29\pm 0.08$		&$2.42\pm 0.01$		&$0.021\pm 0.00012$		&$6046/1.06$			&$2.75$		\\\hline\\


4.IRAS~F00521$^{\rm E}$	&X-1(06)	&$6.42\pm1.65$		&$12.90\pm 4.01$	&$0.72\pm 0.16$			&$2.5^*$	&$0.0051\pm 0.0006$		&$137/1.31$			&$0.50$			\\	
			&X-2(06)	&$7.21\pm1.10$		&$18.95^*$		&$0.53^*$			&$2.5^*$	&$0.0041\pm 0.0008$		&$154/1.31$			&$0.35$		\\	
			&S-1(13)	&$7.26\pm 1.56$		&$42.90\pm 3.09$ &$0.51\pm 0.13$		&$2.44\pm 0.03$		&$0.0041\pm 0.0001$		&$1533/1.06$			&$0.87$		\\\hline\\


5.IRAS~F05189$^{\rm E}$ 	&X-1(01)	&$6.29\pm0.73$		&$--$			&$--$			&$2.20\pm 0.15$	&$0.0023\pm 0.0011$	&$105/0.82$			&$0.67$	\\	
				&X-2(13)	&$5.50\pm0.54$		&$12.40\pm4.35$		&$0.61\pm0.11$		&$2.51\pm 0.20$	&$0.0057\pm 0.0004$	&$274/1.26$			&$0.76$	\\	
				&CA-1(01)	&$6.08\pm1.20$		&$--$			&$--$			&$1.82\pm 0.24$		&$0.0012\pm 0.0001$		&$127/1.04$			&$0.67$	\\	
				&CA-2(02)	&$5.20\pm0.91$		&$--$			&$--$			&$1.43\pm0.22$		&$0.00068\pm 0.00033$		&$106/1.01$			&$0.71$	\\\hline\\


6.{MCG--5--23--16}	&CH-1(00)	&$1.84\pm0.20$		&$--$			&$--$			&$1.66\pm 0.05$	&$0.0129\pm0.0003 $		&$471/1.05$			&$2.51$	\\	
			&CH-2(05)	&$1.44\pm0.25$		&$--$			&$--$			&$1.56\pm 0.03$	&$0.0093\pm0.0004 $		&$329/0.98$			&$5.37$	\\	
			&CH-3(05)	&$1.29\pm0.25$		&$--$			&$--$			&$1.58\pm 0.03$	&$0.011 \pm 0.0004$		&$232/0.97$			&$4.26$	\\	
		
			&X-1(01)	&$1.02\pm0.02$		&$--$			&$--$			&$1.47\pm 0.01$	&$0.012\pm  0.0001$		&$307/1.15$			&$7.76$	\\	
			&X-2(05)	&$1.24\pm0.01$		&$--$			&$--$			&$1.61\pm 0.01$	&$0.0188\pm0.0001 $		&$1148/4.3$			&$9.33$	\\	
			&X-3(13)	&$1.26\pm0.01$		&$--$			&$--$			&$1.78\pm 0.02$	&$0.0366\pm 0.0004$		&$1356/5.10$			&$13.4$	\\	
			&X-4(13)	&$1.25\pm0.01$		&$--$			&$--$			&$1.77\pm 0.01$	&$0.0359\pm 0.0003$		&$1383/5.21$			&$13.4$	\\

			&S-1(05)	&$1.44\pm0.01$		&$--$			&$--$			&$1.85\pm 0.01$	&$0.0271\pm 0.0001 $		&$7764/1.08$			&$9.33$	\\	
			&S-2(13)	&$1.34\pm0.02 $		&$--$			&$--$			&$1.88\pm 0.01$	&$0.0340\pm0.0001 $		&$8995/1.17$			&$10.91$	\\	
			&S-3(13)	&$1.36\pm0.03 $		&$--$			&$--$			&$1.90\pm 0.01$	&$0.0321\pm 0.0001$		&$8574/1.13$			&$12.22$	\\\hline\\


7.{Mkn~348}	&CA-1(10)	&$8.49\pm0.25 $		&$--$		&$--$		&$1.8^* $	&$0.0095\pm 0.0005 $		&$528/1.01$			&$1.99$		\\	
       		
       		&X-1(02)	&$6.62\pm 2.01$		&$8.93\pm1.10$	&$0.89\pm0.05$	&$1.58\pm 0.01$&$0.0107\pm 0.0001$		&$320/1.35$			&$5.12$		\\
       		
		&X-2(13)	&$10.02\pm 1.53 $	&$15^*$		&$0.54\pm0.20$	&$1.80\pm0.02 $	&$0.0041\pm 0.0008$		&$176/1.10$			&$1.51$		\\	
       		
       		&S-1(08)	&$5.42\pm 0.70$		&$6.54\pm2.11 $	&$0.72\pm0.11$	&$1.80\pm 0.01$	&$0.0166\pm 0.0001$		&$6007/1.02$			&$6.45$		\\

\hline

\end{tabular}
	
	{Quantities with a single $^*$ symbol are kept fixed during fitting, mostly due to limited spectral band-pass and/or due to low signal to noise ratio. The errors on parameters quoted in the table are at $90\%$ confidence level. }\\
	{$--$ denotes that these components were not required in the fit.}\\
	{$^{\rm A}$ X=\xmm{},  S=\suzaku{}, CH= \chandra{} HETG, and CA= \chandra{} ACIS. }\\
	{$^{\rm B}$ the value of the power-law component at 1 keV in units of $\rm ph\, keV^{-1}\, cm^{-2}\, s^{-1}$}  \\
	{$^{\rm C}$ The $2-10\kev$ unabsorbed flux.}\\
	{ $^{\rm D}$ Due to the large number of observations, the table entries for CygA's \chandra{} observations have been moved to Table \ref{Table:cygA}.}\\
	{$^{\rm E}$ IRAS F00521= IRAS~F00521--7054, IRAS~F05189=IRAS~F05189--2524  }\\

}
\end{table}



\clearpage
\begin{table}
{\scriptsize
\centering
\setcounter{table}{2}
	\caption{The best fit parameters obtained from X-ray spectral fits. }
	
  \begin{tabular}{lllllllllllllll} \hline\hline 

Source	&obsid(year)$^{\rm A}$	&$\nh$			&$\nh^{\rm pc}$			&$f_{\rm pc}$	&$\Gamma_{\rm Hard}$	&PL-norm$^{\rm B}$	&$\chi^2/{\chi^2_{\nu}}$	&$2-10\kev$ flux$^{\rm C}$	 \\
	  &			&($10^{22}\cmsqi$)	&($10^{22}\cmsqi$)		&		&		&	&			& $10^{-11}\funit$	\\ \hline \\


8. NGC~526A		&X-1(02)	&$1.00\pm 0.05$		&$--$	&$--$	&$1.36\pm 0.02$	&$0.0026\pm0.0001 $		&$221/0.98$			&$1.86$	\\
	  
			&X-2(03)	&$1.03\pm 0.02$		&$--$	&$--$	&$1.43\pm 0.03$	&$0.0039\pm0.0002 $		&$338/1.29$			&$2.51$\\
			&X-3(13)	&$1.05\pm 0.02$		&$--$	&$--$	&$1.42\pm 0.02$	&$0.0043\pm0.0007 $		&$359/1.37$			&$2.81$\\
			&X-4(13)	&$1.06\pm 0.02$		&$--$	&$--$	&$1.45\pm 0.04$	&$0.005\pm 0.0003$		&$377/1.45$			&$3.38$\\
			&CH-1(03)	&$1.03\pm 0.31$		&$--$	&$--$	&$1.22\pm 0.12$	&$0.0046\pm0.0022 $		&$63/0.98$			&$9.92$\\
			&CH-2(03)	&$1.18\pm 0.09$		&$--$	&$--$	&$1.57\pm0.02$	&$0.207\pm 0.0011$		&$303/1.03$			&$1.72$\\
			&CA-1(00)	&$1.58\pm 0.20$		&$--$	&$--$	&$1.7^* $	&$0.007\pm 0.002 $		&$240/1.07$			&$3.64$\\
			&CA-2(00)	&$1.50\pm 0.22$		&$--$	&$--$	&$1.7^* $	&$0.0081\pm 0.0022$		&$292/1.30$			&$17.7$\\ 
			&S-1(11)	&$1.24\pm 0.09$		&$--$	&$--$	&$1.70\pm 0.02$	&$0.0112\pm 0.0001$		&$6906/1.05$			&$4.72$\\ \hline\\


	  9.NGC~1052	&X-1(01)	&$4.84\pm 1.28$		&$33.10\pm8.20 $	&$0.87\pm 0.04$	&$2.05\pm0.01 $	&$0.0047\pm0.0002 $		&$502/0.99$			&$0.91$	\\ 			
			&X-2(06)	&$4.65\pm 0.80$		&$16.08\pm 5.21 $	&$0.69\pm 0.09$	&$1.54\pm0.16 $	&$0.00147\pm 0.0006$	&$1341/0.94$			&$0.74$\\ 
			
			&X-3(09)	&$4.22\pm 0.80$		&$11.30\pm 3.20 $	&$0.75\pm 0.07$	&$1.62\pm 0.14$ &$0.00179\pm 0.0007$	&$1343/0.92$			&$0.83$\\

			&X-4(09)	&$3.76\pm 1.72$		&$8.81\pm 3.31 $	&$0.84\pm 0.08$	&$1.57\pm 0.12$	&$0.00163\pm 0.0005 $	&$1475/0.96$			&$0.83$\\

			&CA-1(05)	&$1.06\pm 0.32$		&$11.65\pm 3.82 $	&$0.82\pm 0.05$	&$1.37\pm 0.10 $&$0.0012\pm 0.0005 $	&$336/1.08$			&$0.93$\\
			
			&S-1(07)	&$3.57\pm 0.28$		&$15.11\pm 4.44 $		&$0.76\pm 0.21$	&$1.62\pm 0.01$	&$0.0018\pm 0.0001$		&$2660/1.02$			&$0.95$\\ \hline  \\


10.NGC~2110	&X-1(03)	&$2.21\pm 0.11$		&$5.98\pm1.26$		&$0.47\pm0.03$	&$1.82\pm0.02$	&$0.0092\pm0.0001$	&$477/1.01$			&$3.46$	\\
		&CA-1(01)	&$1.96\pm 0.30$		&$2.96\pm 1.01$		&$0.67\pm 0.12 $&$1.35\pm 0.10$&$0.0068\pm 0.0015 $	&$635/1.12$			&$5.63$	\\
		&CH-1(01)	&$2.30\pm 0.40$		&$--$			&$--$		&$1.39\pm 0.05$	&$0.0048\pm0.0011 $		&$252/1.03$			&$4.66$	\\
		&CH-2(01)	&$2.85\pm 0.33$		&$--$			&$--$		&$1.74^*$	&$0.0078\pm 0.0008$		&$297/1.26$			&$18.6$	\\
		&CH-3(01)	&$2.16\pm 0.26$		&$--$			&$--$		&$1.31\pm 0.06$	&$0.0054\pm 0.0007 $		&$442/0.98$			&$9.45$	\\
		&CH-4(03)	&$2.46\pm 0.25$		&$--$			&$--$		&$1.34\pm 0.06$	&$0.0047\pm 0.0005$		&$543/1.09$			&$7.25$	\\

		&S-1(05)	&$2.37\pm 0.09$		&$3.22\pm 0.88 $	&$0.65\pm 0.11$	&$1.74\pm 0.01$	&$0.036\pm0.0001 $		&$8007/1.05$			&$14.1$	\\		
		&S-2(12)	&$2.61\pm 0.12$		&$4.03\pm 0.97$		&$0.69\pm0.08 $	&$1.79\pm 0.02$	&$0.047\pm 0.0001$		&$7965/1.05$			&$16.8$	\\
		&S-3(15)	&$2.54\pm 0.21$		&$4.91\pm 1.12$		&$0.75\pm 0.15$	&$1.89\pm 0.03$	&$0.028\pm 0.0001$		&$6815/1.09$			&$9.01$	\\  \hline \\


	  11.{NGC~2992}	&CH-1(10)	&$0.71\pm 0.18$		&$--$	&$--$	&$1.71\pm0.15 $	&$0.0004\pm0.0001 $	&$300/1.38$			&$0.275$			\\
			&X-1(03)	&$0.61\pm 0.01$		&$--$	&$--$	&$1.76\pm0.01 $	&$0.0271\pm0.0005 $	&$1055/1.23$			&$10.4$		\\
			&X-2(10)	&$0.82\pm 0.03$		&$--$	&$--$	&$1.64\pm0.03 $	&$0.0011\pm0.0001 $	&$316/1.29$			&$0.62$		\\
			&X-3(10)	&$0.79\pm 0.03$		&$--$	&$--$	&$1.61\pm0.03 $	&$0.0013\pm0.0001 $	&$302/1.23$			&$0.72$	\\
			&X-4(10)	&$0.75\pm 0.02$		&$--$	&$--$	&$1.56\pm0.01 $	&$0.0025\pm0.0001 $	&$336/1.32$			&$1.41$	\\
			&X-5(10)	&$0.77\pm 0.05$		&$--$	&$--$	&$1.64\pm0.04 $	&$0.0008\pm0.0001$	&$289/1.22$			&$0.47$	\\
			&X-6(10)	&$0.73\pm 0.04$		&$--$	&$--$	&$1.63\pm0.04 $	&$0.0008\pm0.0001$	&$245/1.03$			&$0.48$	\\
			&X-7(10)	&$0.71\pm 0.05$		&$--$	&$--$	&$1.63\pm0.04 $	&$0.00059\pm0.00004$	&$315/1.35$			&$0.38$	\\
			&X-8(10)	&$0.73\pm 0.05$		&$--$	&$--$	&$1.77\pm0.05 $	&$0.00029\pm0.00005 $	&$323/1.42$			&$0.17$	\\
			&X-9(10)	&$0.83\pm 0.04$		&$--$	&$--$	&$1.66\pm0.04 $	&$0.001\pm0.0001 $	&$282/1.17$			&$0.58$	\\
			&X-10(13)	&$0.71\pm 0.03$		&$--$	&$--$	&$1.55\pm0.03 $	&$0.003\pm0.0002$	&$194/0.85$			&$1.73$	\\
			&S-1(05)	&$0.98\pm 0.06$		&$--$	&$--$	&$1.74\pm 0.02$	&$0.0021\pm 0.0001$		&$2987/1.08$			&$1.12$	\\
			&S-2(05)	&$1.06\pm 0.06$		&$--$	&$--$	&$1.79\pm 0.01$	&$0.0025\pm 0.0001$		&$3770/1.15$			&$0.95$	\\
			&S-3(05)	&$0.89\pm 0.07$		&$--$	&$--$	&$1.67\pm 0.01$	&$0.0021\pm 0.0001$		&$4066/1.05$			&$1.09$  \\ \hline  \\


12.{NGC~4258}	&CH-1(07)	&$7.56\pm 0.52$		&$--$	&$--$	&$1.90^* $	&$0.0044\pm0.0003 $	&$146/1.05$			&$1.17$	\\
		&CH-2(07)	&$6.64\pm 0.45$		&$--$	&$--$	&$1.90^* $	&$0.0031\pm0.0002 $		&$217/1.28$			&$1.90$\\

		&CH-3(07)	&$6.86\pm 0.43$		&$--$	&$--$	&$1.90^* $	&$0.0022\pm 0.0010$		&$110/1.22$			&$1.94$\\

		&CA-1(00)	&$8.14\pm 0.47$		&$--$	&$--$	&$1.90^* $	&$0.0026\pm 0.0005$		&$365/1.05$			&$1.65$\\	

		&CA-2(00)	&$9.37\pm 0.66$		&$--$	&$--$	&$1.90^* $	&$0.0033\pm 0.0001$		&$239/1.11$			&$1.31$\\	
		&CA-3(01)	&$8.85\pm 0.35$		&$--$	&$--$	&$1.90^* $	&$0.0033\pm 0.0004$		&$316/0.99$			&$<1.01$\\

		&X-1(00) 	&$9.40\pm 0.80$		&$--$	&$--$	&$1.78\pm0.16 $	&$0.0033\pm0.0011 $	&$184/1.00$			&$1.34$\\
		&X-2(01) 	&$8.00\pm 0.79$		&$--$	&$--$	&$1.67\pm0.16 $	&$0.0023\pm0.0011$	&$189/1.03$			&$1.23$\\
		&X-3(01) 	&$7.95\pm 2.15$		&$--$	&$--$	&$1.52\pm0.35 $	&$0.0015\pm0.0012 $	&$85/0.90$			&$1.02$\\
		&X-4(01) 	&$12.68\pm 1.62$	&$--$	&$--$	&$1.46\pm0.20 $	&$0.0013\pm0.0007$	&$157/0.95$			&$0.85$\\
		&X-5(02) 	&$7.47\pm 0.52$		&$--$	&$--$	&$1.59\pm0.11 $	&$0.0027\pm0.0005 $	&$287/1.32$			&$0.81$\\
		
		&X-6(06) 	&$7.49\pm 0.51$		&$--$	&$--$	&$1.64\pm0.08 $	&$0.0012\pm0.0002$	&$390/1.65$			&$0.60$\\		
		
		&S-1(06) 	&$10.35\pm 0.30$	&$--$	&$--$	&$1.90\pm0.01 $	&$0.0043\pm 0.0001 $		&$4298/1.04$			&$1.28$\\

		&S-2(10) 	&$12.19\pm 1.20$	&$--$	&$--$	&$1.74\pm 0.02$	&$0.0019\pm 0.0001$		&$3376/1.07$			&$0.76$\\		\hline \\


13.{NGC~4507}		&X-1(01)	&$16.81\pm 9.21$	&$31.46\pm3.32 $	&$>0.92 $	&$1.7^* $	&$0.0118\pm 0.0006$		&$404/1.67$			&$5.88$	\\
			&X-2(10)	&$12.67\pm 2.21$	&$60.05\pm5.52 $	&$>0.94 $	&$1.7^*$ 	&$0.0101\pm 0.0011$		&$294/1.48$			&$6.60$\\
			&X-3(10)	&$11.56\pm 1.62$	&$62.21\pm5.62 $	&$>0.94$	&$1.7^*$	 &$0.0103\pm0.0007 $		&$339/1.73$			&$5.49$\\
			&X-4(10)	&$12.65\pm 1.55$	&$59.48\pm6.62$		&$>0.94$	&$1.7^*$	 &$0.0109\pm0.0008 $	&$255/1.30$			&$8.12$\\
			&X-5(10)	&$11.03\pm 3.32$	&$58.55\pm7.21 $	&$>0.94$	&$1.7^*$	&$0.0096\pm0.0012 $	&$216/1.16$			&$6.02$\\
			&X-6(10)	&$9.94\pm 1.21$		&$48.33\pm 7.52$	&$>0.94$	&$1.7^*$ 	&$0.0067\pm 0.0007$		&$241/1.27$			&$8.91$\\

			&CH-1(01)	&$46.45\pm 11.22$	&$187.24\pm 50.22$	&$>0.94$	&$1.7^* $	&$0.021\pm0.001 $		&$250/1.42$			&$3.39$\\		
			
			&CA-1(10)	&$12.59\pm6.56 $	&$47.42\pm30.33 $	&$>0.94$	&$1.7^* $	&$0.019\pm 0.003$		&$406/1.23$			&$9.57$\\ 
			&S-1(07)	&$12.35\pm2.23$		&$61.39\pm 21.29$	&$0.91\pm0.02 $	&$1.90\pm0.02 $	&$0.008\pm 0.001$		&$3660/1.34$			&$4.84$\\	\hline \\

\end{tabular}
	

}
\end{table}

\clearpage
\begin{table}
{\scriptsize
\centering
\setcounter{table}{2}
	\caption{The best fit parameters obtained from X-ray spectral fits. }
  \begin{tabular}{lllllllllllllll} \hline\hline 

Source	&obsid(year)$^{\rm A}$	&$\nh$			&$\nh^{\rm pc}$			&$f_{\rm pc}$	&$\Gamma_{\rm Hard}$	&PL-norm$^{\rm B}$	&$\chi^2/{\chi^2_{\nu}}$	&$2-10\kev$ flux$^{\rm C}$	 \\
	  &			&($10^{22}\cmsqi$)	&($10^{22}\cmsqi$)		&		&		&	&			& $10^{-11}\funit$	\\ \hline \\

14.{NGC~5252}	&CA-1(03)	&$2.84\pm 0.07$		&$--$		&$--$		&$1.4^* $	&$0.0025\pm0.0003 $		&$506/1.01$			&$7.02$			\\

		&CA-2(13)	&$4.51\pm 0.11$		&$--$		&$--$		&$\# $		&$0.0028\pm 0.0002$		&$593/1.12$			&$1.31$		\\

		&CA-3(13)	&$4.58\pm 0.16$		&$--$		&$--$		&$\#$		&$0.00234\pm0.0003 $		&$435/0.97$			&$1.73$		\\
		
		&CA-4(13)	&$3.51\pm 0.10$		&$--$		&$--$		&$\# $		&$0.00247\pm0.0006 $		&$505/0.99$			&$2.78$		\\
		&X-1(03)	&$2.26\pm 0.17$		&$8.51\pm2.01$	&$0.47\pm0.05$	&$1.54\pm0.08 $	&$0.0026\pm 0.0008 $		&$249/1.03$			&$1.38$		\\
		&S-1(12)	&$2.28\pm 0.31$		&$4.47\pm0.23$	&$0.79\pm 0.12$		&$1.59\pm 0.01$	&$0.0027\pm0.0003 $		&$2970/0.98$			&$1.60$		\\  \hline\\


15. {NGC~5506}	&X-1(01)	&$2.77\pm 0.05$		&$--$		&$--$		&$1.73\pm0.02 $	&$0.0173\pm0.0007 $		&$302/1.17$			&$7.24$				\\			
		&X-2(02)	&$2.80\pm 0.05$		&$--$		&$--$		&$1.68\pm0.02 $	&$0.0284\pm0.0011 $		&$268/1.04$			&$12.8$				\\			
		&X-3(04)	&$3.02\pm 0.05$		&$--$		&$--$		&$1.81\pm0.02 $	&$0.0233\pm0.0011 $		&$339/1.31$			&$8.70$				\\			
		&X-4(04)	&$3.03\pm 0.04$		&$--$		&$--$		&$1.79\pm0.03 $	&$0.0217\pm 0.0013 $		&$381/1.47$			&$8.31$				\\			
		&X-5(04)	&$2.97\pm 0.05$		&$--$		&$--$		&$1.84\pm0.02 $	&$0.0212\pm 0.0010 $		&$286/1.11$			&$7.58$				\\			
		&X-6(04)	&$2.92\pm 0.07$		&$--$		&$--$		&$1.82\pm 0.04$	&$0.0346\pm 0.0012 $		&$297/1.15$			&$19.99$				\\				
		
		&X-7(08)	&$2.93\pm 0.05$		&$--$		&$--$		&$1.81\pm0.05 $	&$0.0331\pm 0.0010 $		&$433/1.66$			&$12.30$				\\			

		&X-8(09)	&$2.95\pm 0.04$		&$--$		&$--$		&$1.78\pm0.05 $	&$0.0362\pm 0.0015$		&$519/1.98$			&$13.80$				\\			
		&X-9(15)	&$3.00\pm 0.01$		&$--$		&$--$		&$1.72\pm0.06 $	&$0.0201\pm 0.0021 $		&$602/1.64$			&$7.45$				\\			
		&CH-1(00)	&$2.91\pm 0.08$		&$--$		&$--$		&$1.65\pm 0.02$	&$0.0100 \pm 0.0011 $		&$334/0.96$			&$1.69$				\\			


		&S-1(16)	&$3.12\pm 0.10$		&$--$		&$--$		&$1.95\pm0.01 $	&$0.0391\pm 0.0007$		&$7533/1.09$			&$12.30$				\\			
		&S-2(16)	&$3.15\pm 0.10$		&$--$		&$--$		&$1.94\pm 0.01$	&$0.0409\pm 0.0008$		&$7426/1.05$			&$11.40$				\\			
		&S-3(16)	&$3.16\pm 0.09$		&$--$		&$--$		&$1.96\pm 0.01$	&$0.0407\pm 0.0012$		&$7658/1.09$			&$11.70$		\\\hline \\


	 16.{NGC~6251}	&CA-1(00) &$<0.476$		&$--$		&$--$		&$1.41\pm 0.20$	&$0.0005\pm 0.0002 $		&$66/0.99$			&$0.35$		\\		
		&CA-2(03)	&$0.058\pm 0.021$	&$--$		&$--$		&$1.58\pm 0.08$	&$0.0004\pm 0.0001 $		&$260/0.97$			&$0.25$		\\			
		&X-1(02)	&$0.045\pm 0.010$	&$--$		&$--$		&$1.93\pm 0.03$	&$0.00128\pm 0.00005 $		&$260/1.24$			&$0.37$		\\			
		&S-1(10)	&$0.75\pm0.08 $		&$--$		&$--$		&$1.87\pm0.01 $	&$0.00012\pm0.0002 $			&$1504/1.07$			&$0.19$		\\\hline \\


17.{NGC~6300}	&CA-1(09)	&$19.83\pm 1.19$	&$--$		&$--$		&$1.76^* $	&$0.022\pm 0.008$		&$199/1.32$			&$7.84$		\\			
		&CA-2(09)	&$19.29\pm 0.88$	&$--$		&$--$		&$1.76^* $	&$0.0118\pm 0.0051 $		&$160/0.99$			&$4.98$		\\			
		&CA-3(09)	&$21.08\pm 1.07$	&$--$		&$--$		&$1.76^* $	&$0.028\pm 0.004$		&$224/1.42$			&$9.77$		\\			
		&CA-4(09)	&$19.28\pm 0.82$	&$--$		&$--$		&$1.76^* $	&$0.038\pm 0.011$		&$167/0.93$			&$15.80$		\\			
		&CA-5(09)	&$21.25\pm 0.85$	&$--$		&$--$		&$1.76^* $	&$0.036\pm 0.002$		&$190/1.10$			&$23.00$		\\			
		&X-1(01)	&$19.15\pm 2.51$	&$--$		&$--$		&$1.45\pm 0.12$	&$0.00017\pm 0.00011$		&$53/0.84$			&$0.114$		\\			
		&S-1(07)	&$21.76\pm 1.11$	&$--$		&$--$		&$1.76\pm 0.01$	&$0.0096\pm 0.00011$		&$4721/1.09$			&$3.62$		\\\hline\\


18.{NGC~7172}	&CA-1(00)	&$9.95\pm 0.37$		&$--$		&$--$		&$1.89^* $		&$0.0092\pm0.0003 $		&$807/1.83$			&$5.91$			\\			
	
		&X-1(02)	&$7.56\pm 0.61$		&$6.81\pm4.92$	&$0.55\pm0.35$	&$1.66\pm 0.05 $	&$0.0080\pm 0.0004$		&$221/1.02$			&$4.16$		\\			
		&X-2(04)	&$8.14\pm 0.20$		&$--$		&$--$		&$1.54\pm 0.04 $	&$0.0059\pm0.0005 $	&$360/1.43$			&$4.36$		\\			
		&X-3(07)	&$7.60\pm 0.20$		&$--$		&$--$		&$1.57\pm0.03 $		&$0.0133\pm 0.0010$	&$348/1.39$			&$7.58$		\\			
		&S-1(08)	&$9.57\pm 0.32$		&$--$		&$--$		&$1.89\pm0.02 $		&$0.0254\pm 0.0004$		&$6515/1.05$			&$7.65$		\\ \hline\\


19. {NGC~7314}	&CH-1(02)            	&$0.74\pm 0.11$		&$--$		&$--$		&$1.65\pm 0.10$	&$0.0077\pm0.0012  $		&$323/1.02$			&$3.54$		   \\			
			&CH-2(02)  	&$0.80\pm 0.12$		&$--$		&$--$		&$1.75\pm 0.06$	&$0.0099\pm 0.0011 $		&$543/1.08$			&$2.45$			\\		
			&CA-1(06) 	&$0.82\pm 0.10$		&$--$		&$--$		&$1.66\pm 0.06$	&$0.0197\pm 0.0008 $		&$512/1.19$			&$8.51$			\\		
			&CA-2(06)	&$0.85\pm 0.08$		&$--$		&$--$		&$1.79\pm 0.12$	&$0.0198\pm 0.0016 $		&$437/1.16$			&$8.31$			\\

			&X-1(01)	&$0.73\pm 0.02$		&$--$		&$--$		&$1.93\pm 0.01$	&$0.0141\pm 0.0001 $		&$517/2.02$			&$4.36$			\\		
			&X-2(06)	&$0.73\pm 0.03$		&$--$		&$--$		&$1.77\pm 0.02$	&$0.0043\pm  0.0002$		&$355/1.40$			&$1.65$			\\		
			&X-3(13)	&$0.80\pm 0.01$		&$--$		&$--$		&$1.83\pm 0.03$	&$0.0073\pm 0.0002 $		&$488/1.92$			&$2.57$			\\		
			&X-4(13)	&$0.77\pm 0.01$		&$--$		&$--$		&$1.78\pm 0.01$	&$0.0057\pm  0.0001$		&$392/1.54$			&$2.13$			\\		
			&X-5(16)	&$0.75\pm 0.01$		&$--$		&$--$		&$1.89\pm 0.02$	&$0.0127\pm  0.0001$		&$430/1.70$			&$3.98$			\\		
		
			&S-1(07)	&$0.85\pm 0.05$		&$--$		&$--$		&$1.79\pm 0.01$	&$0.0022\pm  0.0002$		&$4662/1.00$			&$8.91$			\\ 
			&S-2(11)	&$0.86\pm0.07 $		&$--$		&$--$		&$2.04\pm 0.01$	&$0.0065\pm0.0003  $		&$6143/1.09$			&$1.53$			\\ 
			
			\hline\\


20.{NGC~7582}	&CA-1(00)	&$17.95 \pm 2.11 $		&$--$		&$--$		&$1.84\pm0.12 $	&$0.0087\pm 0.0012 $		&$160/1.10$			&$1.38$		     \\
		&CA-2(00)	&$19.90\pm 3.85$		&$--$		&$--$		&$1.45^*$	&$0.0047\pm 0.0011 $		&$65/1.16$			&$1.41$			\\

		&X-1(01)	&$15.50\pm 1.71 $	&$58.36\pm$	&$0.85\pm$	&$1.79\pm 0.01$	&$0.0053\pm  0.0008$		&$241/1.34$			&$0.41$			\\
		&X-2(05)	&$18.21\pm 4.71 $	&$78.43\pm$	&$0.89\pm$	&$2.01\pm 0.03$	&$0.0051\pm 0.0007 $		&$360/1.65$			&$0.23$			\\
		&X-3(07)	&$28.79\pm 10.11$	&$--$		&$--$		&$1.38\pm 0.01$	&$0.0022\pm  0.0002$		&$69/1.12$			&$0.71$		     \\
		&X-4(16)	&$29.75\pm 0.56$	&$--$		&$--$		&$1.41\pm 0.04$	&$0.0036\pm  0.0003$		&$518/2.14$			&$0.93$			\\
		&S-1(07)	&$31.71\pm 3.52$	&$--$		&$--$		&$1.45\pm 0.05$	&$0.0017\pm  0.0002$		&$596/1.10$			&$0.93$			\\
		&S-2(07)	&$41.07\pm 5.52$	&$--$		&$--$		&$1.33\pm 0.04$	&$0.0014\pm  0.0001$		&$645/1.31$			&$0.93$			\\
		&S-3(07)	&$36.66_{-8.38}^{+6.02}$&$--$		&$--$		&$1.33\pm 0.01$	&$0.00058\pm  0.0001$		&$466/1.15$			&$255.1$			\\
		&S-4(07)	&$43.32_{-2.16}^{+4.44}$&$--$		&$--$		&$1.51\pm 0.01$	&$0.00078\pm  0.0001$		&$478/1.15$			&$47.1$			\\ \hline \\


\end{tabular}

}
\end{table}

\clearpage
\begin{table}
{\scriptsize
\centering
	\caption{ The best fit soft and hard X-ray broad band continuum parameters obtained from X-ray spectral fits.} \label{Table:Xray2}
  \begin{tabular}{lllllllllllllll} \hline\hline 

	  Source	&obsid(year)$^{\rm A}$	&APEC1 $k_{\rm B}T$	&APEC1 Norm	&APEC2 $k_{\rm B}T$	&APEC2 Norm	&SXPL $\Gamma$	&SXPL norm	&Pexmon(norm)	&Pexmon $\Gamma$	&Pexmon $R$	&FeK norm	 \\
			&			&$\kev$			&($10^{-4}$)	&$\kev$			&($10^{-4}$)	&		&($10^{-4}$) &&&&($10^{-4}$)	\\ \hline \\

1.Cen~A	&X-1(01) 	&$0.84^*$&$4.01^*$	&$--$	&$--$		&$0.73^*$	&$13.86^*$	&$--$		&$--$	 &$--$		&$3.62\pm0.47$	\\	
      	&X-2(02)	&$\#$	&$\#$		&$--$	&$--$		&$\#$		&$\#$		&$--$		&$--$	 &$--$		&$4.74\pm0.71$			\\	
      	&X-3(13)	&$\#$	&$\#$		&$--$	&$--$		&$\#$		&$\#$		&$--$		&$--$	 &$--$		&$3.79\pm1.06$			\\  	
        &X-4(13)	&$\#$	&$\#$		&$--$	&$--$		&$\#$		&$\#$		&$--$		&$--$	 &$--$		&$5.86\pm1.07$			\\	
        &X-5(14)	&$\#$	&$\#$		&$--$	&$--$		&$\#$		&$\#$		&$--$		&$--$	 &$--$		&$3.63\pm0.67$			\\	
      	&X-6(14)	&$\#$	&$\#$		&$--$	&$--$		&$\#$		&$\#$		&$--$		&$--$	 &$--$		&$4.25\pm0.53$			\\	
        &S-1(05) 	&$0.23^*$&$4.05\pm0.33$	&$0.79^*$&$2.67\pm0.14$	&$1.17^*$	&$7.85\pm0.15$	&$--$		&$--$	 &$--$		&$3.30\pm0.13$			\\  	
      	&S-2(09) 	&$\#$	&$5.79\pm0.70$	&$\#$	&$3.39\pm0.22$	&$\#$		&$9.16\pm0.21$	&$--$		&$--$	 &$--$		&$3.91\pm0.17$			\\  	
        &S-3(09) 	&$\#$	&$6.04\pm0.19$	&$\#$	&$3.44\pm0.52$	&$\#$		&$10.93\pm 0.24$&$--$		&$--$	 &$--$		&$4.16\pm0.20$			\\
	  
	&S-4(09)	&$\#$	&$7.09\pm0.83$	&$\#$	&$3.51\pm0.26$	&$\#$		&$9.92\pm0.24$	&$--$		&$--$	 &$--$		&$4.40\pm0.19$			\\

	  &S-5(13) 	&$\#$	&$4.94\pm1.20$	&$\#$	&$2.92\pm0.43$	&$\#$		&$8.55\pm0.43$	&$--$		&$--$	 &$--$		&$2.94\pm0.34$			\\  	
      	&S-6(14)	&$\#$	&$6.28\pm1.82$	&$\#$	&$3.54\pm0.60$	&$\#$		&$11.25\pm0.58$	&$--$		&$--$	 &$--$		&$7.87\pm0.85$			\\  \hline\\


	  2.{Cyg~A}$^{\rm E}$ &S-1(08)    &$0.23^*$	&$6.01^*$	&$4.17^*$	&$130.11^*$	&$1.89^*$	&$49.00^*$	&$0.0099^{\rm T}$	&$1.89^{\rm T}$	 &$0.91\pm$		&$0.46\pm0.11$ \\	
	  
	  	&X-1(05) &$0.29^*$	&$13.11^*$	&$2.9^*$	&$76.01^*$	&$1.77^*$		&$22.00^*$	&$\#$		&$\#$	 &$\#$		&$0.57\pm0.11$	\\ \hline\\


3.{Fairall 49}	&S-1(07)	&$0.008^*$&$0.55^*$	&$--$	&$--$		&$2.3^*$	&$1.01^*$	&$0.021^{\rm T}$&$2.42^{\rm T}$	 &$0.41\pm0.11$	&$0.26\pm0.11$	\\
	   	&CH-1(02)	&$0.3^*$&$0.041^*$	&$--$	&$--$		&$2.1^*$	&$1.01^*$	&$\#$		&$\#$	 &$\#$		&$--$	\\	
		&CH-2(02)	&$\#$	&$\#$		&$--$	&$--$		&$\#$		&$\#$		&$\#$		&$\#$	 &$\#$		&$--$	\\	
		&X-1(01)	&$0.3^*$&$0.10^*$		&$--$	&$--$	&$2.3^*$	&$1.01^*$	&$\#$		&$\#$	 &$\#$		&$--$	\\	
		&X-2(13)	&$\#$	&$\#$		&$--$	&$--$		&$\#$		&$\#$		&$\#$		&$\#$	 &$\#$		&$0.42\pm0.11$	\\	
		&X-3(13)	&$\#$	&$\#$		&$--$	&$--$		&$\#$		&$\#$		&$\#$		&$\#$	 &$\#$		&$0.33\pm0.11$	\\	\hline\\


4.IRAS~F00521$^{\rm F}$	&S-1(13)	&$0.25^*$	&$0.11^*$	&$--$	&$--$		&$2.44^*$		&$0.15^*$		&$0.0041^{\rm T}$	&$2.48^{\rm T}$	 &$0.65\pm0.21$	&$0.028\pm0.008$\\		
	       		&X-1(06)	&$0.25^*$	&$0.11^*$	&$--$	&$--$		&$2.51^*$		&$0.10^*$		&$\#$		&$\#$	 &$\#$		&$0.057\pm0.010$\\	
			&X-2(06)	&$\#$	&$\#$		&$--$	&$--$		&$\#$		&$\#$		&$\#$		&$\#$	 &$\#$		&$0.068\pm0.010$\\	 \hline\\


5.IRAS~F05189$^{\rm F}$ 	&X-1(01)	&$--$	&$--$		&$--$	&$--$		&$3.31^*$	&$0.39^*$	&$0.0051^*$	&$2.47^*$ &$1.48^*$	&$0.051\pm0.011$	\\	
				&X-2(13)	&$--$	&$--$		&$--$	&$--$		&$\#$		&$\#$		&$\#$		&$\#$	 &$\#$		&$0.082\pm0.012$\\	
				&CA-1(01)	&$--$	&$--$		&$--$	&$--$		&$2.91^*$		&$0.42$		&$\#$		&$\#$	 &$\#$		&$0.049\pm0.025$\\	
				&CA-2(02)	&$--$	&$--$		&$--$	&$--$		&$\#$		&$\#$		&$\#$		&$\#$	 &$\#$		&$0.065\pm0.042$\\ \hline\\


6.{MCG--5--23--16}	&CH-1(00)	&$0.01^*$		&$2.01^*$		&$--$		&$--$		&$1.66^*$	&$2.11^*$	&$0.027^*$	&$1.84^*$&$0.35^*$	&$0.81\pm0.41$	\\	
			&CH-2(05)	&$\#$			&$\#$			&$--$		&$--$		&$\#$		&$\#$		&$\#$		&$\#$	 &$\#$		&$0.98\pm0.41$	\\	
			&CH-3(05)	&$\#$			&$\#$			&$--$		&$--$		&$\#$		&$\#$		&$\#$		&$\#$	 &$\#$		&$0.86\pm0.52$	\\	
		
			&X-1(01)	&$0.05\pm0.01$		&$180\pm23$		&$--$		&$--$		&$1.7^*$		&$3.1\pm0.12$		&$\#$		&$\#$	 &$\#$		&$0.53\pm0.081$	\\	
			&X-2(05)	&$0.07\pm0.01$		&$20\pm9$		&$--$		&$--$		&$1.7^*$		&$2.1\pm0.22$		&$\#$		&$\#$	 &$\#$		&$0.72\pm0.21$	\\	
			&X-3(13)	&$0.05\pm0.02$		&$160 \pm29$		&$--$		&$--$		&$2.1^*$		&$2.9^*$		&$\#$		&$\#$	 &$\#$		&$1.08\pm0.11$	\\	
			&X-4(13)	&$0.06\pm0.01$		&$130\pm12$		&$--$		&$--$		&$\#$		&$\#$		&$\#$		&$\#$	 &$\#$		&$1.10\pm0.12$	\\

			&S-1(05)	&$0.14^*$		&$9\pm4$		&$--$		&$--$		&$1.85^*$	&$0.76\pm0.17$	&$0.027^{\rm T}$	&$1.84^{\rm T}$	&$0.35\pm0.12$	&$0.98\pm0.21$	\\	
			&S-2(13)	&$\#$			&$17\pm6$	&$--$			&$--$		&$\#$		&$0.40\pm0.08$	&$0.034^{\rm T}$	&$1.88^{\rm T}$	&$0.24\pm0.08$	&$0.95\pm0.22$	\\	
			&S-3(13)	&$\#$			&$15\pm6$	&$--$			&$--$		&$\#$		&$0.66\pm0.18$	&$0.032^{\rm T}$	&$1.89^{\rm T}$	&$0.40\pm0.05$	&$0.92\pm0.23$	\\ \hline\\


7.{Mkn~348} &S-1(08)	&$0.008^*$	&$2.0^*$	&$0.92^*$	&$0.11^*$	&$1.81^*$		&$0.25$		&$0.0166^{\rm T}$	&$1.8^{\rm T}$	 &$0.45\pm0.08$	&$0.34\pm0.04$				\\

	&CA-1(10)	&$0.18^*$	&$0.62^*$	&$0.82^*$	&$0.11^*$	&$1.8^*$		&$0.65^*$		&$\#$		&$\#$	 &$\#$		&$0.16\pm0.12$			\\	
       		
       	&X-1(02)	&$0.18^*$	&$0.43^*$	&$0.95^*$	&$0.11^*$	&$1.42^*$		&$0.33^*$		&$\#$		&$\#$	 &$\#$		&$0.21\pm0.08$				\\
       		
       	&X-2(13)	&$\#$		&$\#$		&$\#$	&$\#$		&$\#$		&$\#$		&$\#$			&$\#$	 	&$\#$		&$0.24\pm0.07$				\\	
       		
       	\hline\\

\end{tabular}
	
	{The quantities in this table marked with $^*$ are kept fixed at an average value during fitting, hence no errors are quoted. When the quantities are left free, as required by the data, the errors on the free parameters are quoted. Refer to Section \ref{Sec:spec} for details of the fitting. }\\
	{$\#$ marked quantities are the same as the value just above it, and denotes that we have fixed it to an average value over all the observations for a particular instrument.}
	{$--$ denotes that these were not required in the fit.}
	{$^{\rm A}$ X=\xmm{},  S=\suzaku{}, CH= \chandra{} HETG, and CA= \chandra{} ACIS. }\\
	{$^{\rm B}$ $\nh$ and $\nh^{\rm pc}$ are in units of $10^{22}\cmsqi$.} \\
	{ $^{\rm E}$ Cyg~A Chandra observations and the corresponding best fit parameters are listed in Table \ref{Table:cygA}.}\\
	{$^{\rm F}$ IRAS F00521= IRAS~F00521--7054, IRAS~F05189=IRAS~F05189--2524  }\\
	{$^{\rm T}$  The pexmon normalization and the powerlaw slope ($\Gamma$) are tied to the hard X-ray powerlaw parameters for the respective \suzaku{} observation.}

}
\end{table}


\clearpage
\begin{table}
{\scriptsize
\centering
\setcounter{table}{3}
	\caption{The best fit soft and hard X-ray broad band continuum parameters obtained from X-ray spectral fits.}
  \begin{tabular}{lllllllllllllll} \hline\hline 

	  Source	&obsid(year)$^{\rm A}$	&APEC1 $k_{\rm B}T$	&APEC1 Norm	&APEC2 $k_{\rm B}T$	&APEC2 Norm	&SXPL $\Gamma$	&SXPL norm	&Pexmon(norm)	&Pexmon $\Gamma$	&Pexmon $R$	&FeK norm	 \\
			&			&$\kev$			&($10^{-4}$)	&$\kev$			&($10^{-4}$)	&		&($10^{-4}$)	\\ \hline \\


 8. NGC~526A		&X-1(02)	&$0.90^*$	&$0.20^*$	&$0.19^*$	&$0.46\pm0.11$	&$--$		&$--$		&$--$	&$--$	 &$--$		&$0.19\pm0.12$	\\
	  
			&X-2(03)	&$\#$	&$\#$		&$\#$	&$0.42\pm0.12$	&$--$		&$--$		&$--$	&$--$	 &$--$		&$0.17\pm0.05$\\
			&X-3(13)	&$\#$	&$\#$		&$\#$	&$0.42\pm0.12$	&$--$		&$--$		&$--$	&$--$	 &$--$		&$0.23\pm0.03$\\
			&X-4(13)	&$\#$	&$\#$		&$\#$	&$0.53\pm0.16$	&$--$		&$--$		&$--$	&$--$	 &$--$		&$0.25\pm0.05$\\
			&CH-1(03)	&$\#$	&$0.39^*$	&$\#$	&$0.74\pm0.19$	&$--$		&$--$		&$--$	&$--$	 &$--$		&$0.51\pm0.01$\\
			&CH-2(03)	&$\#$	&$\#$		&$\#$	&$18.0^*$	&$--$		&$--$		&$--$	&$--$	 &$--$		&$0.073\pm0.002$\\
			&CA-1(00)	&$\#$	&$\#$			&$\#$	&$\#$		&$--$		&$--$		&$--$	&$--$	 &$--$		&$0.22\pm0.05$\\
			&CA-2(00)	&$\#$	&$\#$			&$\#$	&$\#$		&$--$		&$--$		&$--$	&$--$	 &$--$		&$0.12\pm0.02$\\ 
			&S-1(11)	&$0.90^*$	&$0.0016^*$	&$--$	&$--$		&$--$		&$--$		&$--$	&$--$	 &$--$		&$0.23\pm0.01$	\\ \hline\\


9.NGC~1052	&X-1(01)	&$0.76^*$	&$0.20^*$		&$--$	&$--$		&$1.55^*$	&$0.91^*$	&$0.00189^*$	&$1.62^*$&$0.29^*$	&$0.17\pm0.07$	\\ 			
			&X-2(06)	&$\#$	&$\#$		&$--$	&$--$		&$\#$		&$\#$		&$\#$		&$\#$	 &$\#$		&$0.12\pm0.04$\\ 
			
			&X-3(09)	&$\#$	&$\#$		&$--$	&$--$		&$\#$		&$\#$		&$\#$		&$\#$	 &$\#$		&$0.11\pm0.03$\\

			&X-4(09)	&$\#$	&$\#$		&$--$	&$--$		&$\#$		&$\#$		&$\#$		&$\#$	 &$\#$		&$0.12\pm0.02$\\

			&CA-1(05)	&$0.80^*$	&$0.32^*$	&$--$	&$--$		&$\#$		&$\#$		&$\#$		&$\#$	 &$\#$		&$0.14\pm0.06$\\
			
			&S-1(07)	&$0.76^*$	&$0.26^*$	&$--$	&$--$		&$\#$		&$\#$		&$0.00189^{\rm T}$	&$1.62^{\rm T}$	 &$0.29\pm0.09$	&$0.13\pm0.06$  \\ \hline  \\


10.NGC~2110	&X-1(03)	&$0.9^*$	&$0.41^*$		&$--$	&$--$		&$1.73^*$	&$0.51^*$	&$0.0365^*$		&$1.74^*$	 &$0.268^*$&$0.61\pm0.01$	\\
		&CA-1(01)	&$0.90^*$	&$0.41^*$		&$--$	&$--$		&$1.73^*$	&$0.51^*$	&$0.0365^*$	&$1.74^*$	&$0.268^*$	&$0.82\pm0.02$	\\
		&CH-1(01)	&$\#$	&$\#$		&$--$	&$--$		&$\#$		&$\#$		&$\#$		&$\#$	 &$\#$		&$0.67\pm 0.13$	\\
		&CH-2(01)	&$\#$	&$\#$		&$--$	&$--$		&$\#$		&$\#$		&$\#$		&$\#$	 &$\#$		&$0.39\pm0.11$	\\
		&CH-3(01)	&$\#$	&$\#$		&$--$	&$--$		&$\#$		&$\#$		&$\#$		&$\#$	 &$\#$		&$0.81\pm 0.08$	\\
		&CH-4(03)	&$\#$	&$\#$		&$--$	&$--$		&$\#$		&$\#$		&$\#$		&$\#$	 &$\#$		&$0.95\pm0.03$	\\

		&S-1(05)	&$0.90^*$	&$0.27^*$&$--$	&$--$		&$1.75^*$	&$0.52^*$	&$0.0365^{\rm T}$	&$1.74^{\rm T}$&$0.268\pm0.091$	&$0.62\pm 0.08$	\\		
		&S-2(12)	&$\#$	&$\#$		&$--$	&$--$		&$\#$		&$\#$		&$0.047^{\rm T}$	&$1.79^{\rm T}$&$0.295\pm0.110$	&$0.80\pm 0.06$	\\
		&S-3(15)	&$\#$	&$\#$		&$--$	&$--$		&$\#$		&$\#$		&$0.028^{\rm T}$	&$1.89^{\rm T}$&$0.588\pm0.071$	&$0.83\pm 0.12$	\\  \hline \\


	  11.{NGC~2992}	&CH-1(10)	&$--$	&$--$		&$--$	&$--$		&$1.51^*$	&$0.86^*$	&$0.0023^*$	&$1.74^*$&$1.36^*$	&$0.25\pm0.04$		\\
			&X-1(03)	&$0.23\pm0.03$&$1.61\pm0.50$	&$--$	&$--$	&$1.51^*$	&$1.41^*$	&$\#$		&$\#$	 &$\#$		&$0.91\pm0.11$		\\
			&X-2(10)	&$0.78^*$&$0.21^*$	&$--$	&$--$		&$1.51^*$	&$1.41^*$	&$\#$		&$\#$	 &$\#$		&$0.21\pm0.03$		\\
			&X-3(10)	&$\#$	&$\#$		&$--$	&$--$		&$\#$		&$\#$		&$\#$		&$\#$	 &$\#$		&$0.21\pm0.02$	\\
			&X-4(10)	&$\#$	&$\#$		&$--$	&$--$		&$\#$		&$\#$		&$\#$		&$\#$	 &$\#$		&$0.28\pm0.02$	\\
			&X-5(10)	&$\#$	&$\#$		&$--$	&$--$		&$\#$		&$\#$		&$\#$		&$\#$	 &$\#$		&$0.24\pm0.02$	\\
			&X-6(10)	&$\#$	&$\#$		&$--$	&$--$		&$\#$		&$\#$		&$\#$		&$\#$	 &$\#$		&$0.25\pm0.02$	\\
			&X-7(10)	&$\#$	&$\#$		&$--$	&$--$		&$\#$		&$\#$		&$\#$		&$\#$	 &$\#$		&$0.23\pm0.02$	\\
			&X-8(10)	&$\#$	&$\#$		&$--$	&$--$		&$\#$		&$\#$		&$\#$		&$\#$	 &$\#$		&$0.19\pm0.02$	\\
			&X-9(10)	&$\#$	&$\#$		&$--$	&$--$		&$\#$		&$\#$		&$\#$		&$\#$	 &$\#$		&$0.19\pm0.02$	\\
			&X-10(13)	&$\#$	&$\#$		&$--$	&$--$		&$\#$		&$\#$		&$\#$		&$\#$	 &$\#$		&$0.35\pm0.02$	\\
			&S-1(05)	&$0.62^*$&$0.22^*$	&$--$	&$--$		&$1.53^*$	&$2.3^*$	&$0.00211^{\rm T}$	&$1.74^{\rm T}$&$1.36\pm0.23$	&$0.24\pm0.02$	\\
			&S-2(05)	&$\#$	&$\#$		&$--$	&$--$		&$\#$		&$\#$		&$0.0024^{\rm T}$	&$1.79^{\rm T}$&$1.36\pm0.21$	&$0.23\pm0.02$	\\
			&S-3(05)	&$\#$	&$\#$		&$--$	&$--$		&$\#$		&$\#$		&$0.0021^{\rm T}$	&$1.67^{\rm T}$	 &$0.98\pm0.09$	&$0.28\pm0.02$\\ \hline  \\


12.{NGC~4258}	&CH-1(07)	&$0.10^*$&$0.01^*$	&$0.45^*$	&$0.10^*$		&$--$		&$--$		&$0.0043^*$	&$1.90^*$	 &$0.165^*$	&$<0.54$	\\
		&CH-2(07)	&$\#$	&$\#$		&$\#$	&$\#$		&$--$		&$--$		&$\#$		&$\#$	 &$\#$		&$<0.08$\\

		&CH-3(07)	&$\#$	&$\#$		&$\#$	&$\#$		&$--$		&$--$		&$\#$		&$\#$	 &$\#$		&$<0.03$\\

		&CA-1(00)	&$\#$	&$\#$		&$0.54\pm0.08$&$0.28\pm0.05$	&$--$		&$--$		&$\#$		&$\#$	 &$\#$		&$--$\\	

		&CA-2(00)	&$\#$	&$\#$		&$1.01\pm0.02$&$0.20\pm0.06$	&$--$		&$--$		&$\#$		&$\#$	 &$\#$		&$<0.01$\\	
		&CA-3(01)	&$\#$	&$\#$		&$1.01\pm0.03$&$0.21\pm0.05$	&$--$		&$--$		&$\#$		&$\#$	 &$\#$		&$--$\\

		&X-1(00) 	&$0.70^*$&$10.01^*$	&$0.21^*$&$0.51^*$		&$2.1^*$		&$0.71^*$	&$\#$		&$\#$	 &$\#$		&$0.07\pm 0.004$\\
		&X-2(01) 	&$\#$	&$\#$		&$\#$	&$\#$		&$\#$		&$\#$		&$\#$		&$\#$	 &$\#$		&$<0.05$\\
		&X-3(01) 	&$\#$	&$\#$		&$\#$	&$\#$		&$\#$		&$\#$		&$\#$		&$\#$	 &$\#$		&$0.05\pm0.004$\\
		&X-4(01) 	&$\#$	&$\#$		&$\#$	&$\#$		&$\#$		&$\#$		&$\#$		&$\#$	 &$\#$		&$<0.083$\\
		&X-5(02) 	&$\#$	&$\#$		&$\#$	&$\#$		&$\#$		&$\#$		&$\#$		&$\#$	 &$\#$		&$0.042\pm0.003$\\
		
		&X-6(06) 	&$\#$	&$\#$		&$\#$	&$\#$		&$\#$		&$\#$		&$\#$		&$\#$	 &$\#$		&$0.027\pm0.002$\\		
		
		&S-1(06) 	&$0.85^*$	&$1.71^*$		&$0.50^*$	&$2.01^*$		&$1.90^*$		&$1.91^*$		&$0.0043^{\rm T}$	&$1.90^{\rm T}$	 &$0.165\pm0.050$	&$0.73\pm0.03$\\

		&S-2(10) 	&$\#$	&$\#$		&$\#$	&$\#$		&$\#$		&$\#$		&$0.0019^{\rm T}$	&$1.74^{\rm T}$	 &$0.201\pm0.091$	&$0.047\pm0.002$	\\		\hline \\


13.{NGC~4507}		&X-1(01)	&$0.18^*$	&$0.71^*$&$0.78^*$	&$0.35^*$		&$1.70^*$		&$1.01^*$		&$0.0088^*$	&$1.90^*$	 &$2.19^*$	&$0.84\pm0.09$	\\
			&X-2(10)	&$\#$	&$\#$		&$\#$	&$\#$		&$\#$		&$\#$		&$\#$		&$\#$	 &$\#$		&$1.91\pm0.30$\\
			&X-3(10)	&$\#$	&$\#$		&$\#$	&$\#$		&$\#$		&$\#$		&$\#$		&$\#$	 &$\#$		&$1.61\pm0.30$\\
			&X-4(10)	&$\#$	&$\#$		&$\#$	&$\#$		&$\#$		&$\#$		&$\#$		&$\#$	 &$\#$		&$2.01\pm0.60$\\
			&X-5(10)	&$\#$	&$\#$		&$\#$	&$\#$		&$\#$		&$\#$		&$\#$		&$\#$	 &$\#$		&$2.31\pm0.60$\\
			&X-6(10)	&$\#$	&$\#$		&$\#$	&$\#$		&$\#$		&$\#$		&$\#$		&$\#$	 &$\#$		&$2.41\pm1.10$\\

			&CH-1(01)	&$\#$	&$\#$		&$\#$	&$\#$		&$\#$		&$\#$		&$\#$		&$\#$	 &$\#$		&$0.32\pm0.19$\\		
			
			&CA-1(10)	&$\#$	&$\#$		&$\#$	&$\#$		&$\#$		&$\#$		&$\#$		&$\#$	 &$\#$		&$2.91\pm1.12$\\ 
			&S-1(07)	&$0.78^*$	&$0.45^*$		&$--$	&$--$		&$1.90^*$		&$1.51^*$		&$0.0088^{\rm T}$	&$1.90^{\rm T}$	 &$2.18\pm0.167$	&$2.92\pm1.22$\\	\hline \\


\end{tabular}

}
\end{table}


\clearpage
\begin{table}
{\scriptsize
\centering
\setcounter{table}{3}
	\caption{The best fit soft and hard X-ray broad band continuum parameters obtained from X-ray spectral fits. }
  \begin{tabular}{lllllllllllllll} \hline\hline 

	  Source	&obsid(year)$^{\rm A}$	&APEC1 $k_{\rm B}T$	&APEC1 Norm	&APEC2 $k_{\rm B}T$	&APEC2 Norm	&SXPL $\Gamma$	&SXPL norm	&Pexmon(norm)	&Pexmon $\Gamma$	&Pexmon $R$	&FeK norm	 \\
			&			&$\kev$			&($10^{-4}$)	&$\kev$			&($10^{-4}$)	&		&($10^{-4}$)	\\ \hline \\

	  14.{NGC~5252}	&CA-1(03)&$0.15^*$	&$2.67\pm0.87$	&$--$	&$--$		&$1.32^*$	&$0.450\pm0.043$&$--$	&$--$	&$--$		&$0.08\pm0.06$\\

		&CA-2(13)	&$\#$		&$\#$		&$--$	&$--$		&$\#$		&$0.965^*$	&$--$	&$--$	&$--$		&$0.16\pm0.06$		\\

		&CA-3(13)	&$\#$		&$\#$		&$--$	&$--$		&$\#$		&$\#$		&$--$	&$--$	&$--$		&$0.19\pm0.07$		\\
		
		&CA-4(13)	&$\#$		&$\#$		&$--$	&$--$		&$\#$		&$\#$		&$--$	&$--$	&$--$		&$0.15\pm0.06$		\\

		&X-1(03)	&$\#$		&$0.46\pm 0.04$			&$0.90^*$	&$0.085^*$	&$2.48^*$		&$0.35^*$		&$--$	&$--$	&$--$		&$0.09\pm0.01$		\\
		&S-1(12)	&$\#$		&$2.23\pm 0.12$			&$0.82^*$	&$0.21^*$		&$1.53^*$		&$0.46^*$		&$--$	&$--$	&$--$		&$0.13\pm0.03$		\\  \hline\\


15. {NGC~5506}	&X-1(01)	&$0.94^*$		&$0.41^*$&$--$		&$--$		&$1.80^*$	&$4.01^*$	&$0.0391^*$	&$1.95^*$	 &$0.475^*$		&$0.43\pm0.12$	\\			
		&X-2(02)	&$\#$	&$\#$		&$--$	&$--$		&$\#$		&$\#$		&$\#$		&$\#$	 &$\#$			&$0.41\pm0.14$	\\			
		&X-3(04)	&$\#$	&$\#$		&$--$	&$--$		&$\#$		&$\#$		&$\#$		&$\#$	 &$\#$			&$0.48\pm0.15$	\\			
		&X-4(04)	&$\#$	&$\#$		&$--$	&$--$		&$\#$		&$\#$		&$\#$		&$\#$	 &$\#$			&$0.72\pm0.12$	\\			
		&X-5(04)	&$\#$	&$\#$		&$--$	&$--$		&$\#$		&$\#$		&$\#$		&$\#$	 &$\#$			&$0.67\pm0.11$	\\			
		&X-6(04)	&$\#$	&$\#$		&$--$	&$--$		&$\#$		&$\#$		&$\#$		&$\#$	 &$\#$			&$0.57\pm0.11$	\\				
		
		&X-7(08)	&$\#$	&$\#$		&$--$	&$--$		&$\#$		&$\#$		&$\#$		&$\#$	 &$\#$			&$0.56\pm0.12$	\\			

		&X-8(09)	&$\#$	&$\#$		&$--$	&$--$		&$\#$		&$\#$		&$\#$		&$\#$	 &$\#$			&$0.46\pm0.11$	\\			
		&X-9(15)	&$\#$	&$21\pm5^{\rm A}$		&$--$	&$--$ 		&$--$		&$--$		&$\#$	&$\#$	 &$\#$		&$0.80\pm0.13$	\\

		&CH-1(00)	&$0.37^*$	&$17.01^*$	&$--$	&$--$		&$1.64^*$		&$0.49^*$		&$\#$		&$\#$	 &$\#$			&$0.71\pm0.11$	\\			


		&S-1(16)	&$0.68^*$	&$0.77^*$	&$--$	&$--$		&$1.95^*$		&$4.5^*$		&$0.0391^{\rm T}$	&$1.95^{\rm T}$	 &$0.475\pm0.081$	&$0.35\pm0.12$	\\			
		&S-2(16)	&$\#$	&$\#$		&$--$	&$--$		&$\#$		&$\#$		&$0.0409^{\rm T}$	&$1.94^{\rm T}$	 &$0.447\pm0.088$	&$0.68\pm0.12$	\\			
		&S-3(16)	&$\#$	&$\#$		&$--$	&$--$		&$\#$		&$\#$		&$0.0407^{\rm T}$	&$1.95^{\rm T}$	 &$0.406\pm0.091$	&$0.39\pm0.12$	\\\hline \\


16.{NGC~6251}	&CA-1(00) 	&$0.54$		&$2.81$		&$--$	&$--$		&$--$		&$--$		&$0.00012$	&$1.87$	 &$0.59$	&$0.25\pm0.21$	\\		
		&CA-2(03)	&$0.66$		&$\#$		&$--$	&$--$		&$--$		&$--$		&$\#$		&$\#$	 &$\#$		&$0.06\pm0.02$	\\			
		&X-1(02)	&$0.56$		&$0.37$		&$--$	&$--$		&$--$		&$--$		&$\#$		&$\#$	 &$\#$		&$0.07\pm0.03$	\\			
		&S-1(10)	&$0.78$		&$0.44$		&$--$	&$--$		&$1.87$		&$4.61$		&$0.00012^{\rm T}$&$1.87^{\rm T}$	 &$0.59\pm0.22$		&$0.01\pm0.01$	\\\hline \\


17.{NGC~6300}	&CA-1(09)	&$0.69^*$	&$0.05^*$		&$--$	&$--$		&$1.86^*$		&$0.51^*$		&$0.0096^*$	&$1.77^*$	 &$0.417^*$	&$0.95\pm0.55$	\\			
		&CA-2(09)	&$\#$	&$\#$		&$--$	&$--$		&$\#$		&$\#$		&$\#$	&$\#$	 &$\#$			&$0.22\pm0.11$	\\			
		&CA-3(09)	&$\#$	&$\#$		&$--$	&$--$		&$\#$		&$\#$		&$\#$	&$\#$	 &$\#$			&$0.68\pm0.52$	\\			
		&CA-4(09)	&$\#$	&$\#$		&$--$	&$--$		&$\#$		&$\#$		&$\#$	&$\#$	 &$\#$			&$0.80\pm0.60$	\\			
		&CA-5(09)	&$\#$	&$\#$		&$--$	&$--$		&$\#$		&$\#$		&$\#$	&$\#$	 &$\#$			&$0.72\pm0.52$	\\			
		&X-1(01)	&$0.69^*$	&$0.005^*$	&$--$	&$--$		&$1.42^*$		&$0.006^*$	&$\#$	&$\#$	 &$\#$			&$0.026\pm0.011$	\\			
		&S-1(07)	&$0.99^*$	&$0.059^*$	&$--$	&$--$		&$1.77^*$		&$0.81^*$		&$0.0096^{\rm T}$	&$1.77^{\rm T}$	&$0.417\pm0.12$	&$0.34\pm0.18$	\\\hline\\


18.{NGC~7172}	&CA-1(00)	&$0.72^*$	&$0.0016^*$	&$--$	&$--$		&$1.70^*$		&$0.081^*$	&$0.0254^*$	&$1.89^*$	 &$0.33^*$		&$0.41\pm0.14$	\\			
	
		&X-1(02)	&$0.70^*$	&$0.041^*$	&$--$	&$--$		&$1.70^*$		&$0.23^*$	&$\#$		&$\#$	 &$\#$			&$0.31\pm0.09$	\\			
		&X-2(04)	&$\#$	&$\#$		&$--$	&$--$		&$\#$		&$\#$		&$\#$		&$\#$	 &$\#$			&$0.27\pm0.08$	\\			
		&X-3(07)	&$\#$	&$\#$		&$--$	&$--$		&$\#$		&$\#$		&$\#$		&$\#$	 &$\#$			&$0.45\pm0.09$	\\			
		&S-1(08)	&$--$	&$--$		&$--$	&$--$		&$1.89^*$	&$0.17^*$	&$0.0254^{\rm T}$	&$1.89^{\rm T}$	 &$0.33\pm0.08$	&$0.42\pm0.09$	\\ \hline\\


19. {NGC~7314}		&CH-1(02)       &$--$	&$--$	&$--$	&$--$		&$1.80^*$		&$3.01^*$	&$0.0022^*$	&$1.79^*$	 &$0.67^*$	&$0.23\pm0.11$   \\			
			&CH-2(02) 	&$--$	&$--$			&$--$	&$--$		&$\#$		&$\#$		&$\#$		&$\#$	 &$\#$		&$0.11\pm0.06$		\\		
			&CA-1(06)	&$--$	&$--$			&$--$	&$--$		&$\#$		&$\#$		&$\#$		&$\#$	 &$\#$		&$0.94\pm0.11$	\\		
			&CA-2(06)	&$--$	&$--$			&$--$	&$--$		&$\#$		&$\#$		&$\#$		&$\#$	 &$\#$		&$0.20\pm0.10$	\\

			&X-1(01)	&$0.29\pm0.02$	&$0.67\pm0.07$	&$--$	&$--$		&$\#$		&$1.3\pm0.02$	&$\#$		&$\#$	 &$\#$		&$0.17\pm0.04$	\\		
			&X-2(06)	&$0.19\pm0.03$	&$0.26\pm0.11$	&$--$	&$--$		&$\#$		&$0.61\pm0.12$	&$\#$		&$\#$	 &$\#$		&$0.19\pm0.03$	\\		
			&X-3(13)	&$0.34\pm0.04$	&$0.15\pm0.01$	&$--$	&$--$		&$\#$		&$1.45\pm0.06$	&$\#$		&$\#$	 &$\#$		&$0.15\pm0.03$	\\		
			&X-4(13)	&$0.27\pm0.03$	&$0.21\pm0.04$	&$--$	&$--$		&$\#$		&$1.15\pm0.07$	&$\#$		&$\#$	 &$\#$		&$0.15\pm0.03$	\\		
			&X-5(16)	&$0.27\pm0.01$	&$0.77\pm0.08$	&$--$	&$--$		&$\#$		&$0.89\pm0.11$	&$\#$		&$\#$	 &$\#$		&$0.17\pm0.03$	\\		
		
			&S-1(07)	&$0.45^*$	&$0.0029^*$	&$--$	&$--$		&$1.79^*$	&$0.86^*$	&$0.0022^{\rm T}$	&$1.79^{\rm T}$&$0.67\pm0.12$	&$0.09\pm0.05$	\\ 
			&S-2(11)	&$\#$		&$\#$		&$--$	&$--$		&$\#$		&$\#$		&$0.0065^{\rm T}$	&$2.04^{\rm T}$&$0.41\pm0.11$	&$0.08\pm0.04$	\\ 
			
			\hline\\


20.{NGC~7582}	&CA-1(00)	&$0.54\pm$	&$0.22\pm$		&$1.20\pm 0.24$	&$0.39\pm 0.08$		&$1.7^*$	&$0.72^*$		&$0.0017^*$	&$1.43^*$	 &$1.59^*$	&$0.53\pm0.12$		     \\
		&CA-2(00)	&$0.99^*$	&$0.27\pm$		&$<0.56$	&$0.011\pm 0.008$		&$\#$		&$\#$		&$\#$		&$\#$	 &$\#$		&$0.77\pm0.52$			\\

		&X-1(01)	&$0.097^*$&$0.86^*$		&$0.59^*$	&$0.33^*$		&$1.5^*$		&$0.91^*$		&$\#$		&$\#$	 &$\#$		&$0.47\pm0.12$			\\
		&X-2(05)	&$\#$	&$\#$		&$\#$	&$\#$		&$\#$		&$\#$		&$\#$		&$\#$	 &$\#$		&$0.77\pm0.12$			\\
		&X-3(07)	&$\#$	&$\#$		&$\#$	&$\#$		&$\#$		&$\#$		&$\#$		&$\#$	 &$\#$		&$0.31\pm0.08$		     \\
		&X-4(16)	&$\#$	&$\#$		&$\#$	&$\#$		&$\#$		&$\#$		&$\#$		&$\#$	 &$\#$		&$0.82\pm0.09$			\\
		&S-1(07)	&$0.81^*$&$0.62\pm 0.08$	&$--$	&$--$		&$1.43^*$		&$0.89^*$		&$0.0017^{\rm T}$	&$1.43^{\rm T}$	 &$1.59\pm0.12$	&$0.37\pm0.12$			\\
		&S-2(07)	&$\#$	&$0.55\pm 0.04$		&$--$	&$--$		&$\#$		&$\#$		&$0.0014^{\rm T}$	&$1.33^{\rm T}$	 &$0.66\pm0.23$	&$0.35\pm0.11$			\\
		&S-3(07)	&$\#$	&$0.53\pm 0.09$		&$--$	&$--$		&$\#$		&$\#$		&$0.0005^{\rm T}$	&$1.33^{\rm T}$	 &$4.19\pm0.72$	&$0.42\pm0.15$			\\
		&S-4(07)	&$\#$	&$0.51\pm 0.12$		&$--$	&$--$		&$\#$		&$\#$		&$0.00078^{\rm T}$	&$1.51^{\rm T}$	 &$4.20\pm0.63$	&$0.39\pm0.11$			\\ \hline \\

\end{tabular}

{$^{\rm A}$ Abundance of {\it Apec} component is low $0.0162\pm 0.0012$ relative to Solar, and hence the normalization is larger. See Appendix A for details.}

}
\end{table}

\clearpage
\begin{table}
	\centering
	\caption{Summary of measurements of $\Delta \nh$ $^{\rm A}$.}\label{Table:NHsumm}

\begin{tabular}{lccc} \hline\hline \\
	Source & Instrument $^{\rm B}$  &  $\Delta \nh$ $^{\rm C}$ &   $\Delta t ^{\rm D}$   \\
	&             & ($\times 10^{22} \cmsqi)$ & (years)      \\ \hline\\
  Cen~A         & S  &  $ -1.05 \pm 0.29 $ &     4.0   \\
  Fairall~49    & X  &  $ +0.68 \pm 0.04    $ &     12.5  \\
  MCG--5-23-16  & X  &  $ +0.35 \pm 0.01 $ &     4.0   \\
  MCG--5-23-16  & S  &  $ -0.10 \pm 0.02 $ &     7.5   \\
  NGC~2992      & X  &  $ +0.22 \pm 0.01 $ &     10.0     \\
  NGC~5252      & CA &  $ +1.67 \pm 0.07 $  &  9.6      \\
                & CA  &  $ -1.07 \pm 0.10 $  &  0.17     \\
  NGC~5506      & X  &  $ +0.25 \pm 0.07 $ &     3.4      \\
  NGC~7582      & X  &  $ +14.3 \pm 2.7  $ &     14.2      \\  \hline\\ \\

\end{tabular}

	{$^{\rm A}$ See Section \ref{subsec:variablesources} for details.}\\
	{$^{\rm B}$ S, X, and CA denote \textit{Suzaku}, \textit{XMM-Newton} EPIC pn, and
\textit{Chandra} ACIS, respectively.}\\

	{$^{\rm C}$ The positive and negative signs indicate increase and decrease in $\nh$ values respectively.}\\

	{$^{\rm D}$ $\Delta t$ are the event durations. However, due to sparse sampling these timescales can be regarded as lower limits on the actual event durations. }\\

\end{table}


\clearpage
\begin{figure*}
  \centering

	\vbox{
	
		\hbox{
	
	\includegraphics[width=9cm,angle=0]{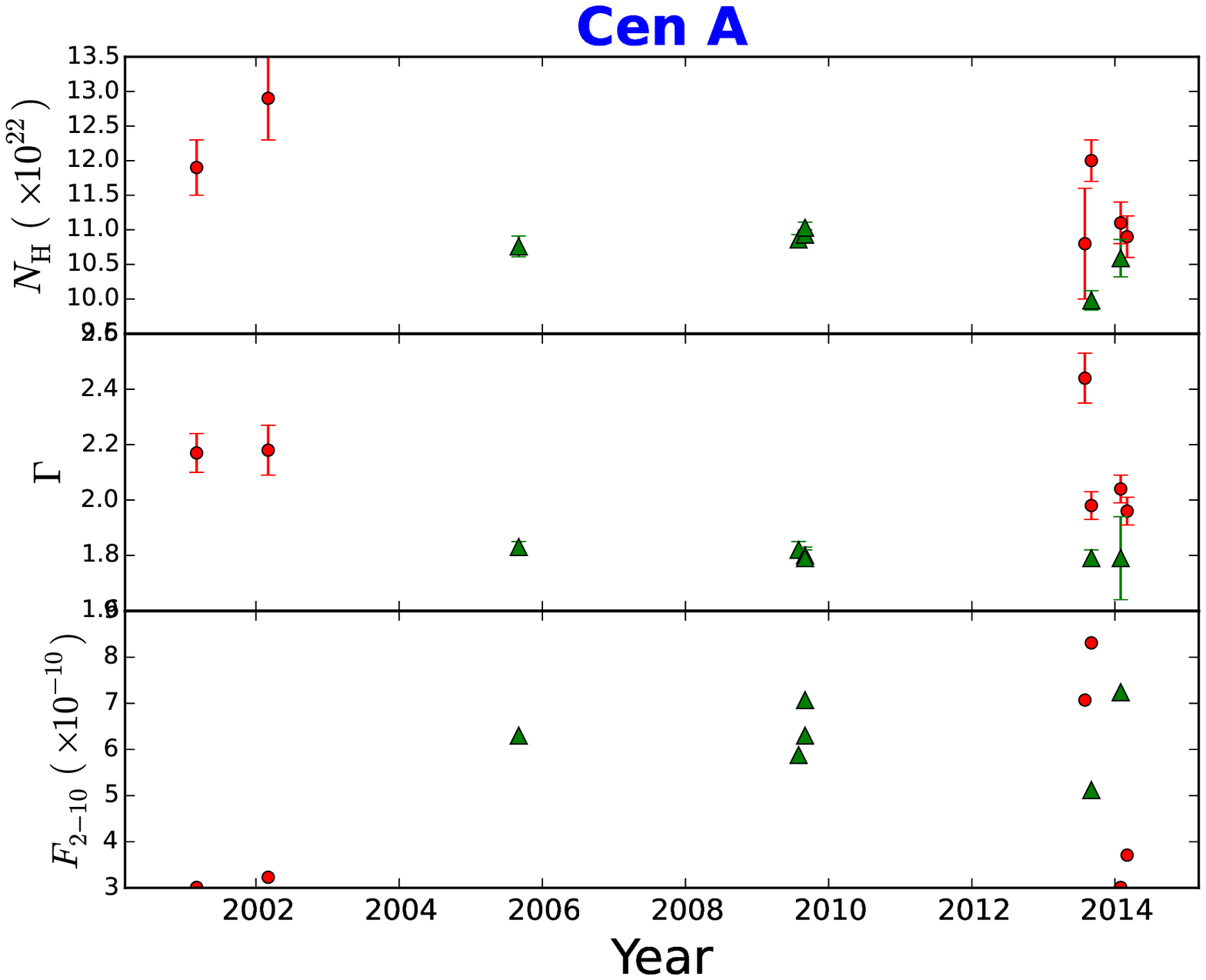} 
	
	\includegraphics[width=9cm,angle=0]{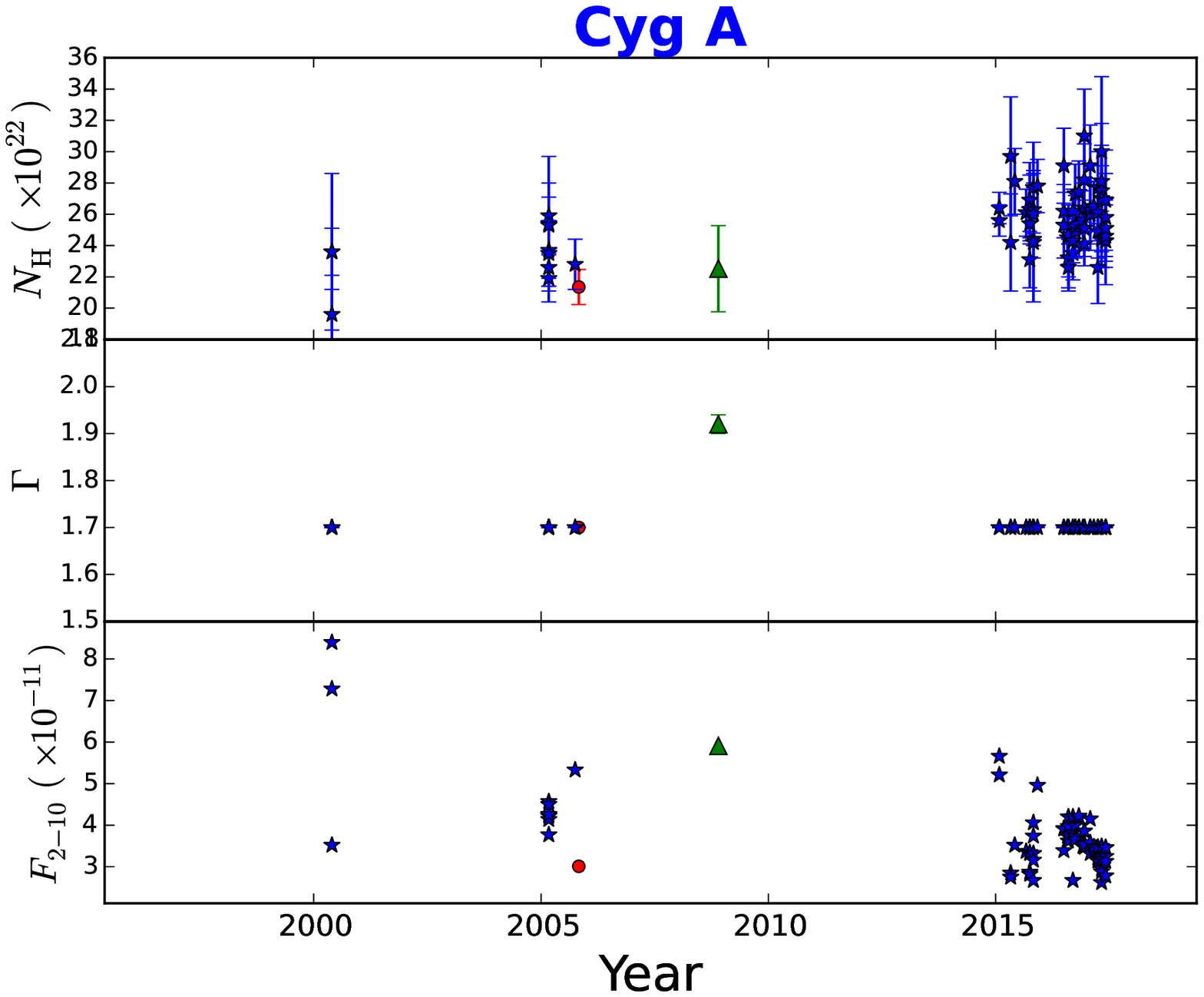} 

		}

	\hbox{
	
	\includegraphics[width=9cm,angle=0]{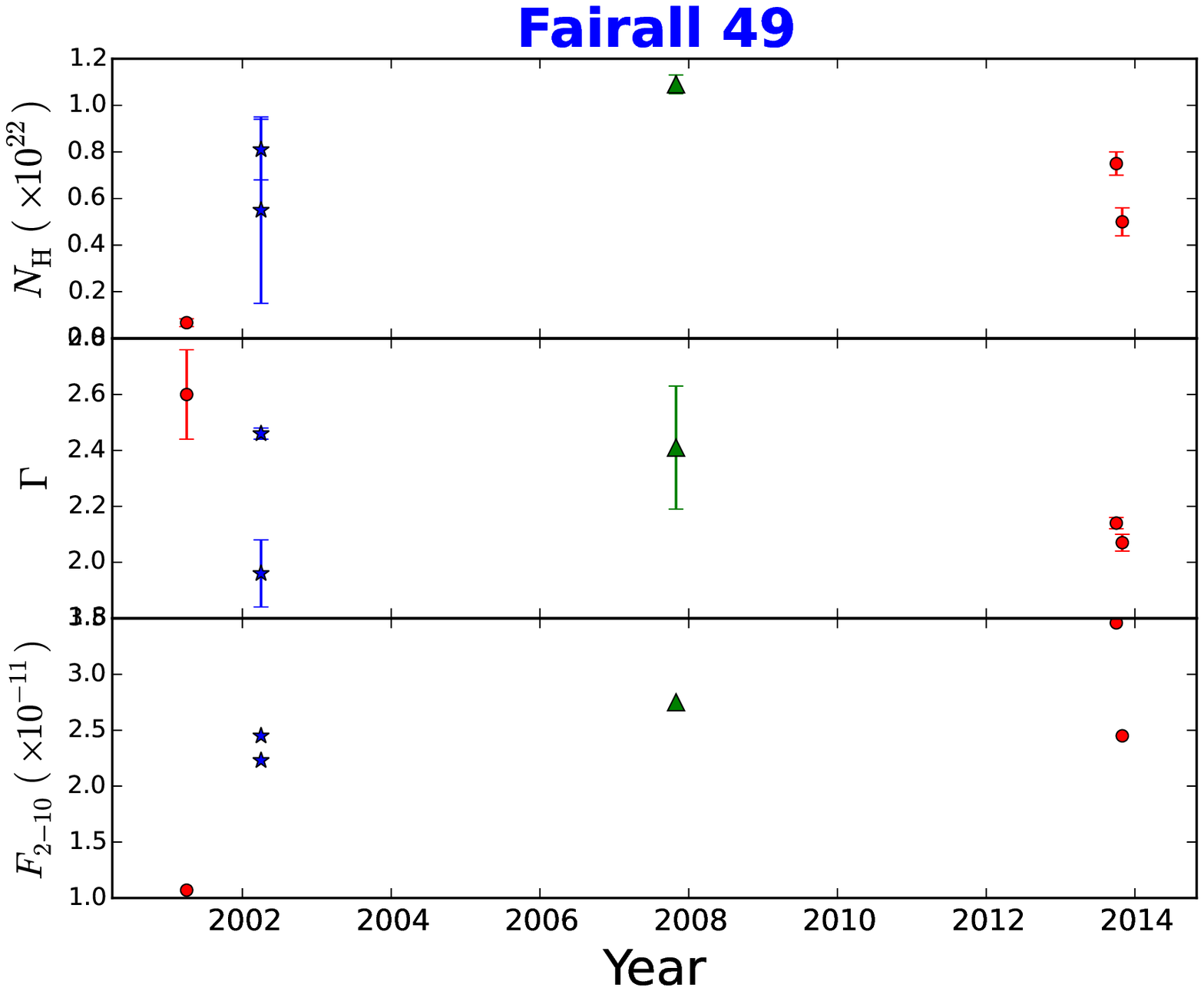}	
	
	\includegraphics[width=9cm,angle=0]{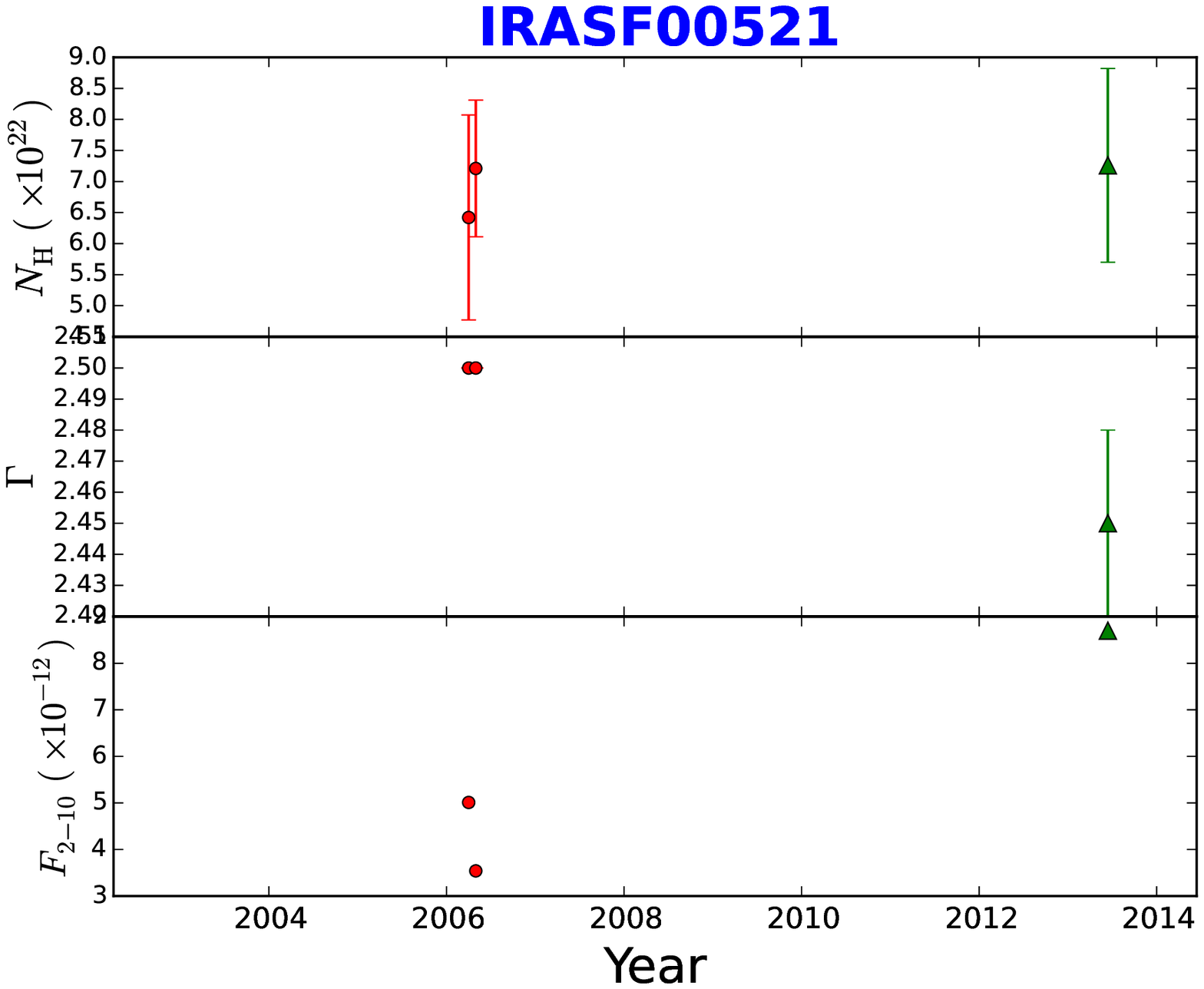}

	}

\hbox{
	
\includegraphics[width=9cm,angle=0]{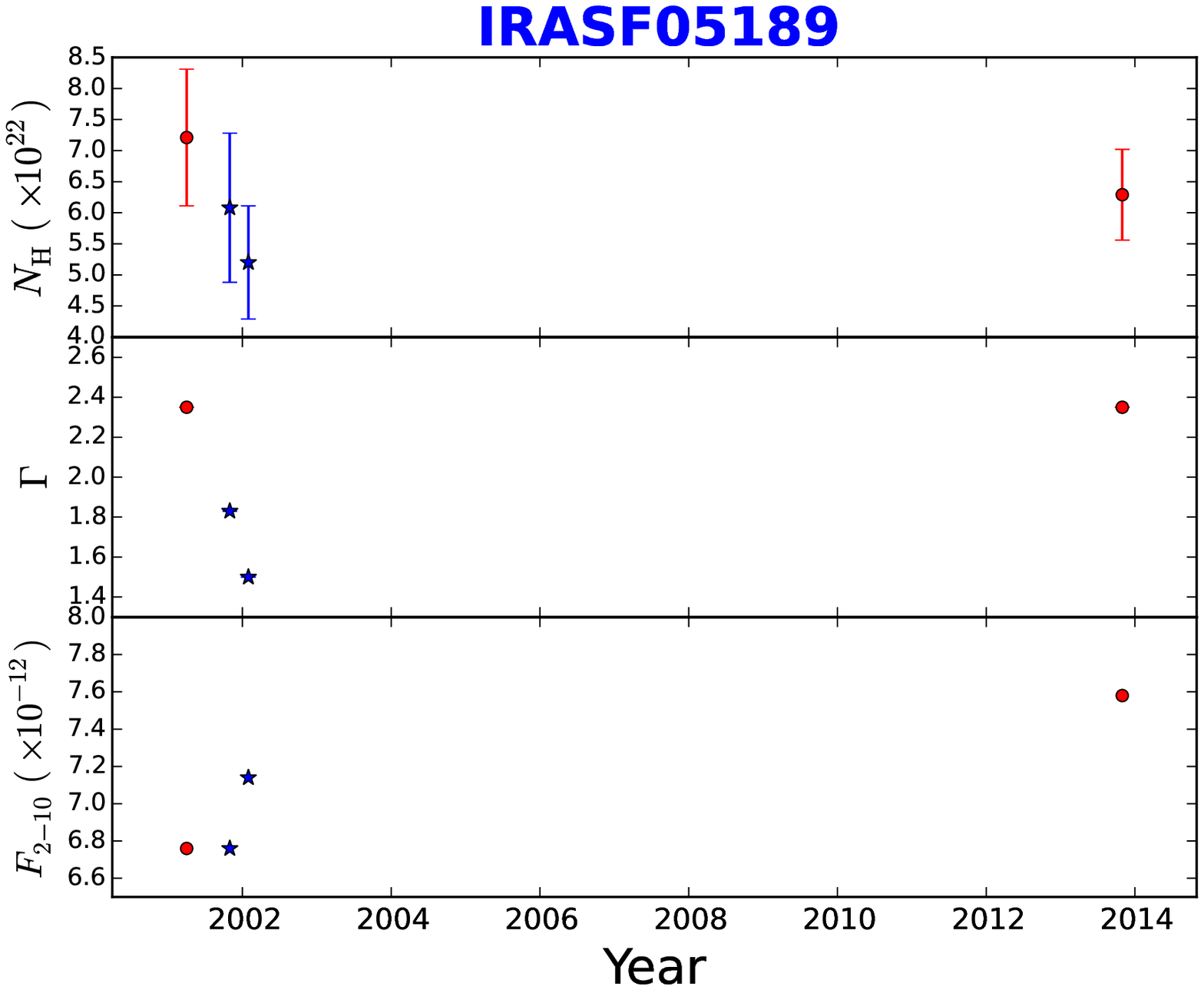} 

\includegraphics[width=9cm,angle=0]{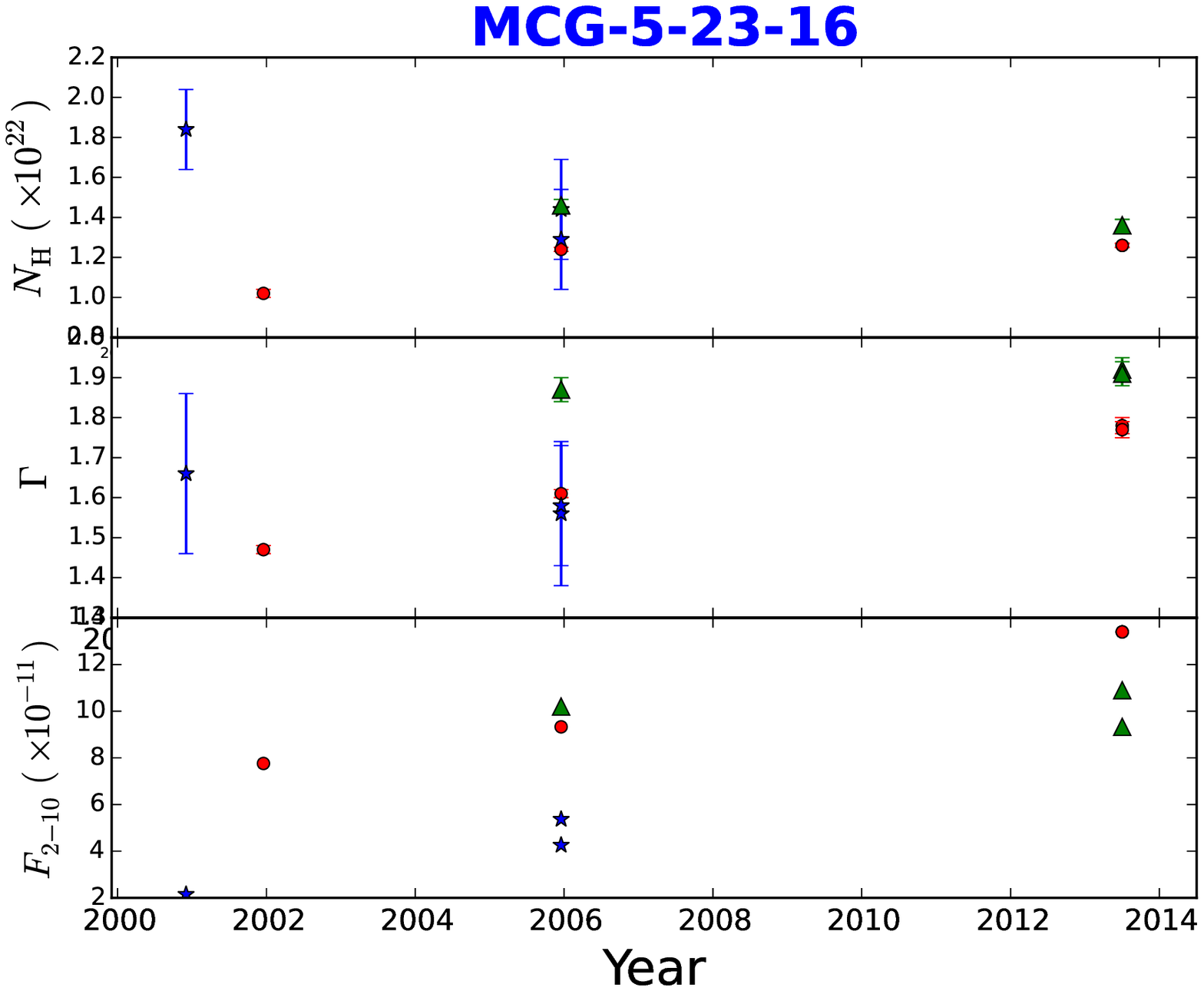}

	}

}

	\caption{The light curves of $\nh$, $\Gamma$, and the $2-10\kev$ unabsorbed flux of the sources in the sample. The red circles, blue stars and the green triangles denote the data points obtained from \xmm{}, \chandra{} and \suzaku{} telescopes respectively.} \label{Fig:lightcurve}
\end{figure*}


\begin{figure*}
  \centering

	\vbox{
	
		\hbox{
	
	\includegraphics[width=9cm,angle=0]{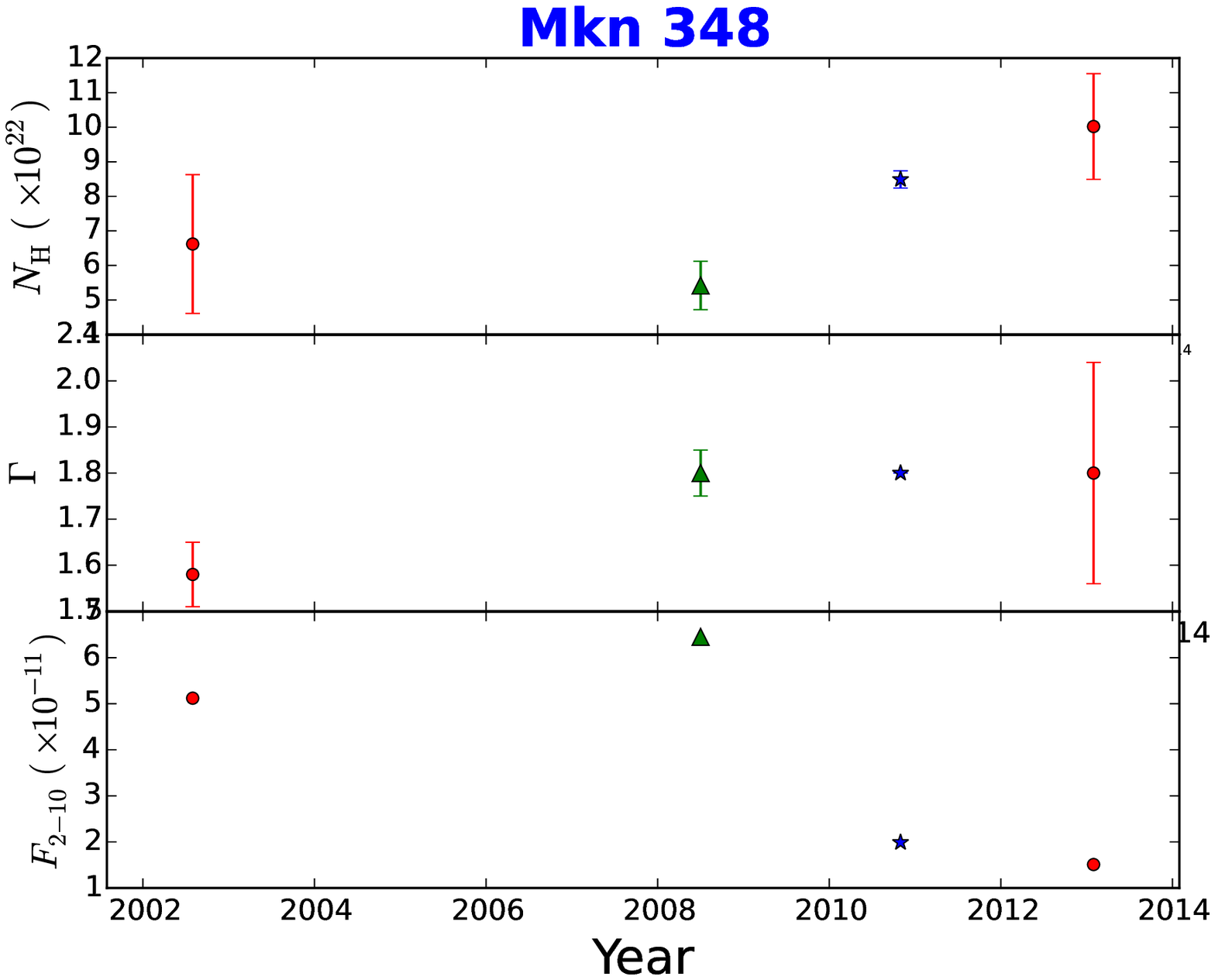} 

	\includegraphics[width=9cm,angle=0]{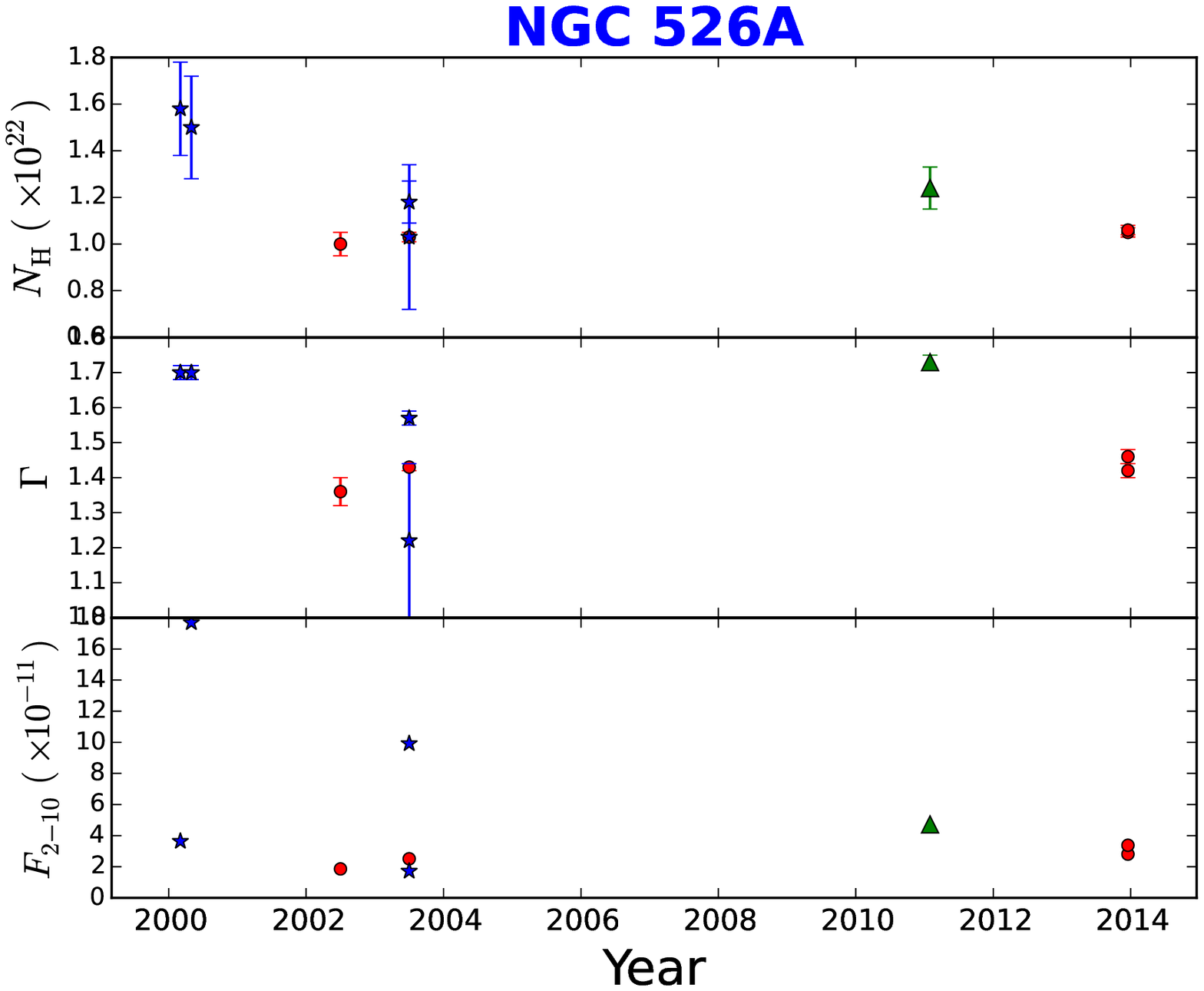}

	}

	\hbox{
	
\includegraphics[width=9cm,angle=0]{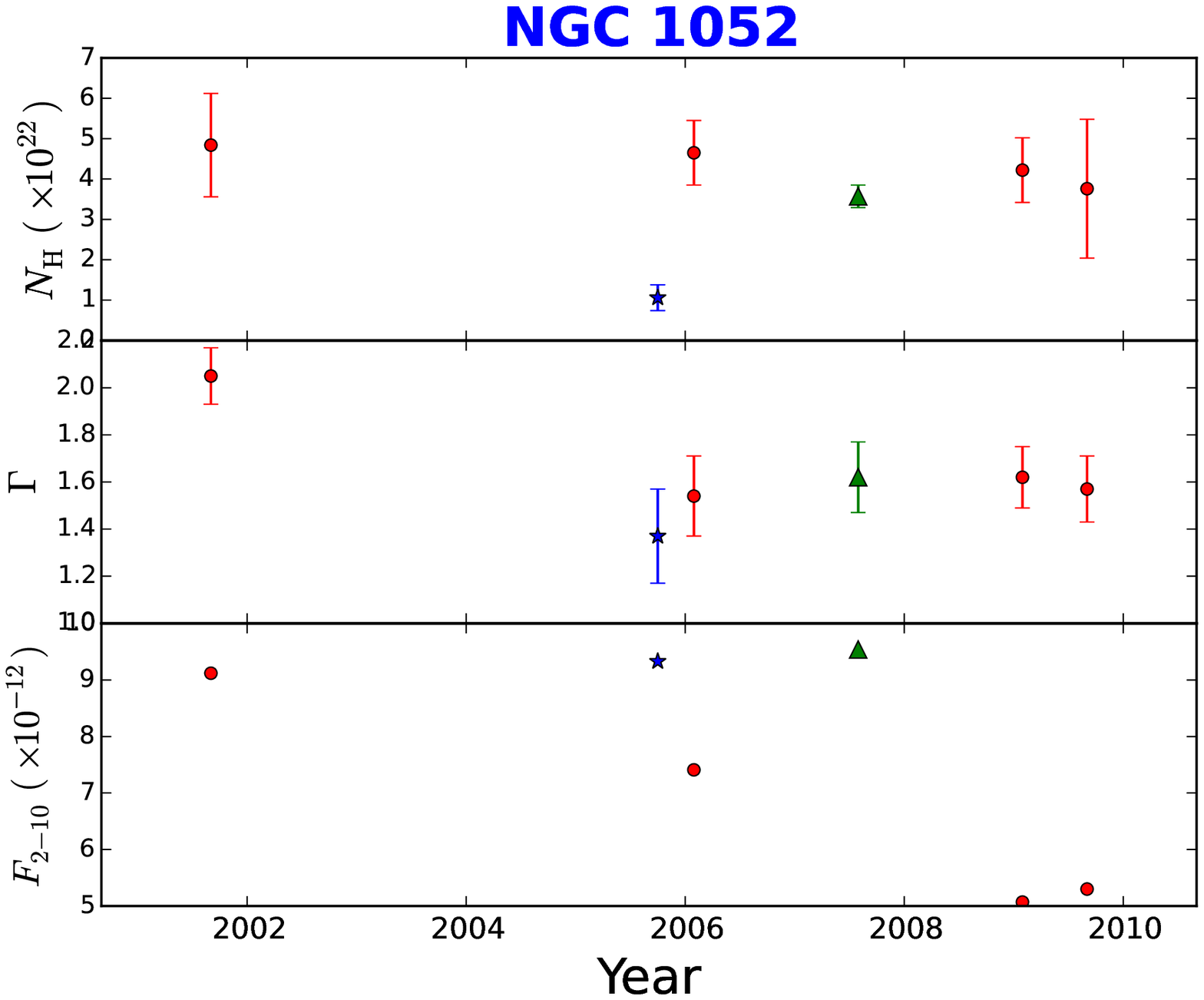} 
	\includegraphics[width=9cm,angle=0]{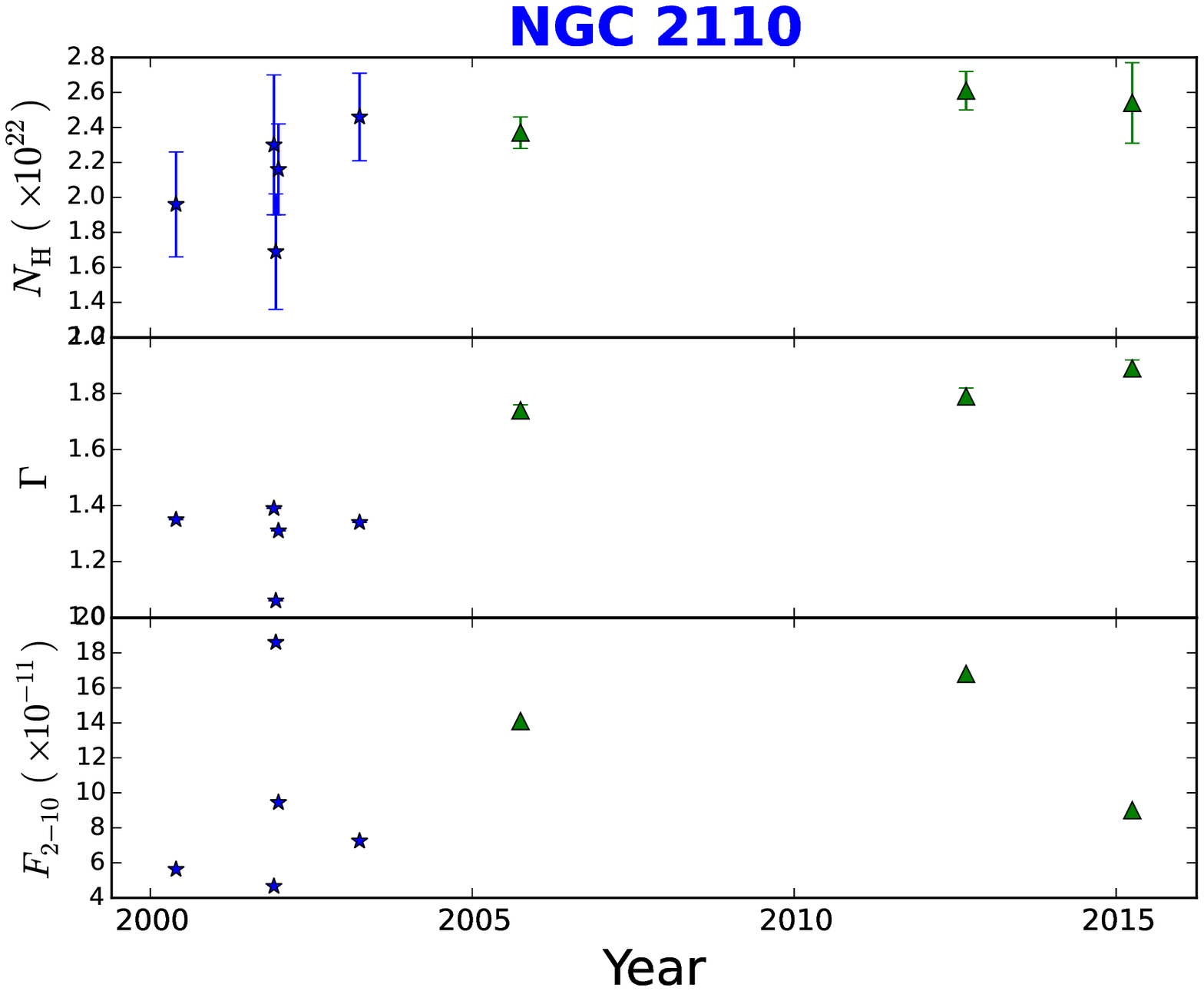}

	}

\hbox{
\includegraphics[width=9cm,angle=0]{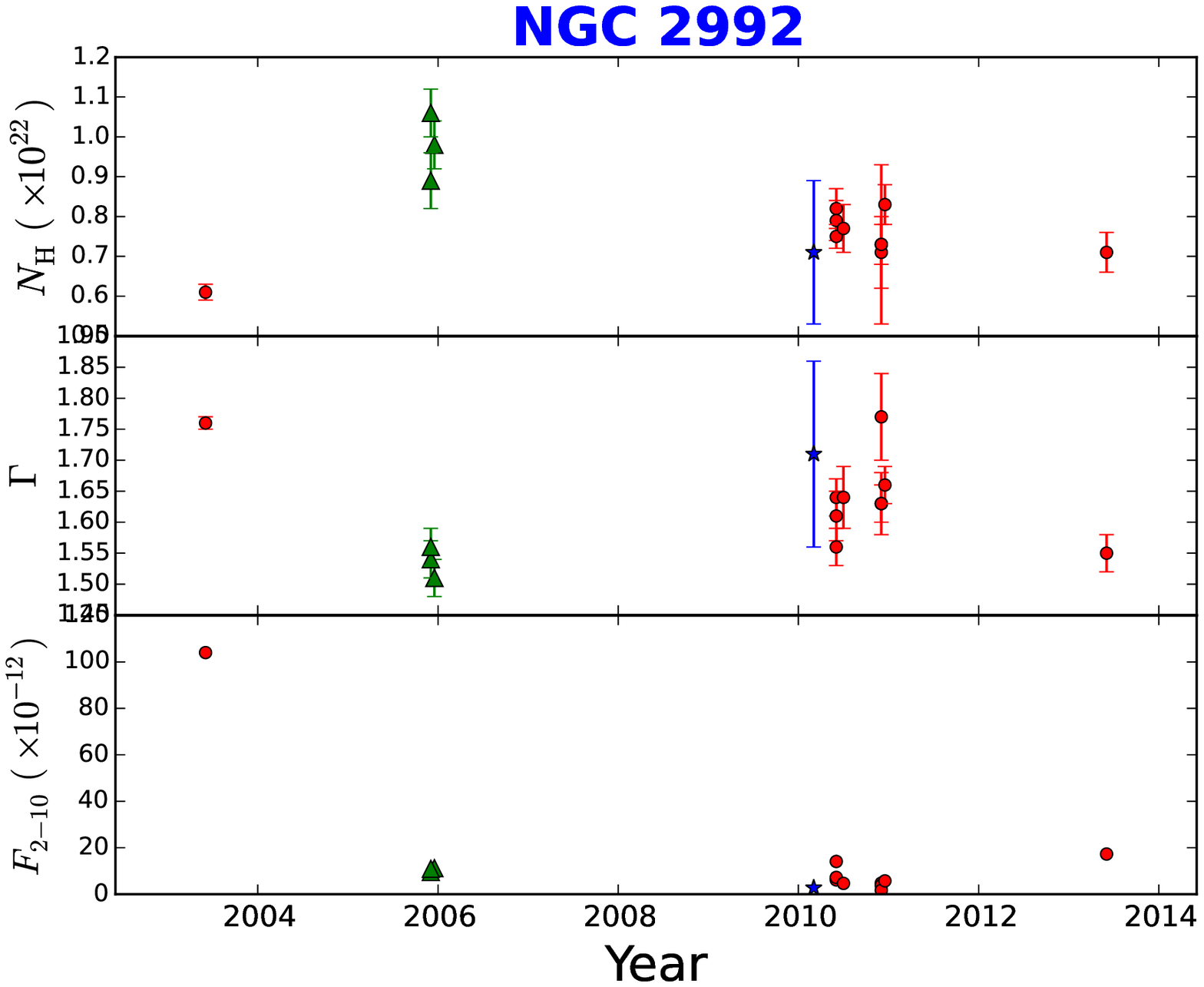}
	
	\includegraphics[width=9cm,angle=0]{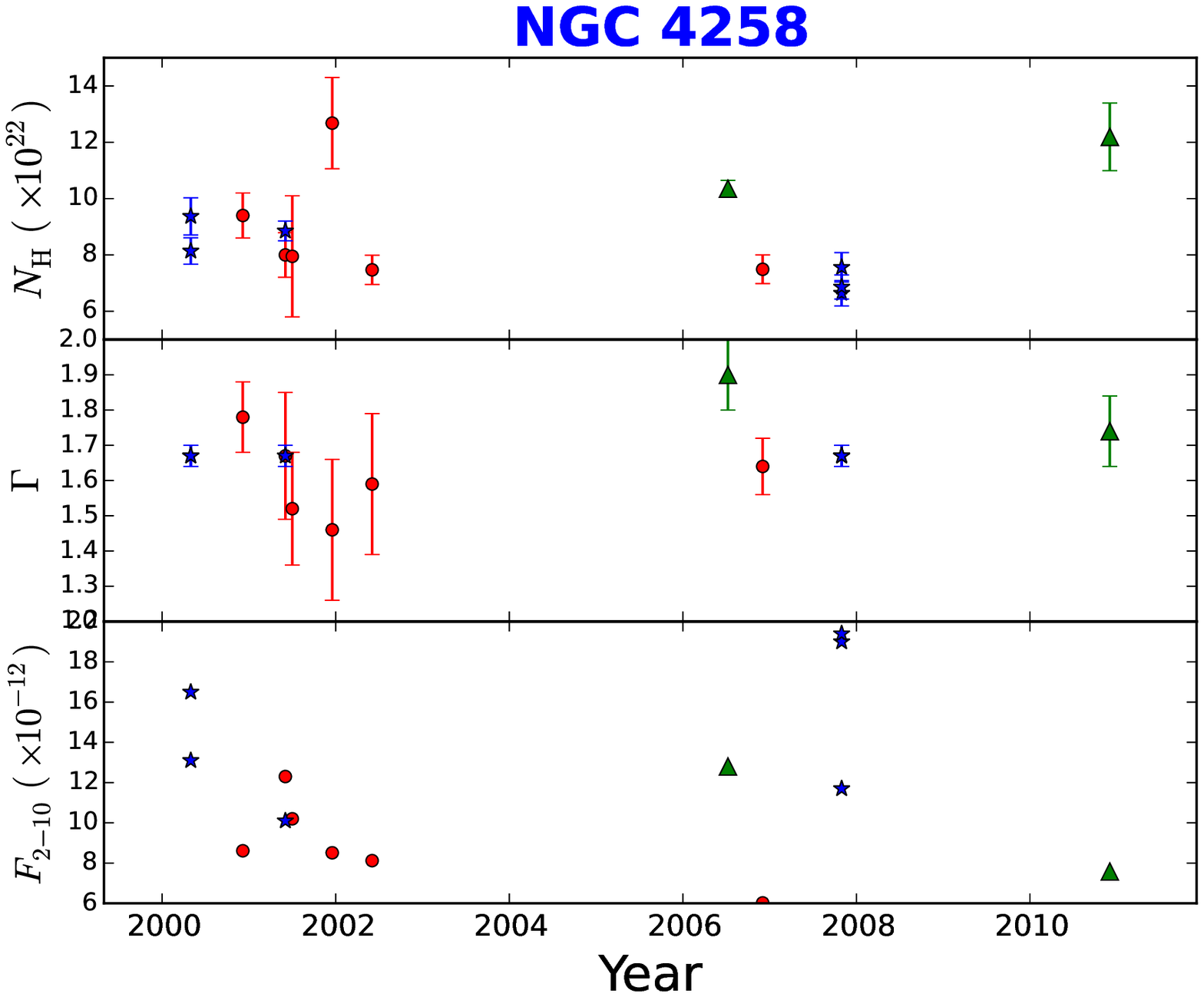}

	}

}

\setcounter{figure}{0}
	\caption{The light curves of $\nh$, $\Gamma$, and the $2-10\kev$ unabsorbed flux of the sources in the sample. The red circles, blue stars and the green triangles denote the data points obtained from \xmm{}, \chandra{} and \suzaku{} telescopes respectively.}

\end{figure*}


\begin{figure*}
  \centering

	\vbox{
	
		\hbox{
	
	\includegraphics[width=9cm,angle=0]{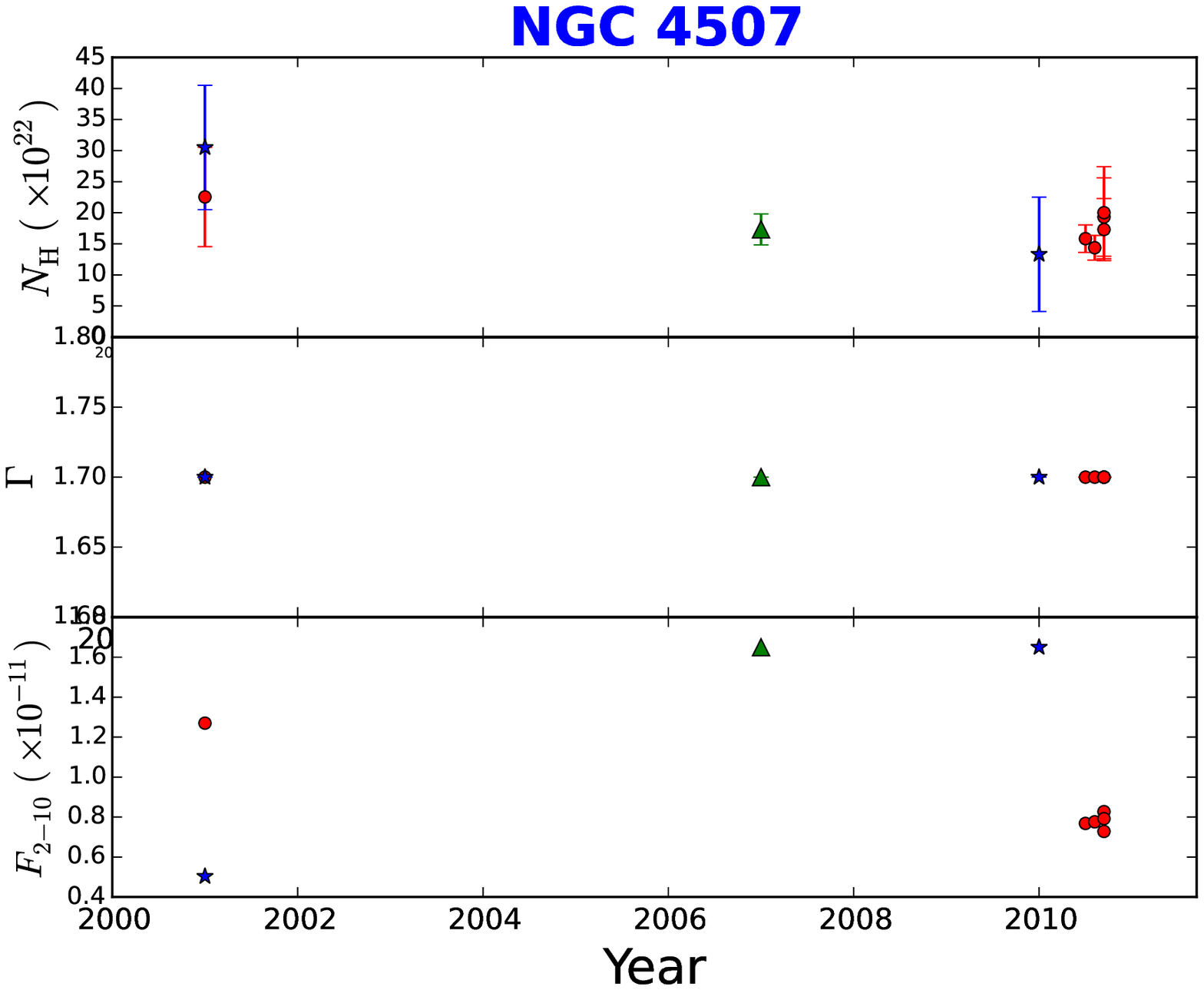}
	\includegraphics[width=9cm,angle=0]{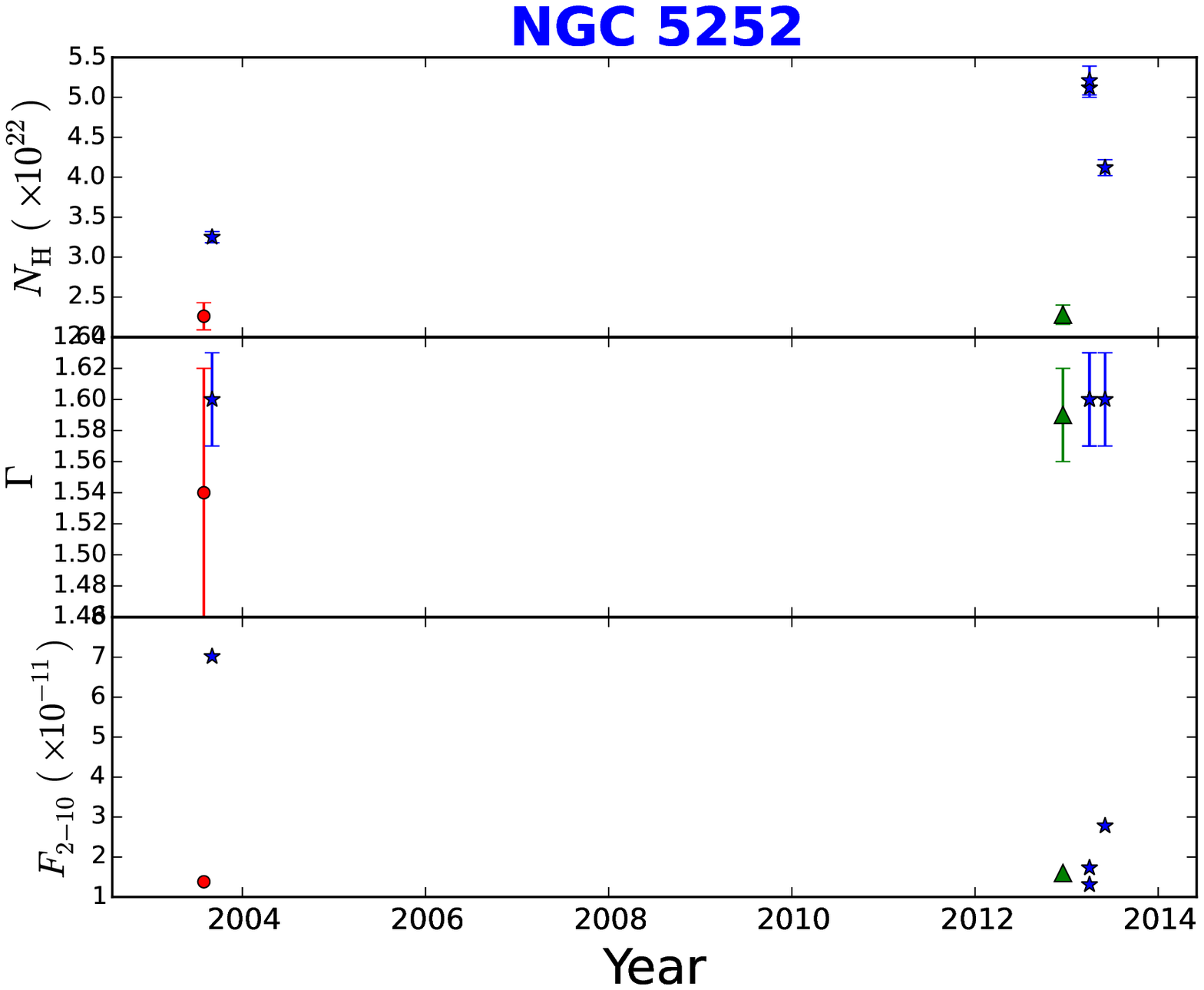} 
	
	}

	\hbox{
	
	\includegraphics[width=9cm,angle=0]{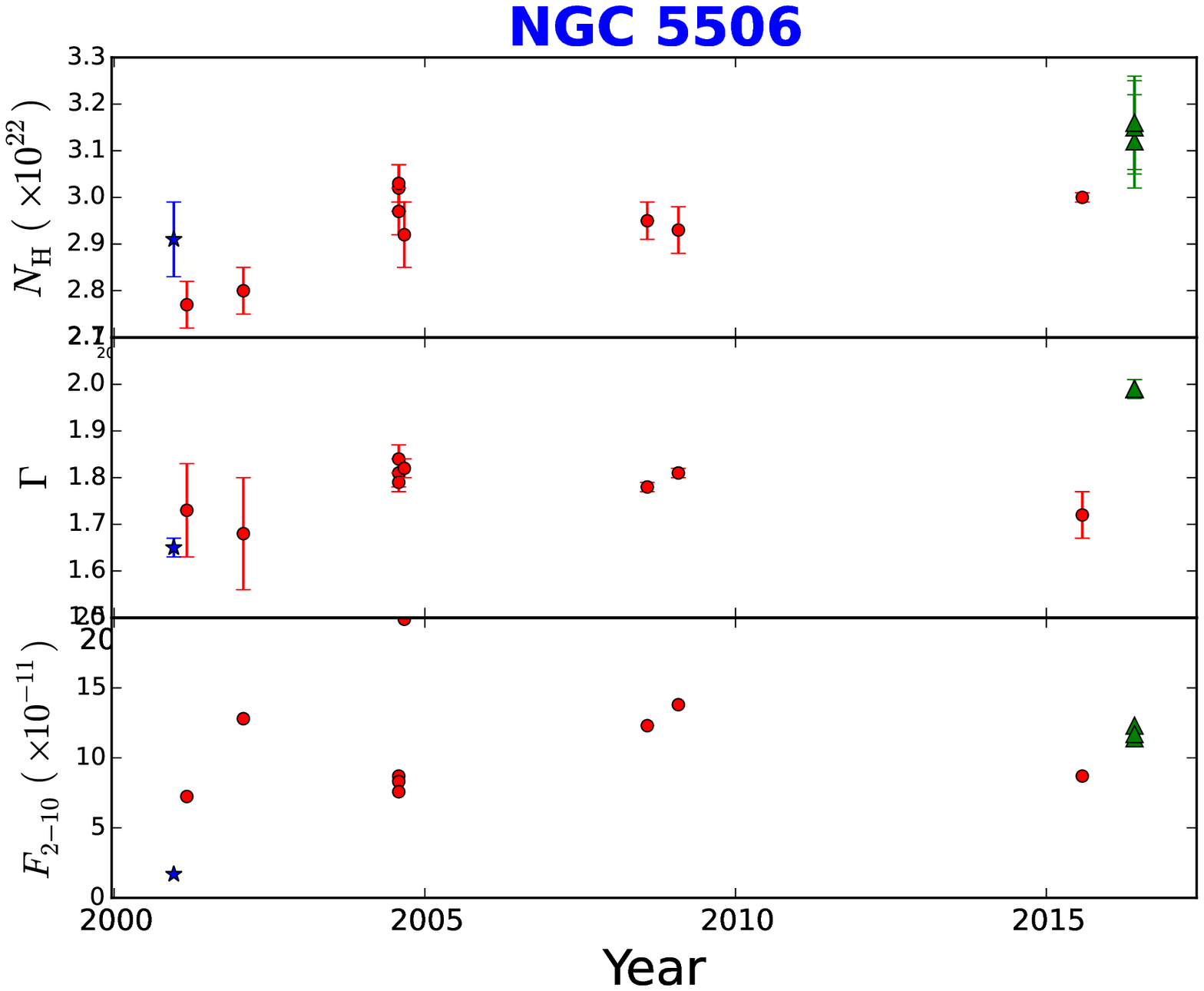}
	
	\includegraphics[width=9cm,angle=0]{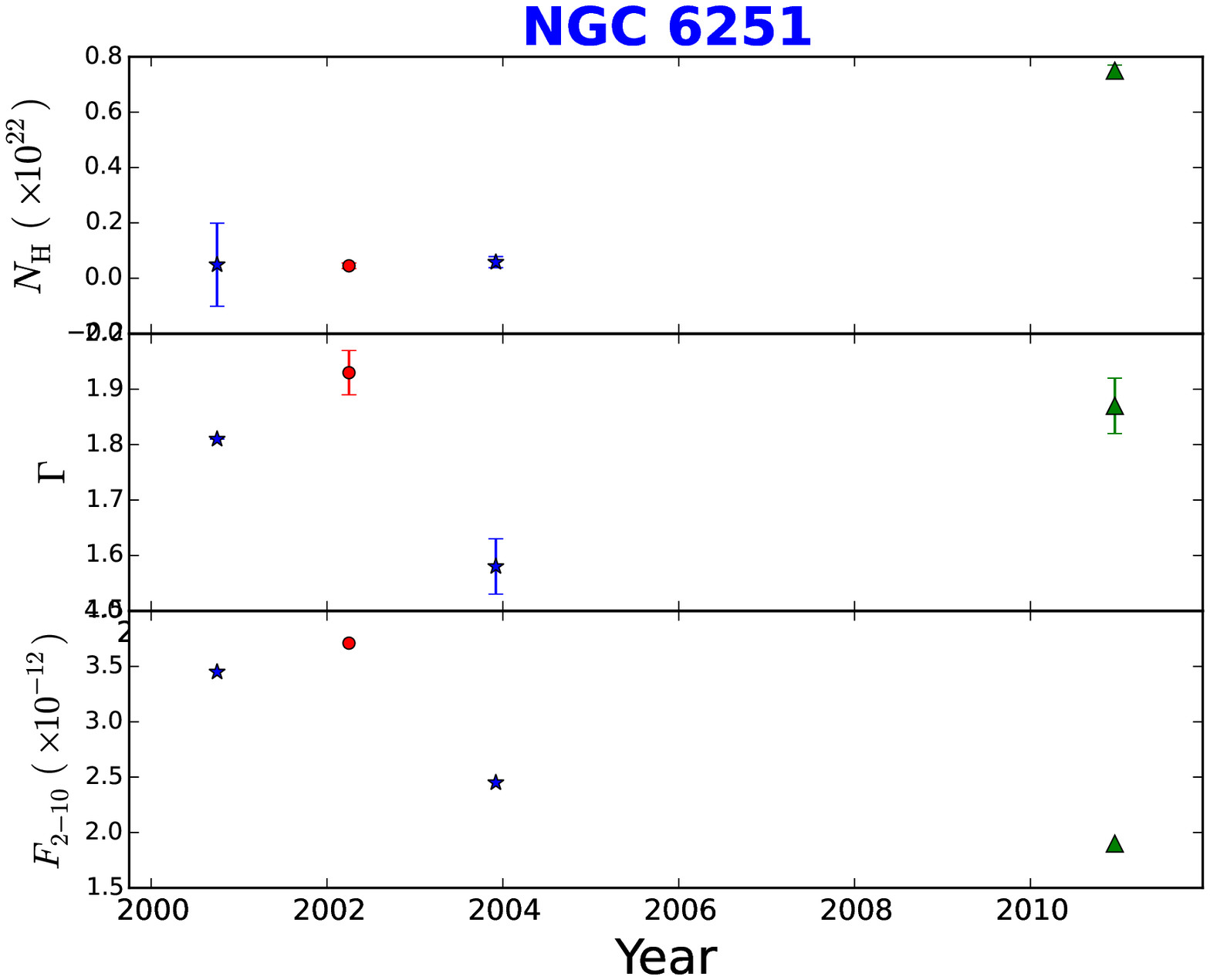}

	}

\hbox{
	
\includegraphics[width=9cm,angle=0]{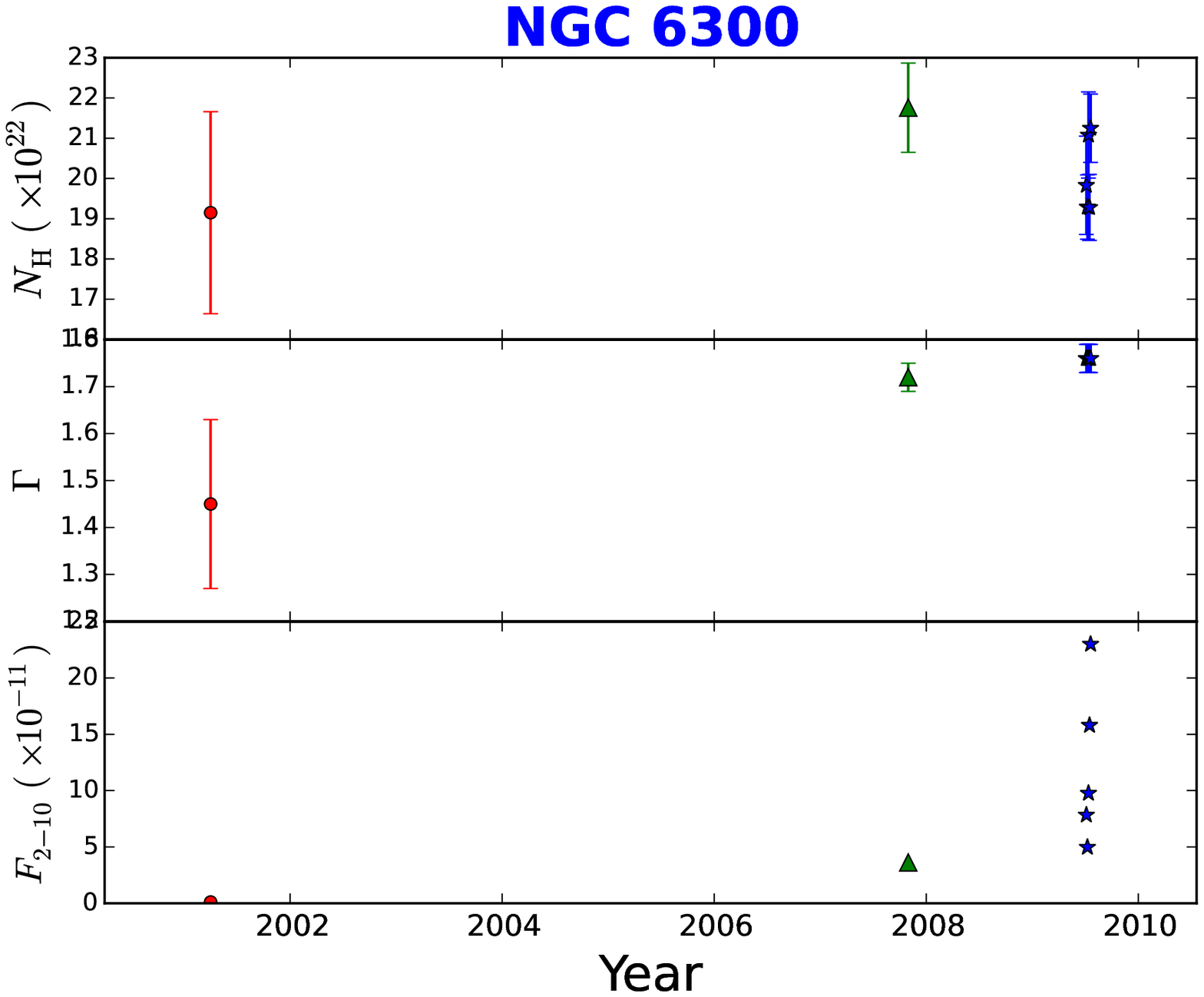}
\includegraphics[width=9cm,angle=0]{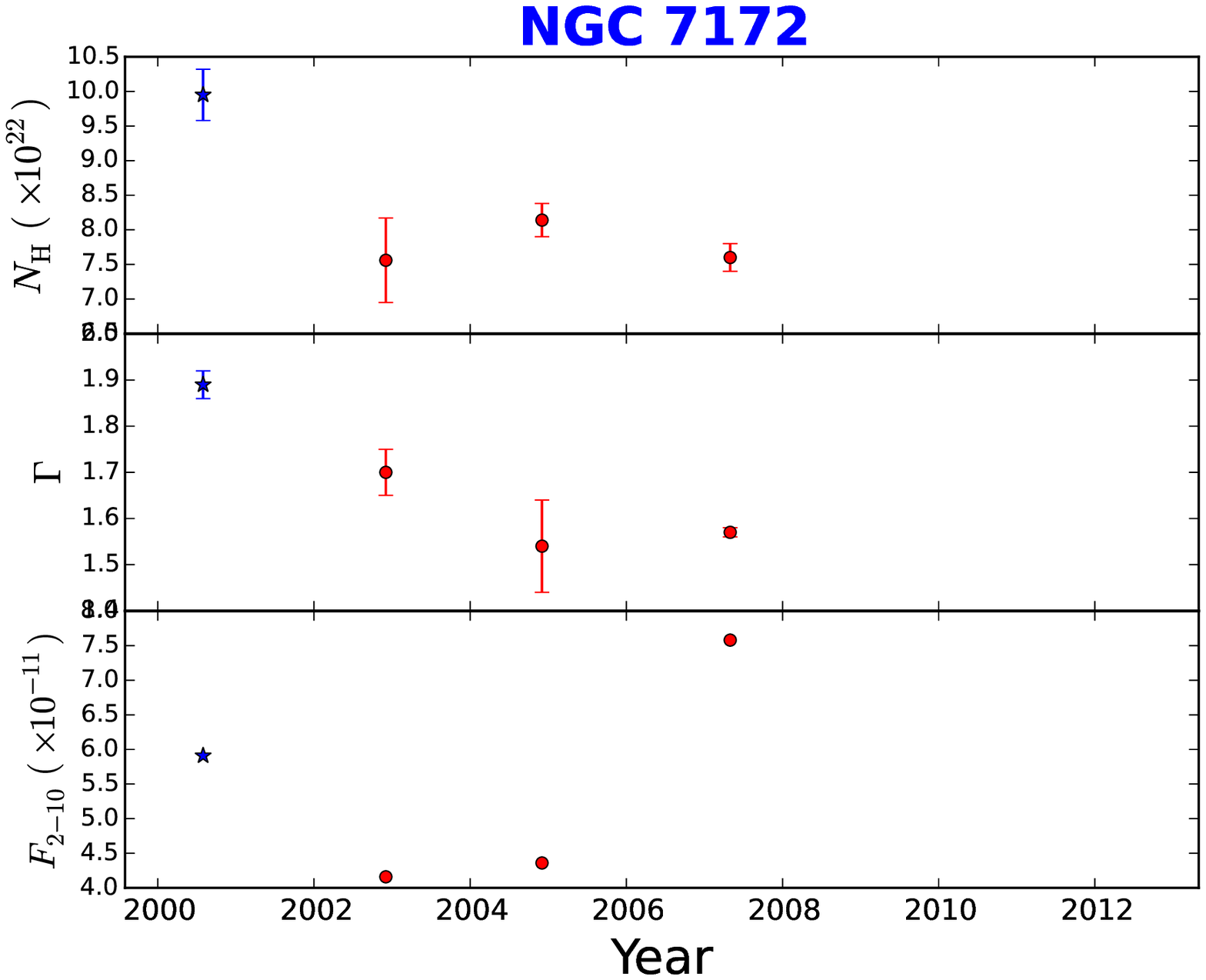} 		 

	}

}

\setcounter{figure}{0}
	\caption{The light curves of $\nh$, $\Gamma$, and the $2-10\kev$ unabsorbed flux of the sources in the sample. The red circles, blue stars and the green triangles denote the data points obtained from \xmm{}, \chandra{} and \suzaku{} telescopes respectively.}

\end{figure*}


\begin{figure*}
  \centering

	\vbox{
	
		\hbox{
	
	\includegraphics[width=9cm,angle=0]{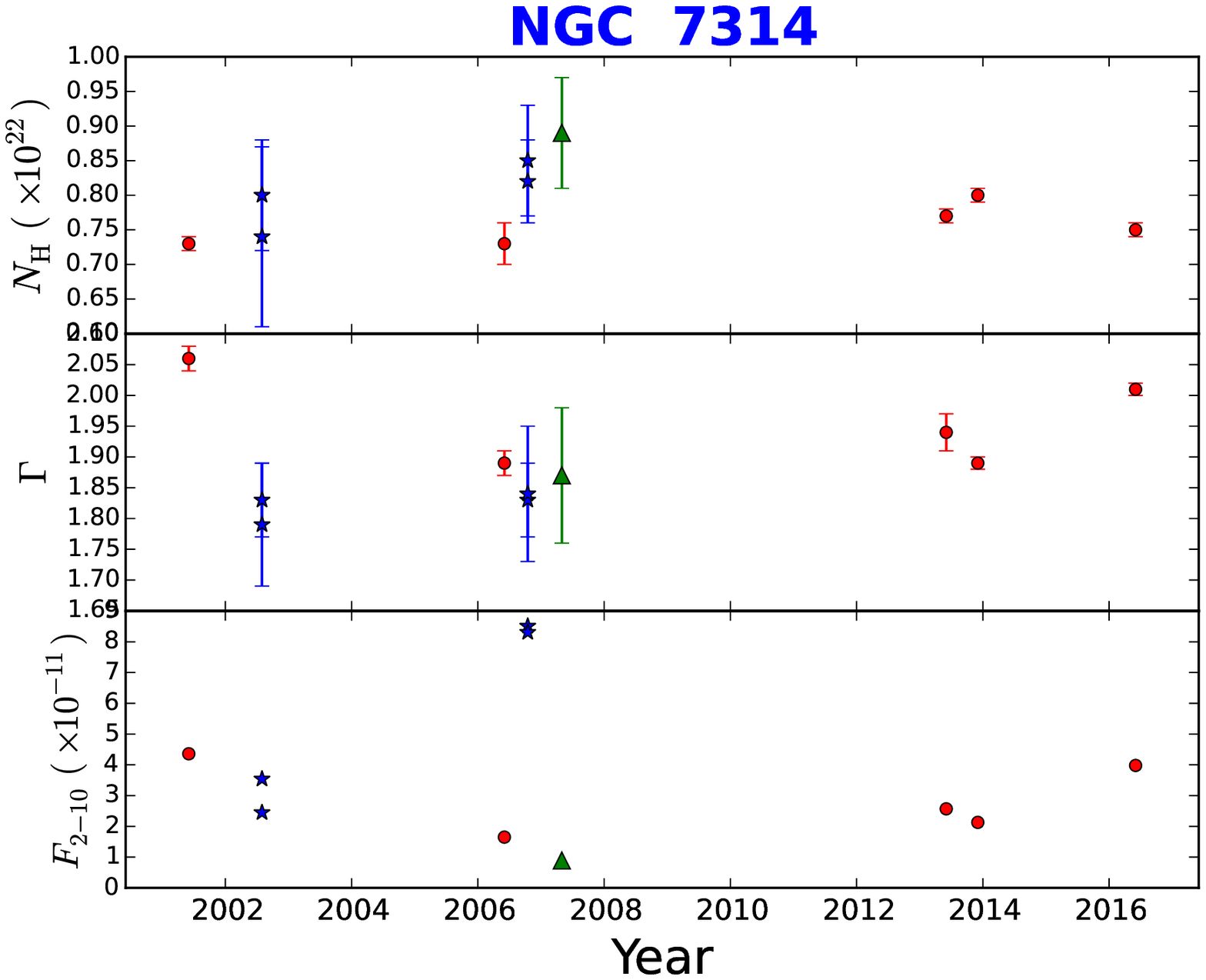}	
	\includegraphics[width=9cm,angle=0]{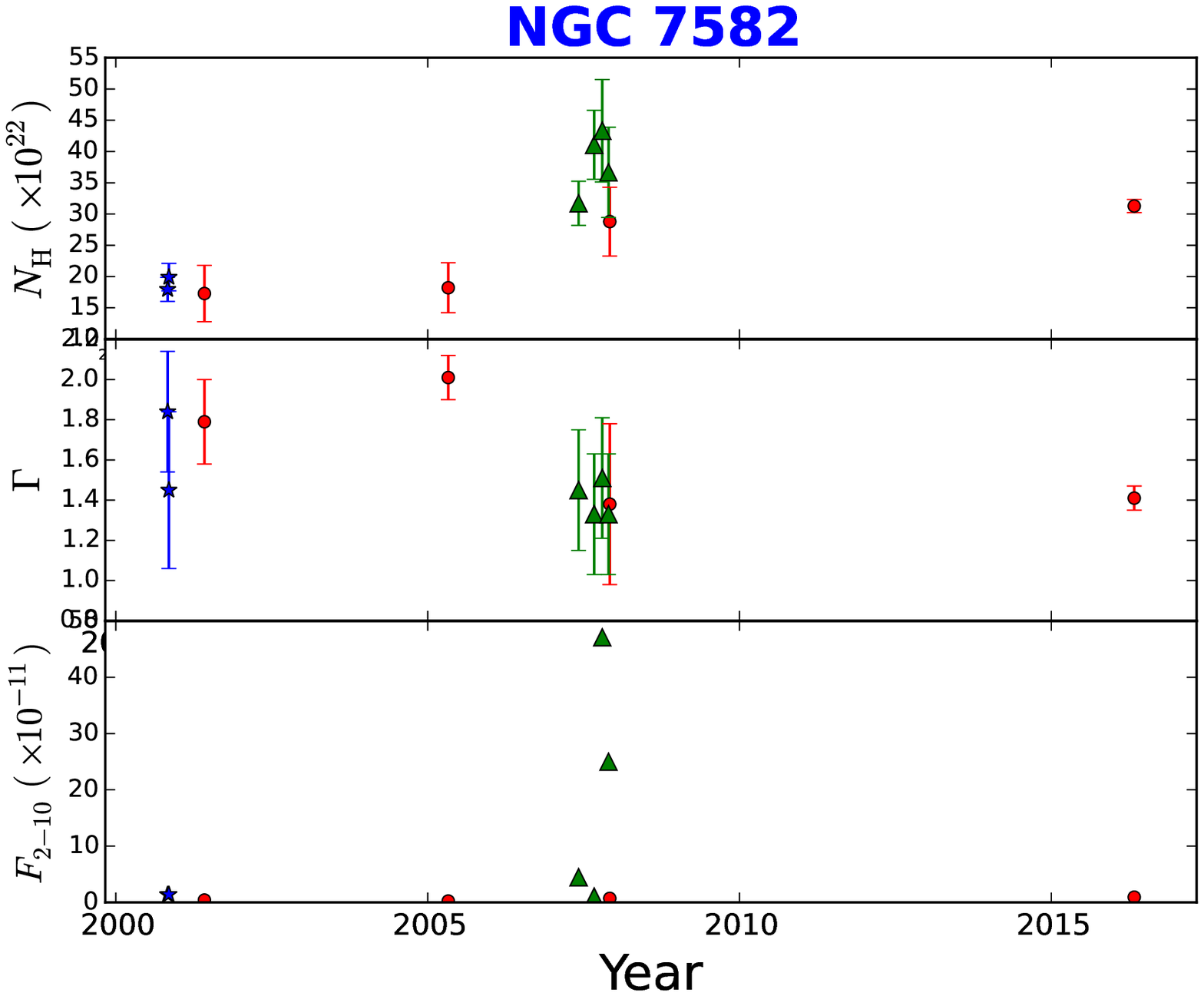} 

	}

}

\setcounter{figure}{0}
	\caption{The light curves of $\nh$, $\Gamma$, and the $2-10\kev$ unabsorbed flux of the sources in the sample. The red circles, blue stars and the green triangles denote the data points obtained from \xmm{}, \chandra{} and \suzaku{} telescopes respectively.}

\end{figure*}


\begin{figure*}
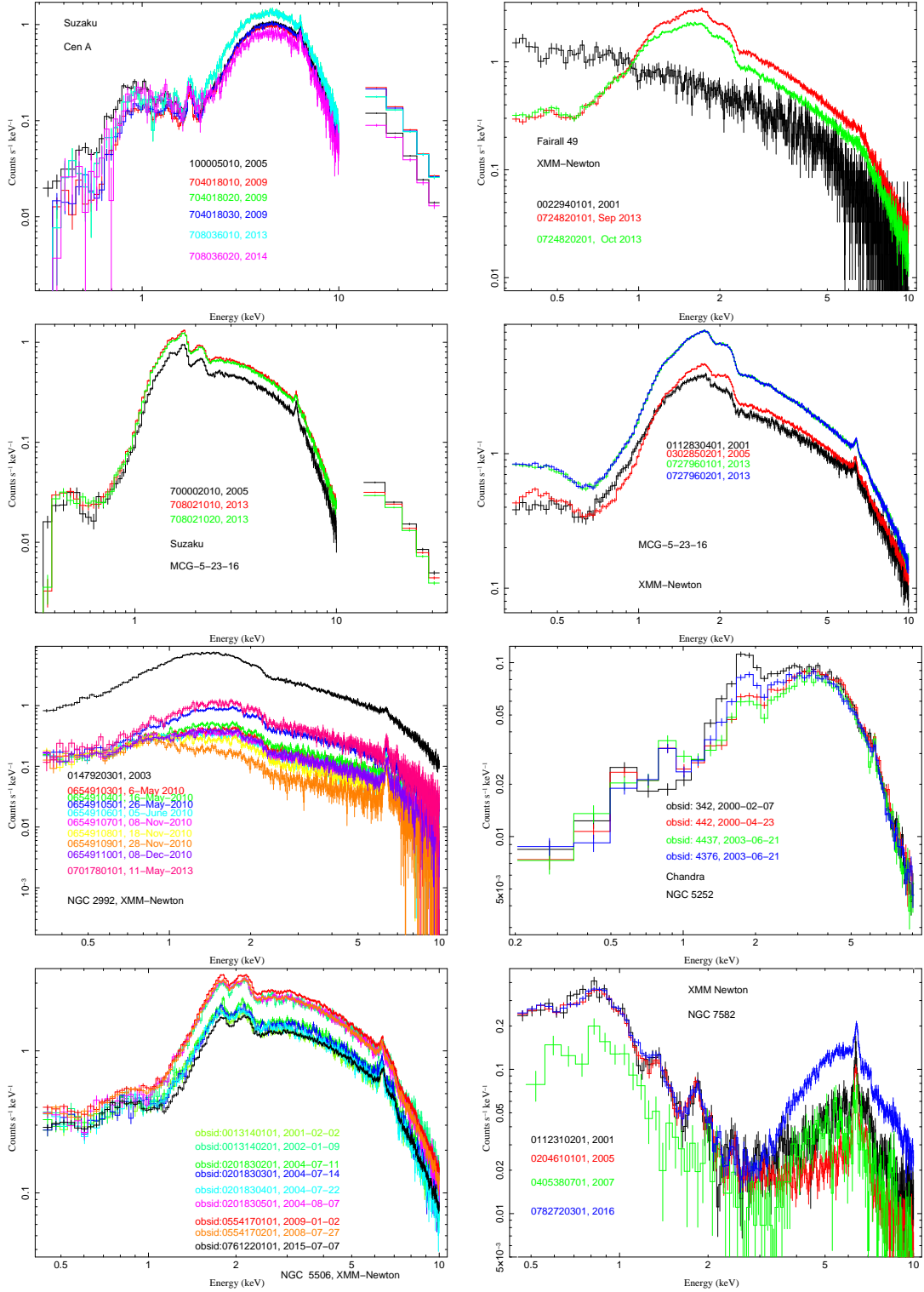

  \centering

	\vbox{	
		\hbox{
	
\includegraphics[width=5.0cm,angle=-90]{Plots_29thOct2017/CenA/CenA_compare_Suzaku.ps} 
\includegraphics[width=5.0cm,angle=-90]{Plots_29thOct2017/Fairall49/Fairall49_XMM_comparison.ps}

	}

	\hbox{	
\includegraphics[width=5.0cm,angle=-90]{Plots_29thOct2017/MCG-5-23-16/MCG5_Suzaku_compare.ps}
\includegraphics[width=5.0cm,angle=-90]{Plots_29thOct2017/MCG-5-23-16/MCG-5-23-16_XMM_comparison.ps}

		}

\hbox{
	
\includegraphics[width=5.0cm,angle=-90]{Plots_29thOct2017/NGC2992/NGC2992_XMM_Compare.ps} 
\includegraphics[width=5.0cm,angle=-90]{Plots_29thOct2017/NGC5252/NGC5252_Chandra_Compare.ps} 
	
	}

	\hbox{
	
\includegraphics[width=5.0cm,angle=-90]{Plots_29thOct2017/NGC5506/NGC5506_XMM_compare.ps} 
\includegraphics[width=5.0cm,angle=-90]{Plots_29thOct2017/NGC7582/NGC7582_comparing_XMM.ps}
	
	}	
	}

	\caption{The overplot of the spectra of the sources listed in Table \ref{Table:obs} and discussed in Section \ref{Sec:results} whose full-covering $\nh$ values have varied between the observations. Here the spectra have been binned by a factor of four for plotting and visual purposes only. The source names, instruments, the observation identifiers (obsid), and dates of observations are written in the individual figures. } \label{Fig:overplot}
\end{figure*}


\begin{figure*}
  \centering

	\includegraphics[width=11cm,angle=-90]{NHCOMP_MAR20.ps}

	\caption{ Mean values of full-covering column density for each object in our
sample are plotted in ranked order. Superimposed are estimates of
X-ray-absorbing column density based on various optical or IR
extinction, assuming the standard Galactic dust/gas conversion: Balmer
	decrements for broad lines or narrow lines \citep[][, respectively]{2001A&A...365...37M,2017ApJ...846..102M}, 9.7 $\mu$m Si line absorption
	\citep{2010ApJS..187..172G}, IR SED modeling \citep{2016A&A...586A..28B}, or
	color-color maps \citep[][]{1994ApJ...436..586M,1994ApJ...433..625M,1996ApJ...459..535S,2014MNRAS.442.2145P}. ``${\Delta}N_{\rm H}$" denotes a detection of variability in the full-covering absorption
component as measured in this paper. "PC" denotes those 11 sources wherein a partial covering component
was detected in most or all of a given sources' observations. The top row denotes the nuclear dust morphology classification
	from \citet{1998ApJS..117...25M}, for those sources included in their sample;
``DC", ``D-[directional]", ``DI", and ``F/W" denote
a dust lane directly crossing the line of sight to the nucleus,
a dust lane just offset from the line of sight to one direction,
irregular dust, and flocculent/wispy, respectively.} \label{Fig:NHcompare}
\end{figure*}


\begin{figure*}
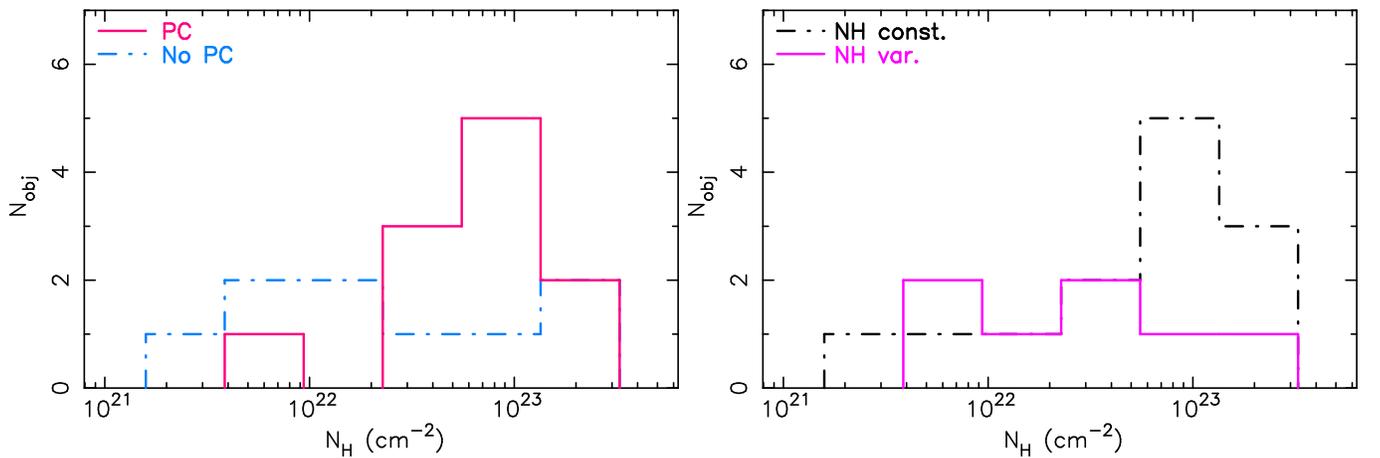

  \centering

	\vbox{
	
		\hbox{
	
	\includegraphics[width=6cm,angle=-90]{hist_apr20_PC.ps}	
	\includegraphics[width=6cm,angle=-90]{hist_apr20_VAR.ps} 

	}

}

	\caption{{\it Left panel:} The histogram plots for the the $\nh$ values of the two sets of sources in our sample, the ones which have been detected with partial covering absorption (red solid line) and the ones which do not have partial absorbers (blue dash-dotted line) {\it Right panel:} Same as left, except that here we consider the sets of sources which have shown $\nh$ variability (pink bold line) and the ones that have not shown $\nh$ variability (black dash-dotted line). } \label{Fig:hist}
\end{figure*}


\clearpage
\appendix

\section{Appendix A: Previous studies and details of analysis of the sources in the sample.}\label{Appendix:sources}
In this section we discuss the previous studies of the sources in the sample in context to the analysis we have carried out and science goals in the paper. We also comment on a few important issues in analysis for each of the sources and the best fit baseline model used in each case. {Here NH22 denotes $\nh/(10^{22}\cmsqi)$} :

\begin{enumerate}[label=\arabic*]

	\item{\bf CenA:} This is the nearest radio-loud galaxy. \textit{RXTE} monitoring
revealed two previous eclipse events by transiting clouds, each
causing total line-of-sight $N_{\rm H}$ to increase by
$8\pm1\times10^{22}$ cm$^{-2}$: one in $\sim$2003--4 and one in
		2010--11, studied by \citet{2011ApJ...733...23R} and \citet{2011ApJ...742L..29R}, respectively.  MKN14 inferred these clouds to reside in the
inner dusty torus.  As described earlier, MKN14 also noted mild,
smooth variations in the baseline level of $N_{\rm H}$, thanks to the
combination of Cen~A's X-ray brightness and sustained, regular
monitoring (and to our knowledge, this is the only such case so far).

		\chandra{} data for the core of Cen A were not usable; the
		pileup was too severe to derive any reliable fluxes or spectral slopes (even with modeling the pileup during spectral fitting). Hence the \chandra{} observations were not used for this source.

		 Following e.g., \citet{2007ApJ...665..209M}, we model the soft band
emission for the \textit{Suzaku} data using two \textsc{apec}
components plus a flat power law, the latter to model blended emission
from point sources, jet components, and diffuse emission.  For
\textit{XMM-Newton} data, using only one \textsc{apec} component
sufficed; adding a second, lower-temperature component yielded no
		improvement to the fit and poor parameter constraints. No strong
evidence for Compton reflection component has been found so
far. Various studies with \textit{RXTE}, \textit{Suzaku},
\textit{INTEGRAL}, and \textit{NuSTAR} have yielded upper limits to
the \textsc{pexrav} reflection fraction $R$ of e.g., 0.28, 0.05, 0.05,
		0.01, and 0.005 by \citet{2011A&A...531A..70B,2007ApJ...665..209M,2001A&A...371..858B,2016ApJ...819..150F, 2011ApJ...742L..29R}, respectively; we thus freeze $R$ at 0 in our fits.

 We find evidence for a partial-covering component in observation S-1,
		consistent with \citet{2007ApJ...665..209M}, and in X-6.  We also modeled
		partial covering components for S-2, S-3, and S-4, following \citet{2011ApJ...743..124F}. However, for S-2, S-3, and S-4, the improvement in fit
when the partial covering component is added is modest, the covering
fractions are quite low, and moreover, values of $N_{\rm H,pc}$ are
poorly constrained in each case, usually falling to the same value as
		$N_{\rm H,full}$. { We detected that NH22 dropped from $11.03\pm0.08$ (S-3) to $9.98\pm 0.14$ (S-5) and hence consider this source as a variable $\nh$ source.}


	\item {\bf CygA:} This is another powerful radio galaxy. \cite{2015ApJ...808..154R} observed the source with \nustar{} and measured a powerlaw $\Gamma=1.7$, a neutral absorption column of $\nh \sim 1.6\times 10^{23} \cmsqi$, a Compton hump at energies $>10\kev$ with a reflection coefficient of $R=1.0$, and a high temperature thermal emission in the soft X-rays modeled by {\it APEC} ($kT\sim 6.4 \kev$). 

	Here, we describe the \chandra{} ACIS observations of Cyg~A.
The active nucleus is embedded in  hot, X-ray emitting gas, and \chandra{} ACIS has been used to image both the active nucleus and the surrounding cluster gas \citep[e.g.,][ and references therein]{2018ApJ...855...71S}. The archive contains dozens of observations aimed at mapping the
cluster emission, with the effect that the nucleus was not observed
consistently, with many observations using different ACIS-I or ACIS-S chips
and having different off-axis angles to the nucleus.

We excluded those observations where the nucleus was located in a chip gap.
Backgrounds were generated using the CIAO "blanksky" tool,
as the cluster gas usually fills the majority of the ACIS chip field of view.
Our model included a hard X-ray power law absorbed by a full-covering
absorber; the photon index of the hard power-law was always poorly
constrained, and we froze it at 1.7.  We also included two
{\it APEC} components and a soft power law whose photon index was
also frozen at 1.7.  The {\it APEC} components typically had
temperatures of 0.73 and 4.0 keV.

We paid attention to those observations where the nucleus was within
16" from the edge of a chip, as dithering could cause some fractions
of photons to fall into the chip gap. However, we did not see any
significant deviation in model parameters in these observations.  We
searched for, but did not detect, any correlation between $N_{\rm H}$
and the off-axis angle of the nucleus, or between $N_{\rm H}$ the
temperature of either thermal component; there was also no correlation
with the ACIS chip used.

		We find a mean value (standard deviation) in $N_{\rm H, full}$ of $ (25.6\pm 2.1) \times 10^{22}\cmsqi$.
Given the variation in observations from one obsid to the next
(e.g., different offset pointings, the
nucleus falling on different chips),
we conservatively avoid concluding that measured
variations in $N_{\rm H, full}$ from one obsid to the next
is intrinsic to the source, although we can rule out the presence of
		strong systematic trends of order $10^{24}$ cm$^{-2}$. See Table \ref{Table:cygA} for details of the best fit parameters.

The best fit model used is {\it tbabs*(apec+apec(2)+powerlaw(1)+ztbabs*(pexmon+powerlaw(2)+zgauss))}. We do not detect significant variability in $\nh$ for this source.


\item {\bf Fairall 49:} Also known as IRAS 18325-5926. The source was studied by \citet{2004MNRAS.347..411I} who found a relatively steep powerlaw slope $\Gamma=2.2$, and a neutral absorption column of $\nh=1.62\times 10^{22} \cmsqi$. \citet{2016ApJS..225...14K} detected a partial covering absorption with covering fraction of $f_{\rm PC}=0.22$ and column density of $\nh=3.2\times 10^{22}\cmsqi$, in addition to a full covering absorption, with \suzaku{} observations. {We detected partial covering in addition to full covering absorption in all the observation, except for CH-2, where the low SNR did not allow us to constrain $\nh$ of the partial absorber. We detected significant variability in $\nh$ for this source with \xmm{} observations, with variations of $\nh$ by almost an order of magnitude.} The best fit model used is {\it tbabs*(apec(1)+powerlaw(1)+ztbabs*zpcfabs*(pexmon+powerlaw(2)+zgauss))}


\item {\bf IRAS~F00521--7054:} \citet{2014ApJ...795..147R} modeled the X-ray spectrum by a steep powerlaw of $\Gamma=2.2-2.3$, obscured by a neutral column of $\nh\sim 10^{22.9}\cmsqi$, in addition to a blurred ionised disk reflection component with a reflection coefficient of $R=2.7$. In our study, we too detected steep powerlaw slope pegged at $\Gamma=2.5$. The best fit model is {\it tbabs*(apec+powerlaw(1)+ztbabs*zpcfabs*(pexmon+powerlaw(2)+zgauss))}. We do not detect any significant variability in $\nh$ for the source, as also concluded by a recent study by \citet{2019MNRAS.484.2544W} using \nustar{} observations.\\


\item {\bf IRAS~F05189--2524:} A detailed study of this source has been carried out by \citet{2009ApJ...691..261T}, where the authors have found interesting variability in spectral shape and flux between different observations taken between 2001 and 2006 with \chandra{}, \xmm{} and \suzaku{} satellites. The source has shown an increase in obscuring column by a factor of 10 from 2002 (\chandra{}) to 2006 (\suzaku{}) observations, although the intrinsic unabsorbed $2-10\kev$ flux is constant over time. This source has been studied as a part of a sample of ULIRGs by \citet{2010ApJ...725.1848T} using \chandra{} observations. They found that for different observations the hard X-ray powerlaw varied from $\Gamma=1.4-2.3$ (between 2001 and 2002 observations). They also detected a non-variable partial covering absorber of the source with $\nh\sim 7 \times 10^{22}\cmsqi$ and a covering fraction of $f^{\rm pc} \sim 0.96$. A more recent study by \citet{2015ApJ...814...56T} using \nustar{} data revealed a powerlaw $\Gamma=2.5$ and two partial covering absorbers of $\nh$ of $5.2 \times 10^{22}\cmsqi$ and $9.3 \times 10^{22} \cmsqi$ with covering fractions of $98 \pm 0.2\%$ and $74\pm 1.2\%$ respectively. Another study by \citet{2017ApJ...837...21X} using \nustar{} and \xmm{} observations have constrained the powerlaw slope $\Gamma=2.29$ and Pexrav reflection component $R=1.48$. In our study we too required a partially covering neutral intrinsic absorber, but only for the observation X-2. The best fit model used is: {\it tbabs*(powerlaw(1)+apec+ztbabs*zpcfabs*(zgauss+pexmon+powerlaw(1)+zgauss(2)))}. We do not detect any significant variability in $\nh$ for the source. \\


\item {\bf MCG--5--23--16:} \citet{2017ApJ...836....2Z} observed this source using \nustar{} for half mega seconds, with a focus on the relativistic reflection and primary coronal emission. The authors detected the presence of reflection features with a reflection coefficient of $R=0.84$. \citet{2007ApJ...670..978B} carried out a time resolved spectral study of this source and found that there is an absorption variability intrinsic to the source. 
	
	{In our study we found that the X-ray spectra of this source are quite complex for all the observations. In particular the high SNR long exposure \xmm{} observations X-2, X-3 and X-4 ($\sim 130\ks$ each) exhibited several discrete spectral features which were visible after the `basline' model fit. We found that in the cases of X-2, X-3 and X-4, the FeK$\alpha$ emission line and the compton reflection hump coule not be modeled simultaneously, using the {\it pexmon} parameter values obtained from \suzaku{} observations. We therefore thawed the {\it pexmon} normalization and this improved the fit by $\dc=500$. But the FeK emission line was not fully modeled, so we thawed the Fe abundance assuming that the reprocessing media for FeK emission and the Compton hump are different, which again improved the fit by $\dc=300$ and the best fit Fe abundance $\sim 0.15-0.19$ times solar abundance. We used a {\it zgauss} model to describe the higher ionization Fe emission lines. The best fit model used is: {\it tbabs*(apec+powerlaw(1)+ztbabs*(pexmon+powerlaw(2)+zgauss))}. We detect significant variability in $\nh$ for the source using \xmm{} observations. }


\item {\bf Mkn~348:} \citet{2014MNRAS.437.2806M} studied this Seyfert 2 galaxy with \suzaku{} and \xmm{}. The authors detected variability in the X-ray spectral curvature which they concluded could be due to changes in column density of neutral and ionized absorbers. They obtained a powerlaw photon index of $\Gamma=1.72$, a neutral absorber column density of $\nh=4.50\times 10^{22}\cmsqi$, and a reflection coefficient of $R=0.24_{-0.04}^{+0.04}$. {In our analysis we found that the spectra need a partial covering absorber in addition to a full covering absorption, in all but one observation (CA-1, possibly due to low SNR)}. The best fit model used is {\it tbabs*(apec+apec(2)+powerlaw(1)+zpcfabs*ztbabs*(zgauss+pexmon+powerlaw(1)))}. We detected variability in $\nh$ between 2008 observation with \suzaku{} and 2010 observation with \chandra{}. However, the variations are within the uncertainties of cross-instrument-calibration as discussed in Section \ref{subsec:crossinstrument}, and so we have not considered this as one of the variable $\nh$ sources.


\item {\bf NGC~526A:} This is a Seyfert 1.9 galaxy, studied by \citet{2001A&A...379...46L} using \sax{} X-ray telescope in the energy band $0.1-150 \kev$. The authors detected a relatively flat powerlaw slope of $\Gamma=1.6$. Although the source flux varies strongly between the observations, the powerlaw slope remains constant over time. The reflection component detected is weak ($R\sim 0.7$). In our study we found that the best fit baseline model for this source is {\it tbabs $\times$ (apec(1) + apec(2)+ powerlaw(1)+ ztbabs $\times$ (powerlaw(2) + pexmon + zgauss ))}. {The spectra did not require any soft X-ray powerlaw  for any of the observations. We do not detect any significant variability in $\nh$ for the source.}


\item {\bf NGC~1052:} Identified as a low ionization nuclear emission region (LINER) by \citet{1980A&A....87..152H}, the galaxy hosts a low luminosity AGN (LLAGN) of a luminosity $L_{\rm 1-100 GHz}=4.4\times 10^{40} \lunit$ \citep{1984ApJ...284..531W}. The compact core of the AGN has a very flat X-ray powerlaw slope. The source is absorbed by an intrinsic neutral column of $\nh \sim (0.6-0.8)\times 10^{22} \cmsqi$. In our analysis we found that the \chandra{} observations are piled up. We used the best fit model {\it tbabs*(apec+apec(2)+powerlaw(2)+ztbabs*zpcfabs*(zgauss+pexmon+powerlaw(1)))}. We detected variability in $\nh$ between 2005 observation with \chandra{} and 2007 observation with \suzaku{}. However, due to uncertain cross-instrument-calibration we have not considered this as one of the variable $\nh$ cases.


\item {\bf NGC~2110:} \citet{2015MNRAS.447..160M} studied this Seyfert 2 galaxy using \nustar{}, \xmm{}, \chandra{} etc. No detectable contribution from Compton reflection has been found. The powerlaw slope $\Gamma=1.64$ and neutral absorption measured is $\nh=4.3\times 10^{22}\cmsqi$. In our study, the best fit model used is: {\it tbabs $\times$ (apec(1) + powerlaw(1)+ ztbabs $\times$ zpcfabs $\times$ (powerlaw(2) + pexmon + zgauss ))}. {For the \xmm{} observation the frozen {\it pexmon} parameters (obtained from the \suzaku{} fits) were over-predicting the FeK$\alpha$ emission line, likely pointing towards a different origin of the emission line as that of the Compton hump. We thawed the Fe abundance in the pexmon fit and fitted the data. The fit improved by $\dc=81$, with a best fit Fe abundance of the {\it pexmon} model $\rm Fe_{\rm abund}=0.55\pm 0.07$ relative to Solar. This has also been observed in MCG-5-23-16. The powerlaw slope $\Gamma$ could not be constrained for the observation CH-2(01) and hence frozen to 1.74 (the $\Gamma$ estimated from S-1(05)). We do not detect any significant variability in $\nh$ for the source.}


\item {\bf NGC~2992:} \citet{2010ApJ...713.1256S} studied the variability of the source in X-rays. This source is known to exhibit X-ray flaring on timescales of days to weeks. They measured a powerlaw $\Gamma=1.83$, the neutral absorption column density $\nh=6.45\times 10^{21}\cmsqi$, the reflection fraction $R=0.40$. 
	
	{In our analysis we detected large changes in flux (almost an order of magnitude) between the \xmm{} observation in 2003 X-1 and the rest of them in 2010 (X-2 to X-10). The X-1 observation is in an unusually high flux state of the source with both the soft and the hard X-ray bands orders of magnitude higher than the other observations. The soft X-ray band in X-1 is dominated by the nuclear emission from the AGN. In our analysis we therefore could not use averaged values for X-1. From X-1 to X-2: NH22 increased from $0.60\pm0.01$
to $0.82 \pm 0.03$ from 2003 to 2010, and hence we regard this source as a $\nh$ variable source. The best fit model used is {\it tbabs*(apec+powerlaw(1)+ztbabs*(zgauss+pexmon+powerlaw(2)++zgauss(3)))}. }


	\item {\bf NGC~4258:} \citet{2009ApJ...691.1159R} have studied this low-luminosity AGN, also classified as Seyfert-2, with \xmm{} and \suzaku{}. The authors conclude that the circumnuclear environment of this AGN is very clean and lacks Compton-thick obscuring torus. They obtained a best fit powerlaw $\Gamma=1.75$, and an absorption column of $\nh=9.2\times 10^{22}\cmsqi$. In our analysis we found that this source has spatially extended structures in soft X-ray emission as viewed with XMM EPIC-pn camera. The best fit model we used is {\it tbabs*(apec+apec(2)+powerlaw(1)+ztbabs*(zgauss+pexmon+powerlaw(2)))}. We do not detect any significant variability in $\nh$ for the source.


	\item {\bf NGC 4507:} The \chandra{} HETG and \xmm{} data of this source has been studied by \citet{2004A&A...421..473M} where they detect a Compton-thin absorption of column density of $\nh=4\times 10^{23}\cmsqi$. This source has also been studied as a part of a sample using \suzaku{} data by \citet{2016ApJS..225...14K}, and the authors found that the source required a Compton-thin absorber, a partial covering absorber and a neutral reflection component of $R=0.43\pm 0.07$. The powerlaw slope estimated by the authors are $\Gamma=1.79$. In our study, the best fit model used is {\it tbabs*(apec+apec(2)+powerlaw(1)+zgauss(2)+ztbabs*zpcfabs*(zgauss+pexmon+powerlaw(2)))}. We do not detect any significant variability in $\nh$ for the source.


	\item {\bf NGC 5252:} \citet{2010A&A...516A...9D} studied this source with \chandra{}. The intrinsic powerlaw slope and absorption column density obtained are $\Gamma=1.4-1.5$, and $\nh=10^{22}\cmsqi$ respectively. No mention is made of a possible Compton hump. 
		
		{In our fits to the \textit{Chandra}-ACIS data, $\Gamma_{\rm Hard}$ is
very poorly constrained, but we obtained best fits when it is frozen
at 1.4 (i.e., forcing $\Gamma_{\rm Hard}$ to 1.66, as measured by Kawamuro et al.\ (2016) with
\textit{Suzaku}, yielded high values of $\chi^2_{\nu} \sim 1.2$).
This value consistent with previous fits to \textit{Chandra}
data done by Dadina et al.\ (2010), who measured $\Gamma = 1.4-1.5$.
Traditionally, such flat values of $\Gamma_{\rm Hard}$ culd be
attributed to unmodeled Compton Reflection Hump emission, but we did
not find any mention in the literature of a Compton Hump, and our fits
to the \textit{Suzaku} data did not require one.}

{In our study, the best fit model used is tbabs*(apec+apec(2)+powerlaw(1)+zpcfabs*ztbabs*(zgauss+pexmon+powerlaw(2))).
We obtained good fits in the soft band by using a single \textsc{apec}
component with a common temperature and normalization for all four
spectra.  $\Gamma_{\rm Soft}$ was poorly constrained, but an average
value of 1.32 allowed for good fits in all four spectra.  However, the
soft power-law normalization was found to vary: Satisfactory fits to
CA2, CA3, and CA4 were obtained with an average normalization of
$9.65\times10^{-5}$, but for CA1, which occurred ten years earlier,
this value yielded a poor fit.  CA1 required a normalization of
$4.50\pm0.43 \times10^{-5}$ (and in the MultiNest runs, this parameter
is left free in CA1).}

{Partial coverers were required in the \textit{XMM-Newton} and
\textit{Suzaku} data only.  We tried to insert partial coverers into the
\textit{Chandra} fits, but could obtain reasonable constraints on
those parameters.  We note, though, that through using various values
of photon index or requiring/lacking partial coverers, that the
sequence of $\nh$ values in the ACIS data always consistently followed $\nh$(CA1) $<$ $\nh$(CA2) $\sim$ $\nh$(CA3) $>$ $\nh$(CA4), and $\nh$(CA1) $<$ $\nh$(CA4).}


\item {\bf NGC~5506:} \citet{2015MNRAS.447.3029M} studied this source using \nustar{} observations. They found that the spectrum is well fitted by $\Gamma=1.9$ intrinsically absorbed by $\nh=3.10\times 10^{22}\cmsqi$, and a distant reflection component and narrow Fe emission lines. \citet{2010MNRAS.406.2013G} studied this source using \xmm{} and measured a $\Gamma=1.9$. 
	
	In our study, we used the best fit model {\it tbabs*(ztbabs(2)*(apec+powerlaw(1))+zgauss(3)+ztbabs*(pexmon+powerlaw(1)+ \\ zgauss+zgauss(2)))}. Note that we required two different full covering absorbers. One that absorbs the central AGN and the other which absorbs both the AGN and the outer parts of the galaxy (stellar emission) with a lower absorbing column density. The outer absorption column density is $\nh \sim 10^{20}\cmsqi$, and the inner absorption column $\nh \sim 10^{22} \cmsqi$ . Most of the \chandra{} observations are piled-up and the pile-up could not be modeled using the pile-up kernel alone, hence we have reported only one \chandra{}-HETG observation which has lower pile-up. We detect significant variability of $\nh$ in this source, with a gradual increase in column density from 2002-2004. The \xmm{} observation of this source X9, has a long exposure ($\sim 132\ks$) with high SNR. The soft X-ray spectra ($<1.3\kev$) could be modeled by a single {\it Apec} model component with a higher normalization but a lower abundance of elements, $Z=0.0162\pm 0.0012$ relative to Solar, as also detected in previous studies \citep{2003A&A...402..141B}. We did not require any soft X-ray powerlaw for this observation. The measured value of $\nh=(3.00\pm 0.01)\times 10^{22}\cmsqi$ for this observation is consistent with those detected for observations from 2004-2015 indicating a constant absorption for that period of time.


\item {\bf NGC~6251:} This is a low excitation radio loud galaxy. \citet{2011ApJ...741L...4E} studied the \suzaku{} observation and detected a powerlaw of $\Gamma=1.82$ and two thermal plasma components, $kT=0.89\kev$, and $kT=2.63\kev$, and no reflection component in the hard X-rays. The authors however could not rule out a Compton-thick obscuration. In our analysis the best fit model used is {\it tbabs*(apec+powerlaw(1)+ztbabs*(pexmon+powerlaw(2)+zgauss))}. {The \chandra{}-ACIS observation has no sufficient SNR to constrain the column density. We do not detect any significant variability in $\nh$ for the source.}


\item {\bf NGC 6300:} \citet{2004ApJ...617..930M} studied the Seyfert 2 galaxy using \xmm{} and obtained a powerlaw $\Gamma=1.83\pm 0.08$ and a Compton-thin absorber of column density $\nh=2.2\times 10^{23} \cmsqi$. The authors could model the soft emission using a powerlaw. The relative reflection strength producing the Compton hump is estimated to be $R=1.1_{-0.6}^{+1.1}$ using {\it Pexrav} model. In our study we used the best fit model: {\tt tbabs*(apec+powerlaw(1)+ztbabs*(zgauss+pexmon+powerlaw(2)))}. We do not detect any significant variability in $\nh$ for the source.


\item {\bf NGC 7172:} \citet{2006ApJ...645..928A} studied the variability of this source in a sample of Seyfert 2 galaxies, and measured a $\Gamma=1.55\pm0.07$, and an absorption column of $\nh=(8.3\pm0.4)\times 10^{22}\cmsqi$. \citet{2001MNRAS.324..521A} have studied this source and detected a flatter slope of $\Gamma=1.64$ and a neutral absorption of $\nh=9.0\times 10^{22}\cmsqi$. However, they could not rule out the possiblity of an alternative scenario of a steeper slope of $\Gamma=1.78$ and a reflection component of $R=1.2_{-0.9}^{+0.7}$. In our work we found that the \chandra{} observations are mostly piled up. The best fit model we have used is {\it tbabs*(powerlaw(1) + apec(2) + ztbabs * zpcfabs * (zgauss+pexmon+powerlaw(2)))}. We do not detect any significant variability in $\nh$ for the source.


\item {\bf NGC7314:} \citet{2011A&A...535A..62E} studied the source using \xmm{} and \suzaku{} and found that the source shows rapid short term variability. No Compton reflection was detected. A powerlaw slope of $\Gamma=2.14$ was measured along with a neutral absorber column of $\nh=2.9\times 10^{21}\cmsqi$. \citet{2005ApJ...625L..31D} studied this source as a part of a small sample of obscured NLSy1 galaxies and found a reflection component of $R=2.83$, $\Gamma=2.19$, and no intrinsic absorption.In our study, we used the best fit model {\it tbabs*(apec+powerlaw+ztbabs*(zgauss+pexrav))}. We do not detect any significant variability in $\nh$ for the source.


\item {\bf NGC 7582:} This is a star-burst dominated galaxy with a Seyfert 2 nucleus at the centre. \citet{2015ApJ...815...55R} studied the source using \nustar{} and found that the source is variable with strong reflection features. The obscuring torus is patchy with a covering fraction of $80-90\%$ with a column density of $3.6\times 10^{24}\cmsqi$. Another full covering absorber was also needed with a column density of $\sim 3-12\times 10^{23}\cmsqi$. The authors modeled the Compton hump with a Pexrav model with a best fit reflection $R=4.3$, which is much higher than expected. The authors suggest that the geometry of the reflecting material is not that of a flat disk. The powerlaw photon index obtained by them is $\Gamma=1.78\pm 0.07$. \citet{2009ApJ...695..781B} studied this source with \suzaku{}. The source is characterized by very rapid changes of the column density of an inner absorber, which makes the authors conclude the presence of complex absorbing system and not just a simple torus followed from the unification model.

	{In our study the best fit model used is {\it tbabs*(apec + apec(2) + apec(3) + powerlaw + zpcfabs * ztbabs * (zgauss + pexrav + zgauss(2) + zgauss(3) + zgauss(4))) }. We detected variability of $\nh$ in this source both using \xmm{} and \suzaku{} observations. However, the \suzaku{} estimated values of $\nh$ are different with respect to each other at only 90\% error margins obtained using ISIS, and do not qualify our `variability' criteria described in Section \ref{subsec:variablesources}. Due to high SNR in the \xmm{} observations, we detected discrete emission lines at $2.5\kev$ (Sulphur K-$\alpha$) and $6.4\kev$, $7.12\kev$ and $7.5\kev$ indicative of FeK$\alpha$ and higher ionization of Fe emission lines. The observations X-1 and X-2 required a partial covering absorption, while X-3 and X-4 did not require them. When we used a {\it zpcfabs} model for X-3 and X-4 and froze the parameter values to those obtained in X-1, we found that fit worsened by $\dc=200$. When left free the value of $\nh^{\rm pc}$  went to zero with an improvement in statistics. We required three Apec models for this source, possibly because this source is dominated by stellar emission from different regions in the galaxy (with different temperatures and emissivity) in the soft X-rays. }

\end{enumerate}


\clearpage
\section{Appendix B: CygA \chandra{} observations}
In this section we list the best fit parameters obtained from the fits to the \chandra{} observation of Cyg~A (See Table \ref{Table:cygA}). 

\begin{table}

{\footnotesize
\centering
\setcounter{table}{0}
	\caption{The Chandra observations of Cyg-A and the best fit absorption column densities, Apec temperatures and best fit statistic. The powerlaw slope has been fixed at $\Gamma=1.7$ in all cases.}\label{Table:cygA}
  \begin{tabular}{lllllllllllllll} \hline\hline

	  Chandra obs-id	&Date of obs	&$\nh$	&Apec-KT	&Apec-KT		&Fluxes$^a$		&$\chi^2/\chi^2_{\nu}$ \\ 
				&		&$(\times 10^{22}\cmsqi)$	&($\kev$)	&($\kev$)&$\times 10^{-11}\funit$	& \\ \hline \\


	  359& 2000-03-08 &	$23.6^{+5.0}_{-4.1}$   	 	& 1.1 &4.6  & 8.40 		& $25.93/0.960$	\\
	  360 &2000-05-21 &	$23.6^{+1.5}_{-1.4}$	   	& 0.26 &1.9 & 3.52      	&  $375.92/0.989$	\\
	  1707& 2000-05-26&	$19.6^{+1.6}_{-1.4}$       	& 0.28 &2.0 & 7.28      	&$170.69/0.938$		\\
	  6225&2005-02-15 &	$21.9^{+1.5}_{-1.8}$       	& 0.92 &3.9 & 3.77      	&$298.73/1.205$	\\
	  5831& 2005-02-16 &	$22.6^{+1.2}_{-1.1}$       	& 0.78 &3.3 & 4.22      	&$448.74/1.039$	\\
	  6226 &2005-02-19 &	$23.7^{+1.8}_{-1.6}$      	& 1.1 &3.5  & 4.23       	&$270.43/1.104$ \\
	  6250	&2005-02-21 &	$25.4^{+4.3}_{-3.5}$      	& 0.43& 3.3 & 4.57      	&$80.97/0.976$ \\ 
	  5830 	&2005-02-22 &	$25.9^{+2.1}_{-2.0}$      	& 0.43& 4.0 & 4.14      	&$265.78/1.089$\\
	  6229 &2005-02-23 &	$25.3^{+1.8}_{-1.8}$	  	& 0.69 &4.1 & 4.26      	&$228.11/0.939$	\\
	  6228 &2005-02-25 &	$23.5^{+2.1}_{ -2.0}$     	& 0.85 &2.7 & 4.50      	&$156.82/0.866$ \\
	  6252 	&2005-09-07 &	$22.8^{+1.6}_{-1.5}$      	& 0.93 &4.0 & 5.33      	&$350.30/1.052$ \\
	  17505 &2015-01-05 &	$26.4^{+1.0}_{ -0.8}$     	& 0.82 &4.9 & 5.21      	&$471.77/1.051$\\
	  17145 &2015-01-10 &	$25.6^{+1.0}_{ -0.9}$ 	  	& 0.97 &5.1 & 5.66      	&$420.56/0.960$	\\
	  17530 &2015-04-19 &	$24.2^{+3.1}_{ -2.6}$     	& 0.52 &4.7 & 2.85      	&$116.75/0.965$\\
	  17650 &2015-04-22 &	$29.7^{+3.8}_{-2.9}$      	& 1.06 &7.0 & 2.75      	&$161.40/1.121$	\\
	  17144 &2015-05-03 &	$28.1^{+2.1}_{-1.9}$      	& 0.74 &4.7 & 3.52      	&$236.37/0.879$\\
	  17528 &2015-08-30 &	$26.1^{+1.5}_{ -1.4}$     	& 0.88 &3.7 & 3.37      	&$390.46/1.004$\\
	  17143 &2015-09-03 &	$23.1^{+1.8}_{ -1.7}$	  	& 0.68 &2.7 & 3.32      	&$  228.42/1.053$ \\
	  17524 &2015-09-08 &	$26.9^{+2.4}_{ -2.2} $ 	  	& 0.66 &2.6 & 2.87      	&$196.28/ 1.175$	\\
	  18441 &2015-09-14 &	$26.3^{+2.2}_{-2.2}$  	  	& 0.86 &3.5 & 2.82      	&$186.22/ 1.029$	\\
	  17526 &2015-09-20 &	$25.4^{+1.1}_{-1.3}$ 	  	& 0.35 &3.3 & 3.34      	&$397.67/1.063$	\\
	  17527 &2015-10-11 &	$26.3^{+2.3}_{-2.3}$      	& 0.92 &4.5 & 3.74      	&$159.48/0.886$\\
	  18682 &2015-10-14 &	$24.4^{+3.3}_{-2.3}$      	& 0.79 &7.2 & 3.32      	&$142.93/0.928$\\
	  18641 &2015-10-15 &	$26.0^{+2.8}_{-2.7}$      	& 2.0 &4.0  & 3.16      	&$121.57/0.921$\\
	  18683 &2015-10-18 &	$24.2^{+3.8}_{-2.7}$     	& 0.69 &4.3 & 2.67		&$98.54/1.263$\\
	  17508 &2015-10-28 &	$27.7^{+2.9}_{-2.6}$      	& 1.0 &4.4  & 4.06      	&$144.79/0.894$	\\
	  18688 &2015-11-01 &	$27.8^{+1.7}_{-1.6}$      	& 0.54 &1.3 & 4.96      	&$344.60/0.999$	\\
	  18871 &2016-06-13 &	$29.1^{+2.4}_{-2.5}$      	& 0.44 &2.7 & 3.92      	&$200.73/0.947$	\\
	  17133 &2016-06-18 &	$25.3^{+2.1}_{-1.9}$      	& 0.85 &3.5 & 3.39      	&$298.38/1.033$	\\
	  17510 &2016-06-26 &	$26.2^{+1.7}_{-1.6}$      	& 1.03 &4.4 & 3.91	     	&$311.58/0.898$	\\
	  17509 &2016-07-10 &	$24.5^{+1.3}_{-1.2}$      	& 0.70 &3.6 & 3.74      	&$414.29/0.989$	\\
	  17518 &2016-07-16 &	$23.2^{+1.2}_{-1.4}$      	& 0.52 &4.0 & 3.62      	&$382.97/0.930$	\\
	  17521 &2016-07-20 &	$23.2^{+1.9}_{-1.8}$      	& 0.43 &2.0 & 3.78      	&$297.22/1.126$	\\
	  18886 &2016-07-23 &	$24.7^{+1.9}_{-1.8}$      	& 0.69 &2.6 & 4.20      	&$234.27/1.014$	\\
	  17138 &2016-07-25 &	$22.6^{+1.5}_{-1.6}$      	& 0.89 &2.9 & 3.97      	&$320.98/1.180$	\\
	  17513 &2016-08-15 &	$26.2^{+1.4}_{-1.4}$      	& 0.76 &4.0 & 4.20	     	&$428.77/1.033$	\\
	  17516 &2016-08-18 &	$24.3^{+1.4}_{-1.3}$      	& 0.67 &3.2 & 3.82      	&$418.26/1.023$	\\
	  17523 &2016-08-31 &	$23.5^{+1.7}_{ -1.6}$    	& 1.04 &4.8 & 2.67		&$258.93/0.838$	\\
	  17512 &2016-09-15 &	$24.9^{+1.2}_{-1.2}$      	& 0.82 &3.8 & 3.73      	&$445.67/0.975$	\\
	  17139 &2016-09-16 &	$27.3^{+1.9}_{-1.7}$      	& 0.54 &2.2 & 4.00      	&$361.65/1.005$	\\
	  17517 &2016-09-17 &	$25.2^{+2.0}_{-1.9}$      	& 1.11 &4.1 & 3.65      	&$253.62/0.998$	\\
	  19888 &2016-10-01 &	$25.4^{+2.1}_{-1.9}$      	& 0.43 &5.5 & 4.17      	&$194.71/0.969$	\\
	  17140 &2016-10-02 &	$27.4^{+2.0}_{-1.8}$      	& 0.44 &2.2 & 4.22      	&$304.83/0.929$	\\
	  17507 &2016-11-12 &	$25.6^{+1.9}_{ -1.7}$    	& 1.06 &4.3 & 3.49		&$319.56/1.055$	\\
	  17520 &2016-12-06 &	$31.0^{+3.0}_{-2.6}$      	& 0.74 &10.9& 3.58      	&$209.14/0.890$	\\
	  19956 &2016-12-10 &	$26.4^{+1.6}_{-1.4}$      	& 0.71 &5.0 & 3.47      	&$391.13/0.959$	\\
	  17514 &2016-12-13 &	$24.1^{+1.4}_{ -1.3}$    	& 0.72 &2.8 & 3.54		&$407.98/0.990$	\\
	  17529 &2016-12 15 &	$25.1^{+1.8}_{ -1.7}$    	& 0.98 &1.7 & 3.86		&$314.61/0.948$	\\	
	17519 &2016-12-19 &	$28.2^{+2.3}_{ -2.0}$    	& 0.17 &4.3 & 3.86		&$7300.66/1.051$	\\
	17135 &2017-01-20 &	$29.1^{+2.6}_{ -2.4}$    	& 0.70 &3.8 & 4.15		&$182.04/0.910$	\\
	17136 &2017-01-26 &	$26.0^{+2.3}_{ -2.0}$    	& 0.12 &3.2 & 3.58		&$216.93/1.028$	\\
	19996 &2017-01-28 &	$26.3^{+2.0}_{ -1.8}$    	& 0.54 &4.4 & 3.32		&$282.59/1.075$	\\ \hline \\

\end{tabular}                                             
   
   {$^a$ The $2-10\kev$ unasborbed power law fluxes.}
	                                                  
}                                                         
\end{table}                                               

\begin{table}

{\footnotesize
\centering

\setcounter{table}{0}
	\caption{The Chandra observations of Cyg-A and the best fit absorption column densities, Apec temperatures and best fit statistic. The powerlaw slope has been fixed at $\Gamma=1.7$ in all cases.}

   \begin{tabular}{lllllllllllllll} \hline\hline

	  Chandra obs-id	&Date of obs	&$\nh$				&Apec-KT	&Apec-KT	&Fluxes		&$\chi^2/\chi^2_{\nu}$  \\
				&		&$(\times 10^{22}\cmsqi)$	&($\kev$)	&($\kev$)	&$\times 10^{-11}\funit$	& \\ \hline \\

	  19989 &2017-02-12 &	$26.5^{+1.6}_{-1.6}$     	& 0.68 &3.9 & 3.48		&373.15/1.014	\\
	  17515 &2017-03-21 &	$24.9^{+1.7}_{-1.5}$      	& 0.7 &4.2  & 3.23       	&295.30/0.937	\\
	  20043 &2017-03-25 &	$27.7^{+2.3}_{-2.1}$      	& 1.0 &6.9  & 3.14       	&194.67/0.801	\\
	  20044 &2017-03-26 &	$22.6^{+2.3}_{-2.1}$      	& 0.12 &2.4 & 3.47      	&140.45/1.025	\\
	  17137 &2017-03-29 &	$26.1^{+2.0}_{-2.0}$      	& 1.4 &5.0  & 3.20       	&215.37/1.002	\\
	  17522 &2017-04-08 &	$27.5^{+1.6}_{ -1.5}$    	& 0.72 &5.9 & 3.49		&336.64/0.879	\\
	  20059 &2017-04-19 &	$30.0^{+4.8}_{ -2.7}$    	& 0.27 &4.3 & 2.95		&114.31/1.099	\\
	  17142 &2017-04-20 &	$27.0^{+3.4}_{ -2.7}$    	& 0.54 &5.0 & 2.88		&174.94/1.346	\\
	  17525 &2017-04-22 &	$24.4^{+1.8}_{-2.3}$      	& 1.22 &4.1 & 2.62      	&191.00/1.151	\\
	  20063 &2017-04-22 &	$28.1^{+3.7}_{-2.6}$      	& 0.88 &6.3 & 3.18      	&193.57/1.030	\\
	  17511 &2017-05-10 &	$24.3^{+2.8}_{-2.2}$      	& 0.44 &2.4 & 3.46      	&131.70/0.808	\\
	  20077 &2017-05-13 &	$24.6^{+2.}_{-1.9}$       	& 0.21 &4.2 & 3.25      	&272.28/1.072	\\
	  20048 &2017-05-19 &	$26.9^{+3.2}_{ -2.7}$    	& 0.70 &3.5 & 2.78		&119.64/0.989	\\
	  17134 &2017-05-20 &	$25.1^{+1.8}_{-1.7}$      	& 1.1 &4.3  & 3.46       	&270.15/0.986	\\
	  20079 &2017-05-21 &	$25.8^{+2.8}_{-2.3}$      	& 0.74 &2.8 & 3.13      	&229.10/1.046	\\ \hline \\

\end{tabular}

}
\end{table}



\section{Appendix C: The spectral overplots}

\clearpage

\begin{figure*}                                             
  \centering

		\hbox{

\includegraphics[width=5.2cm,angle=-90]{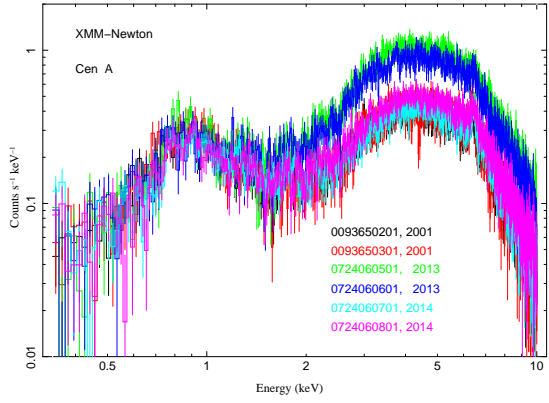} 

	}	
	
\setcounter{figure}{0}
	\caption{The overplot of the spectra of Cen~A. Here the spectra have been binned by a factor of four for plotting and visual purposes only.} \label{Fig:CenA}
\end{figure*}


\begin{figure*}
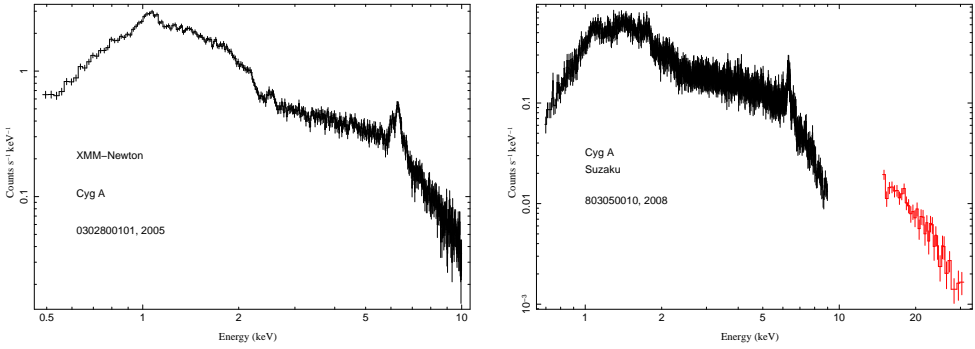
                                            

  \centering

	\hbox{

\includegraphics[width=4.5cm,angle=-90]{Plots_29thOct2017/CygA/CygA_XMM_comparison.ps} 
\includegraphics[width=4.5cm,angle=-90]{Plots_29thOct2017/CygA/CygA_Suzaku_comparison.ps}
		
	}

\setcounter{figure}{1}
	\caption{Same as Fig. \ref{Fig:CenA}, except for the source which is Cyg~A. } \label{Fig:CygA}
\end{figure*}


\begin{figure*}
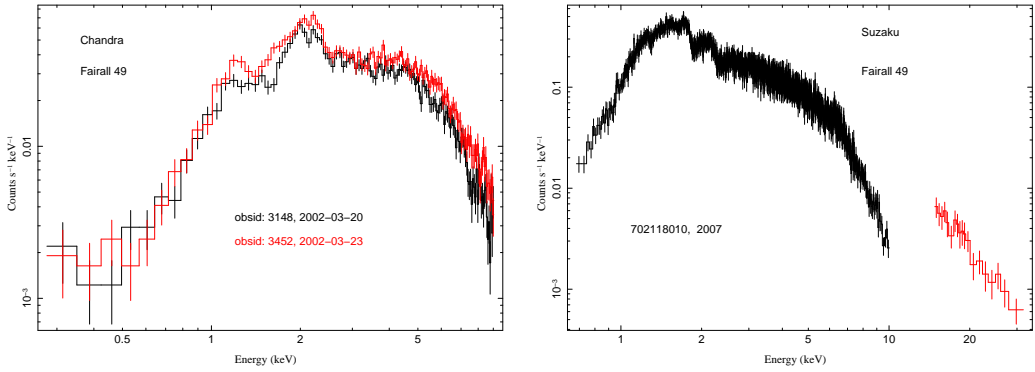
                                                    

  \centering

	\hbox{

\includegraphics[width=4.8cm,angle=-90]{Plots_29thOct2017/Fairall49/Fairall49_Chandra_Compare.ps}
\includegraphics[width=4.8cm,angle=-90]{Plots_29thOct2017/Fairall49/Fairall49_Suzaku_compare.ps}	
	}	

\setcounter{table}{2}
\caption{Same as Fig. \ref{Fig:CenA}, except for the source which is Fairall 49  } \label{Fig:Fairall49}
\end{figure*}


\begin{figure*}
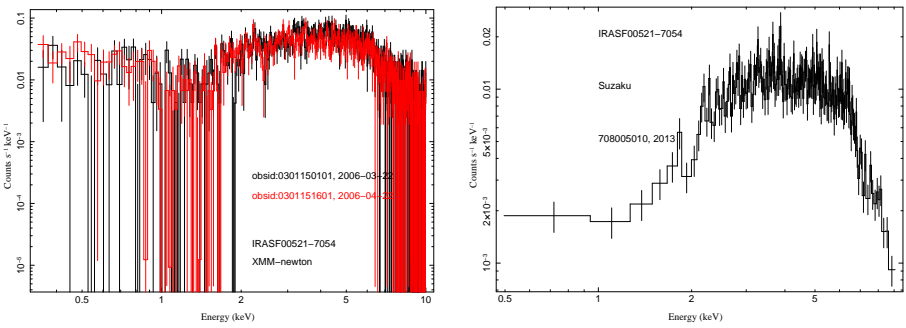
                                                 

  \centering

	\hbox{

\includegraphics[width=4.2cm,angle=-90]{Plots_29thOct2017/IRASF00521/IRASF00521_XMM_comparison.ps} 
\includegraphics[width=4.2cm,angle=-90]{Plots_29thOct2017/IRASF00521/IRASF00521_Suzaku_comparison.ps}
	}

\caption{Same as Fig. \ref{Fig:CenA}, except for the source which is IRASF00521. } \label{Fig:IRASF00521}
\end{figure*}


\clearpage

\begin{figure*}
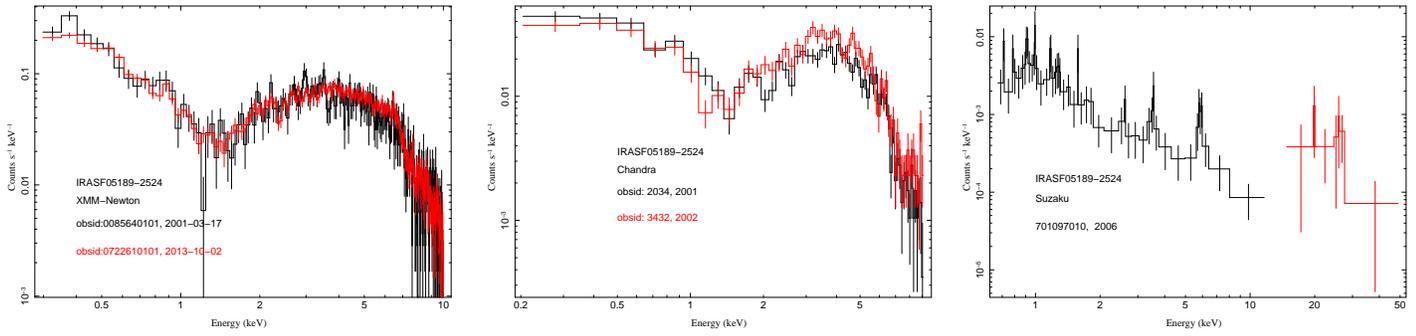
                                          

  \centering

		\hbox{

\includegraphics[width=4.3cm,angle=-90]{Plots_29thOct2017/IRAS05189/IRAS05189_XMM_comparison.ps} 
\includegraphics[width=4.3cm,angle=-90]{Plots_29thOct2017/IRAS05189/IRASF05189_Chandra_Compare.ps}
\includegraphics[width=4.3cm,angle=-90]{Plots_29thOct2017/IRAS05189/IRASF05189_Suzaku_compare.ps}
	
	}

	\caption{Same as Fig. \ref{Fig:CenA}, except for the source which is IRASF05189} \label{Fig:IRASF05189}
\end{figure*}


\begin{figure*}                                          
  \centering

	\hbox{

\includegraphics[width=5cm,angle=-90]{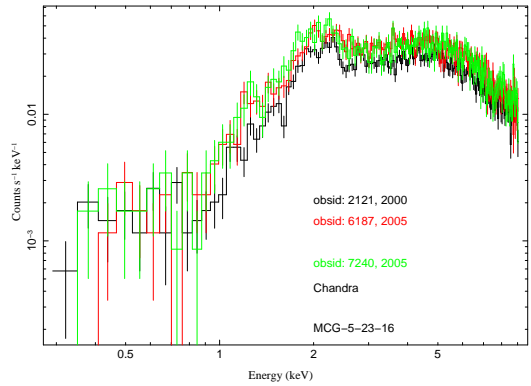}

	}

	\caption{Same as Fig. \ref{Fig:CenA}, except for the source which is MCG-5-23-16.} \label{Fig:MCG5}
\end{figure*}


\begin{figure*}
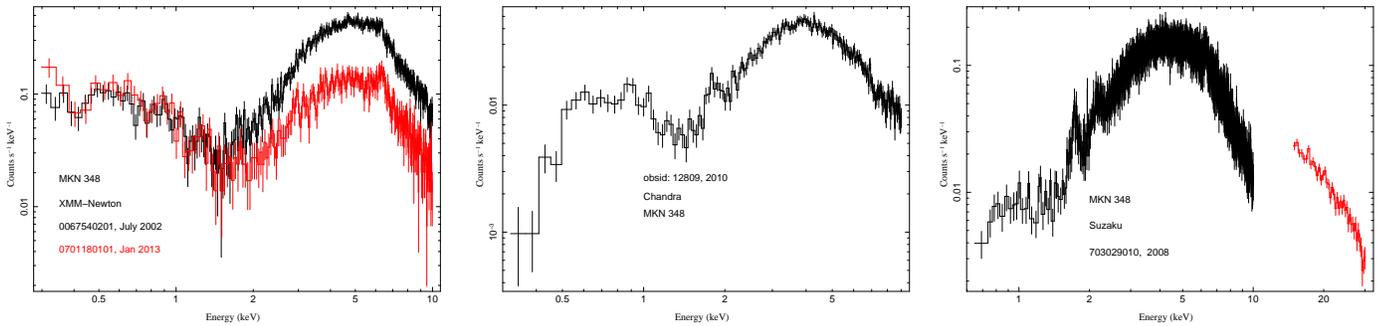
                                              
  \centering

	\hbox{

\includegraphics[width=4.2cm,angle=-90]{Plots_29thOct2017/MKN348/MKN348_XMM_comparison.ps} 
	\includegraphics[width=4.2cm,angle=-90]{Plots_29thOct2017/MKN348/MKN348_Chandra_Compare.ps}
\includegraphics[width=4.2cm,angle=-90]{Plots_29thOct2017/MKN348/MKN348_Suzaku_compare.ps}	
	}

\caption{Same as Fig. \ref{Fig:CenA}, except for the source which is MKN 348. } \label{Fig:MKN348}
\end{figure*}


\begin{figure*}                                             

  \centering

	\hbox{

\includegraphics[width=4.2cm,angle=-90]{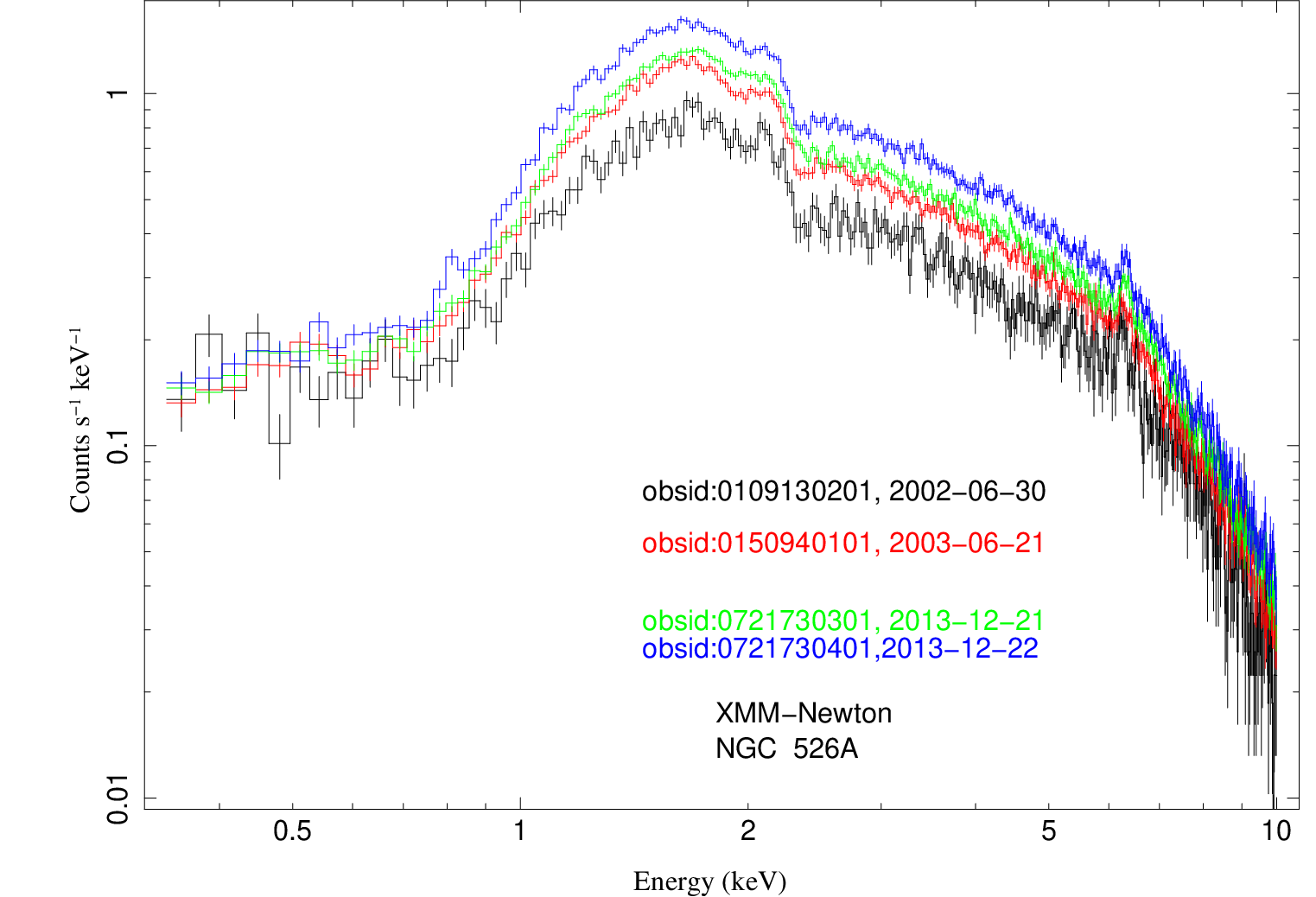} 
\includegraphics[width=4.2cm,angle=-90]{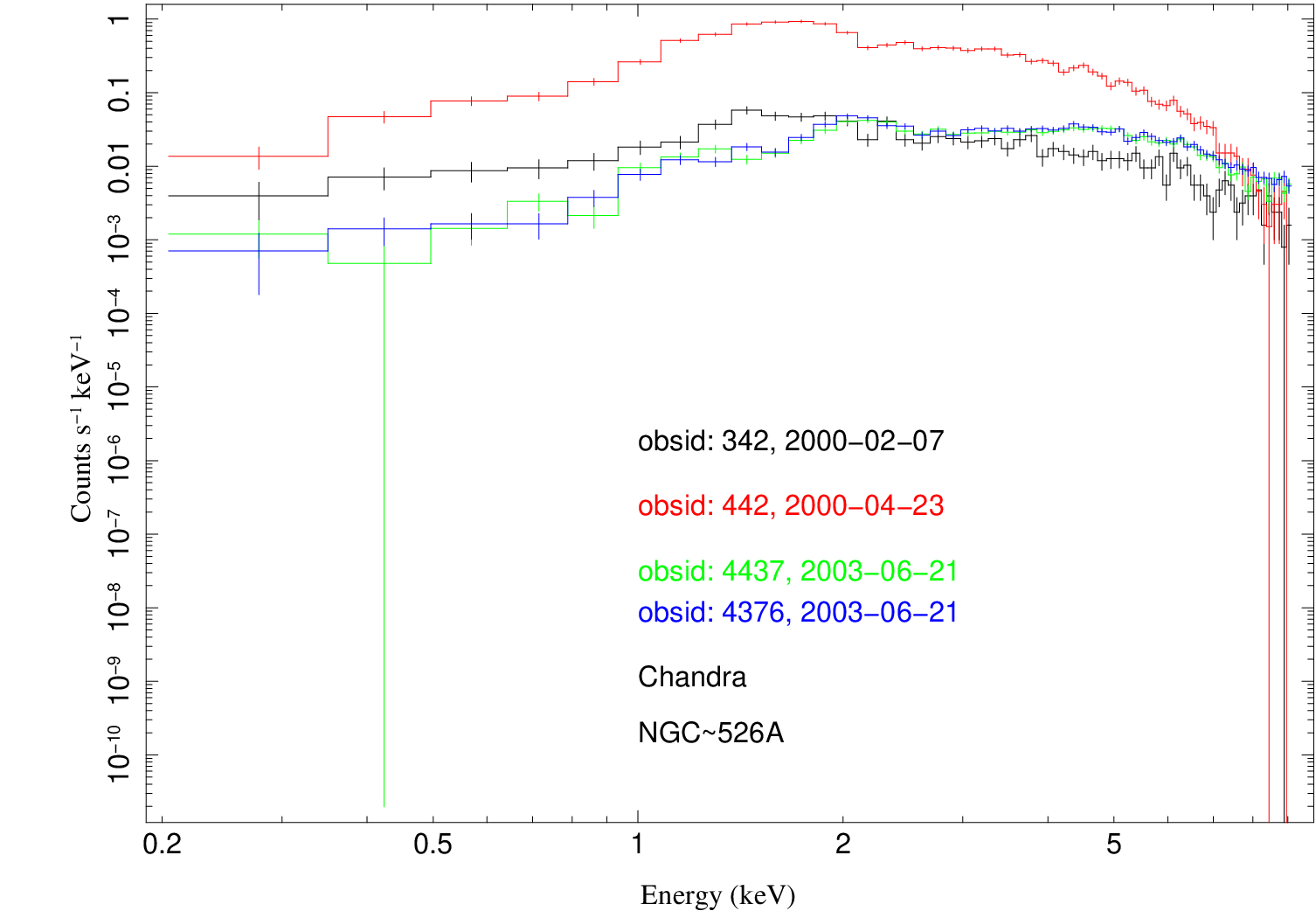}
\includegraphics[width=4.2cm,angle=-90]{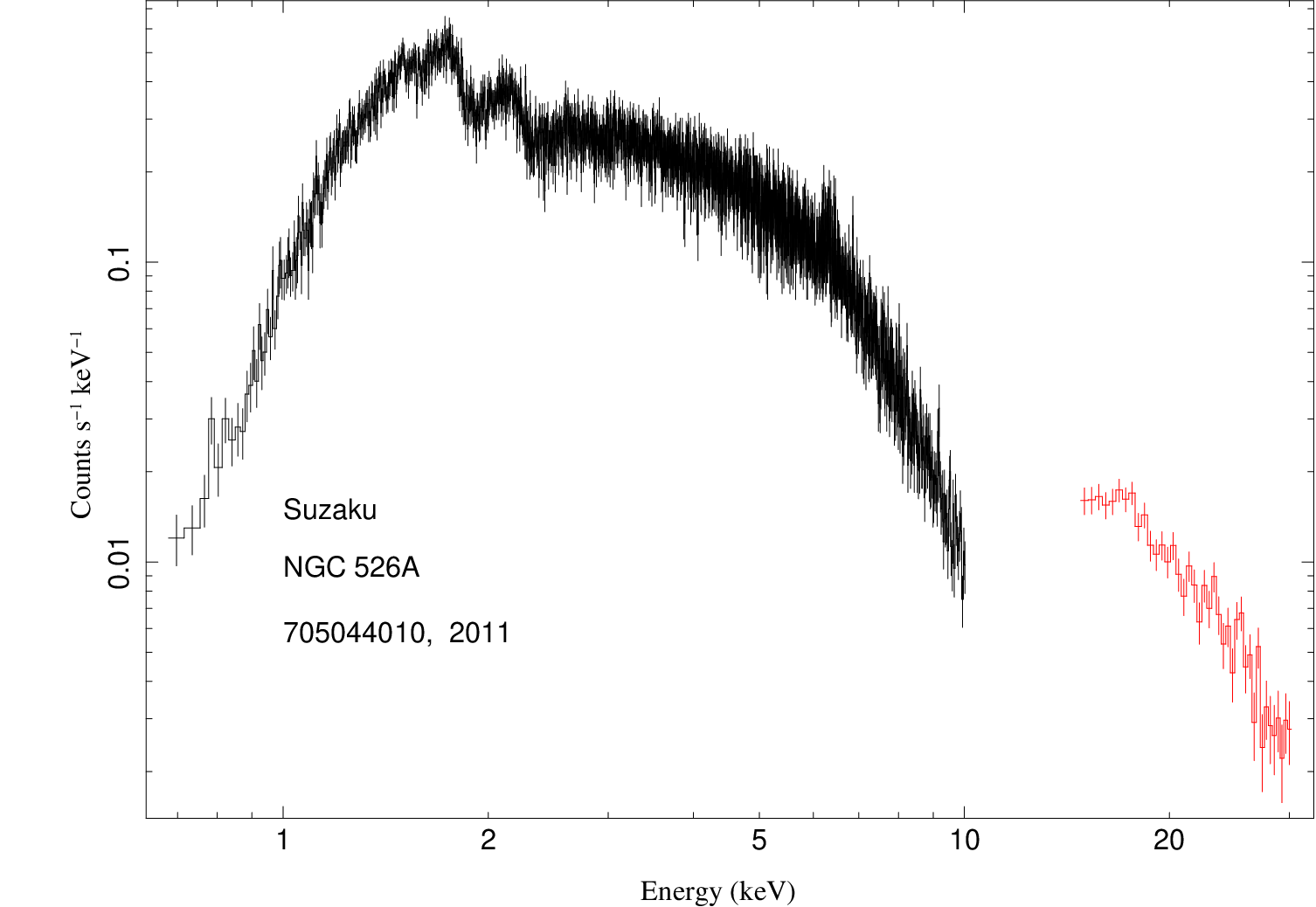}	
	}

\caption{Same as Fig. \ref{Fig:CenA}, except for the source which is NGC 526A} \label{Fig:NGC526A}
\end{figure*}


\clearpage

\begin{figure*}                                     

  \centering

		\hbox{

\includegraphics[width=4.5cm,angle=-90]{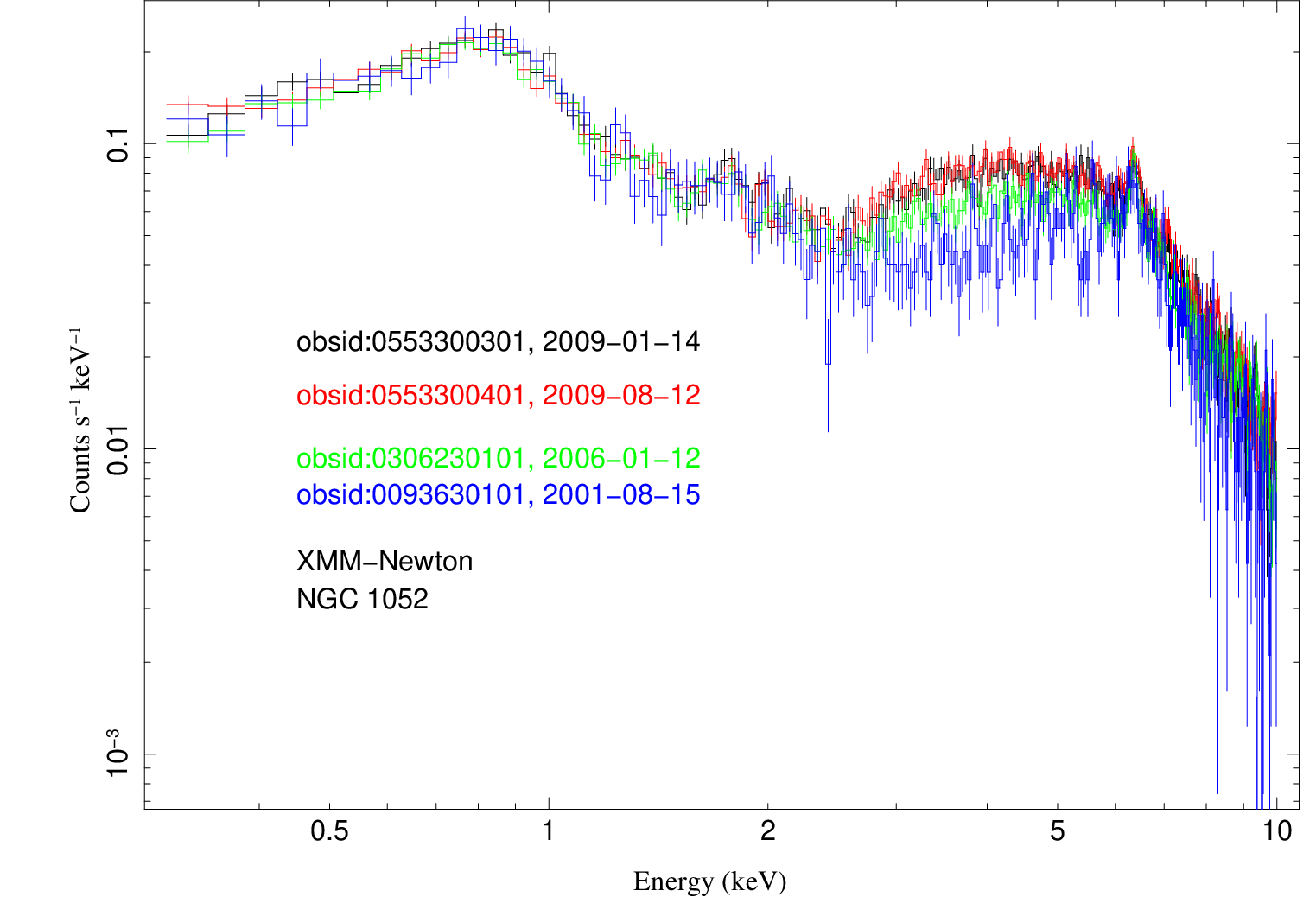} 
\includegraphics[width=4.5cm,angle=-90]{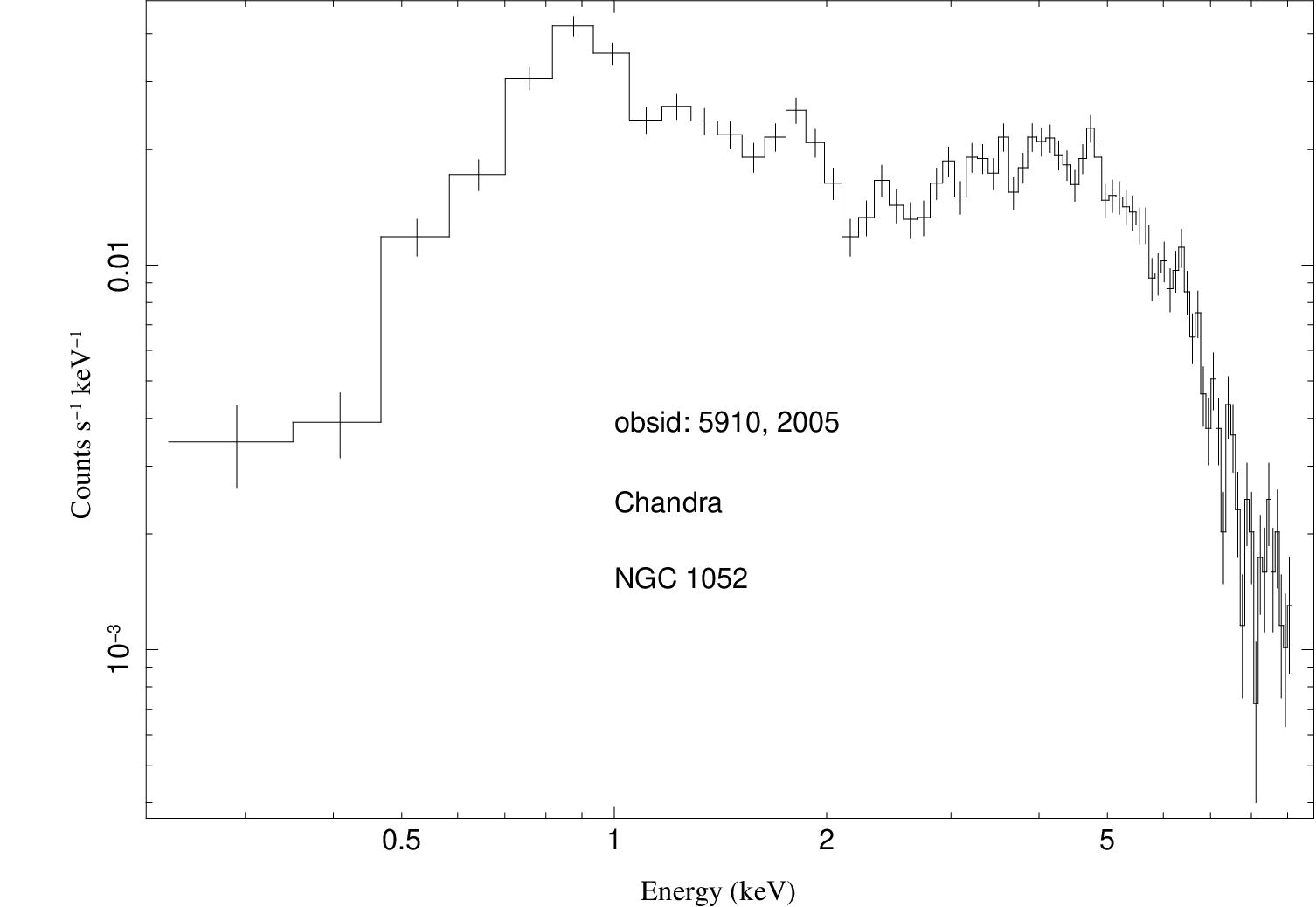}
\includegraphics[width=4.5cm,angle=-90]{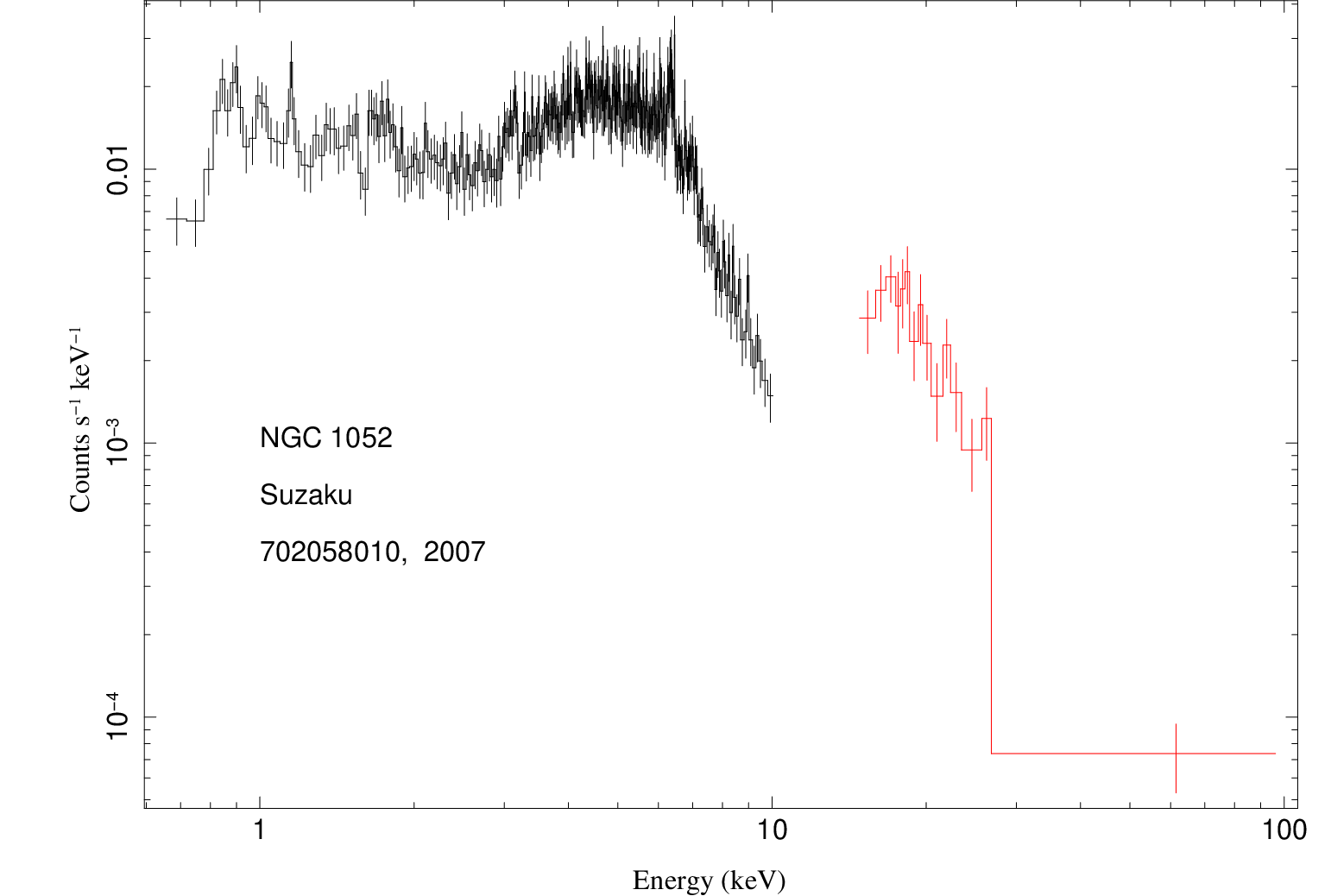}
	
	}

	\caption{Same as Fig. \ref{Fig:CenA}, except for the source which is NGC 1052.} \label{Fig:NGC1052}
\end{figure*}


\begin{figure*}                                          

  \centering

	\hbox{

\includegraphics[width=4.2cm,angle=-90]{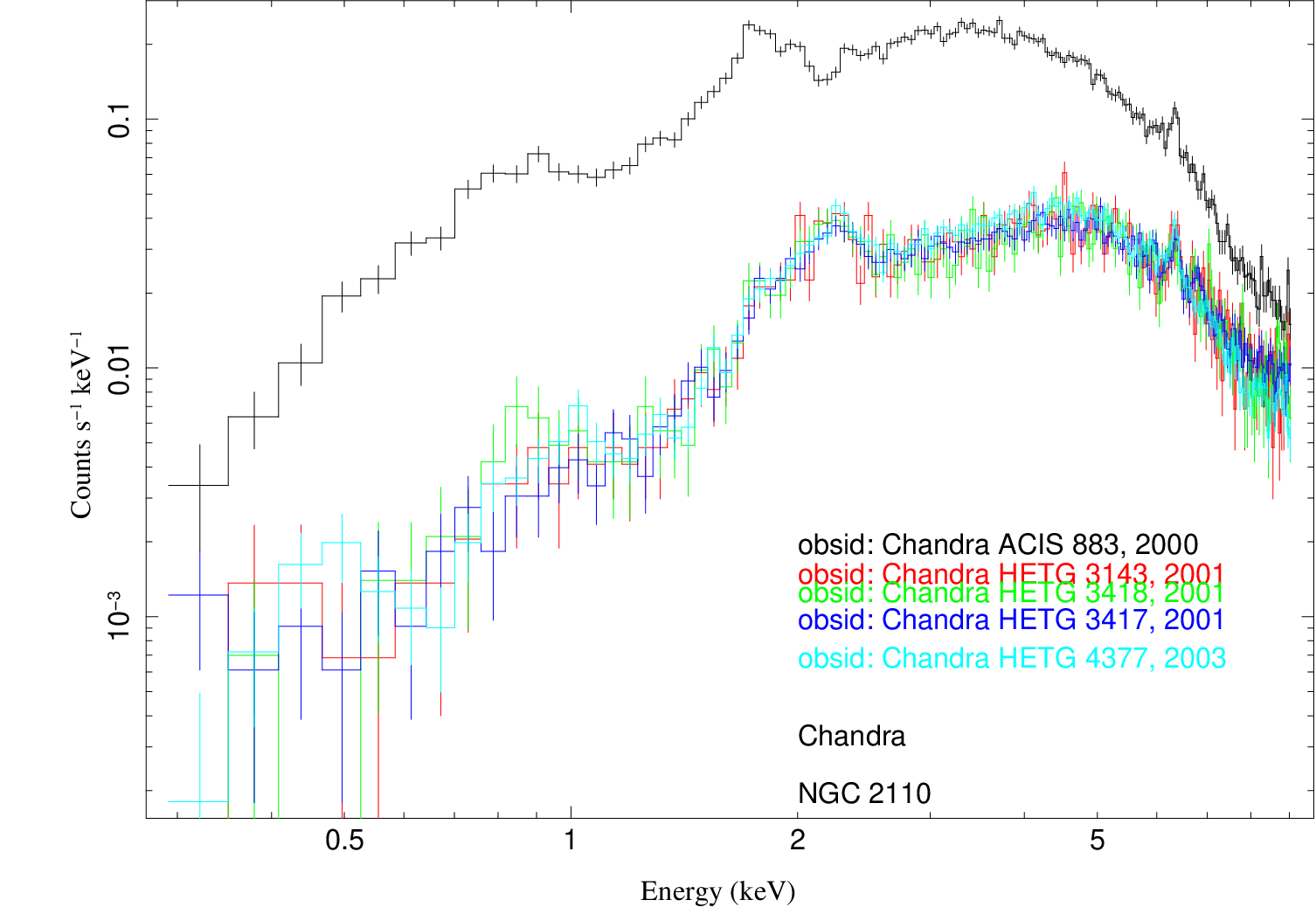} 
\includegraphics[width=4.2cm,angle=-90]{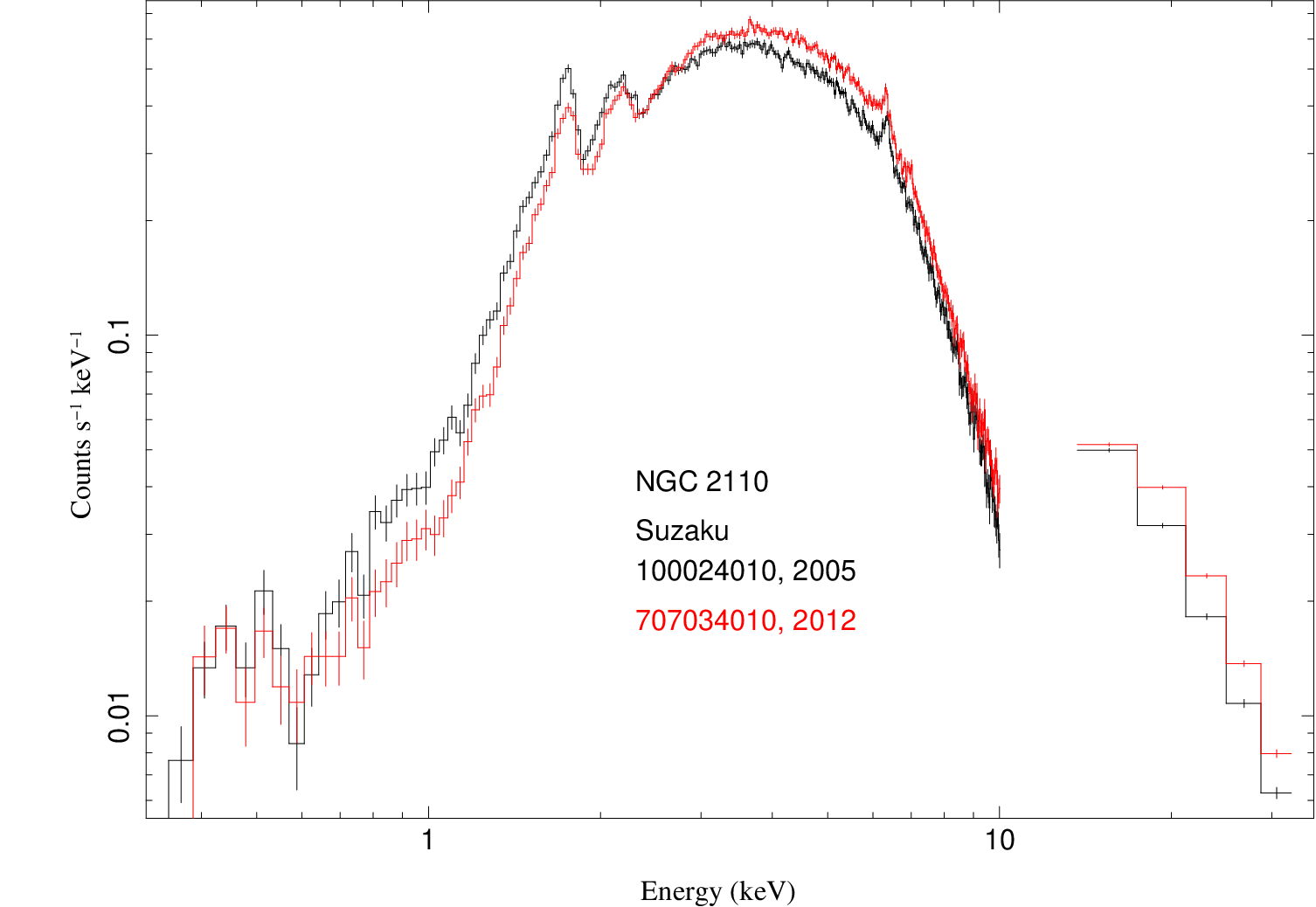}
	
	}

	\caption{Same as Fig. \ref{Fig:CenA}, except for the source which is NGC 2110. } \label{Fig:NGC2110}
\end{figure*}


\begin{figure*}                                            

  \centering

	\hbox{

\includegraphics[width=4.2cm,angle=-90]{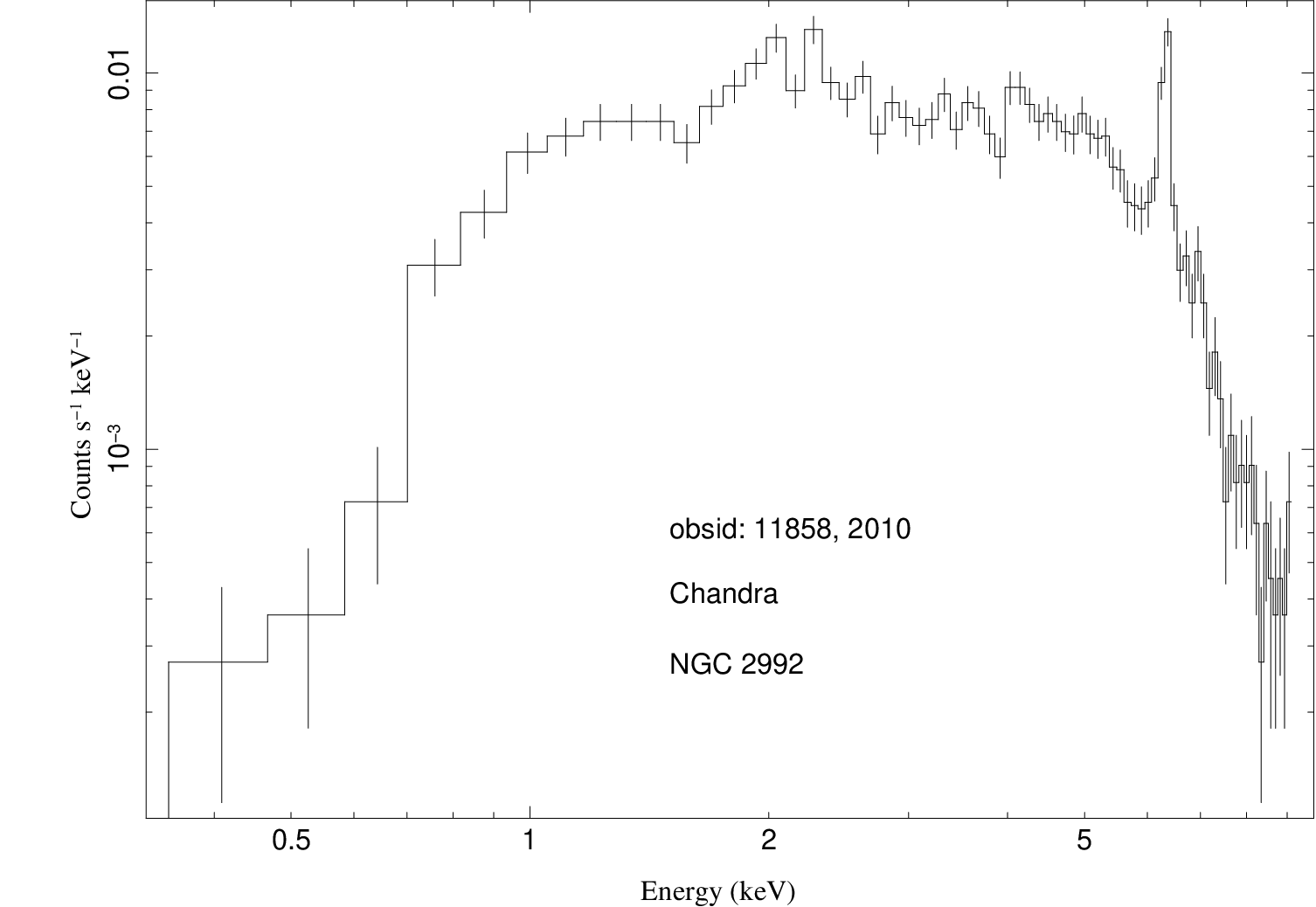}
\includegraphics[width=4.2cm,angle=-90]{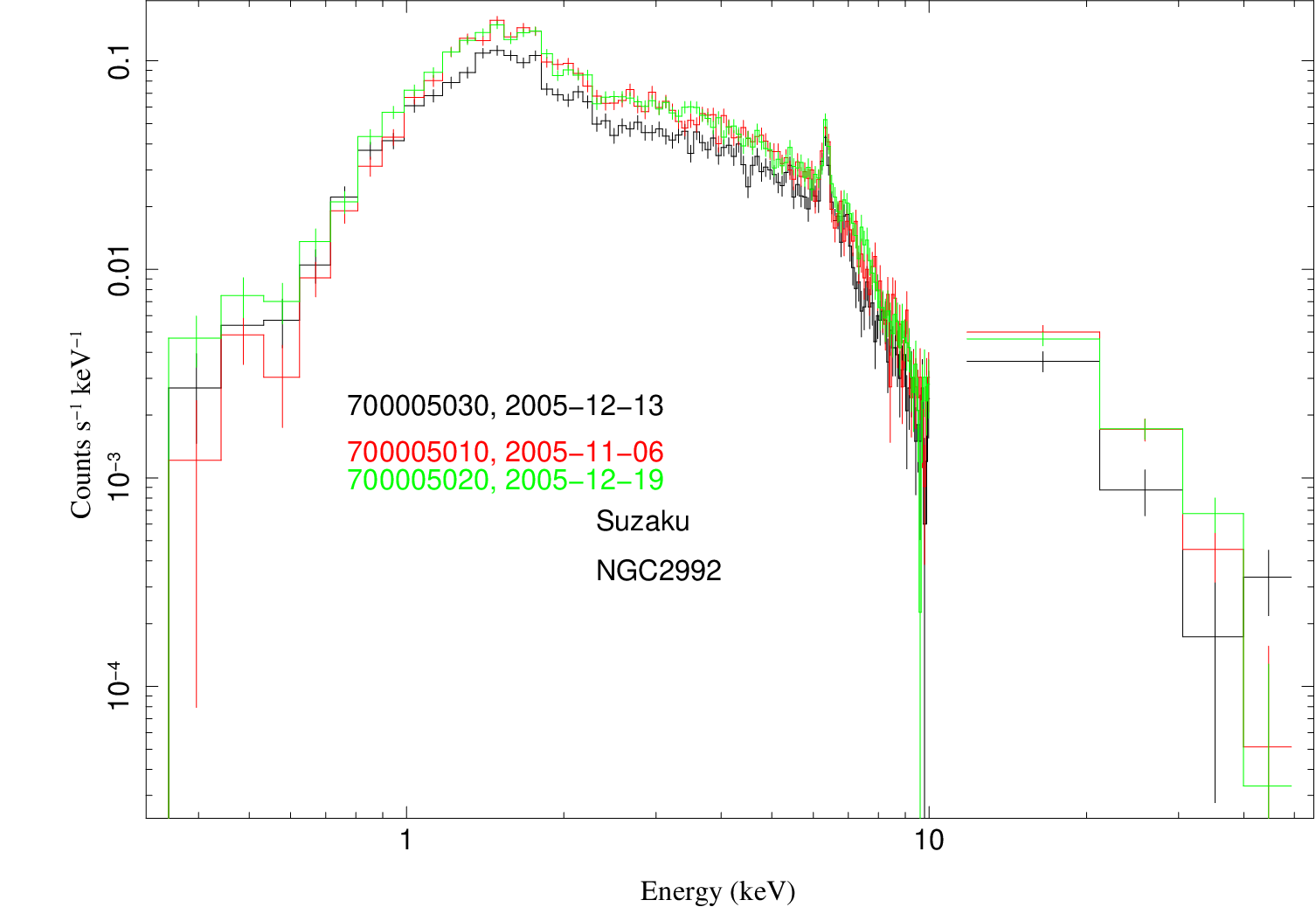}	
	}

\caption{Same as Fig. \ref{Fig:CenA}, except for the source which is NGC 2992. } \label{Fig:NGC2992}
\end{figure*}


\begin{figure*}                                          

  \centering

	\hbox{

\includegraphics[width=4.2cm,angle=-90]{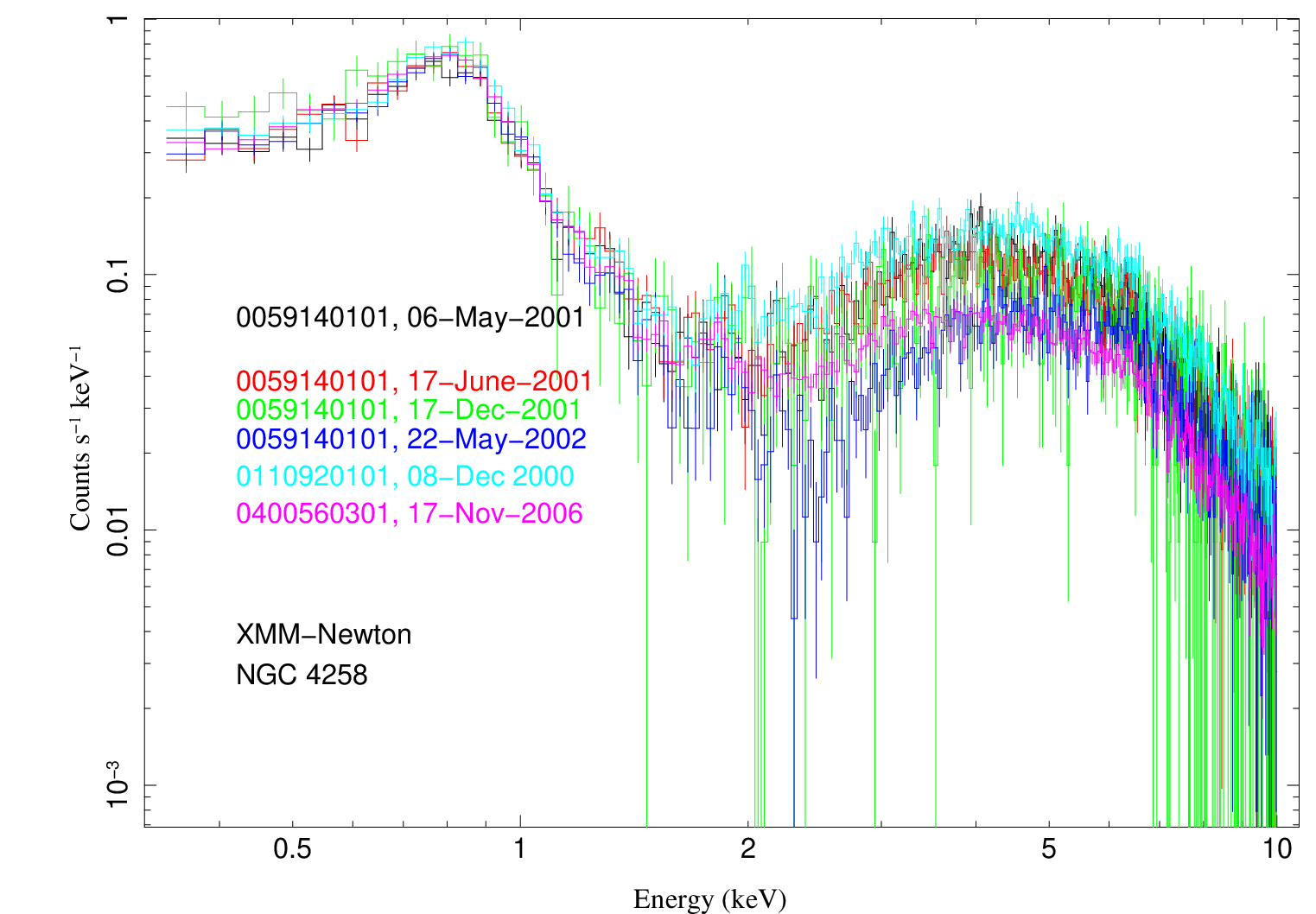} 
\includegraphics[width=4.2cm,angle=-90]{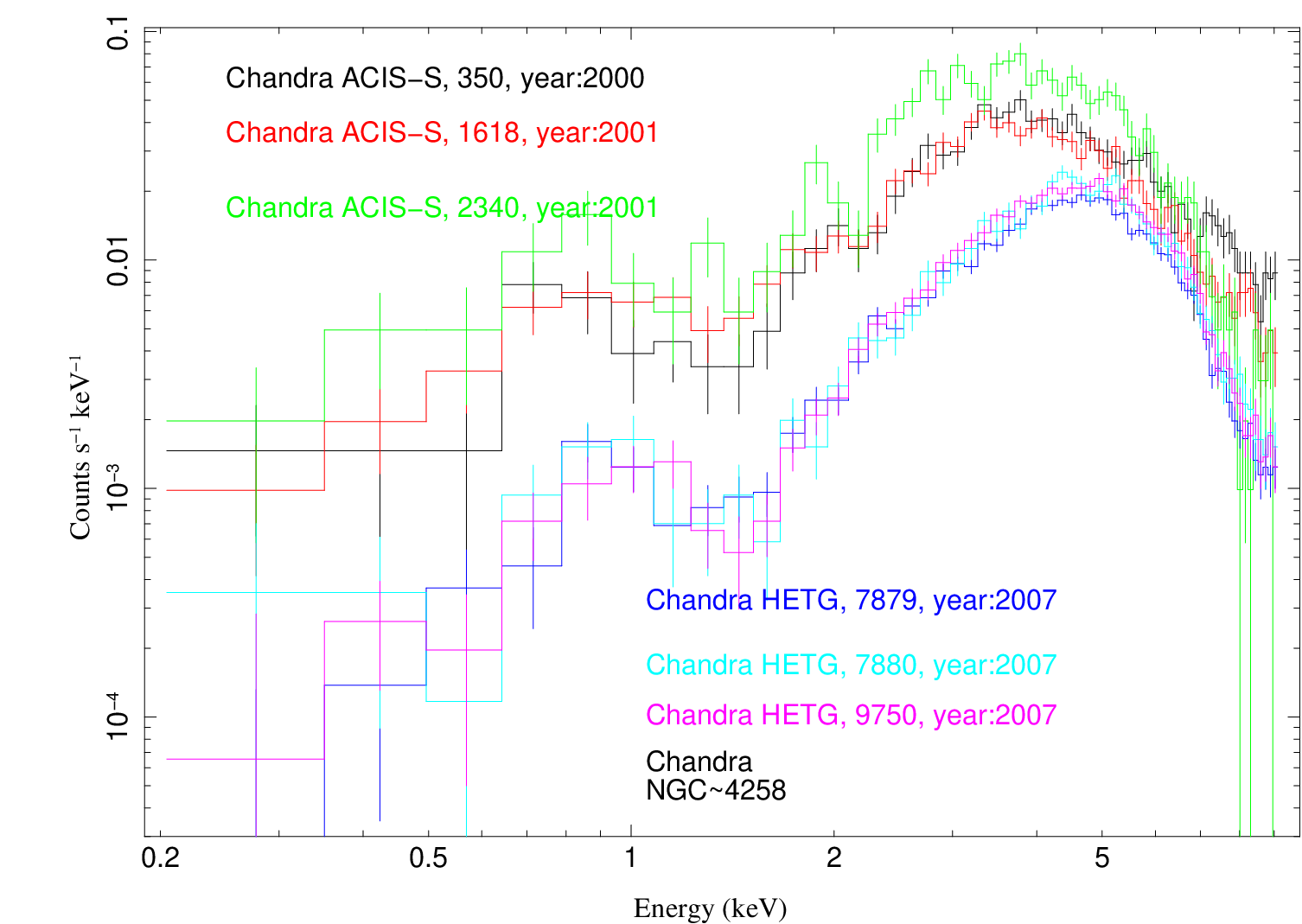}
\includegraphics[width=4.2cm,angle=-90]{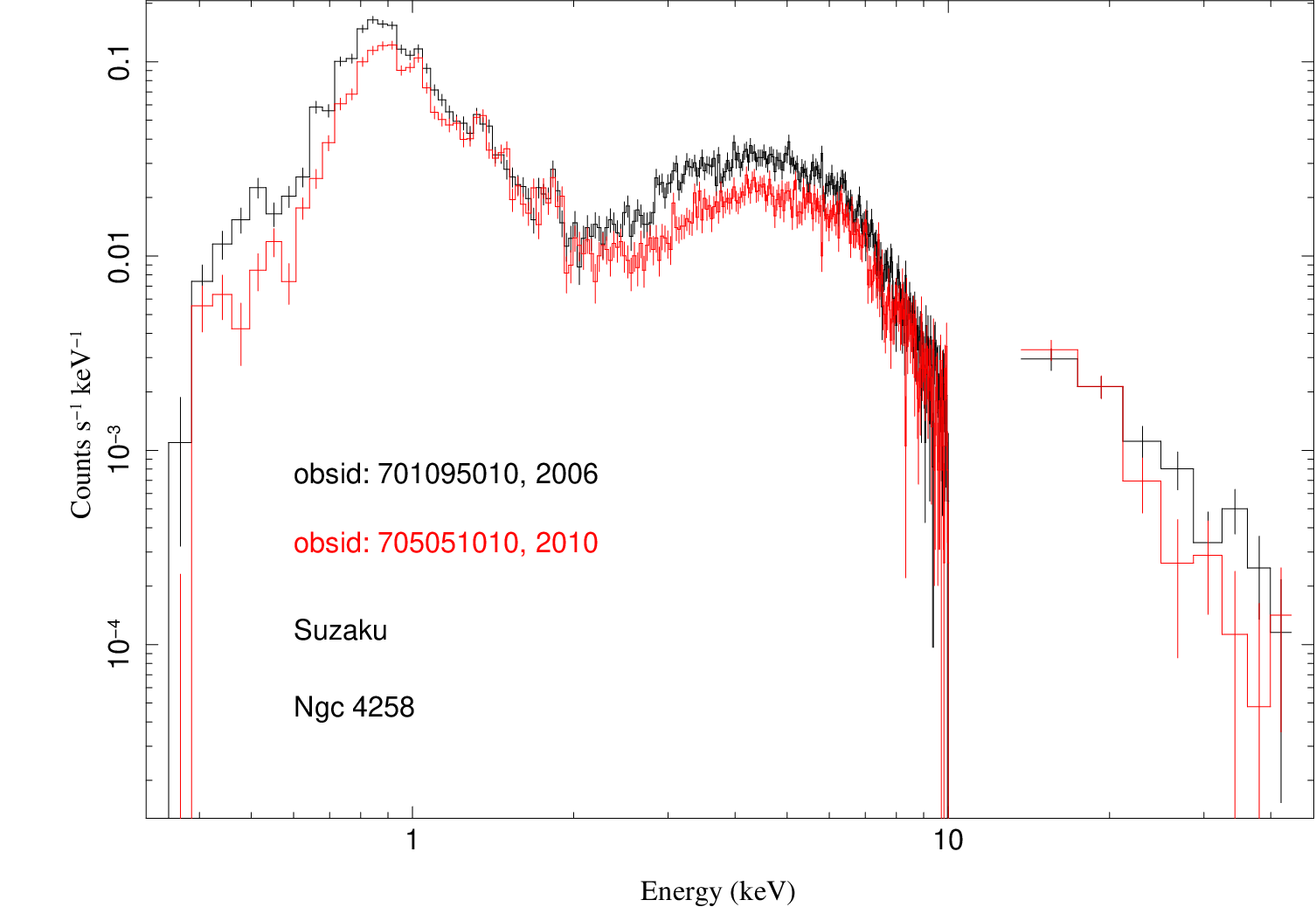}	
	}

\caption{Same as Fig. \ref{Fig:CenA}, except for the source which is NGC 4258} \label{Fig:NGC4258}
\end{figure*}


\clearpage

\begin{figure*}                                        

  \centering

		\hbox{

\includegraphics[width=4.2cm,angle=-90]{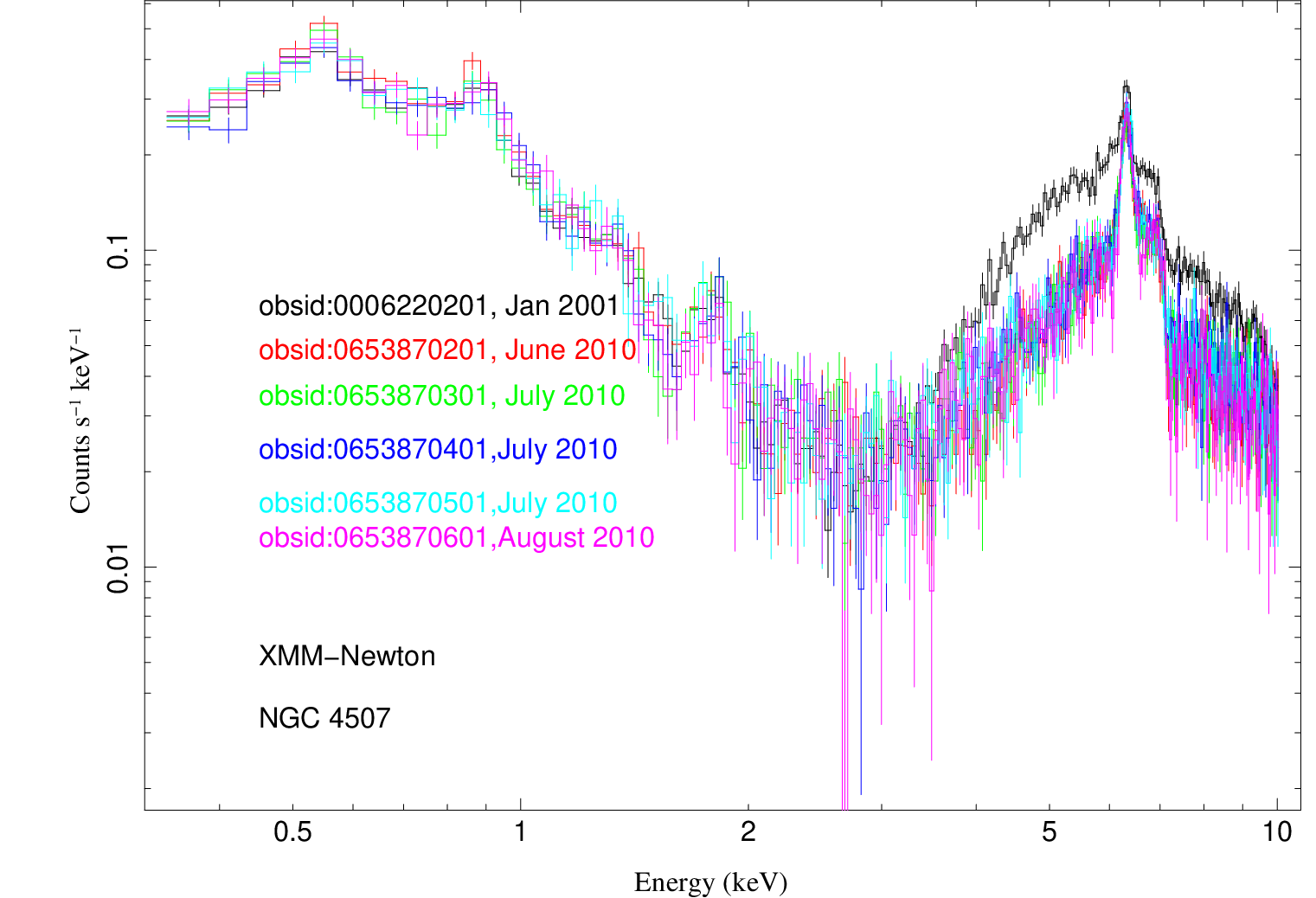} 
\includegraphics[width=4.2cm,angle=-90]{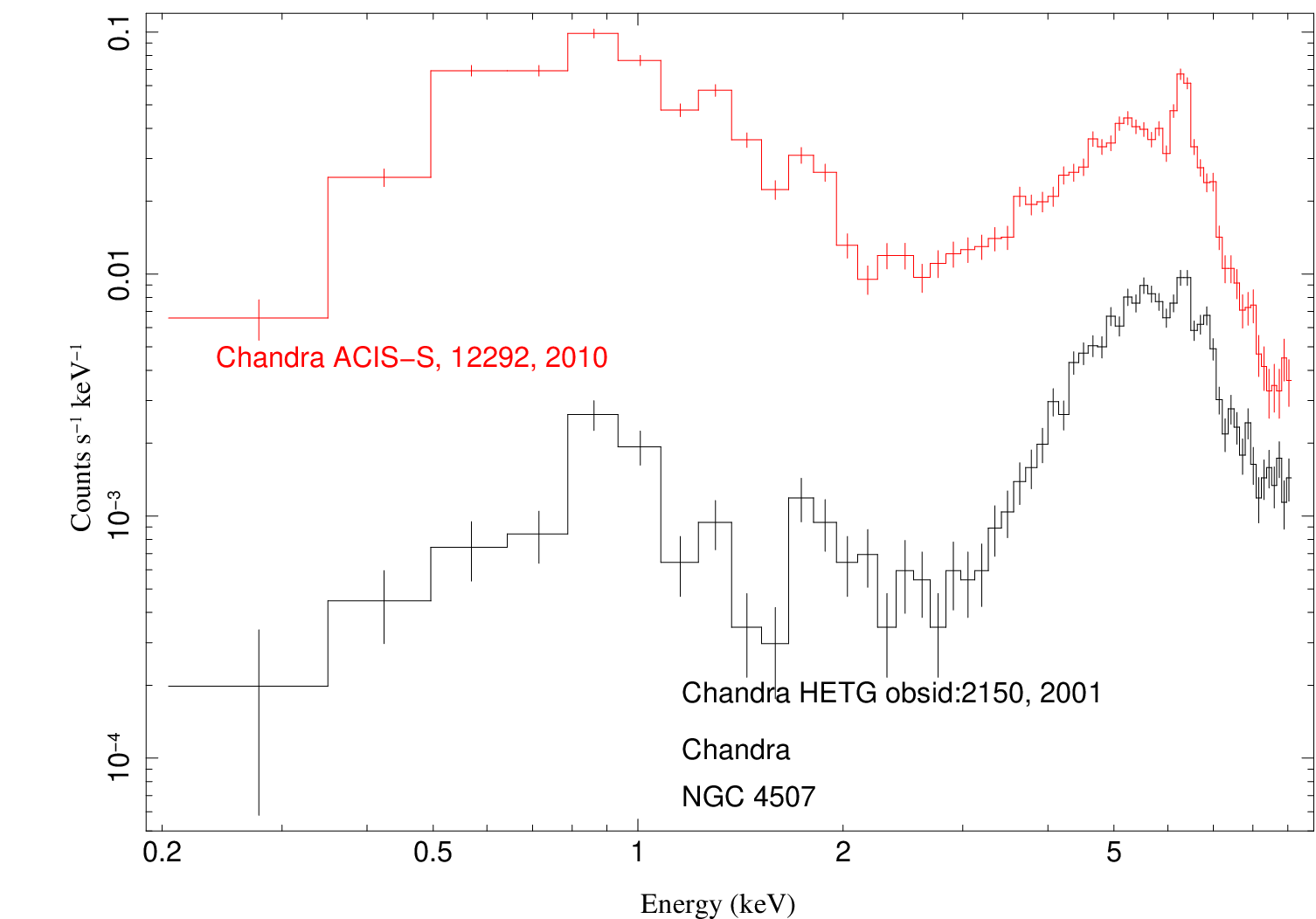}
\includegraphics[width=4.2cm,angle=-90]{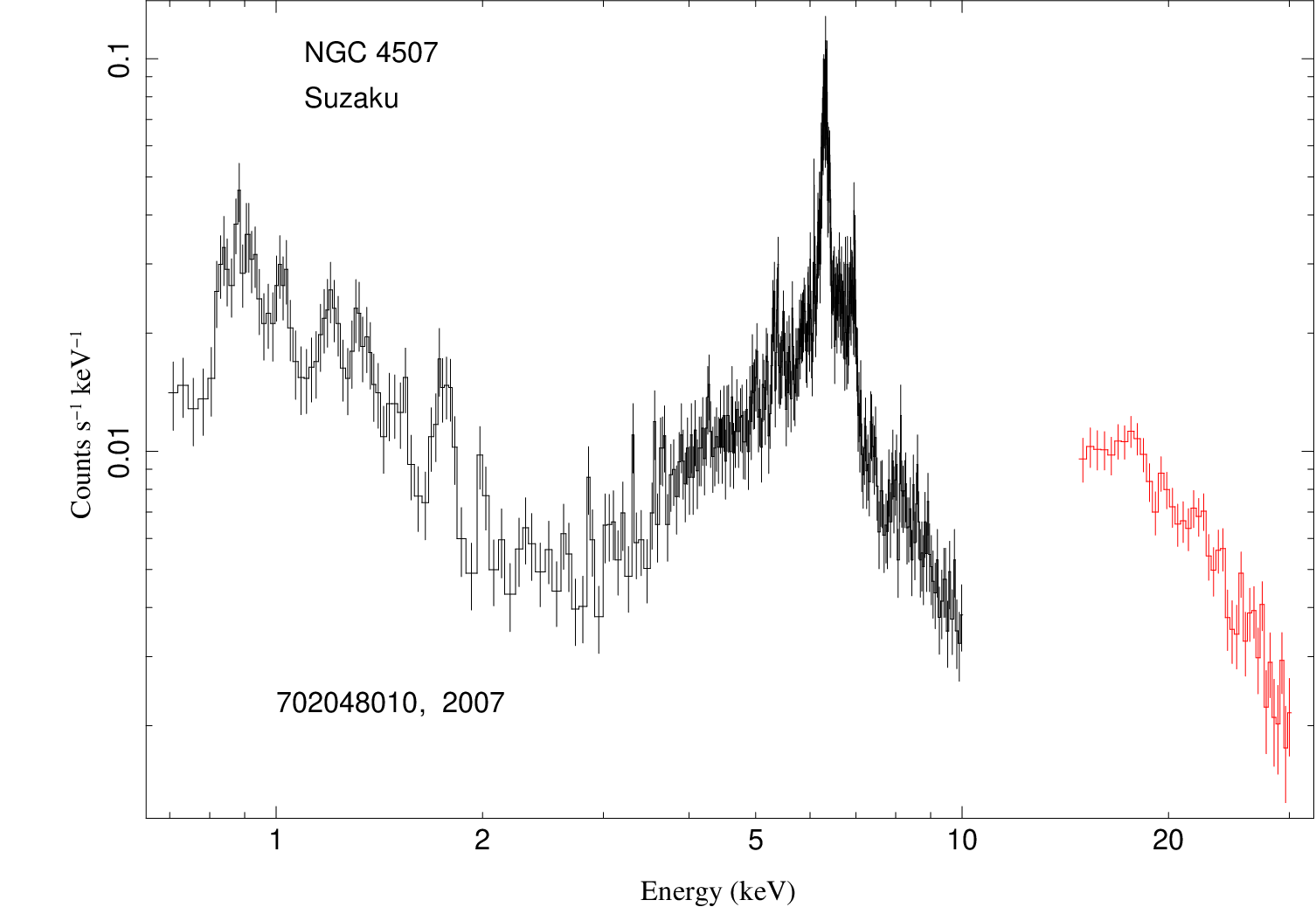}
	
	}

	\caption{Same as Fig. \ref{Fig:CenA}, except for the source which is NGC 4507} \label{Fig:NGC4507}
\end{figure*}


\begin{figure*}                                      

  \centering

	\hbox{

\includegraphics[width=4.2cm,angle=-90]{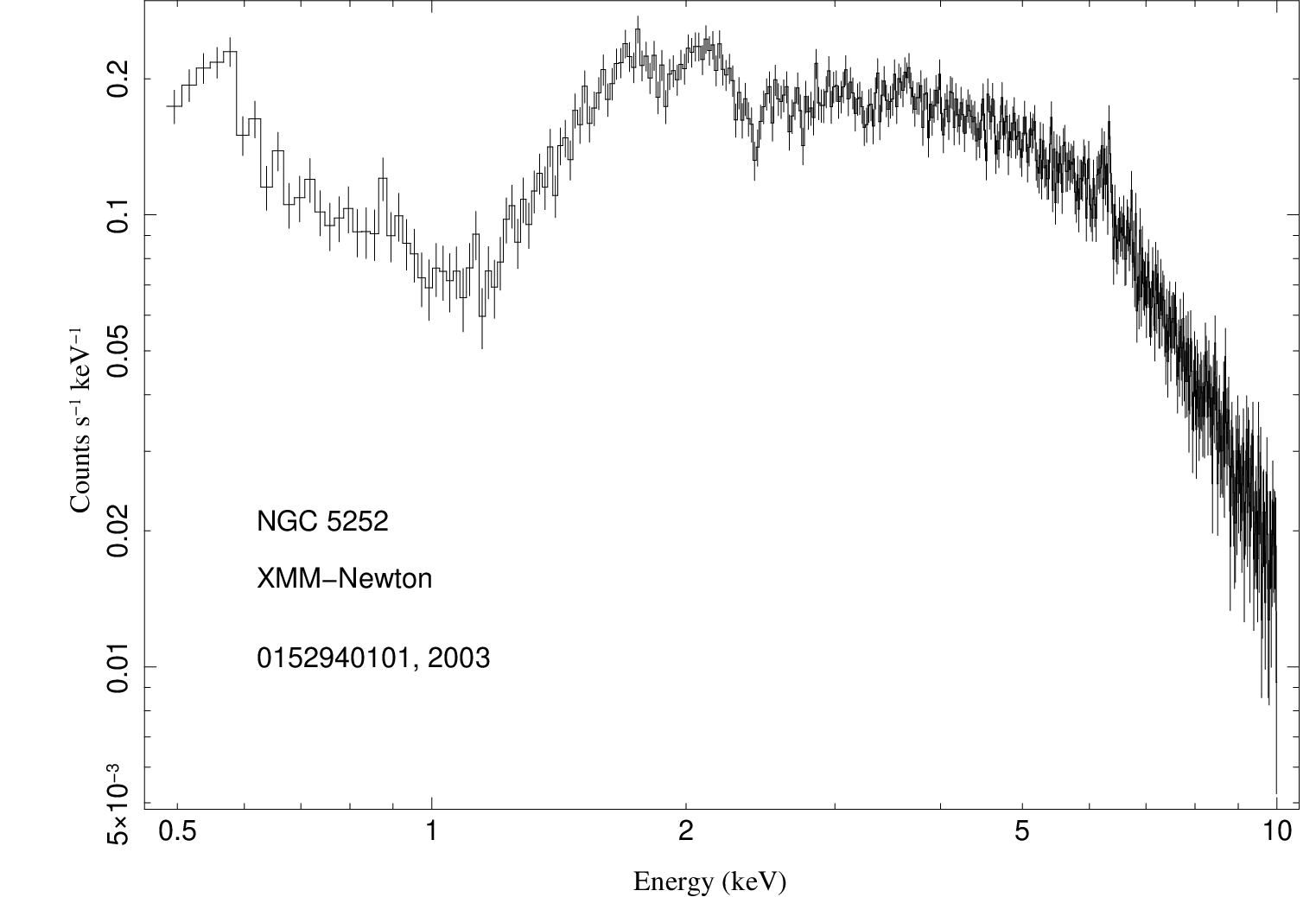} 
	\includegraphics[width=4.2cm,angle=-90]{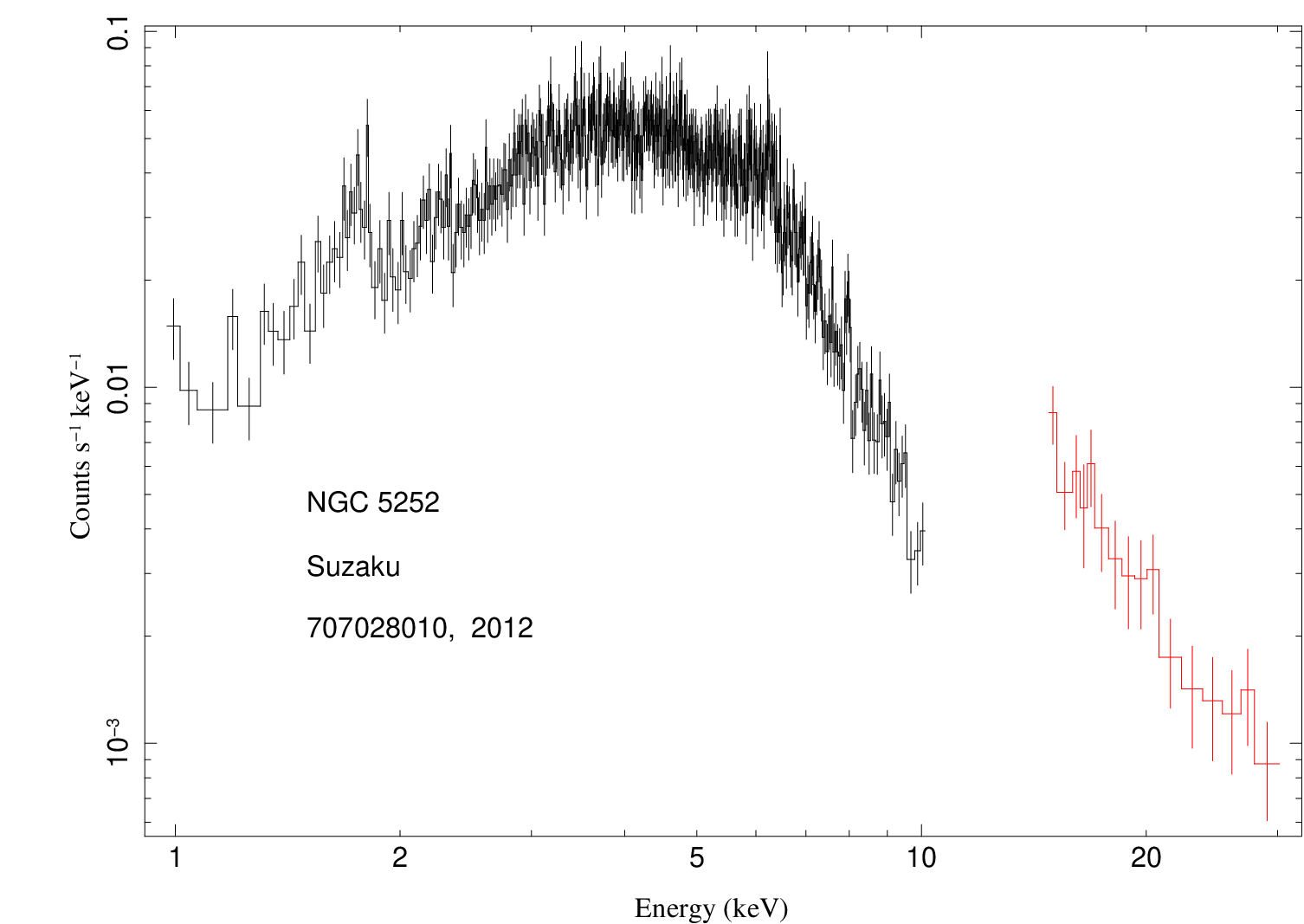}
	
	}

	\caption{Same as Fig. \ref{Fig:CenA}, except for the source which is NGC~5252 } \label{Fig:NGC5252}
\end{figure*}


\begin{figure*}                                       

  \centering

	\hbox{

\includegraphics[width=4.2cm,angle=-90]{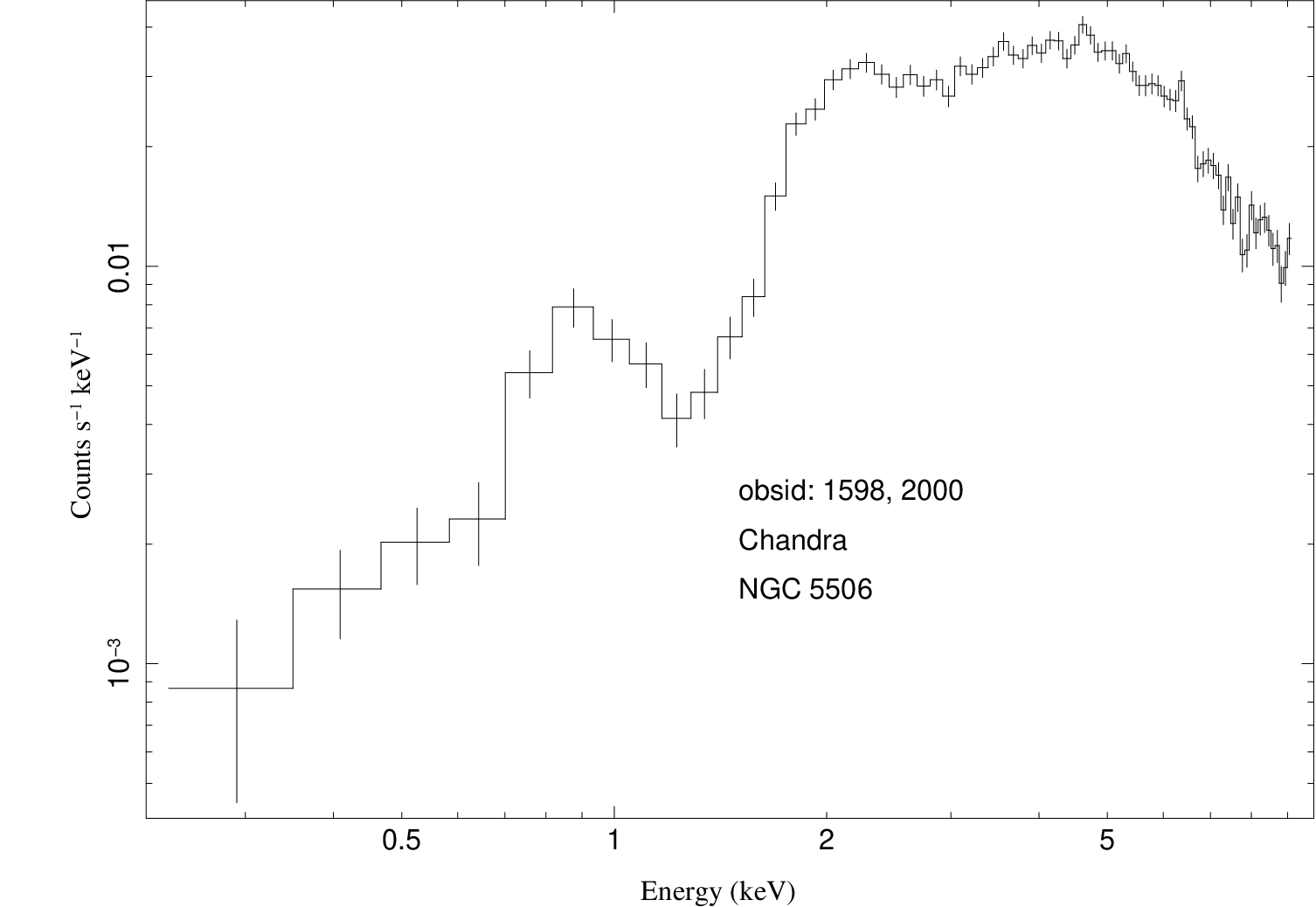}
\includegraphics[width=4.2cm,angle=-90]{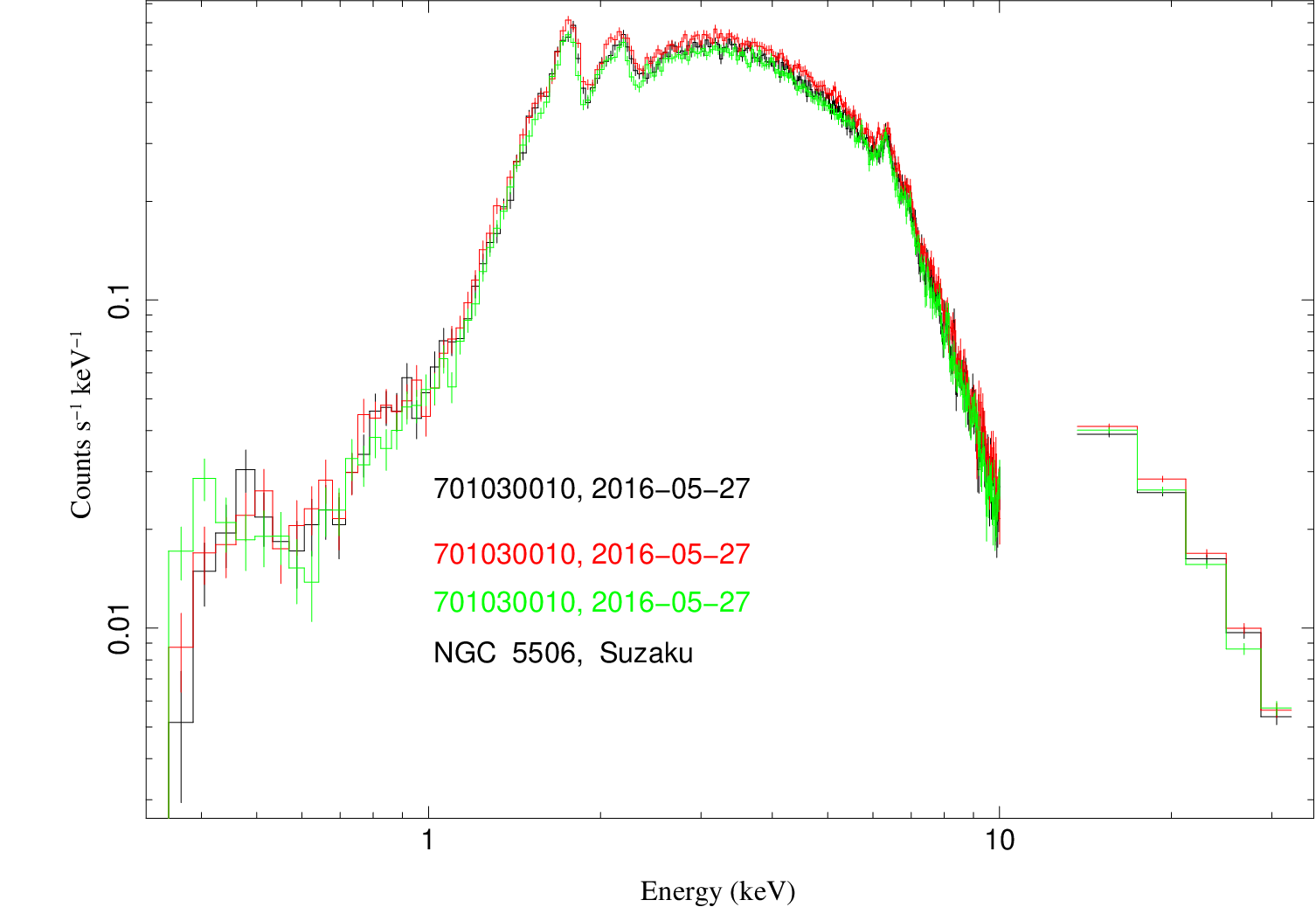}	
	}

\caption{Same as Fig. \ref{Fig:CenA}, except for the source which is NGC 5506 } \label{Fig:NGC5506}
\end{figure*}


\begin{figure*}                                             

  \centering

	\hbox{

\includegraphics[width=4.2cm,angle=-90]{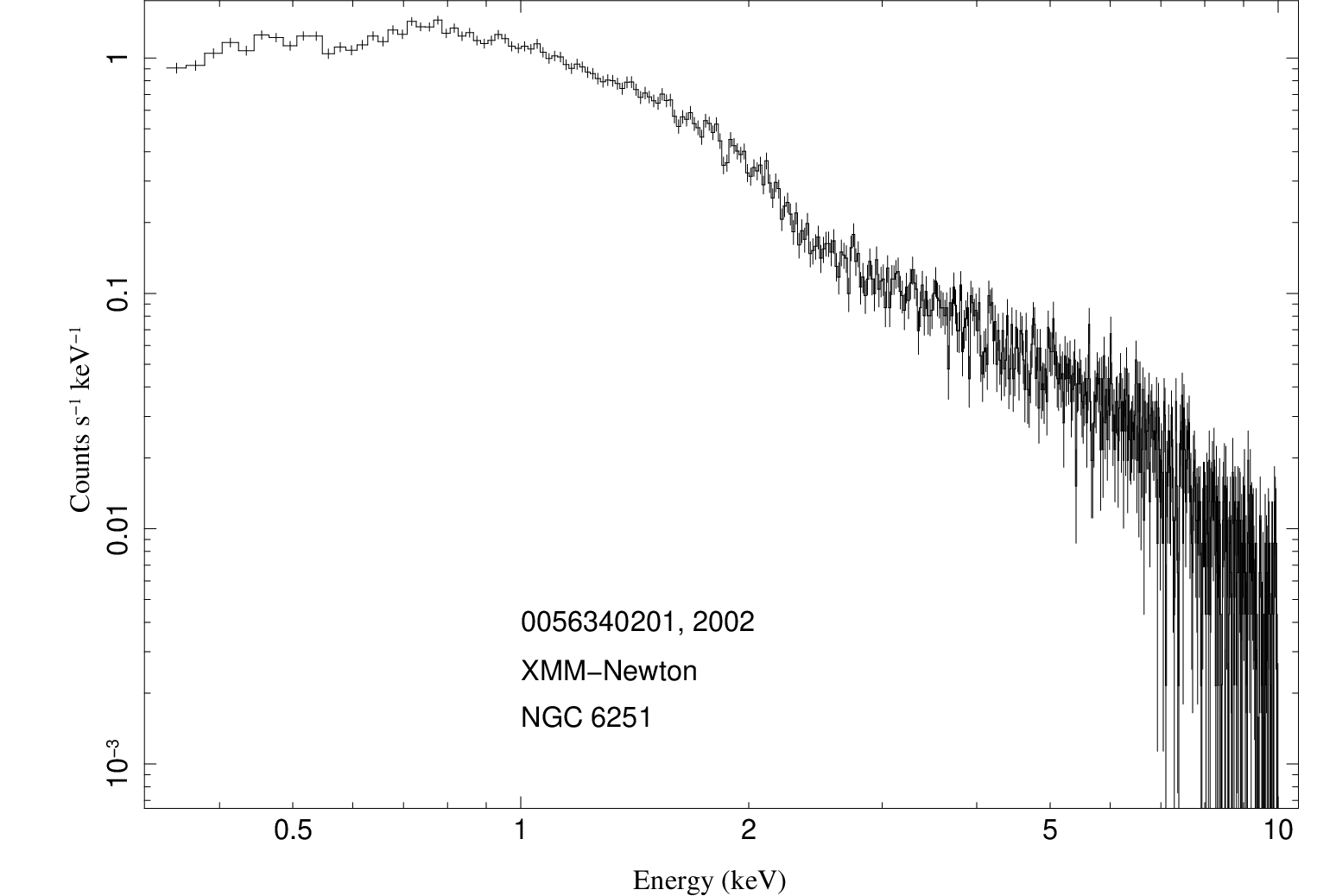} 
\includegraphics[width=4.2cm,angle=-90]{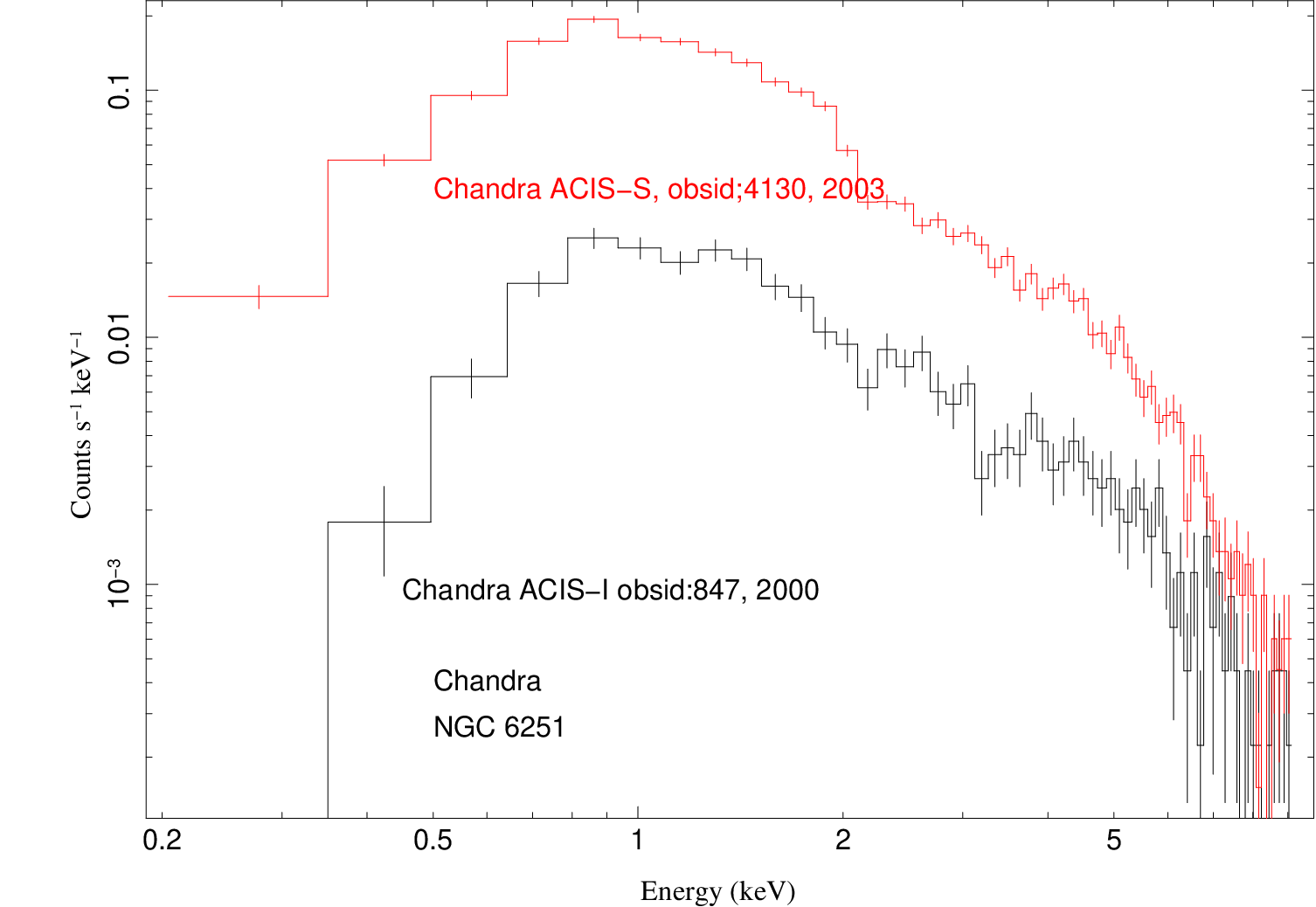}
\includegraphics[width=4.2cm,angle=-90]{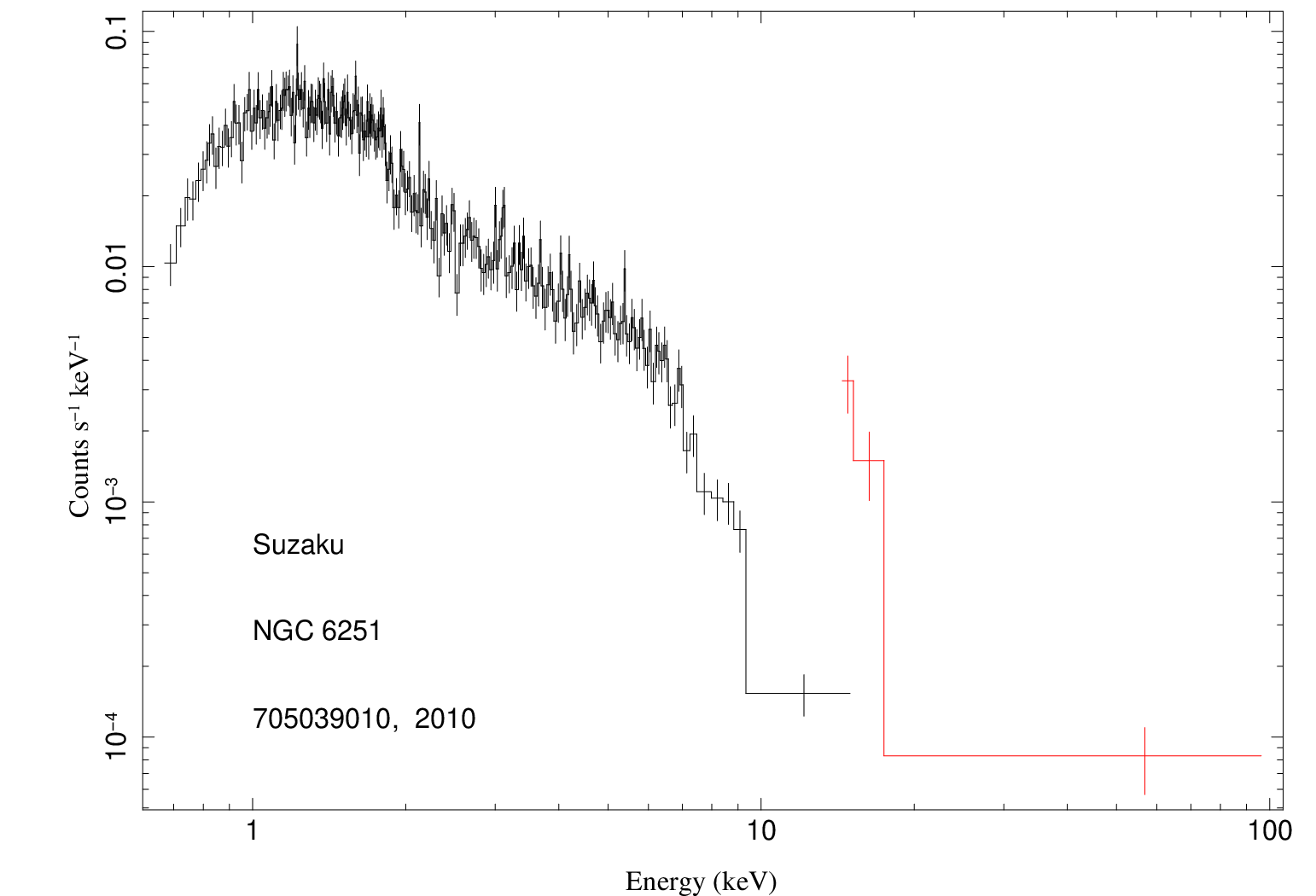}	
	}

\caption{Same as Fig. \ref{Fig:CenA}, except for the source which is NGC 6251} \label{Fig:NGC6251}
\end{figure*}


\clearpage

\begin{figure*}                                        

  \centering

		\hbox{

\includegraphics[width=4.2cm,angle=-90]{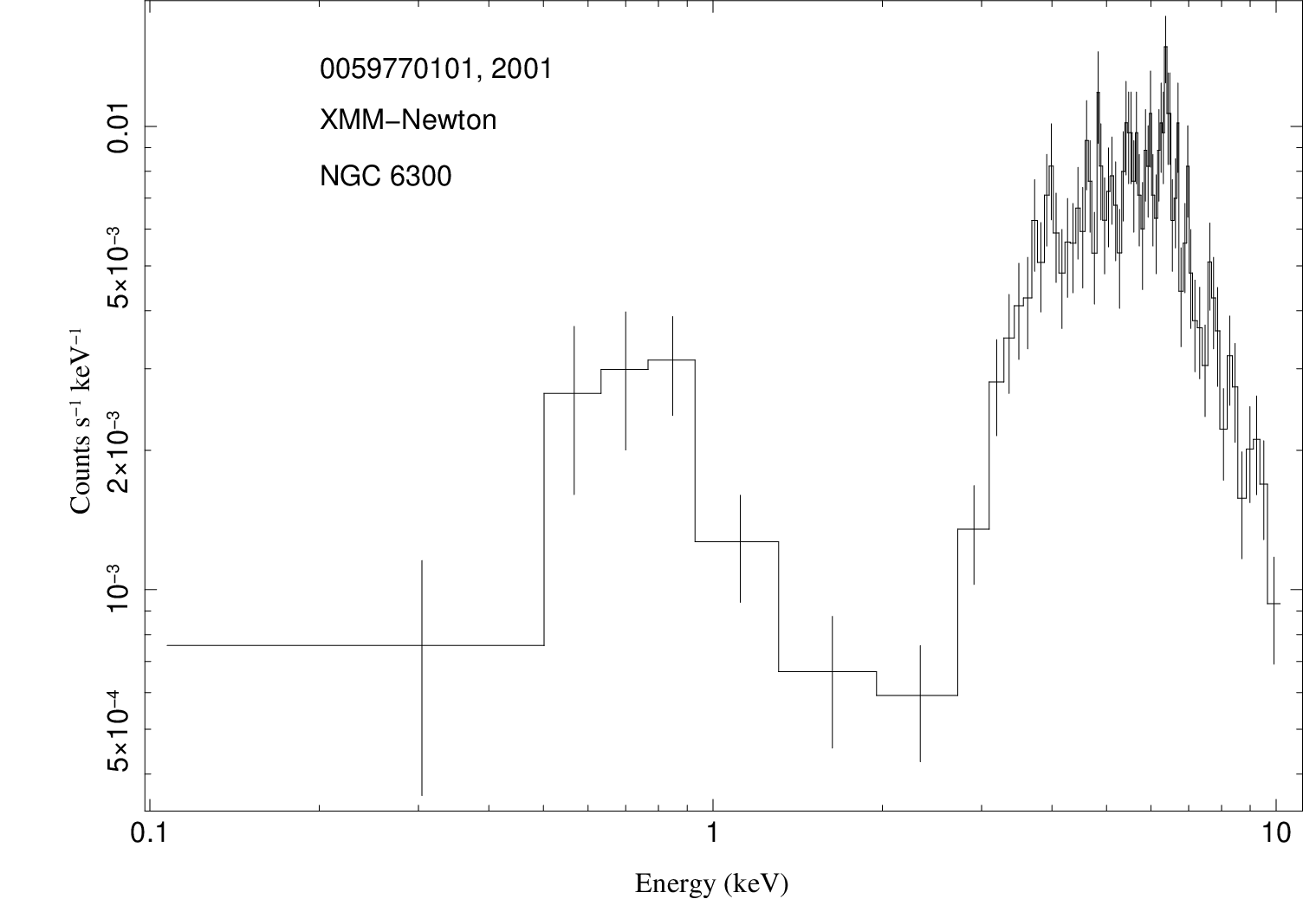} 
\includegraphics[width=4.2cm,angle=-90]{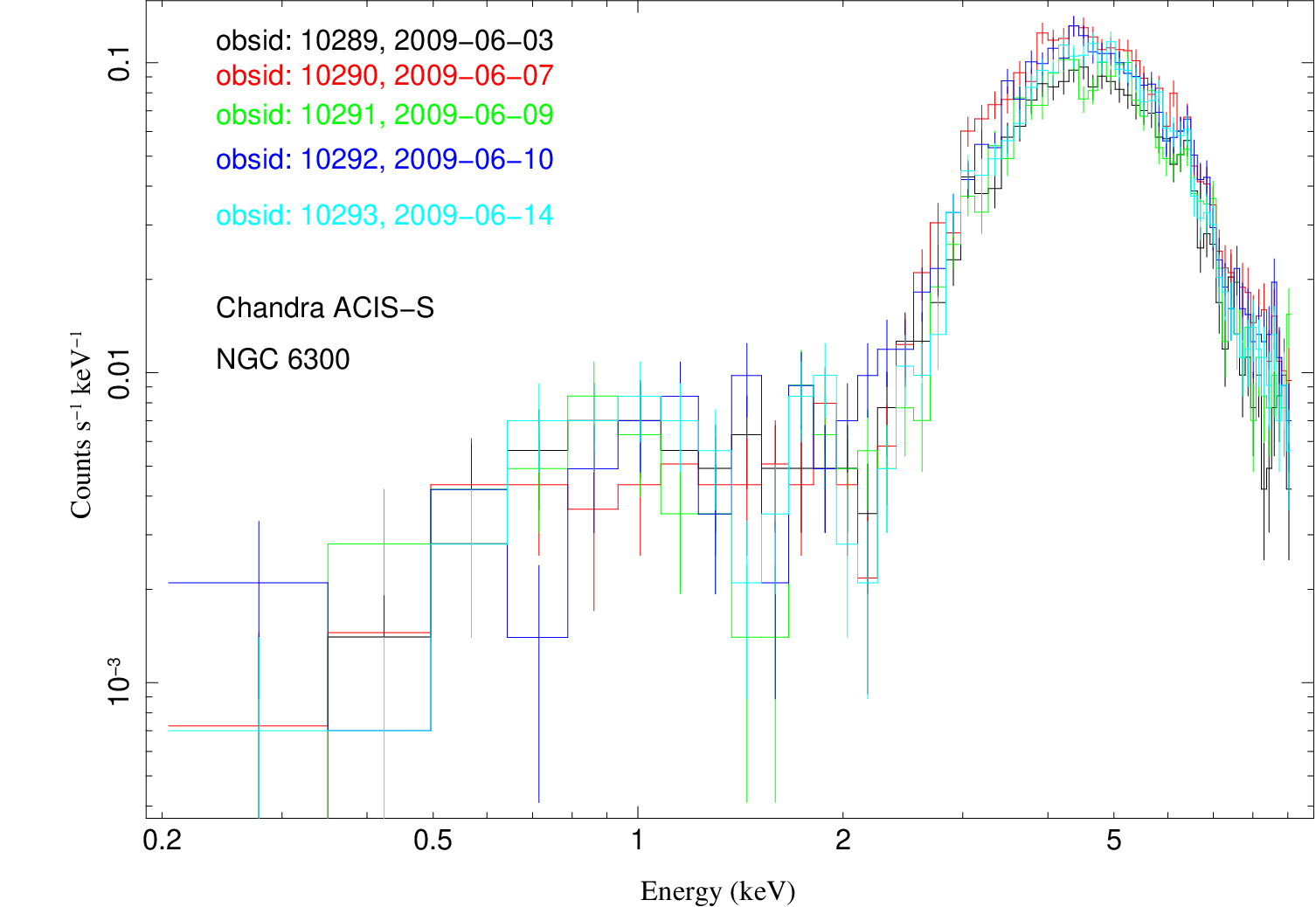}
\includegraphics[width=4.2cm,angle=-90]{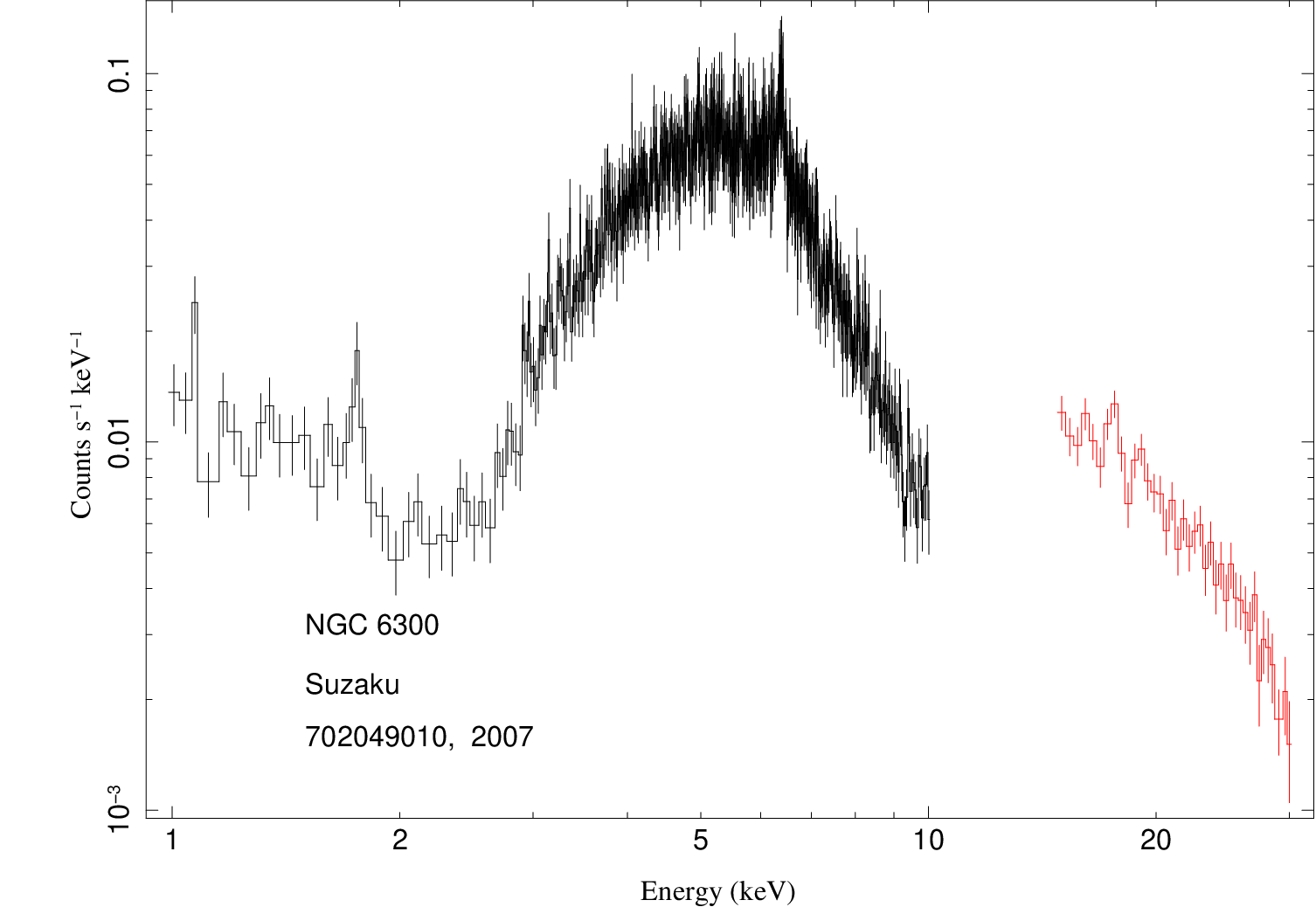}
	
	}

	\caption{Same as Fig. \ref{Fig:CenA}, except for the source which is NGC 6300} \label{Fig:NGC6300}
\end{figure*}


\begin{figure*}                                     

  \centering

	\hbox{

\includegraphics[width=4.2cm,angle=-90]{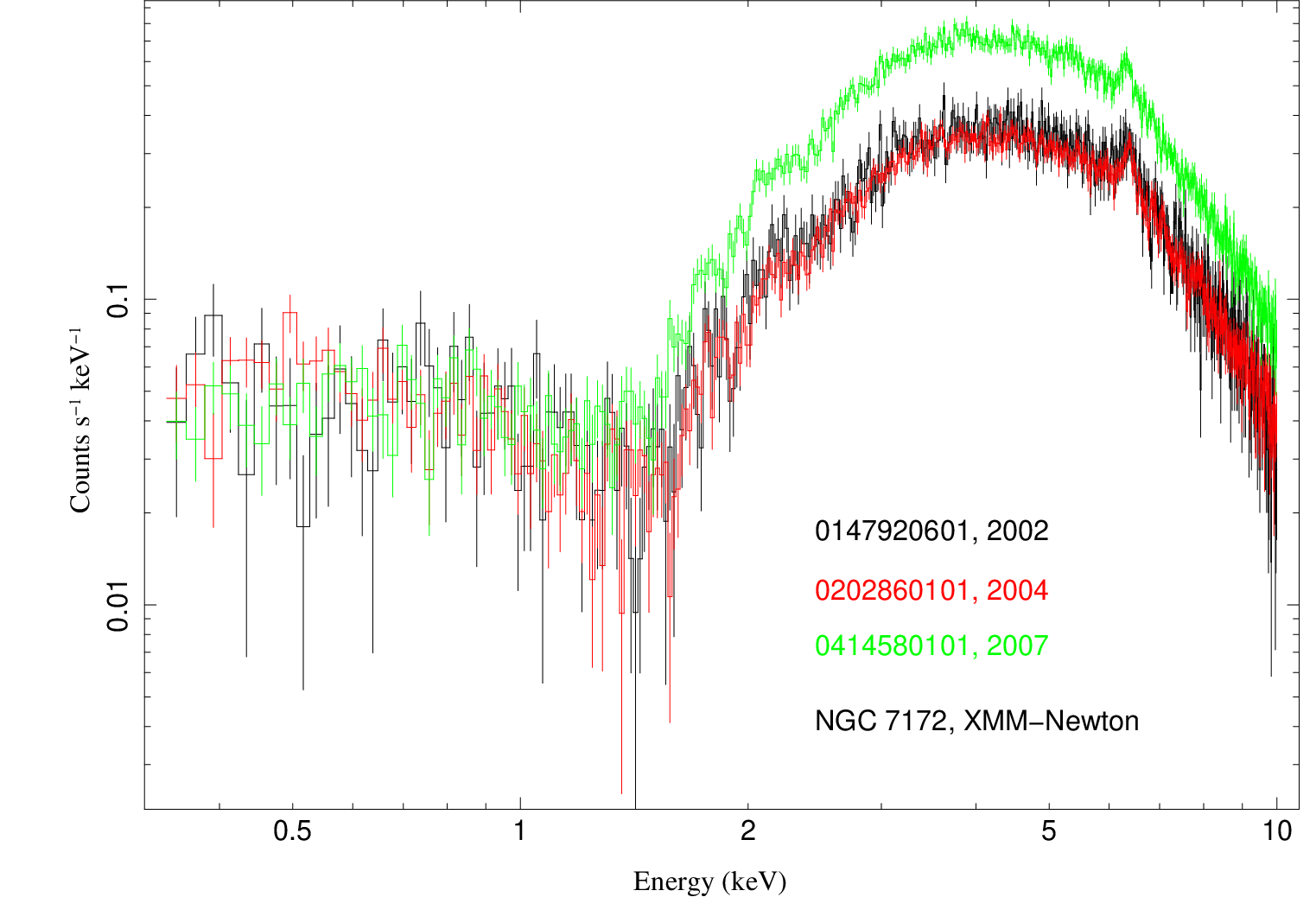} 
\includegraphics[width=4.2cm,angle=-90]{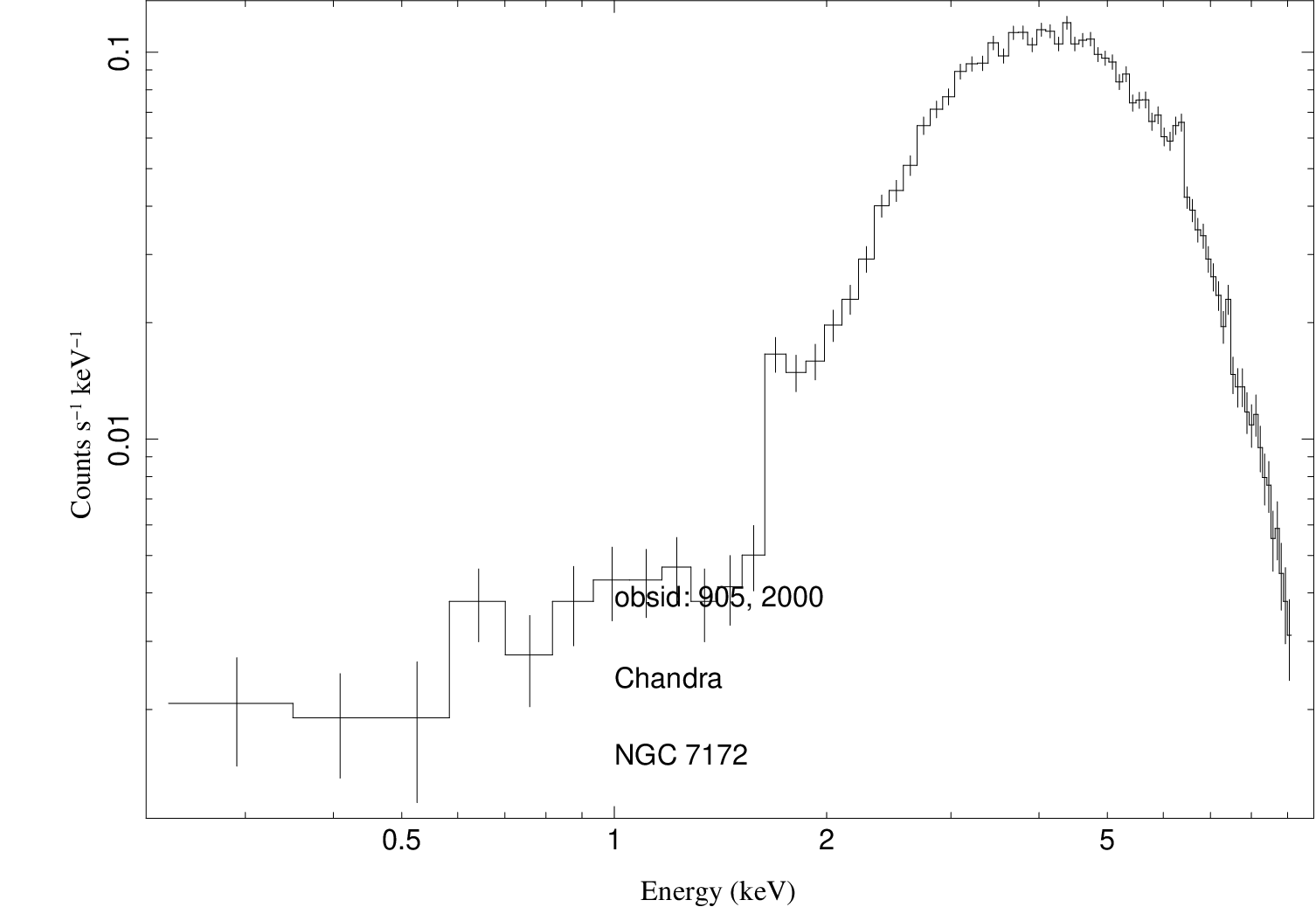}
	\includegraphics[width=4.2cm,angle=-90]{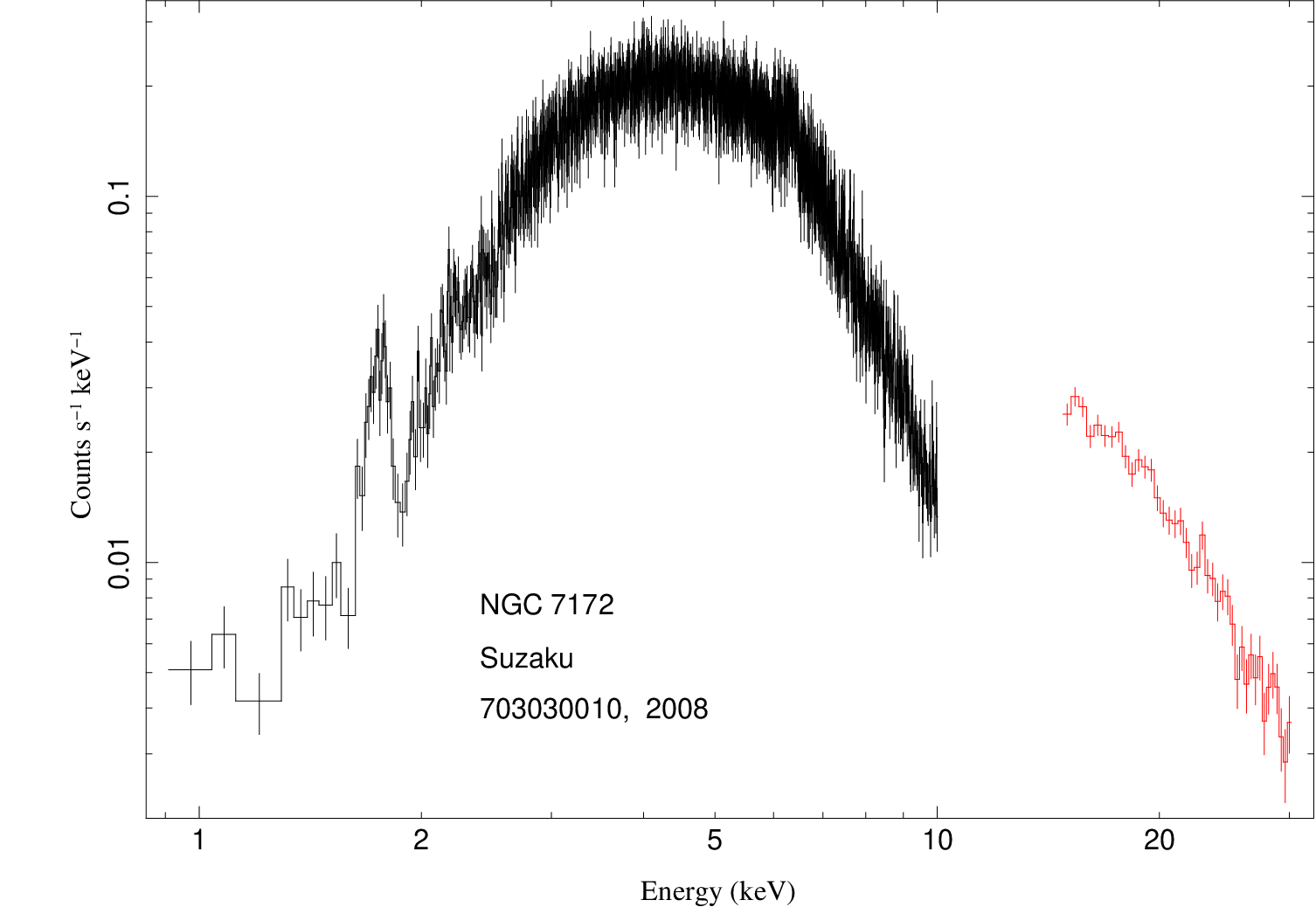}
	
	}

	\caption{Same as Fig. \ref{Fig:CenA}, except for the source which is NGC~7172 } \label{Fig:NGC7172}
\end{figure*}


\begin{figure*}                                       

  \centering

	\hbox{

\includegraphics[width=4.2cm,angle=-90]{Plots_29thOct2017/NGC7314/NGC7314_XMM_comparison.ps} 
\includegraphics[width=4.2cm,angle=-90]{Plots_29thOct2017/NGC7314/NGC7314_Chandra_Compare.ps}
\includegraphics[width=4.2cm,angle=-90]{Plots_29thOct2017/NGC7314/NGC7314_Suzaku_compare.ps}	
	}

\caption{Same as Fig. \ref{Fig:CenA}, except for the source which is NGC 7314. } \label{Fig:NGC7314}
\end{figure*}


\begin{figure*}
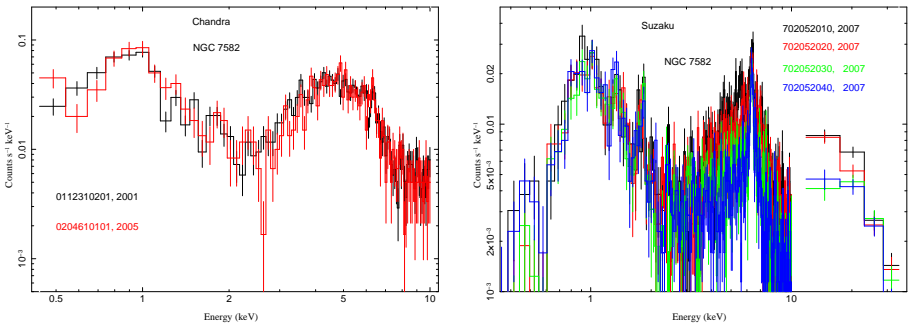
                                         
  \centering

	\hbox{

\includegraphics[width=4.2cm,angle=-90]{Plots_29thOct2017/NGC7582/NGC7582_comparing_Chandra.ps}
\includegraphics[width=4.2cm,angle=-90]{Plots_29thOct2017/NGC7582/NGC7582_comparison_Suzaku.ps}	
	}

\caption{Same as Fig. \ref{Fig:CenA}, except for the source which is NGC 7582. } \label{Fig:NGC7582}
\end{figure*}


\section{Appendix D: Bayesian X-ray Analysis (BXA) simulations}\label{Sec:BXA}

As discussed in Section 4, we ran the MultiNest nested-sampling
algorithm (Skilling 2004; Feroz et al.\ 2009) via the Bayesian X-ray
Analysis (BXA) and PyMultiNest packages (Buchner et al.\ 2014) for
XSPEC version 12.10.1f to assess fit parameter distributions and
explore potential model degeneracies between $\nh$ and other
model parameters.  Due to the long computational times required, it
was not feasible to run MultiNest on all observations for a given
source/instrument combination.  Instead we ran MultiNest on select
cases of interest to make sure that deviating parameter values for a
particular observation were not the result of e.g., a low exposure
time in that observation (as we had initially suspected, but then
ruled out, for Cen~A S5).

We caution the reader, however, about the limitations of such
simulations, as they are intended for exploring fit parameter
distributions for an assumed model.  They are not intended for
determining goodness-of-fit of that model to the observed data.
In addition, a couple assumptions implicit in these simulations are
worth noting.  When we specify a model as input for these simulations,
we implicitly assume that the input model is accurate in terms of
containing the proper components (we assume we're not missing spectral
components that are intrinsically present in the source, nor have we
added components to our model that are intrinsically lacking in the
source).  We also assume that each spectral component intrinsically
present in the real data follows the equation-based model components
we use (e.g., we assume that the real data's primary continuum indeed
follows a strict power law). If the real data has, for example, some
very mild continuum curvature such that our modeling cannot
signficantly reject a strict power law, then best-fit parameters could
differ slightly between in the observed/modeled spectrum and the
simulated spectra.

We display the resulting parameter posterior contour plots in Figures \ref{Fig:BXA_CenA} through \ref{Fig:BXA_ngc7582X4}.



\begin{figure*}                                         
  \centering

	\includegraphics[width=16cm,angle=0]{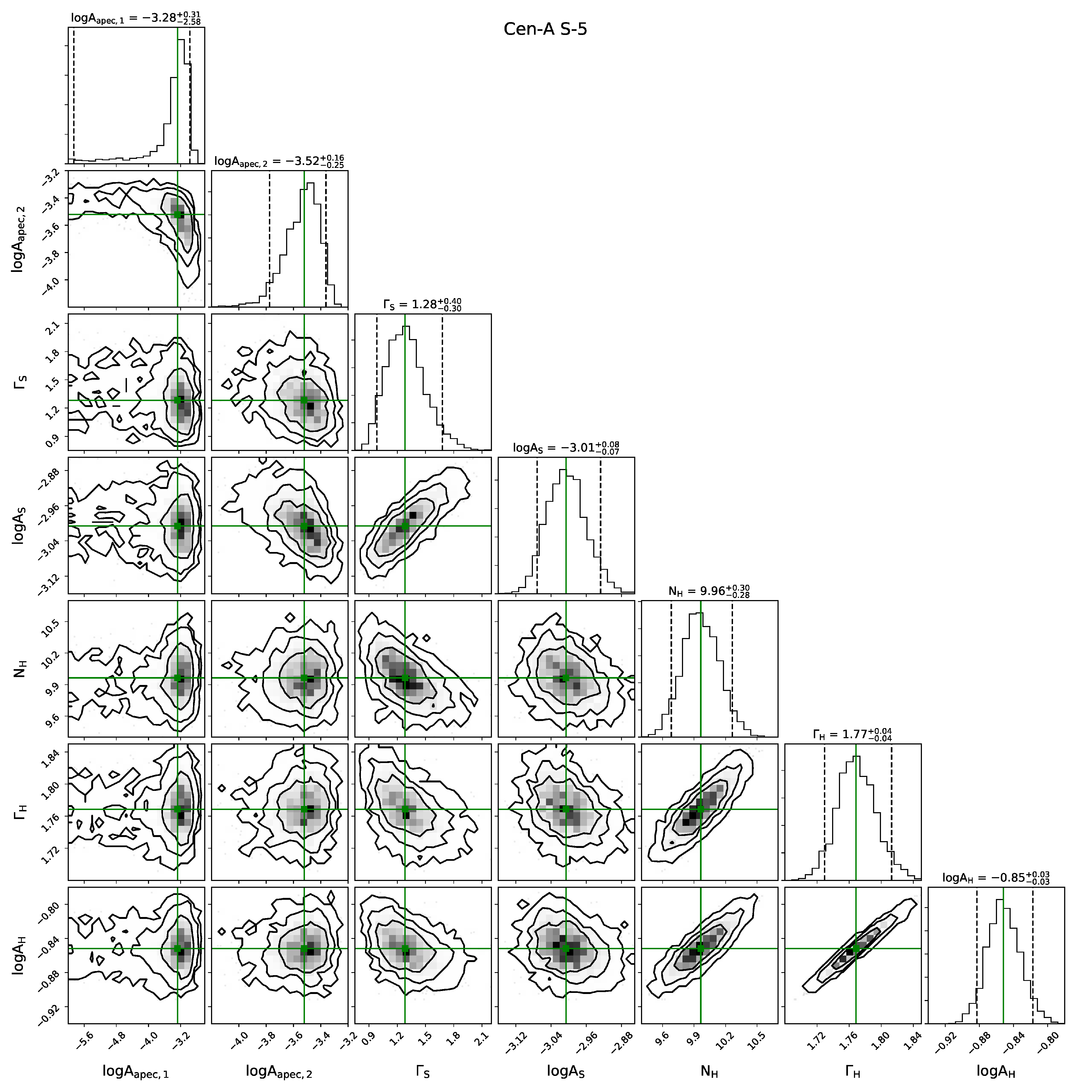}

\setcounter{figure}{0}
\caption{ Parameter posterior distributions for the MultiNest run of
Cen~A, observation S-5. Green solid lines denote the median
posterior value; dashed black lines denote the 5 and 95$\%$-tile
values.  The solid black contours denote the 68, 90, and 99$\%$ confidence
levels. Here, $A_{\rm apec1, 2}$ denote \textsc{apec} normalizations,
$A_{\rm S, H}$ denote the 1 keV normalization of the soft- and
hard-band power laws, $A_{\rm HXD}$ refers to the HXD-PIN/XIS0
cross-instrument constant. } \label{Fig:BXA_CenA}
\end{figure*}


\begin{figure*}                                         
  \centering

	\includegraphics[width=16cm,angle=0]{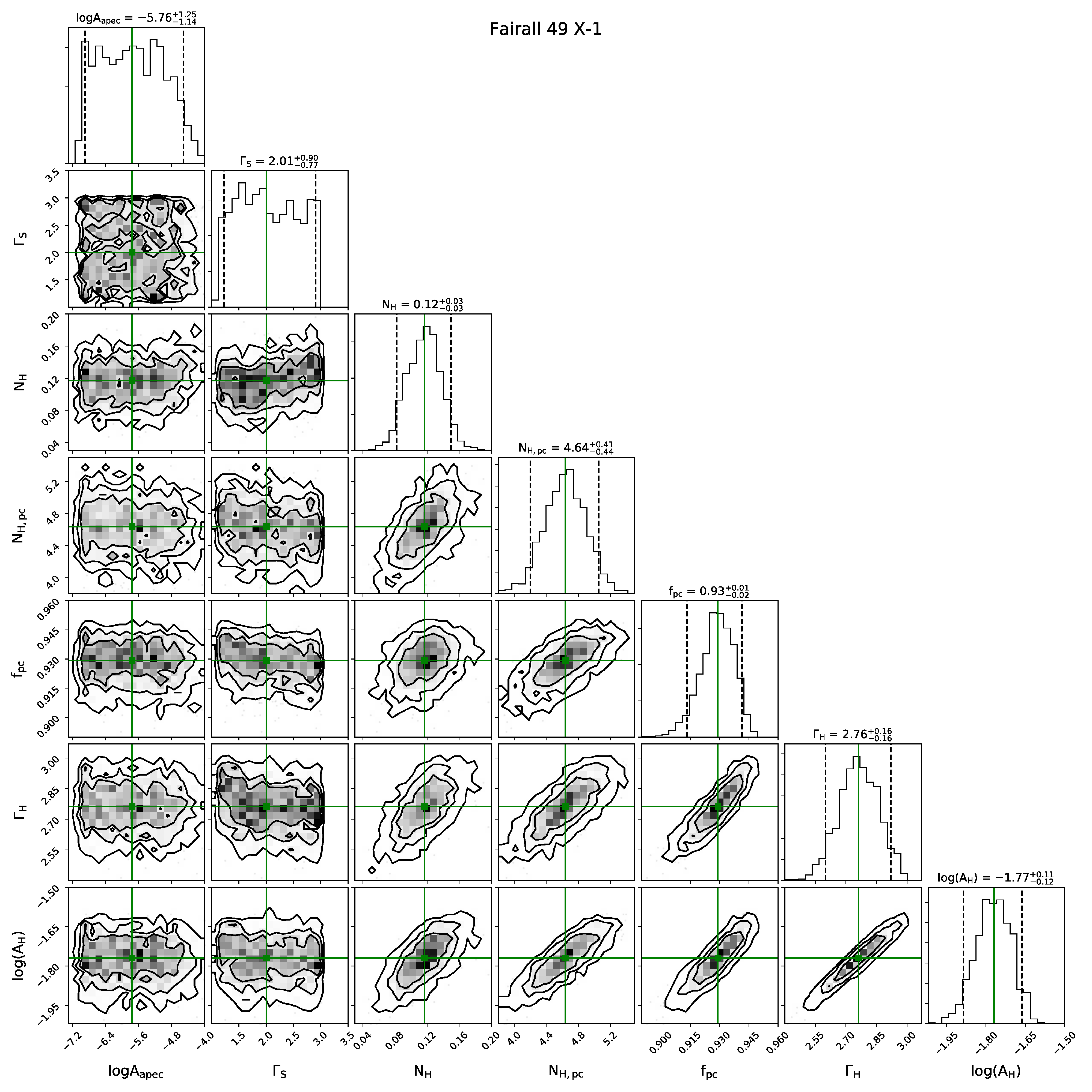}

\caption{Same as Fig.~\ref{Fig:BXA_CenA}, but for Fairall49, X1. Here, $N_{\rm H, pc}$ and $f_{\rm pc}$ denote partial-covering absorber parameters. } \label{Fig:BXA_F49X1}
\end{figure*}


\begin{figure*}                                         
  \centering

	\includegraphics[width=16cm,angle=0]{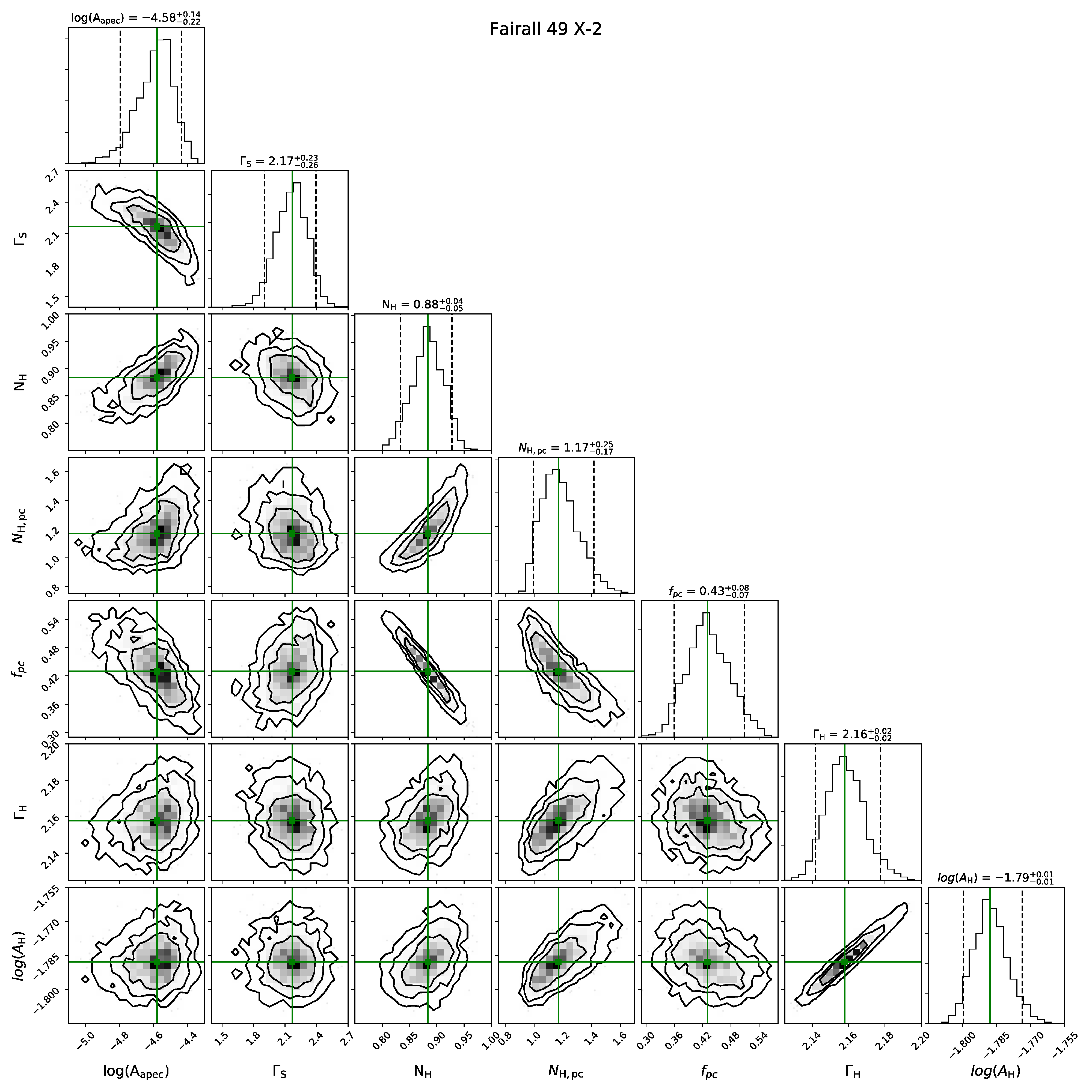}

\caption{Same as Fig.~\ref{Fig:BXA_CenA}, but for Fairall 49, X2 } \label{Fig:BXA_F49X2}
\end{figure*}


\begin{figure*}                                         
  \centering

	\includegraphics[width=16cm,angle=0]{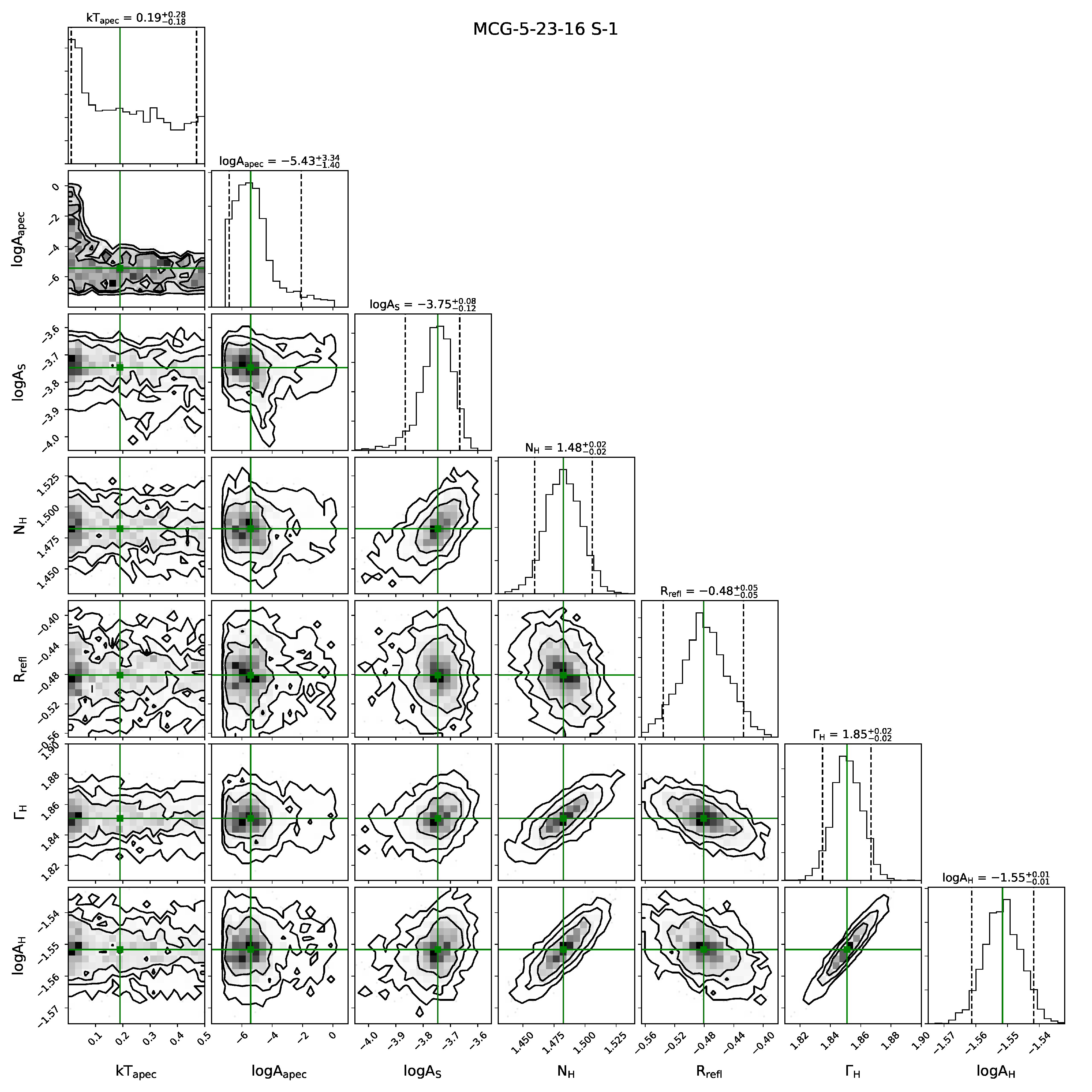}

\caption{Same as Fig.~\ref{Fig:BXA_CenA}, but for MCG-5-23-16, S-1 } \label{Fig:BXA_mcg5S1}
\end{figure*}


\begin{figure*}                                         
  \centering

	\includegraphics[width=16cm,angle=0]{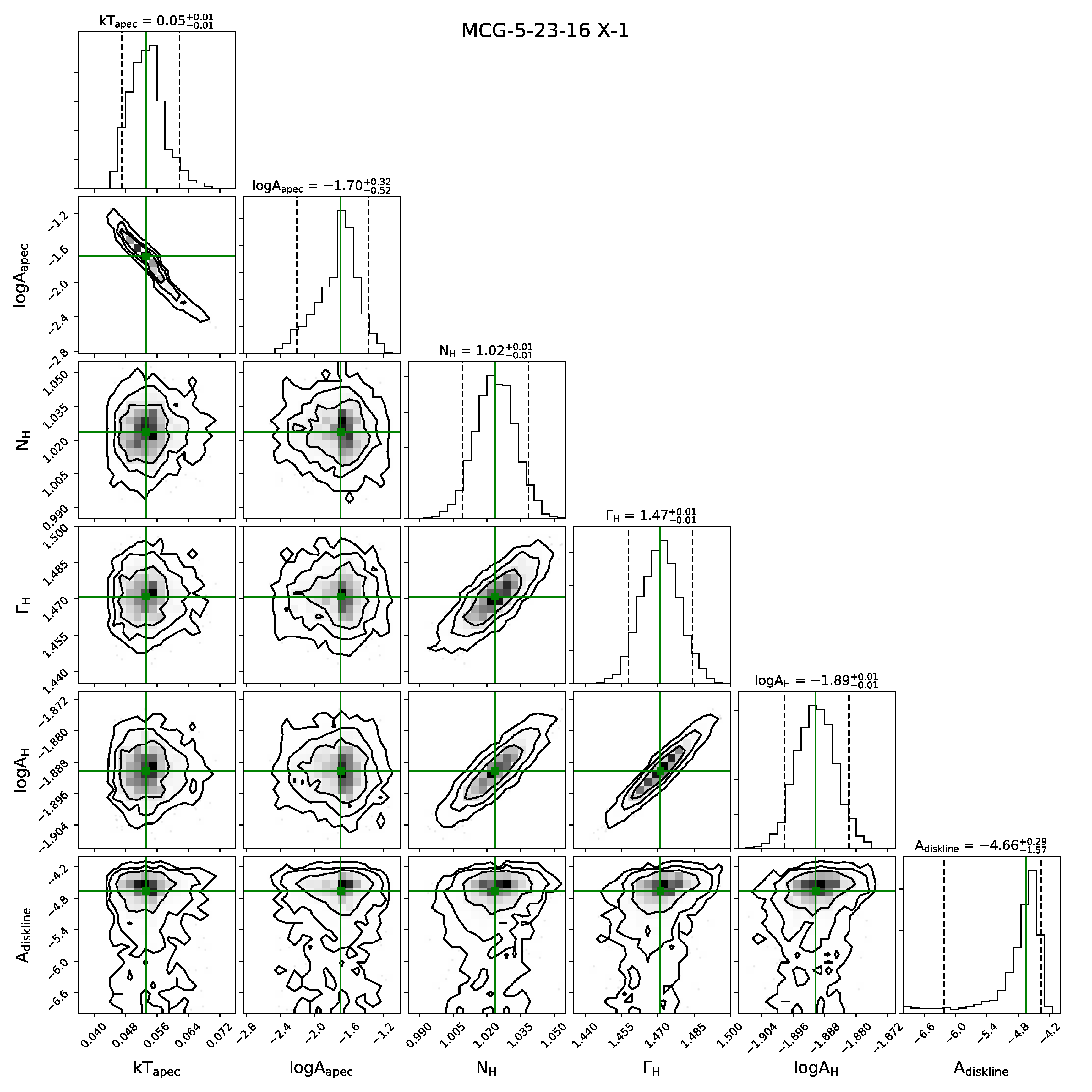}

\caption{Same as Fig.~\ref{Fig:BXA_CenA}, but for MCG-5-23-16, X1 } \label{Fig:BXA_mcg5X1}
\end{figure*}


\begin{figure*}                                         
  \centering

	\includegraphics[width=16cm,angle=0]{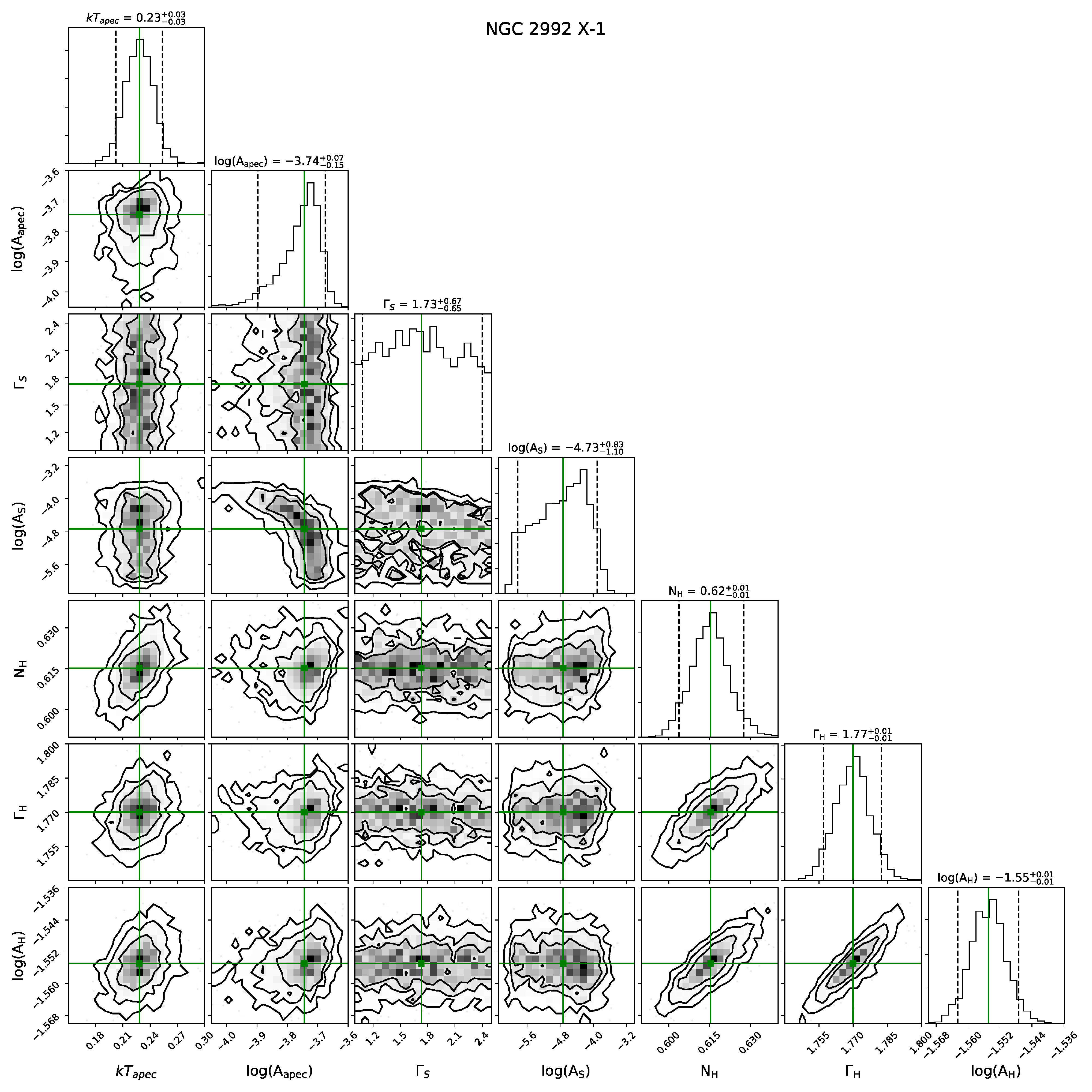}

\caption{Same as Fig.~\ref{Fig:BXA_CenA}, but for NGC~2992, X1 } \label{Fig:BXA_ngc2992X1}
\end{figure*}


\begin{figure*}                                         
  \centering

	\includegraphics[width=16cm,angle=0]{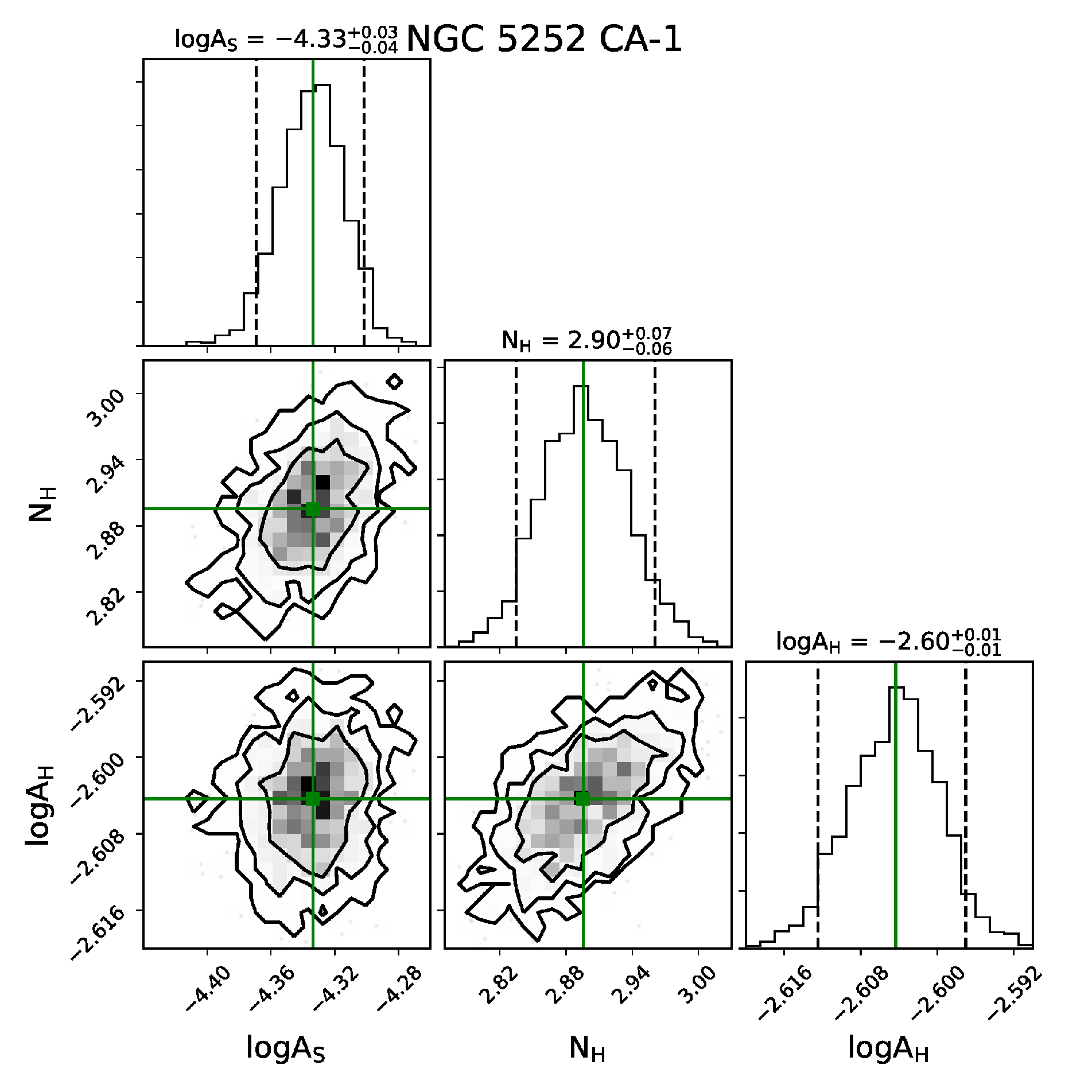}

\caption{Same as Fig.~\ref{Fig:BXA_CenA}, but for NGC~5252, CA-1 } \label{Fig:BXA_ngc5252CA1}
\end{figure*}


\begin{figure*}                                         
  \centering

	\includegraphics[width=16cm,angle=0]{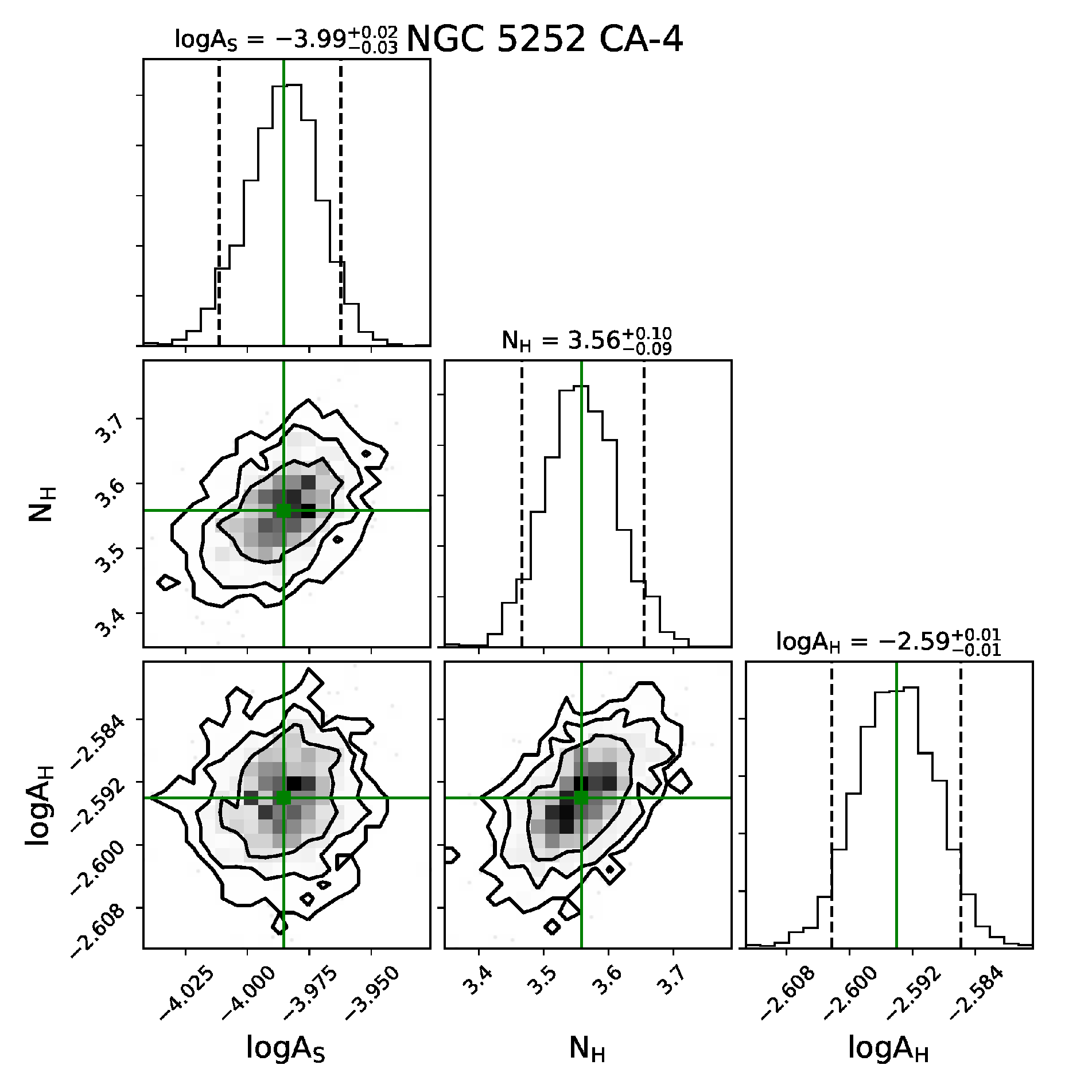}

\caption{Same as Fig.~\ref{Fig:BXA_CenA}, but for NGC~5252, CA4 } \label{Fig:BXA_NGC5252CA4}
\end{figure*}


\begin{figure*}                                         
  \centering

	\includegraphics[width=16cm,angle=0]{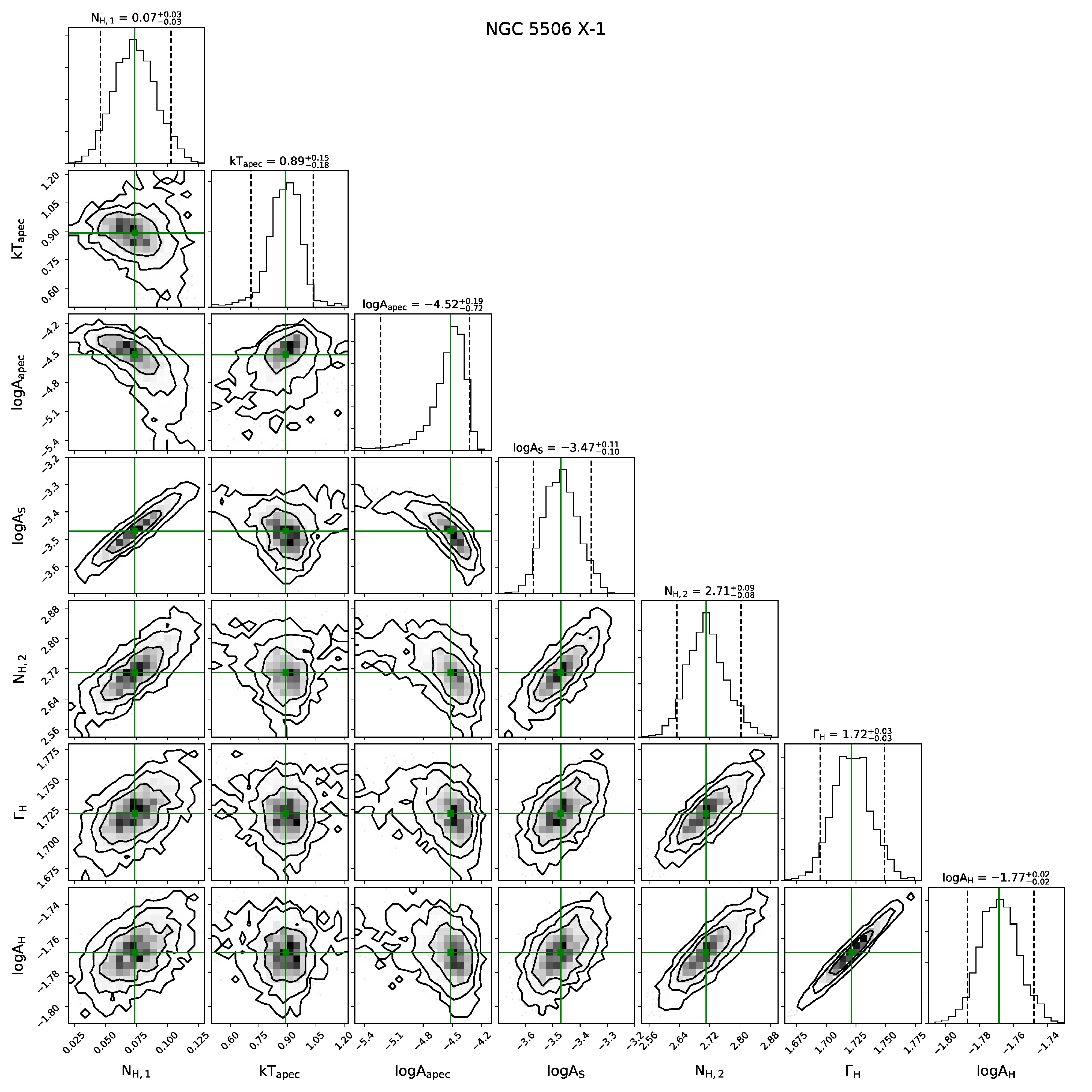}

\caption{Same as Fig.~\ref{Fig:BXA_CenA}, but for NGC~5506, X1. Here, $N_{\rm H,1}$ refers to the component absorbing the soft-band emission components,
as discussed in Appendix A; $N_{\rm H,2}$ refers to the variable full-covering absorber absorbing the hard-band
components. } \label{Fig:BXA_ngc5506X1}
\end{figure*}


\begin{figure*}                                         
  \centering

	\includegraphics[width=16cm,angle=0]{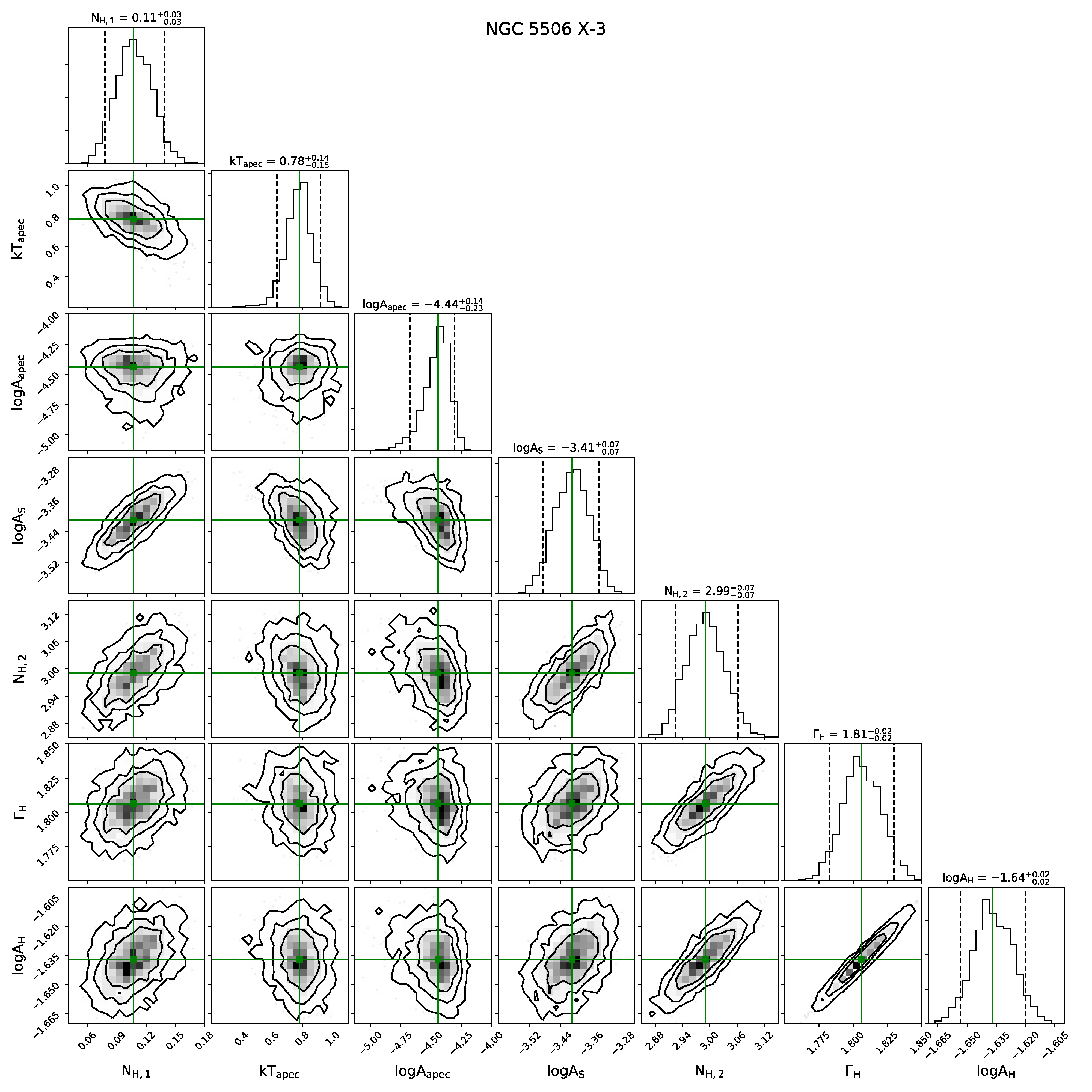}

	\caption{Same as Fig.~\ref{Fig:BXA_ngc5506X1}, but for NGC~5506 X3. } \label{Fig:BXA_ngc5506X3}
\end{figure*}


\begin{figure*}                                         
  \centering

	\includegraphics[width=16cm,angle=0]{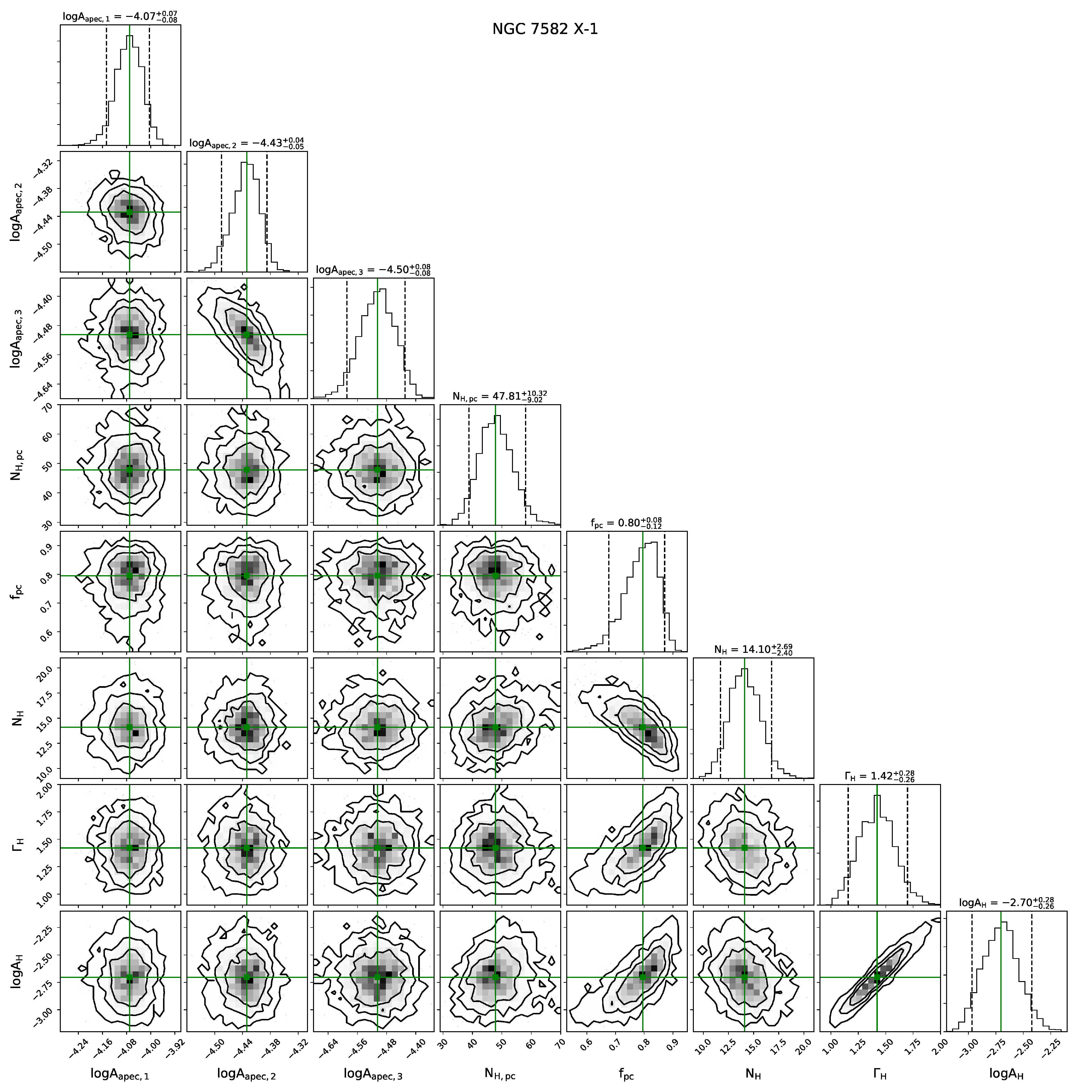}

\caption{Same as Fig.~\ref{Fig:BXA_CenA}, but for NGC~7582, X1 } \label{Fig:BXA_ngc7582X1}
\end{figure*}


\begin{figure*}                                         
  \centering

	\includegraphics[width=16cm,angle=0]{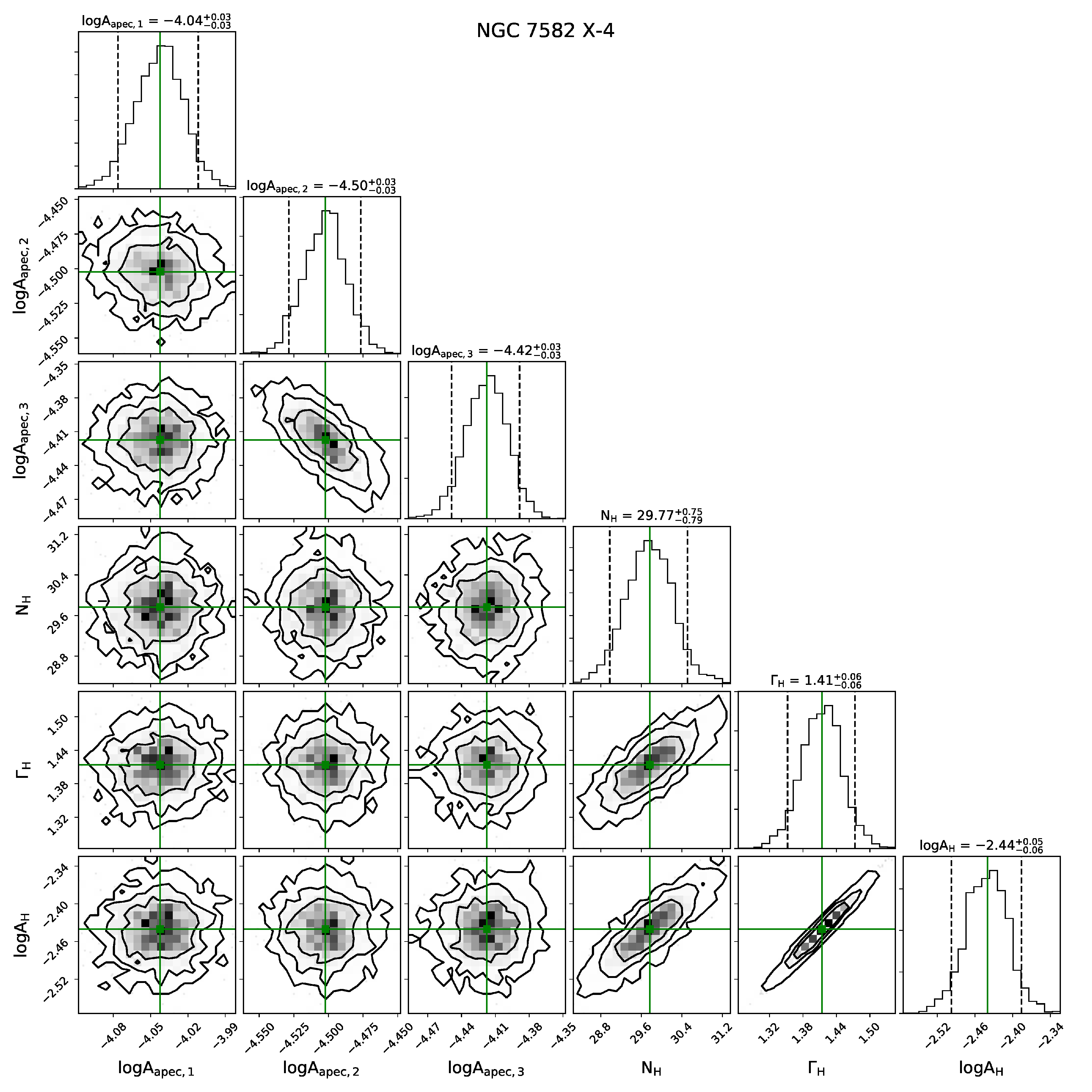}

\caption{Same as Fig.~\ref{Fig:BXA_CenA}, but for NGC~7582 X-4, and which lacks a partial-covering component.} \label{Fig:BXA_ngc7582X4}
\end{figure*}


\clearpage
\bibliographystyle{apj}

\bibliography{mybib}

\end{document}